\documentclass[12pt,twoside]{article}
\usepackage{amsmath}
\usepackage{amssymb}
\usepackage{latexsym}
\usepackage{color}
\input BoxedEPS
\SetRokickiEPSFSpecial  %% for dvips by Tom Rokicki, for VMS
\HideDisplacementBoxes

\newcommand{\xdot}{\dot{X}}

\newcommand{\brr}{({\bf r}-{\bf r}')}

\newcommand{\varep}{\varepsilon^{ijk}}
\newcommand{\rd}[1]{\mathop{\mathrm{d}#1}}
\newcommand{\tr}{\mathop{\mathrm{tr}}}

\newcommand{\fract}[2]{{\textstyle\frac{#1}{#2}}}
\newcommand{\grad}{\vec\nabla}
\newcommand{\dtr}{\rd{\bf r}}

\newcommand{\CS}{Chern-Simons}
\newcommand{\Cpr}{Clebsch pa\-ra\-me\-ter\-iza\-tion}
\newcommand{\pr}{parameterization}
\newcommand{\BI}{\mathrm{Born\mbox{\scriptsize -}Infeld}}
\newcommand{\balpha}{\mbox{\boldmath$\alpha$}}
\newcommand{\bcP}{\mbox{\boldmath$\cal P$}}
\newcommand{\rmd}{{\rm d\null}}
\newcommand{\dotrho}{\dot{\rho}}
\newcommand{\dotv}{\dot{v}}

\newcommand{\dotth}{\dot{\theta}}
\newcommand{\dotpsi}{\dot{\psi}}
\newcommand{\sql}{\sqrt{2\lambda}}

\newcommand{\tn}[2]{\theta^{#1#2}}
\newcommand{\en}[2]{\eps^{#1#2}}
\newcommand{\vth}{\vec\theta}
\newcommand{\vom}{\vec\omega}
\newcommand{\on}[2]{\omega_{#1#2}}

\newcommand{\rfq}[1]{~(\ref{#1})}
\newcommand{\gk}{{\mathfrak g}}
\newcommand{\co}{coordinate}
\newcommand{\vx}{{\bf x}}
\newcommand{\fvx}[1]{f^{#1}(\vx)}
\newcommand{\pa}[1]{\frac\partial{\partial #1}}
\newcommand{\pas}[1]{\partial_{#1}}

\renewcommand{\Re}{\operatorname{Re}}
\renewcommand{\Im}{\operatorname{Im}}

\newcommand{\paj}{\partial_j}

\newcommand{\mn}{{\mu\nu}}
\newcommand{\tvr}{(t,\vrr)}
\newcommand{\tvx}{(t,\vx)}
\newcommand{\tvX}{(t,\vX)}
\newcommand{\tvk}{(t,\vk)}

\newcommand{\han}[1]{\hat A_{#1}}
\newcommand{\ha}{\hat A}

\newcommand{\tenshf}{\skew2\hat{\mathsf F}}

\newcommand{\vha}{\skew3\hat{\bf A}}

\newcommand{\vD}{{\bf D}}

\newcommand{\vX}{{\bf X}}

\renewcommand{\Re}{\operatorname{Re}}
\renewcommand{\Im}{\operatorname{Im}}
\newcommand{\fkf}{{\vec{\mathfrak f}}}
\newcommand{\vcr}{{\bf r}}
\newcommand{\vchi}{{\vec \chi}}
\newcommand{\rD}[1]{\mathop{\mathrm{D}#1}}
\newcommand{\vv}{{\bf v}}
\newcommand{\vk}{{\bf k}}
\newcommand{\vj}{{\bf j}}
\newcommand{\vrr}{{\bf r}}
\newcommand{\rX}{{\rm X}}
\newcommand{\rk}{{\rm k}}
\newcommand{\rJ}{{\rm J}}

\newcommand{\ct}{coordinate transformation}
\newcommand{\df}{diffeomorphism}

\newcommand{\nc}{noncommutative}
\newcommand{\ncg}{noncommuting}
\newcommand{\sw}{Seiberg-Witten}

\def\beq{\begin{equation}}
\def\eeq{\end{equation}}
\def\bi{\begin{itemize}}
\def\ei{\end{itemize}}
\def\beqar{\begin{eqnarray}}
\def\eeqar{\end{eqnarray}}

\def\half {{\textstyle{1\over 2}}}
\def\Tr {{\rm Tr}}
\def\tr {{\rm tr}}

\def\del {{\partial}}

\def\d {{\partial}}

\def\vf {{\varphi}}
\def\dd {{\,d}}
\def\bw{{\bar w}}
\def \P {{\cal P}}
\def \La {{\cal L}}

\def\appendixa{
 \vskip 1cm
 \noindent
 {\large \bf A. Sidebar on canonical formalism determined by a Lagrangian}
 \vskip 0.5cm

 \setcounter{equation}{0}
 \def\theequation{A.\arabic{equation}}
 }
 \def\appendixb{
 \vskip 1cm
 \noindent
 {\large \bf  B. Sidebar on Clebsch parameterization and the Chern-Simons term}
 \vskip 0.5cm
 \par
 \setcounter{equation}{0}
 \def\theequation{B.\arabic{equation}}
 }

\newcommand{\numeq}[2]{\begin{equation}
#2
\label{#1}
\end{equation}}
\newcommand{\refeq}[1]{(\ref{#1})}

\let\vec\boldsymbol
\let\eps\varepsilon

\numberwithin{equation}{section}
\renewcommand{\theequation}{\thesubsection.\arabic{equation}}

\begin{document}
%\pagenumbering{roman}
 
\title{Perfect Fluid Theory and its Extensions}
\author{R. Jackiw$^1$ \negthickspace \footnote{jackiw@lns.mit.edu}
\enspace, V.P. Nair$^2$\footnote{vpn@sci.ccny.cuny.edu}\enspace, S.-Y. Pi$^3$\footnote{soyoung@buphy.bu.edu}\enspace, 
A.P. Polychronakos$^2$\footnote{alexios
@sci.ccny.cuny.edu} \\
\\
{\small\itshape $^1$Department of Physics}\\[-1ex]
{\small\itshape Massachusetts Institute of
Technology} \\[-1ex]
{\small\itshape  Cambridge, MA 02139}\\
{\small\itshape $^2$Department of Physics}\\[-1ex]
{\small\itshape CCNY-CUNY}\\[-1ex]
{\small\itshape New York, NY 10031}\\[.3ex]
{\small\itshape $^3$Department of Physics}\\[-1ex]
{\small\itshape Boston University}\\[-1ex]
{\small\itshape Boston, MA 02215}\\[10ex]
%{\small\itshape BUHEP-0407}
}

\date{\small MIT-CTP-3509\\
BUHEP-04-07}

\maketitle
\begin{abstract}
We review the canonical theory for perfect fluids, in Eulerian and Lagrangian formulations. The theory is related to a description of extended structures
in higher dimensions. Internal symmetry and supersymmetry degrees of freedom are incorporated. Additional miscellaneous subjects
that are covered include physical topics concerning quantization, as well as mathematical issues of volume preserving diffeomorphisms and
representations of Chern-Simons terms (= vortex or magnetic helicity).
\end{abstract}
\newpage
\tableofcontents
\pagestyle{myheadings} \markboth{R.Jackiw}{Perfect Fluid Theory and its Extensions}
\thispagestyle{empty}
\newpage

\section{INTRODUCTION}

The dynamics of fluids is described by a classical field theory, whose origins lie in the nineteenth century, like Maxwell electrodynamics with which it shares some antecedents. The electromagnetic theory has enjoyed much development: it passed into quantum physics, and the resulting quantum electrodynamics served as a model for its non-Abelian generalization, Yang-Mills theory, which today is at the center of a quantum field theoretic description for fundamental physics. We believe that fluid dynamics can undergo a similar evolution and play a similar generative role in physics.

Modern (quantum) field theory has expanded concepts and calculational possibilities beyond what was familiar to (classical) field theorists. One learned about higher and unexpected symmetries, which also facilitate partial or complete
integrability of the relevant differential equations. Topological and geometric
concepts and structures, like solitons and instantons, were recognized as encoding
crucial dynamical information about the models. New entities like Pontryagin
densities and Chern-Simons terms entered into the description of kinematics and
dynamics. Degrees of freedom were enlarged and unified by new organizing
principles based on non-Abelian and supersymmetries. Indeed application of field
theory to particle physics has now evolved to a study of extended structures like
strings and membranes, whose mathematical description bears some similarity to
the theory of fluids.

The novelties introduced in particle field theories can be also developed for the non-particle field theory of fluid dynamics. Correspondingly, fluid
dynamics can illuminate some aspects of particle physics, especially as concerns the
extended structures that these days are the focus of attention for particle physicists.

We begin by addressing the question of why fluid mechanics would be interesting in its own right.
Fluid mechanics, for most physical situations, is obtained from an underlying particle description by suitable averages of  Boltzmann-type equations.
Recall that the Boltzmann equation for the single particle distribution function
$f (X, {\bf P}, t) $ is given by
\beq
{\partial f \over \partial t} +{{\bf P} \over m} \cdot {\partial f\over \partial {\bf X}}
+ {\bf F}\cdot {\partial f \over \partial {\bf P}} = {\cal C}(f)
\label{intro1}
\eeq
where ${\bf X}, {\bf P}$, refer to the coordinates and momenta, the phase space variables,
of a single particle of mass $m$; ${\bf F}$ is the force acting on the particle.
$C(f)$ is the collision integral, which takes into account particle interactions.
In the special
case of the collisonless limit, {\it i.e.}, with
${\cal C}=0$, the
Boltzmann equation (\ref{intro1}) is the
equation for the distribution function for single particles obeying the
standard classical equations of motion.
Solving the Boltzmann equation is not very easy. The equilibrium distribution function
is a solution of the equation; in particular, ${\cal C}(f) =0$ for the equilibrium solution.
The general strategy which has been used for solving
equation (\ref{intro1}) is to seek a perturbative
solution of the form $f= f^{(0)}+ f^{(1)}+...$, where
$f^{(0)}=n_p$ is the equilibrium distribution, appropriately chosen for
bosons and fermions. Transport coefficients and fluid equations of motion
can then be obtained from the perturbative corrections.

This approach has the virtue of simplicity and does capture many of the
general features of the problem of deriving coarse-grained dynamics, fluid mechanics
in particular, from the underlying particle dynamics.
However, there are two  limitations: First because we are only taking care of
the single-particle distributions, and second
because the treatment is essentially classical.

Regarding the first point, one
could indeed try to be more general by starting with the completely
general
$N$-particle Liouville equation for the phase space distribution $\rho
({\bf X}_n, {\bf P}_n)$, $n =1, 2, ..., N$.
The one-particle distribution function is then given by
$\int d\mu_{N-1}~\rho
({\bf X}_n, {\bf P}_n)$, the two-particle distribution is given by
$\int d\mu_{N-2}~\rho
({\bf X}_n, {\bf P}_n)$, etc., where $d\mu_{N}$ denotes the phase space volume for
$N$ particles.
The Liouville equation then leads to a hierarchy
of kinetic equations,
the so-called BBGKY (Bogolyubov-Born-Green-Kirkwood-Yvon)
hierarchy,
involving higher and higher correlated $n$-particle distribution
functions. (For the one-particle distribution function, we get the
Boltzmann equation, but with the collision integral
given in terms of the two-particle distribution function.)
To be able to solve this infinite hierarchy of equations, one needs to truncate it,
very often at just the single particle distribution
function.  Therefore, even though a more general formulation is
possible,
the feasibility of solving these equations limits the kinetic
approach to dilute systems near equilibrium, where the truncation can be
justified.

Regarding the question of quantum corrections, the needed formalism
is that of the Schwinger-Dyson equations with a time-contour, the
so-called Schwinger-Bakshi-Mahanthappa-Keldysh approach \cite{keldysh}.
The Green's functions are defined by the generating functional
\beq
Z[\eta ]= {{\Tr \rho_0 ~{\cal T}_C \exp (iI_{int}+i\phi\cdot \eta )}\over
\Tr \rho_0},\label{intro2}
\eeq
where the time-integral goes from $-\infty$ to $\infty$, folds back and
goes from $\infty$ to $-\infty$; ${\cal T}_C$ denotes
ordering along this time-contour. $\phi_\mu$ generically represents
fields of interest, $\eta^\mu$ is a
source function and
$\rho_0$ is the thermal density matrix.
One can represent $Z[\eta ]$ as a functional integral.
\beq
Z[\eta ] =\int d\mu [\phi ] ~\exp (iI_C (\phi ) +
i\phi \cdot \eta )
\label{intro3}
\eeq
The action is again defined on the time-contour. The Green's
functions for which some of the fields are on the forward
time line and some are on the reverse time line will represent
the effect of statistical distributions.
One has to solve the hierarchy of
coupled Schwinger-Dyson equations which follow from
(\ref{intro3}), again
truncating them
at a certain level to get a description of nonequilibrium phenomena.
In practice, one has to carry out semiclassical expansions to simplify
these to the point where a solution can be found and
again we have a formalism of limited validity.

This discussion shows that
we should expect that the regime of
validity of the fluid dynamical equations derived within kinetic theory or
within the time-contour approach is
rather limited, basically a semiclassical regime for dilute systems not too far
from equilibrium.
However, fluid dynamical equations can also be derived
from very general principles, showing that they have a much wider regime
of validity, and, indeed in practice, we apply them over such a wider
range. This is the `universality' of fluid dynamics.
It is this property which shows the value of a study of fluid mechanics in its
own right, rather than its derivation from an underlying description 
in some approximation.
It seems that a
reconsideration of the whole setting and development of the
ideas of fluid mechanics in the context of modern concepts in
particle physics is entirely appropriate.
The novelties introduced in particle field theories can be generalized to
the non-particle field theory of fluid dynamics. Correspondingly, fluid
dynamics can illuminate some aspects of particle physics, especially as concerns the
extended structures that these days are the focus of attention for particle physicists.

In this Introduction, we shall review Lagrange's and Euler's description of fluid kinematics and dynamics, and we shall describe the mapping between the two. We shall discuss the Hamiltonian (canonical) formulation, together with the associated {Poisson} brackets. We shall also present
(configuration space) Lagrangians for fluid motion; in the Eulerian case a Clebsch
parameterization is needed. Both nonrelativistic and relativistic systems in various
spatial dimensions are treated. These are standard topics and they are well known \cite{LanLif}.
However, in our presentation of this familiar material we shall approach the subject
with an eye towards the various enhancements of fluid mechanics, which comprise
our current research and which are reviewed in the remainder of this article.
 
\subsection{Lagrange and Euler descriptions of a fluid and the relationship between them.}
\setcounter{equation}{0}

The Lagrange description of a fluid focuses on the coordinates of the individual fluid particles. These satisfy a Newtonian equation of motion (in the
nonrelativistic case). On the other hand, in Euler's formulation, the fluid is described by a density  $ \rho$ and $\bf
\mbox{velocity} \  \bf v$, which are linked by a continuity equation, while Euler's equation describes the dynamics. Euler's
method is akin to a classical field theory in physical space-time. 

In fact one may exemplify the two approaches, and the relation between them, already for a single particle, carrying mass $m$ and located on the coordinate ${\bf X} (t)$, whose time evolution  is governed by a force $\bf F$.
\begin{equation}
{\bf \ddot{X}} (t) = \frac{1}{m} {\bf F} ({\bf X} (t))
\label{eqone1}
\end{equation}
(Over-dot denotes differentiation with respect to explicit time-dependence.) This is the ``Lagrange" description (of course it is Newton's). Next, we introduce the Eulerian (single) particle density by
\begin{equation}
\rho (t, {\bf r}) = m \delta ({\bf X} (t) - {\bf r}).
\label{eqone2}
\end{equation}
The delta function follows the dimensionality of space so that the volume integral of $\rho$ is $m$.
Differentiation (\ref{eqone2}) with respect to $t$ leaves 
\begin{eqnarray}
\dot{\rho} (t, {\bf r}) &=& m \frac{\partial}{\partial X^i} \delta({\bf X} (t) - {\bf r}) \dot{X}^i (t) \nonumber\\
&=& -\frac{\partial}{\partial r^i} \bigg(\dot{X}^i (t) m \delta ({\bf X}(t)-{\bf r}) \bigg) = -\frac{\partial}{\partial r^i} \bigg[v^i (t, {\bf r}) \rho (t, \bf r) \bigg],
\label{eqone3}
\end{eqnarray}
where the Eulerian velocity $\bf v$ is given by
\begin{equation}
{\bf v} (t, {\bf r}) = {\bf \dot{X}} (t)~ \ \ \mbox{with} ~{\bf r} = {\bf X} (t).
\label{eqone4}
\end{equation}
Note that the velocity function ${\bf v} (t, {\bf r})$ is only defined  at the point ${\bf r} = {\bf X} (t)$; its value at other points being undetermined and irrevelent.
Evidently the continuity equation is satisfied as a consequence of the above definitions.
\begin{equation}
\dot{\rho} (t, {\bf r}) + {\bf \nabla} \cdot {\bf j} (t, {\bf r}) =0
\label{eqone5}
\end{equation}
\begin{equation}
{\bf j} (t, {\bf r})= {\bf v} (t, {\bf r}) \rho (t, {\bf r}) = {\bf \dot{X}} (t) m \delta ({\bf X} (t) - {\bf r})
\label{eqone6}
\end{equation}
To arrive at the dynamical Euler equation, we differentiate $\bf j$ with respect to time.
From (\ref{eqone6}) it follows that
\begin{subequations}
\begin{eqnarray}
\rho (t, {\bf r})  {\bf \dot{v}} (t, {\bf r}) +  {\bf v} (t, {\bf r}) \dot{\rho} (t, {\bf r}) =
\qquad \qquad \qquad \qquad \
\nonumber\\ 
{\bf \ddot{X}} (t) m \delta ({\bf X}(t)-{\bf r}) + {\bf \dot{X}} 
(t) m \frac{\partial}{\partial X^j} \delta ({\bf X} (t) - {\bf r}) \dot{X}^j (t).
\label{eqone7a}
\end{eqnarray}
Use of the continuity equation on the left side and Newton's equation (\ref{eqone1}) on the right leaves
\begin{eqnarray}
\rho (t, {\bf r}) {\bf \dot{v}} (t, {\bf r}) - {\bf v} (t, {\bf r}) \nabla \cdot \big({\bf v} (t, {\bf r})\rho (t, {\bf
r})\big)=\qquad \qquad \quad \nonumber\\ 
{\bf F} ({\bf X} (t)) \delta ({\bf X} (t) - {\bf r}) -
\frac{\partial}{\partial r^j} \bigg({\bf \dot{X}} (t) \dot{X}^j (t) m \delta ({\bf X} (t) -
{\bf r})\bigg).
\label{eqone7b}
\end{eqnarray}
Cancelling common terms gives Euler's equation (for a single-particle ``fluid"!).
\begin{equation}
\dot{v}^i (t, {\bf r}) + v^j (t, {\bf r}) \partial_j v^i (t, {\bf r})= \frac{1}{m} F^i ({\bf r})
\label{eqone7c}
\end{equation}
\end{subequations}
[$\partial_j$ denotes a derivative with respect to the components of a spatial vector, which can be ${\bf X} \ \mbox{or}\ {\bf
r}$, or below,
${\bf x}$. If context does not determine unambiguously which vector is involved, it will be specified explicitly, as in (1.1.7a,b).]
Strictly speaking, (1.17c) only holds at the point ${\bf r} = {\bf X}$. Suitable continuations of the
function ${\bf v}(t, {\bf r} )$ away  from the point ${\bf r}={\bf X}$ can be found that make (1.1.7c)
hold everywhere. Such continuations represent the velocity field of a fictitious accompanying
fluid, of which $\bf X$ is one particle.
 For a complete dynamical description an appropriate expression for the force $\bf F$ still needs to be
given.

 The same development holds for a collection of $N$ particles: (\ref{eqone1}) becomes replaced by
\begin{subequations}\label{eqone8ab}
\begin{equation}
{\bf \ddot{X}}_n (t) = \frac{1}{m_n} {\bf F} \bigg({\bf X}_1(t),...,  {\bf X}_n (t) \bigg).
 \label{eqone8a}
\end{equation}
The particle label $n$ ranges from $1$ to $N$. We shall take the particles to be identical. 
Therefore the mass does not carry the $n$ label, and  the force ${\bf F}_n$ has a functional
dependence on  ${\bf X}_n$ independent of $n$, and is symmetric under exchange of the
remaining $N-1$ particle coordinates.
\begin{equation}
 {\bf \ddot{X}}_n (t) = \frac{1}{m} {\bf F} \bigg({\bf X}_n (t); \{ {\bf X}_k (t), k \ne n \}\bigg)
 \label{eqone8b}
\end{equation}
\end{subequations}
The Eulerian mass density, velocity and current are defined as
\begin{equation}
\rho (t, {\bf r}) = m \sum^N_{n=1} \delta ({\bf X}_n (t) - {\bf r}),
\label{eqone9}
\end{equation}
\begin{equation}
{\bf j} (t, {\bf r}) = {\bf v} (t, {\bf r}) \rho (t, {\bf r}) = m \sum^N_{n=1} {\bf \dot{X}}_n (t) \delta ({\bf X}_n (t) - {\bf r}).
\label{eqone10}
\end{equation}
Similarly to the single-particle case, the function ${\bf v(t,p)}$ is defined only at the points ${\bf r}= {\bf X}_n (t)$.
Evidently the continuity and Euler equations continue to hold.

For the true fluid formulation, we promote the discrete particle label $n$ to a continuous label
${\bf x}$ and the Lagrange coordinate ${\bf X}_n (t)$ becomes ${\bf X} (t, {\bf x})$. Frequently
$\bf x$ is specified by the statement that it describes the fluid coordinate $\bf X$ at initial time
$t=0$, {\it i.e.}
\begin{equation}
{\bf X} (0,{\bf x}) = \bf x.
\label{eqone11}
\end{equation}
Thus $\bf x$ is the comoving coordinate. Dynamics is again Newtonian.
\begin{equation}
{\bf \ddot{X}} (t, {\bf x}) =\frac{1}{m} {\bf F} ({\bf X} (t, {\bf x}))
\label{eqone12}
\end{equation}
The density and velocity are now defined by
\begin{eqnarray}
\rho (t, {\bf r}) &=& \rho_0 \int d  x \ \delta({\bf X} (t, {\bf x})- {\bf r}), 
\label{eqone13}\\
{\bf j} (t, {\bf r}) &=& {\bf v} (t, {\bf r}) \rho \ (t, {\bf r}) = \rho_0 \int d  x \ {\bf \dot{X}} (t, {\bf x}) \delta({\bf X} (t, {\bf x}) - {\bf r}).
\label{eqone14}
\end{eqnarray}
The integration is over the entire relevant volume, be it 1-dimensional, 2-dimensional or 3-dimensional. (The dimensionality of the measure will be specified only when formulas are dimension specific.) $\rho_0$ is a background mass density, so that the volume integral of $\rho$ is the total mass.
In the above, we assume that the particles that are nearby in space have similar velocities and thus the function of discrete points ${\bf v} (t, {\bf X}_n)$ goes over to a smooth continuous function
${\bf v}(t,{\bf r})$. This is the fundamental assumption in classical fluids. The more general case 
calls for a phase space density (Boltzmann) description of the collection of particles, which takes
us away from the realm of classical Lagrange fluids.

The continuity and Euler equations follow as before, with the latter reading
\begin{equation}
\dot{v}^i (t, {\bf r}) +  v^j (t, {\bf r}) \partial_j v^i (t, {\bf r}) =  \frac{1}{m} F^i ({\bf r}).
\label{eqone15}
\end{equation}
The determination of the force term in the above equation requires some discussion. For
external forces, ${\bf F}({\bf r})$ is simply a known function of space. For internal particle forces,
however, the force could be a nonlocal function depending on the full distribution of particles in
space. We shall postulate that interparticle forces are short-range and thus depend only on the
distribution of nearby particles. For such forces, $\bf F$ would depend only on the density of
particles and its derivatives at the point $\bf r$. Under additional assumptions of isotropicity,
$\bf F$ will be proportional to the gradient of the density, $\vec \nabla \rho$. In
this case, a standard argument shows that the right hand side of (1.1.15) should be set to
$-\frac{1}{\rho} \vec \nabla P$, where $P$ is the pressure [2]. Thus once an equation of state is
given, {\it i.e.} once the dependence of
$P$ on $\rho$ is known, we have a self-contained system of equations: the continuity and 
Euler equations.
\begin{equation}
\dot{\rho} (t, {\bf r}) + {\bf \nabla} \cdot \bigg({\bf v} (t, {\bf r}) \rho (t, {\bf r}) \bigg) = 0
\label{eqone16}
\end{equation}
\begin{equation}
{\bf \dot{v}} (t, {\bf r}) + \bigg({\bf v} (t, {\bf r}) \cdot {\bf \nabla} \bigg) {\bf v} (t, {\bf r}) = -\frac{1}{\rho} {\bf \nabla} P (\rho)
\label{eqone17}
\end{equation}
A fluid obeying these equations is called ``perfect".
Later this formalism will be generalized to account for an internal symmetry. For that purpose, it will be useful to delineate explicitly the effect of the $\delta$ functions in (\ref{eqone13}) and ((\ref{eqone14}). In the course of the $\bf x$ integral, $\bf x$ becomes evaluated at a function ${\boldsymbol\chi} (t, {\bf r})$, which is inverse to ${\bf X} (t, {\bf r})$.
\begin{subequations}\label{eighteen}
\begin{eqnarray}
{\bf X} \bigg(t, {\boldsymbol \chi}(t, {\bf r}) \bigg) &=& {\bf r} \label{eqone18a}\\
{\boldsymbol \chi} \bigg(t, {\bf X}(t, {\bf x}) \bigg) &=& {\bf x} \label{eqone18b}
\end{eqnarray}
\end{subequations}
(Uniqueness is assumed.) Thus the $ \bf x$ integration sets $\bf X$ equal to $\bf r$ and also there is a Jacobian, $det \frac{\partial X^i}{\partial x^j}$ for the $\bf X \to r$ transformation.

Consequently from (\ref{eqone13}) and (\ref{eqone14}) it follows that
\begin{eqnarray}
\rho &=& \rho_0 \frac{1}{| det \frac{\partial X^i}{\partial x^j}|_{{\bf x} =\boldsymbol \chi}},  \label{eqone19}\\
{\bf v}&=& {\bf \dot{X}} |_{{\bf x} = \bf \boldsymbol \chi}. \label{eqone20}
\end{eqnarray}
In other words $\bf X$ effects a diffeomorphism ${\bf x} \to {\bf X}\equiv {\bf r}$, while $\boldsymbol \chi$ acts similarity for ${\bf r}\to {\bf x}$. The interchange between a dependent variable $\bf X$ and an independent variable $\bf r$ is called a hodographic transformation.
\subsection{Lagrangian, Hamiltonian formulations and symmetries of dynamics}\label{Sec1.2}
\subsubsection*{(i) Lagrangian and Hamiltonian functions for Lagrange fluid mechanics}
\addcontentsline{toc}{subsubsection}{(i) Lagrangian and Hamiltonian functions for Lagrange fluid
mechanics}
\setcounter{equation}{0}

Since the Lagrange method is essentially Newtonian, Lagrangian and Hamiltonian formulations for (\ref{eqone12}) are readily constructed provided the force is derived from a potential
$\mathcal{V}(\bf X)$.
\begin{subequations}
\begin{eqnarray}
L_L = \int d x \bigg(\frac{1}{2} m {\bf \dot{X}}^2 - \mathcal{V} ({\bf X}) \bigg) \label{eqone21a}\\
{\bf F}({\bf X}) = -{\boldsymbol \nabla} \mathcal{V} (\bf X) \label{eqone21b}
\end{eqnarray}
\end{subequations}
\begin{subequations}
\begin{eqnarray}
H_L = \int d x \bigg(\frac{1}{2 m} {\bf P}^2 + \mathcal{V} (\bf X) \bigg) \label{eqone22a}\\
{\bf P} = m \bf \dot{X}
\label{eqone22b}
\end{eqnarray}
\end{subequations}
A canonical formulation for the Lagrange description of fluid motion follows in the usual way, so that bracketing
with $H_L$ and using conventional Poisson brackets for $\bf X$ and $\bf P$ reproduces the equation of motion
(\ref{eqone12}).
\\

\subsubsection*{(ii) Diffeomorphism symmetry of Lagrange fluid mechanics}
\addcontentsline{toc}{subsubsection}{(ii) Diffeomorphism symmetry of Lagrange fluid mechanics}
In the discrete antecedent to the continuum formulation there is the obvious freedom of renaming the {\it n} label. The continuum version of this freedom
manifests itself in that the Lagrange formulation of fluid dynamics enjoys invariance against
volume-preserving diffeomorphisms of the continuous label $\bf x$.

An infinitesimal diffeomorphism of $\bf x$, generated by a infinitesimal function ${\bf f} ({\bf x})$ reads
\begin{subequations}
\begin{equation}
\delta_f {\bf x} = -{\bf f} ({\bf x}),
\label{eqone23a}
\end{equation}
and this is volume-preserving when $\bf f$ is transverse.
\begin{equation}
{\bf \nabla} \cdot {\bf f} = 0
\label{eqone23b}
\end{equation}
\end{subequations}
Provided the Lagrange coordinate transforms as a scalar
\begin{equation}
\delta_f {\bf X} (t, {\bf x}) = {\bf f} ({\bf x}) \cdot {\bf \nabla} {\bf X} (t, {\bf x}),
\label{eqone24}
\end{equation}
$L_L$ is invariant.
\begin{eqnarray}
\delta_f L_L &=& \int d x \bigg( m \dot{X}^i f^j \partial_j \dot{X}^i -\frac{\partial}{\partial X^i}
\mathcal{V}({\bf X}) f^j \partial_j X^i \bigg) \nonumber\\ &=& \int d x f^j \partial_j
\bigg(\frac{m}{2} {\bf \dot{X}}^2 - \mathcal{V}({\bf X})\bigg)
\label{eqone25}
\end{eqnarray}
The last expression vanishes after partial integration by virtue of  (\ref{eqone23b}). (Surface terms are ignored.) Noether's theorem then gives the constant of motion associated with the relabeling symmetry
(volume-preserving diffeomorphism).
\begin{equation}
\mathcal{C}_f = \int d x  \dot{X}^i f^j \partial_j X^i
\label{eqone26}
\end{equation}

Since $\bf f$ is an arbitrary transverse function it can be stripped away from
(\ref{eqone26}). Explicit formulas reflect the spatial dimensionality of the system. In
three dimensions we can present the transverse $f^i$ as $\varepsilon^{ijk} \partial_j
\Tilde{f}^k,
\Tilde{f}^k$ arbitrary, leading to conserved local vector quantities.
\begin{equation}
\mathcal{C}^i_{(3)} = \varepsilon^{ijk} \partial_j \dot{X}^\ell \partial_k X^\ell
\label{eqone27}
\end{equation}
In two dimensions $f^i$ involves a scalar function, $f^i = \varepsilon^{ij} \partial_j f$, and for planar systems the conserved quantity is a local scalar.
\begin{equation}
\mathcal{C}_{(2)} = \varepsilon^{ij} \partial_i \dot{X}^k \partial_j X^k
\label{eqone28}
\end{equation}
Finally, in one dimension, a ``transverse" function is constant, so the constant of motion for lineal systems remains integrated.
\begin{equation}
\mathcal{C}_{(1)} = \int d x^1 \dot{X} \partial_{x^1} X = \int d X \dot{X}
\label{eqone29}
\end{equation}

\subsubsection*{(iii) Hamiltonian function for Euler fluid mechanics}
\addcontentsline{toc}{subsubsection}{(iii)  Hamiltonian function for Euler fluid mechanics}
A Hamiltonian for the continuity and Euler equations (\ref{eqone16}), (\ref{eqone17}) is obtained from $H_L$
(\ref{eqone22a}) by transforming to Eulerian variables. First, however, to ensure proper dimensionality the kinetic term is
divided by the spatial volume so that $m$ is replaced by $\rho_0$. Also to reproduce the pressure form of the forces, we
write $
\mathcal{V}({\bf X})$ as a function of the Jacobian, $det \frac{\partial X^i}{\partial x^j}$. Thus $H_L$ reads
\begin{equation}
H_L = \int d  x \bigg( \frac{1}{2} \rho_0 {\bf \dot{X}}^2 + \mathcal{V} (| det \frac{\partial X^i}{\partial x^j} | /\rho_0) \bigg).
\label{eqone30}
\end{equation}
The transformation to Eulerian variables is effected by multiplying (\ref{eqone30}) by unity, in the form $\int d r \delta ( {\bf X} - {\bf r})$, and interchanging orders of integration. In this way $H_L$ becomes $H_E$.
\begin{equation}
H_E = \int d  r \bigg( \frac{1}{2} \rho {\bf v}^2 + \frac{\rho}{\rho_0} \mathcal{V} (\frac{1}{\rho}) \bigg)
\label{eqone31}
\end{equation}
Agreement with the pressure expression for the force, as in (\ref{eqone17}), is achieved when we identify
\begin{equation}
P (\rho) = -\frac{1}{\rho_0}\ \mathcal{V}' (1/\rho)
\label{eqone32}
\end{equation}
(Dash denotes derivative with respect to argument.)
In the subsequent we drop the subscript $E$ on $H$ and rename $\frac{\rho}{\rho_0} \mathcal{V}(\frac{1}{\rho})$ as $V(\rho)$. Thus the Euler Hamiltonian reads
\begin{equation}
H = \int d  r \bigg(\frac{1}{2} \rho  {\bf v}^2 + V(\rho) \bigg),
\label{eqone33}
\end{equation}
and the pressure is Legendre transform of $V$.
\begin{equation}
P (\rho) = \rho V' (\rho) - V (\rho)
\label{eqone34}
\end{equation}
Conventional nomenclature for $V' (\rho)$ is enthalpy and $\sqrt{P' (\rho)} = \sqrt{\rho V'' (\rho)}$ is the sound speed $s$.

Bracketing the Hamiltonian with $\rho$ and  $\bf v$ should generate equations (\ref{eqone16}) and (\ref{eqone17}). To
verify this, we need to know the brackets of $\rho, \bf v$ with each other. These may be obtained from the canonical
brackets in the Lagrange formulation, the only non-vanishing one being
\begin{equation}
\{\dot{X}^i ({\bf x}), X^j ({\bf x'})\} = \frac{1}{\rho_0} \delta^{ij} \delta ({\bf x} -{\bf x'}).
\label{eqone35}
\end{equation}
Using the definitions of $\rho$ and $\bf j$ in terms of $\bf X$ and $\bf \dot{X}$, 
eqs. (\ref{eqone13}) and (\ref{eqone14}), as well as the canonical brackets
(\ref{eqone35}), determines the brackets of $\rho$ and ${\bf j}$.
\begin{eqnarray}
\{\rho({\bf r}), \rho ({\bf r}')\} &=&0 \label{eqone36}\\
\{j^i ({\bf r}), \rho ({\bf r'})\}&=& \rho ({\bf r}) \partial_i \delta \brr \label{eqone37}\\
\{j^i ({\bf r}),j^j ({\bf r}')\} &=& j^j ({\bf r}) \partial_i \delta \brr + j^i ({\bf r'}) \partial_j
\delta \brr \label{eqone38}
\end{eqnarray}
Since ${\bf j}={\bf v} \rho$ these in turn imply that the brackets for $\rho$ and $\bf v$ take the
form \cite{ArnKhe}
\begin{eqnarray}
\{v^i({\bf r}), \rho({\bf r'})\} = \partial_i \delta \brr, \qquad \label{eqone39}\\
\{v^i({\bf r}), v^j({\bf r'})\} = -\frac{\omega_{ij} ({\bf r})}{\rho({\bf r})} \delta \brr,
\label{eqone40}
\end{eqnarray}
where
\begin{equation}
\omega_{ij} ({\bf r}) = \partial_i \ v_j {(\bf r)}-\partial_j v_i ({\bf r})
\label{eqone41}
\end{equation}
is called the fluid vorticity.
[In the above expressions, (\ref{eqone35})-(\ref{eqone41}), all quantities are at equal
times, so the time argument is omitted.] Of course the Jacobi identity is satisfied by
the brackets.

It is now straight forward to verify from (\ref{eqone36}), (\ref{eqone39})-(\ref{eqone41})  that bracketing with $H$
reproduces the  equations of motion (\ref{eqone16}) and (\ref{eqone17}).
\begin{eqnarray}
\dot{\rho} = \{H, \rho\} = -\vec \nabla \cdot ({\bf v} \rho )
\label{eqone42}\\
\dot{\bf v} = \{H, {\bf v}\} = - ({\bf v} \cdot \vec \nabla) {\bf v} -
\vec \nabla V' (\rho)
\label{eqone43} 
\end{eqnarray}

These equations may also be presented as continuity equations for a (nonrelativistic) energy momentum tensor. 
The energy
density
\begin{equation}
\mathcal{E} = \frac{1}{2} \rho {\bf v}^2 + V = T^{00},
\label{eqone44}
\end{equation}
together with the energy flux
\numeq{eqone45}{
T^{jo} = \rho v^j (\fract 12 {\bf v}^2 +  V'),
}
obey
\numeq{eqone46}{
\dot{T}^{oo} + \partial_j T^{jo} = 0 . 
}
Similarly the momentum density, $\cal P$,  which in the nonrelativistic theory coincides with
the current,
\begin{subequations}\label{conteqsII}
\numeq{eqone47a}{
{\cal P}^i = \rho v^i  = T^{oi},
}
and the stress tensor $T^{ij}$
\numeq{eqone47b}{
T^{ij} = \delta^{ij} (\rho V' - V) + \rho v^i v^j = \delta^{ij} P +  \rho v^i v^j,
}
\end{subequations}
satisfy
\numeq{eqone48}{
\dot{T}^{oi} + \partial_j T^{ji} = 0 . 
}
Note that $T^{oi} \neq T^{io}$ because the theory is not Lorentz invariant, but
$T^{ij}=T^{ji}$ because it is invariant against spatial rotations. ($T^{\mu\nu}$ is
not, properly speaking, a ``tensor'', but an energy-momentum ``complex''.) Because $\vec{\cal P} =
\bf j$, the current algebra (\ref{eqone38}) is also the momentum density algebra.
\\

\noindent{\bf (iv) Symmetries of Euler fluid mechanics} \label{sec:1.2.iv}
\addcontentsline{toc}{subsubsection}{(iv)  Symmetries of Euler fluid mechanics}

The above continuity equations and other specific properties of the energy-momentum tensor allow constructing constants of motion,
which reflect symmetries of theory. The Hamiltonian = energy, 
\begin{equation}
E = \int d  x \mathcal{E} \qquad \mbox{(time-translation)},
\label{eqone49}
\end{equation}
is constant as a consequence of time-translation invariance, while the constancy of the
momentum,
\begin{equation}
{\bf P} = \int d r \ \vec{\cal P} =  \int d r ~ {\bf j} \qquad
\mbox{(space-translation)},
\label{eqone50}
\end{equation}
follows from space-translation invariance. The index symmetry of $T^{ij}$ is a consequence of rotational invariance and ensures that the angular
momentum,
\begin{equation}
M^{ij} = \int d  r\ (r^i \mathcal{P}^j - r^j \mathcal{P}^i) \qquad \mbox{(spatial rotation)},
\label{eqone51}
\end{equation}
is constant. The identity of the momentum density and the current density allows construction of the Galileo
boost constant of motion,
\begin{equation}
{\bf B} = t\ {\bf P} - \int d r\ {\bf r} \rho \qquad \mbox{(velocity boost)},
\label{eqone52}
\end{equation}
whose time-independence signals invariance against velocity boosts.
Finally, the total number
\begin{equation}
N = \int dr \rho \qquad \mbox{(number)},
\label{eqone53}
\end{equation}
is also conserved, as a consequence of the continuity equation (\ref{eqone16}).
Upon bracketing with each other, the constants (\ref{eqone49})-(\ref{eqone53}) form the extended Galileo Lie algebra, 
and they generate Galileo transformations on $\rho$ and
$\bf v$. Brackets with $N$ vanish; $N$ provides the central extension for the Galileo algebra in the
$\{ {\bf B}, \bf P\}$ bracket.

Further constants of motion are present for specific forms of $V$. In particular if
\begin{subequations}
\begin{equation}
2 T^{00} = \delta_{ij}\ T^{ij},
\label{eqone54a}
\end{equation}
which in $d$ spatial dimensions requires that
\begin{equation}
V(\rho) =\lambda\ \rho^{1+\frac{2}{d}},
\label{eqone54b}
\end{equation}
\end{subequations}
two more constants exist. They are the dilation,
\begin{equation}
D = 2 t H - \int d r\ {\bf r} \cdot \bf \vec{\cal P} \qquad \mbox{(dilation)},
\label{eqone55}
\end{equation}
and the special conformal generator.
\begin{equation}
K = t^2 H - t D - \fract 12 \int d r\ {\bf r}^2 \rho \qquad \mbox{(conformal transformation)}
\label{eqone56}
\end{equation}
The latter two, together with $H$, form the $SO (2,1)$ Lie algebra of the non-relativistic conformal group,
which they generate by bracketing. Together with remaining Galileo group elements this is called the Schr\"{o}dinger group
\cite{Salm}.

Finally, in any dimension when 
\begin{equation}
V(\rho) = \lambda/ \rho,
\label{eqone57}
\end{equation}
we are dealing with a Chaplygin gas and, as we shall discuss below, this model
supports remarkable higher symmetries related to relativistic extended objects \cite{Chp}.

What is the Eulerian image for the volume-preserving diffeomorphism symmetries of the Lagrange
formulation discussed in Section 1.2 (ii)? First we note that the Eulerian variables $\rho$ and $\bf v$ do not
respond to the volume-preserving transformations on the Lagrange variable $\bf X$.
This is seen from (\ref{eqone13}), (\ref{eqone14}), (\ref{eqone24}).
Also the associated constants of motion, summarized by
(\ref{eqone26})-(\ref{eqone29}) do not possess in general simple Eulerian
counterparts. But some do.

Let us begin with the 3-dimensional quantity in (\ref{eqone27}) and integrate it over a surface
$\bar{\bf S}$ in the $\bf X$ parameter space bounded by a closed curve $\bf
\partial \bar{S}$. This gives the constant 
\begin{subequations}
\begin{equation}
A = \int_{\bf \bar{S}} d S^i \varepsilon ^{ijk} \partial_j  (\partial_k X^\ell \dot{
X}^\ell)
 = \oint_{\partial \bf \bar{S}} \ d X^i \ \partial_i \  X^\ell   \dot{X}^\ell.
\label{eqone58a}
\end{equation}
Upon performing the diffeomorphism ${\bf x} \to {\boldsymbol \chi}$, (\ref{eqone58a}) becomes
expressed in terms of Euler variables as
\begin{equation}
A = \oint_{\partial {\bf S}} d {\bf r} \cdot {\bf v} = \int_{\bf S} d{\bf S} \cdot \vec \omega,
\label{eqone58b}
\end{equation}
\end{subequations}
where $\bf \partial \bar{S}$ is the (time-dependent) image of $\bf \partial S$ under the diffeomorphism. The above quantity, known as
the velocity circulation or vorticity flux, is therefore constant; this is Kelvin's  theorem. [One can establish the result directly for
(\ref{eqone58b}) from the equations of motion for Euler variables, provided one takes into account
the time-dependence of the contour $
\partial {\bf S}$.] Similarly conserved velocity circulation exists also in 2-dimensional
fluid mechanics; of course the integration surface or contour then lie in the plane.

Additional important constants arise from the volume integrals of the  $C^i_{(d)}, d =
1,2,3$, in (\ref{eqone27})-(\ref{eqone29}). In three dimensions, we begin with (\ref{eqone27})
contracted with $\xdot^m \partial_i X^m$ and integrated over volume.
\begin{subequations}
\begin{eqnarray}
C_{(3)} &=& \int d^3 x \ \varep (\xdot^m \partial_i X^m) \partial_j (\xdot^n \partial_k X^n) \nonumber\\
&=&  \int d^3 x \ \xdot^m \partial_j \xdot^n \ \varep \ \partial_i X^m \partial_k X^n
\label{eqone59a}
\end{eqnarray}
We use the identity $\varep \partial_i X^m \partial_k X^n = \varepsilon^{m\ell n}
\frac{\partial x^j}{\partial X^\ell} \ det \ \frac{\partial X^p}{\partial x^q}$, and multiply
$C_{(3)}$ by unity in the form $\int d^3 r \ \delta({\bf X} - {\bf r})$ so that $C_{(3)}$
becomes
\begin{eqnarray}
C_{(3)} &=& \int d^3 r \ d^3 x \ \xdot^m\  \frac{\partial \xdot^n}{\partial x^j} \
\varepsilon^{m \ell n} \ \frac{\partial x^j}{\partial X^\ell} \ det \ \frac{\partial
X^p}{\partial x^q} \ \delta({\bf X} - {\bf r}) \nonumber \\
&=& \int d^3 r {\bf v} \cdot (\vec \nabla \times {\bf v}) = \int d^3 r {\bf v} \cdot
\vec \omega,
\label{eqone59b}
\end{eqnarray}
\end{subequations}
(apart from an irrelevant factor of $\pm \rho_0$). Conservation of $C_{(3)}$, called the vortex
helicity, also follows when the Euler equation (\ref{eqone17}) is applied to (\ref{eqone59b}).

An expression like $C_{(3)}$, with $\bf v$ an arbitrary 3-vector, is a well known mathematical entity, called the Chern-Simons term. In the last 
twenty years, Chern-Simons terms have come to play a significant role in physics and
mathematics \cite{DesJac}. We shall have more to say about them, incarnated in various
contexts, but always in odd-dimensional space. Indeed the one-dimensional constant
(\ref{eqone29}), when written in terms of Euler variables reads
\begin{equation}
C_{(1)} = \int d^1 r v.
\label{eqone60}
\end{equation}
(Here ``$r$" is not a radial coordinate, but lies on the real line.) This is  just a
1-dimensional Chern-Simons term. (We shall further discuss Chern-Simons terms in Sidebar B, below.)

Finally we turn to the planar constant $C_{(2)}$ (\ref{eqone28}). Taking it to the
$M^{th}$ power, multiplying by $\delta ({\bf X} - {\bf r})$ and integrating over $\bf x$
and $\bf r$ gives
\begin{subequations}\label{261}
\begin{equation}
C^M_{(2)} = \int d^2x \ d^2 r (\varepsilon^{ij} \partial_i \xdot^k \partial_j X^k)^M \
\delta ({\bf X} - {\bf r}).
\label{eqone61a}
\end{equation}
The 2x2 matrix identity  $\varepsilon^{ij} \frac{\partial X^k}{\partial x^j} = \varepsilon^{mk} \frac{\partial x^i}{\partial
X^m}  det \ \frac{\partial X^p}{\partial x^q}$ converts the above to
\begin{eqnarray}
C^M_{(2)} &=& \int d^2x \ d^2 r \bigg(\frac{\partial \xdot^k}{\partial x^i} \ \varepsilon^{mk}\ \frac{\partial x^i}{\partial X^m}
\  det\ \frac{\partial X^p}{\partial x^q} \bigg)^M
\delta({\bf X}-{\bf r}) \nonumber\\ &=& \int d^2r \rho \bigg(\frac{\partial}{\partial
r^m}
\ v^k \ \varepsilon^{mk} / \rho \bigg)^M  = \int d^2 r \rho \
\bigg(\frac{\omega}{\rho} \bigg)^M.
\label{eqone61b}
\end{eqnarray}
\end{subequations}
apart from irrelevant factors. Here $\omega$ is the planar
vorticity
$\varepsilon^{ij} \partial_i v^j$. Thus in the plane there is a denumarbly infinite set
of particle relabeling constants. Again, time-independence of $C^N_{(2)}$ can be
established directly from (\ref{eqone61b}) with the help of the continuity and Euler
equations (\ref{eqone16}), (\ref{eqone17}).

The brackets of the above volume-integrated relabeling constants with the Euler
Hamiltonian vanish, because they are time-independent. But this does not depend
on the specific form of the Hamiltonian, since the fundamental brackets
(\ref{eqone36})-(\ref{eqone41}) give vanishing brackets for $C_{(d)}$ with $\rho$
and
$\bf v$. This is as it should be, because we have already remarked that the Eulerian
variables do not respond to the relabeling transformations. Quantities whose bracket
vanishes with all the elements of a (bracket) algebra, here $\rho$ and $\bf v$, are
called Casimir invariants. So we see that the relabeling symmetry gives rise to
Casimir invariants in the Euler formulation for fluids. This has a profound impact on
the possibility of constructing a Lagrangian for Eulerian fluids.

Finally, let us remark that even though the volume-preserving diffeomorphism transformations
do not act on the Eulerian $\rho$ and $\bf v$, there remains in the formalism a related structure: 
The brackets (\ref{eqone38}) of  currents (equivalently, momentum densities) present a local
realization of the full (not merely volume-preserving) diffeomorphism algebra. For if we define for
arbitrary $\bf f$
\begin{equation}
j_{\bf f} = \int d   r\ {\bf f} ({\bf r}) \cdot {\bf j} ({\bf r}),
\label{eqone62}
\end{equation}
then (\ref{eqone38}) implies
\begin{equation}
\{j_{{\bf f}_1}, j_{{\bf f}_2} \} = j_ {{\bf f}_{12}},
\label{eqone63}
\end{equation}
where ${\bf f}_{12}$ is the Lie bracket of ${\bf f}_1$ and ${\bf f}_2$
\begin{equation}
 f^i_{12} = f^j_1 \partial_j f^i_2 - f^j_2 \partial_j f^i_1.
\label{eqone64}
\end{equation}
{\bf ({\rm v}) Lagrange function for Eulerian fluid mechanics}\\ 
\addcontentsline{toc}{subsubsection}{(v) Lagrange function for Eulerian fluid mechanics}

While constructing the Euler Hamiltonian is straightforward, for example by
transforming the Lagrange Hamiltonian, as in (\ref{eqone30})-(\ref{eqone33}), an
analogous construction for the Euler Lagrangian is problematic. First, the equations
for the variables $\rho$ and $\bf v$ are first order in time, so the Lagrangian should
reproduce this. Second, time derivatives in the Lagrangian determine the canonical,
bracket structure, which ultimately should reproduce
(\ref{eqone36})-(\ref{eqone41}). However, direct transcription of
$L_L$ (\ref{eqone21a}) to $L_E$, analogous to the passage from $H_L$ in
(\ref{eqone22a}) to $H_E$ (\ref{eqone33}), would yield $L_E = \int d r \, (\frac{1}{2} \, \rho
{\bf v}^2 - V (\rho))$, which contains no time derivatives. So something else must be done.
Moreover, as we shall now explain, the presence of the Casimirs $C_{(d)}$ and $N$ poses obstructions to the construction of a
Lagrangian, which must be overcome.
 
 Before proceeding, we present a Sidebar on the relation between Lagrangians and
the canonical bracket structure.\\
\newpage
\hrule \vspace{-8pt}
{\appendixa
\addtocontents{toc}{\protect\hrulefill\par\vspace{-10pt}}
\addcontentsline{toc}{section}{\quad \ A. Sidebar on canonical formalism
determined by a Lagrangian}
\subsubsection*{(a) Easy case}\label{suba}
\addcontentsline{toc}{subsubsection}{(a) Easy case}

 We begin with a Lagrangian that is first order in time.  This entails no loss of
generality because all second order Lagrangians can be converted to first
order by the familiar Legendre transformation that produces a Hamiltonian:
$H(p,q)=p\dot q-L(\dot q,q)$, where $p\equiv  \partial L/\partial\dot q$. The 
equations of motion gotten by taking the Euler-Lagrange derivative with respect to
$p$ and $q$ of the
 Lagrangian
$L(\dot p,p;\dot q,q)\equiv p\dot q - H(p,q)$ coincide with  the ``usual''
equations of motion 
 obtained by taking the $q$ Euler-Lagrange derivative of $L(\dot q,q)$.  
[In fact $L(\dot p,p;\dot q,q)$ does not depend on~$\dot p$.] Moreover, some
Lagrangians possess only a first-order formulation (for example, Lagrangians for
Schr\"odinger or Dirac fields; also the Klein-Gordon Lagrangian in light-cone
coordinates is first order in the light-cone ``time'' derivative).

Denoting all variables by the generic symbol $\xi^i$, the most general first order
Lagrangian is
\begin{equation} L=a_i(\xi)\dot\xi^i-H(\xi).
\label{lag}
\end{equation}
  Note that  although we shall ultimately be interested in fields defined on
space-time, for present didactic purposes  it  suffices to consider
variables~$\xi^i(t)$ that are functions only of time.  The Euler-Lagrange equation 
that is implied by (\ref{lag}) reads
\begin{equation} f_{ij}(\xi)\dot\xi^j = \frac{\partial H(\xi)}{\partial
\xi^i}\label{E-L}
\end{equation}
 where
\begin{equation} f_{ij}(\xi)=  \frac{\partial a_j(\xi)}{\partial \xi^i}- \frac{\partial
a_i(\xi)}{\partial
\xi^j}.\label{fij}
\end{equation}
The first term in \refeq{lag} determines the canonical 1-form: $a_i(\xi)\dot\xi^i
\rd t = a_i(\xi)\rd{\xi^i}$, while $f_{ij}$ gives the symplectic 2-form: 
$\rd{a_i(\xi)\rd{\xi^i}} = \fract12 f_{ij} (\xi) \rd{\xi^i}\rd{\xi^j}$.

 To set up a canonical formalism, we proceed directly.  We \emph{do not} make
the frequently heard statement that ``the canonical momenta
$ \partial L/\partial\dot\xi^i = a_i(\xi)$ are constrained to depend on the
coordinates $\xi$'', and we \emph{do not}  embark on Dirac's method for
constrained systems \cite{FadJac}.

In fact, if the matrix $f_{ij}$ possesses the inverse $f^{ij}$  there are no
constraints.  Then (\ref{E-L}) implies
\begin{equation}
\dot\xi^i = f^{ij}(\xi)  \frac{\partial H(\xi)}{\partial \xi^j}.\label{eqm}
\end{equation}
 When one wants to express this equation of motion by bracketing with
the Hamiltonian
\begin{equation}
\dot\xi^i=\{H(\xi),\xi^i\} = \{\xi^j,\xi^i\}\frac{\partial H(\xi)}{\partial \xi^j},
\end{equation}
 one is led to postulating the fundamental bracket as
\begin{equation}
\{\xi^i,\xi^j\}=-f^{ij}(\xi).\label{fundbrack}
\end{equation}
 The bracket between functions of $\xi$ is then defined by
\begin{equation}
\{F_1(\xi),F_2(\xi)\} = -\frac{\partial F_1(\xi)}{\partial\xi^i} f^{ij}
\frac{\partial F_2(\xi)}{\partial\xi^j}.
\label{eqone71}
\end{equation}
 One verifies that (\ref{fundbrack}), (\ref{eqone71}) satisfy the Jacobi identity by
virtue of  the Bianchi identity when $f$ is given by (\ref{fij}). 

\begin{equation}
\frac{\partial}{\partial \xi^i} \ f_{j k} \ + \frac{\partial}{\partial \xi^j} \ f_{k i} \ +
\frac{\partial}{\partial \xi^k} \ f_{ij} = 0
\label{aeight}
\end{equation}
\noindent
\subsubsection*{(b) Difficult case}\label{subb}
\addcontentsline{toc}{subsubsection}{(b) Difficult case}
When $f_{ij}$ is singular, we may still proceed in the following manner \cite{bacjac}.
Let us suppose that $f_{ij}$ possesses $N$ zero modes $p^i_{(n)}$
\begin{eqnarray}
p^i_{(n)} f_{ij} = 0\nonumber \\
n=1, ..., N.
\label{neweq}
\end{eqnarray}
If we use a rank $N$ projection operator $P^{\ j}_i$ that satifies
\begin{equation}
P^{\ j}_i P^{\ k}_j = P^{\ k}_i \quad , \  P^{\ j}_i f_{j k} = 0,
\label{neweq2}
\end{equation}
it is possible to find an inverse for $f_{ij}$ on the projected subspace. Namely, the ``inverse" $f^{ij}$
is uniquely determined by
\begin{eqnarray}
f_{ik} f^{k j}&=& \delta^{j}_{i} - P^{\ j}_i \nonumber\\
f^{k j}&=& -f^{jk}, \quad f^{i k} P^{\ j}_k = 0.
\label{neweq3}
\end{eqnarray}
Once $f^{ij}$ is constructed, we define the Poisson bracket for functions of $\xi^i$ by (\ref{eqone71}).

It still remains to verify the Jacobi identity. An easy computation shows that (\ref{eqone71})
continues to satisfy that identity provided
\begin{equation}
P^{\ j}_i \frac{\delta}{\delta \xi^j} F_\ell = 0.
\label{anine}
\end{equation}
Hence we use the brackets (\ref{eqone71}) only between functions
$F_\ell(\xi)$ that satisfy the admissability criterion  (\ref{anine}).

\subsubsection*{(c) Obstructions to a canonical formalism}  
\addcontentsline{toc}{subsubsection}{(c) Obstructions to a canonical formalism}
Our problem in connection with Eulerian fluid mechanics is in fact the 
inverse of what has been summarized above.  From (\ref{eqone36}), (\ref{eqone39})
and (\ref{eqone40}), we know the form of
$f^{ij}$ and that the Jacobi identity holds.  We then wish to determine the
inverse $f_{ij}$, and then
$a_i$ from (\ref{fij}). Since we  know the Hamiltonian from~(\ref{eqone33}),
construction of  the Lagrangian~(\ref{lag}) should follow immediately.

However, an obstacle arises: Since there exist a Casimir invariants $C(\xi)$ whose
brackets with all the $\xi^i$ vanish, then
\begin{equation} 0 = \{\xi^i,C(\xi)\} = -f^{ij}\frac{\partial}{\partial \xi^j}C(\xi)\ . 
\end{equation}
That is, $f^{ij}$ has zero modes $\frac{\partial}{\partial \xi^j} C(\xi)$, and the  
inverse to $f^{ij}$, namely the symplectic 2-form
$f_{ij}$, does not exist.  In that case, something has to be done to neutralize the Casimirs.

\subsubsection*{(d) Canonical transformations}
\addcontentsline{toc}{subsubsection}{(d) Canonical transformations}
It is interesting, and will be later useful, to develop the theory further into a discussion of
canonical transformations \cite{VarMech}. A canonical trasfromation is a
transformation on the phase space coordinates $\xi^i$, given infinitesimally by a
vector function,
\begin{equation}
\delta\, \xi^i = - v^i (\xi),
\label{xiten}
\end{equation}
which leaves the symplectic 2-form $f_{ij}$ invariant.
\begin{subequations}\label{right}
\begin{equation}
\delta f_{ij} = v^m \, \partial_m \, f_{ij} + \partial_i \, v^m \, f_{mj} + \partial_j \, v^m \,
f_{im} = 0
\label{fifteena}
\end{equation}
[The exrpession in (\ref{fifteena}) is the Lie derivative of $f_{ij}$ with respect to the
vector field $v^m$.] Use of the Bianchi identity (\ref{aeight}) allows casting
(\ref{fifteena}) into
\begin{equation}
\partial_i (v^m f_{mj}) - \partial_j (v^m f_{mi}) = 0.
\label{aelevenb}
\end{equation}
\end{subequations} 
This shows that the quantity $v^m f_{mi}$ may be presented as
\begin{equation}
v^m f_{mi} = \frac{\partial G (\xi)}{\partial \xi^i}.
\label{atwelve}
\end{equation}
$G(\xi)$ is called a ``generator" for the vector field $v^m$. Conversely, when $f_{mi}$ possesses an
inverse,  given any function $G(\xi)$ on phase space we can define a vector field
$v^m$. Eq. (\ref{aelevenb}) is the general requirement for $v^i$ to define a
canonical transformation. Eq. (\ref{atwelve}) is a necessary and sufficient condition,
locally on phase space. If the phase space has non-trivial topology, one can have
more general solutions to the condition (\ref{right}).

The change of a function $F(\xi)$ on phase space, due to a canonical transformation, is given by
\begin{eqnarray}
\delta F &=& - \frac{\partial F}{\partial \xi^m} \, v^m(\xi) \nonumber\\
&=&- \frac{\partial G}{\partial \xi^i} \, f^{i m} \, \frac{\partial F}{\partial \xi^m} = \{G, F\}
\label{athirteen}
\end{eqnarray}
where (\ref{fundbrack}) and  (\ref{eqone71}) and (\ref{atwelve}) have been used. This shows that
indeed
$G$ generates the infinitesimal canonical transformation by Poisson bracketing.

Note that the bracket (\ref{athirteen}) has been established without the inverse 2-form $f^{ij}$.
This enjoys an advantage over explicit evaluations relying on the methods presented above in 
(a), and especially in (b), where a projected inverse is employed. Moreover we see
that the admissability criterion (\ref{anine}) is automatically satisfied by any generator $G$ that
solves (\ref{atwelve}). This follows immediately from (\ref{neweq}).
}
\vspace{6pt}
\hrule 
\vspace{12pt}
\setcounter{equation}{44}
To construct a Lagrangian for Euler fluids, which leads to the correct equations of motion (\ref{eqone16}), (\ref{eqone17}) and brackets
(\ref{eqone36})-(\ref{eqone41}), we must neutralize the Casimirs. In three and one spatial dimensions, we must neutralize
the velocity Chern-Simons terms (1.2.59) and (\ref{eqone60}), and also the total
number
$N$.  In two dimensions the Casimirs which must be neutralized comprise the infinite tower
(\ref{261}), as well as $N$.

In three dimensions this is achieved in the following manner, based on an idea of C.C. Lin
\cite{Lin}. We use the Clebsch parameterization for the vector field $\bf v$ \cite{Cle}. 
Any three-dimensional vector, which involves three functions, can be presented as
\begin{equation}
\bf v = \grad \theta+\alpha\grad \beta, \label{Cpar}
\end{equation}
 with three suitably chosen scalar functions $\theta,\alpha$, and $\beta$. 
This is called the Clebsch parameterization, and $(\alpha,\beta)$ are called
Gaussian potentials. In this parameterization, the vorticity reads
\begin{equation}
\vec \omega = \grad\alpha \times \grad \beta,
\label{eqone66} 
\end{equation}
 and the Lagrangian is taken as
\begin{equation}L =  -\int \rd{^3} r \rho(\dot\theta+\alpha\dot\beta) - H |_{{\bf
v}=\grad\theta+\alpha\grad\beta}, \label{takenas}
\end{equation}
 with $\bf v$ in $H$ expressed as in (\ref{Cpar}).  Thus the canonically conjugate pairs are
$(\rho,\theta)$ and $(\rho\alpha,\beta)$ replacing $\rho$ and $\bf v$.  The phase space ($\rho,
\theta, \alpha, \beta$) is $4$-dimensional, corresponding to the four observables $\rho$ and $\bf v$,
and a straightforward calculation shows that the Poisson brackets (\ref{eqone36}), (\ref{eqone39})-(\ref{eqone41}) are
reproduced with $\bf v$ constructed by (\ref{Cpar}).  

But how has the obstacle presented by the Casimirs been overcome?  Let us
observe that in the Clebsch parameterization~$C_3$ is given by
\begin{equation}
 C_3 =  \int \rd{^3} r \epsilon^{ijk}\,\partial_i\theta
\,\partial_j\alpha\,\partial_k\beta, \label{cpars}
\end{equation}
 which is just a surface integral
\begin{equation} 
C_3= \int \rd {\vec S}\cdot (\theta\vec \omega).
\label{three5}
\end{equation}
 In this form, $C_3$ has no bulk contribution, and presents no obstacle to
constructing a symplectic $2$-form and a canonical $1$-form in terms of
$({\rho},\theta,\alpha, \beta)$, which are defined in the bulk, that is, for all
finite~$\bf r$. Moreover, the brackets with $N$ are no longer universally
vanishing. Specifically with the new dynamical variable $\theta$ we find
\begin{equation}
\{N, \theta\} = -1.
\label{ntheta70}
\end{equation}
Note that the response of $\theta$ to a finite boost with velocity $\bf u$
\begin{equation}
\theta (t, {\bf r}) \to \theta (t, {\bf r}-{\bf u} t) + {\bf u} \cdot {\bf r} -\frac{\bf u^2}{2} t,
\label{ntheta71}
\end{equation}
contains the 1-cocycle ${\bf u} \cdot {\bf r} -\frac{\bf u^2}{2} t$, as is familiar 
from representations of the Galileo group.

A ``derivation" of (\ref{takenas}) can be given, based on ideas of Lin and earlier ones
of Eckart \cite{Eck}. We begin with $L_E$, the Euler transcription of the Lagrange Lagrangian,
mentioned earlier and containing no time derivatives. This is supplemented with
constraints, enforced by Lagrange multipliers that ensure various continuity
equations.
\begin{equation}
L = \int \rd{^3} r \bigg(\frac{1}{2} \rho {\bf v}^2 - V (\rho) + \theta (\dot{\rho} + {\boldsymbol
\nabla}
\cdot ({\bf v} \rho)) - \rho \alpha (\dot{\beta} + {\bf v} \cdot {\boldsymbol \nabla} \beta) \bigg)
\label{three6}
\end{equation}

The first two terms in the integrand reproduce $L_E$; the first multiplier, $\theta$, enforces the matter/current continuity equation. The
second continuity equation for $\beta$, enforced by the multiplier $\rho \alpha$, is physically obscure. (Lin argues that it
reflects the conservation of ``initial data" for fluid motion.) Varying $\bf v$ and eliminating it gives (\ref{takenas}). 

Let us carry out all the variations and re-derive the two Eulerian equations. For greater generality, useful for relativistic
kinematics, we shall take an arbitrary kinetic energy $\rho T ({\bf v})$, rather than the above non
relativistic case $T({\bf v}) =
\frac{1}{2} {\bf v}^2$, and define the momentum $\bf p $ be the derivative of $T$.
\begin{equation}
{\bf p} \equiv \frac{\partial T ({\bf v})}{\partial  \bf v}
\label{three7}
\end{equation}
Varying $\theta$ gives the continuity equation (\ref{eqone16}); varying $\alpha$ gives
\begin{subequations}\label{1.2.78}
\begin{equation}
\dot{\beta} + {\bf v} \cdot {\boldsymbol \nabla} \beta = 0,
\label{three8}
\end{equation}
and varying $\beta$ gives a similar continuity equation for $\alpha$. 
\begin{equation}
\dot{\alpha} + {\bf v} \cdot \vec \nabla \alpha = 0
\label{three9}
\end{equation}
\end{subequations}
Next we vary $\bf v$ to find
\begin{equation}
\rho {\bf p} - \rho ({\boldsymbol \nabla} \theta + \alpha {\boldsymbol \nabla} \beta) = 0.
\label{three10}
\end{equation}
Thus in the general case it is $\bf p$ (rather than $\bf v$) that is given by the Clebsch parameterization. With
(\ref{three10}) the Lagrangian (\ref{three6}) is rewritten, apart from a total time derivative, as 
\begin{equation}
L = \int d^3 r \bigg(\rho T ({\bf v}) - V (\rho) - \rho (\dot{\theta} + \alpha \dot{\beta}) - \rho {\bf v} \cdot {\bf p}
\bigg).
\label{three11}
\end{equation}
Let $h({\bf p})$ be the Lagrange transform of $T({\bf v})$
\begin{subequations}
\begin{eqnarray}
 h({\bf p}) &=& {\bf p} \cdot {\bf v} - T({\bf v}),  \label{three12a} \\
  \frac{\partial  h({\bf p})}{\partial \bf p}&=& \bf v, 
\label{three12b}
\end{eqnarray}
\end{subequations}
and $L$ becomes
\begin{equation}
L = \int d^3 r \bigg(- \rho (\dot{\theta} + \alpha \dot{\beta}) - \rho h ({\bf p}) -
V(\rho)\bigg). 
\label{three13}
\end{equation}
The remaining variable to vary is $\rho$.
\begin{subequations}
\begin{equation}
\dot{\theta} + \alpha \dot{\beta} + h({\bf p}) + V'(\rho) =0
\label{three13a}
\end{equation}
Differentiating with respect to $r^i$ converts this to
\begin{equation}
\partial_i \dot{\theta} + \partial_i \alpha \dot{\beta} + \alpha  \partial_i \dot{\beta} = -\frac{\partial  h({\bf p})}{\partial  p^j} \partial_i
p^j - V''(\rho) \partial_i \rho.
\label{three13b}
\end{equation}
\end{subequations}
But according to (\ref{three10}) $\partial_i p^j = \partial_j p^i + \partial_i \alpha \partial_j \beta - \partial_j \alpha \partial_i
\beta$. Thus the first term on the right in (\ref{three13b}) is 
\begin{equation}
-v^i \partial_j p^i - \partial_i \alpha v^j \partial_j \beta + v^j \partial_j \alpha \partial_i \beta = -v^j \frac{\partial p^i}{\partial
v^k}
\partial_j v^k + \partial_i \alpha \dot{\beta} - \dot{\alpha} \partial_i \beta,
%\label{three16} 
\nonumber
\end{equation}
where we have used (1.2.74) and (\ref{three10}). Rearranging (\ref{three13b}) leads to 
\begin{subequations}
\begin{equation}
\partial_i \dot{\theta} + \dot{\alpha} \partial_i \beta + \alpha \partial_i \dot{\beta} + v^j \partial_j p^i = -V''(\rho) \partial_i
\rho,
\label{three14a}
\end{equation}
or
\begin{equation}
\dot{p}^i + v^j \partial_j p^i = -V'' (\rho) \partial_i \rho.
\label{three14b}
\end{equation}
\end{subequations}

With Newtonian kinematics $p^i=v^i$, and Euler's equation is regained. With arbitrary kinematics  $p^i \equiv
\frac{\partial  T({\bf v})}{\partial v^i}$. Also, since $\delta p^i = \frac{\partial  p^i}{\partial v^j}
\delta v^j = \tau^{ij} \delta v^j$ where
$\tau^{ij}=
\frac{\partial^2 T}{\partial v^i \partial v^j}$, the above becomes
\begin{subequations}
\begin{equation}
\tau^{ik} ( \dot{v}^k + v^j \partial_j v^k) = -V''(\rho) \partial_i \rho,
\label{three15a}
\end{equation}
or
\begin{equation}
\dot{v}^i + v^j \partial_j v^i = - (\tau^{-1})^{ij} V''(\rho) \partial_j \rho,
\label{three15b}
\end{equation}
\end{subequations}
whenever the inverse to $\tau^{ij}$ exists. For Newtonian kinematics $\tau^{ij} = \delta^{ij}$.

Note that the free Euler equation, (\ref{three15b}) with $V'' =0$, is insensitive to the form of the kinetic term $T({\bf v})$
(provided
$\tau^{ij}$ possess an inverse),  and can be solved together with the continuity equation
(\ref{eqone16}). The general solution in the non-interacting case is presented in terms of posited
initial data.
\begin{eqnarray}
\rho(t = 0, {\bf r}) &=& \rho_0 ({\bf r}) \label{eule82}\\
{\bf v} (t = 0, {\bf r}) &=& {\bf v}_0 ({\bf r}) \label{eule83}
\end{eqnarray}
Define the quantity ${\boldsymbol \chi} (t, {\bf r})$ by the equation
\begin{equation}
{\bf r} = t {\bf v}_0 \big({\boldsymbol \chi} (t, {\bf r}) \big) + {\boldsymbol \chi} (t, {\bf r}).
\label{eule84}
\end{equation}
Then (\ref{eqone16}) and (\ref{eqone17}) (with ${\boldsymbol \nabla} P = 0$) are solved by 
\begin{eqnarray}
\rho (t, {\bf r}) &=& \rho_0 ({\boldsymbol \chi}) | det \frac{\partial \chi^i}{\partial r^j} | ,\label{rhoray}\\
{\bf v} (t, {\bf r}) &=& {\bf v}_0 ({\boldsymbol \chi}). \label{rhoray2}
\end{eqnarray}
This result is verified by differentiation; alternatively it may be derived from (\ref{eqone13}), (\ref{eqone14}), with
${\bf X} (t, {\bf x})$ taken in the absense of forces to be a linear function of $t$.

In one dimension, we parameterize the
velocity as a derivative of a potential
$\theta$,
\begin{equation}
v = \theta',
\label{three21}
\end{equation}
and the phase space consist of $(\rho, \theta)$. The Casimir (\ref{eqone60}) again becomes a surface term 
(only endpoints contribute) and is neutralized in the bulk. The two variables are the conjugate pair $(\rho, \theta)$, which capture the
two degrees of freedom $(\rho, v)$. Of course lineal fluids possesses no vorticity, so
the velocity bracket (\ref{eqone40}) vanishes, while (\ref{eqone39}) is verified. [In
an alternative approach to lineal fluids, we replace $\theta(r)$ by $\frac{1}{2} \int d r'
\varepsilon(r-r') v(r')$ and the canonical 1-form $\int d r \theta \dot{\rho}$ is
$-\frac{1}{2} \int d r d r' \rho(r)
\varepsilon(r-r') \dot{v}(r')$,where
$\varepsilon$ is the $\pm 1$ step function. Evidently this leads to a spatially non-local, but otherwise completely satisfactory canonical
formulation for fluids on a line.] 

Two dimensions presents the additional problem that the number of physical variables is three: $(\rho,
{\bf v})$, but an odd number cannot form a symplectic structure. At the same time there is an infinite number of
Casimirs, \negthinspace (\ref{261}). One may then conclude, heuristically, that it should be possible
to neutralize an infinite number of non-local Casimirs by suppressing one local degree of freedom,
thereby decreasing the effective variables from three to two, an even number with which one can
build a symplectic structure. But it is not known how to effect this suppression;
rather the Lin/Clebsch method is adopted, which increases the degrees of freedom to
four and the Lagrangian (\ref{takenas}) is used in two dimensions as well. 

Finally note that the Lagrangian in (\ref{takenas}), apart from a total time derivative, can also be written as
\begin{eqnarray}
L &=& - \int d r [\rho (\dot{\theta} + \alpha \dot{\beta}) + \rho {\bf v} \cdot ({\boldsymbol \nabla} \theta + \alpha {\boldsymbol \nabla}
\beta)] + L_E \nonumber \\
&=& -\int d r j^\mu (\partial_\mu \theta + \alpha \partial_\mu \beta) + L_E.
\label{eule88}
\end{eqnarray}
Although we are dealing with nonrelativistic dynamics, we have used covariant notation for the canonical 1-form, with
$j^\mu = (c\rho, \rho {\bf v})        $ and $\partial_\mu = (\frac{1}{c} \frac{\partial}{\partial t}, {\boldsymbol \nabla})$. This
formulation becomes our starting point for the relativistic generalization, discussed in Section 1.4 below. (That is why we
have introduced the velocity of light $c$ in the above definitions; of course it disappears in the nonrelativistic theory, where
it has no role.)\\

\hrule

\vspace{8pt}
{\appendixb
%\addtocontents{toc}{\protect\hrulefill\par}
\addcontentsline{toc}{section}{\quad \ B. Sidebar on Clebsh parameterization and the Chern-Simons term}
We elaborate on the \Cpr\ for a vector field \cite{Cle}, which was
presented for the velocity vector in~(\ref{Cpar}). Here we shall use the notation
of electromagnetism and discuss the \Cpr\ of a vector potential $\bf A$, which
also leads to the magnetic field ${\bf B} =\grad \times \bf A$. (Of course the
same observations apply when the vector in question is the velocity field $\bf
v$, with $\grad\times{\bf v}$ giving the vorticity.) 

The familiar \pr\ of a three-component vector employs a scalar
function~$\theta$  (the ``gauge'' or ``longitudinal'' part) and a two-component
transverse vector
${\bf A}_T$: ${\bf A}=
\grad\theta + {\bf A}_T$, $\grad\cdot {\bf A}_T = 0$. This decomposition is
unique and invertible (on a space with simple topology). In contrast, the \Cpr\
uses three scalar functions, $\theta$, $\alpha$, and~$\beta$,
\begin{equation}
{\bf A} = \grad\theta + \alpha \grad \beta,
\label{threescalar}
\end{equation}
which are not uniquely determined by $\bf A$ (see below). The
associated magnetic field reads 
\begin{equation}
{\bf B} = \vec \nabla \times {\bf A}=\grad\alpha\times\grad\beta.
\label{assmag}
\end{equation}
Repeating the above in form notation, the 1-form $A=A_i \rd r^i$ is presented as 
\begin{equation}
A= \rd \theta + \alpha \rd \beta,
\label{1-form}
\end{equation}
and the 2-form is
\begin{equation}
\rd A= \rd \alpha \rd \beta. 
\label{2-form}
\end{equation}
Darboux's theorem \cite{Darb} ensures that the \Cpr\ is attainable locally in space [in
the form (\ref{1-form})]. Additionally, an explicit construction of $\alpha$,
$\beta$, and $\theta$ can be given by the following procedure \cite{Lam}.

Solve the equations
\begin{subequations}\label{solve}
\begin{equation}
\frac{\rd x}{B_x} =\frac{\rd y}{B_y} =\frac{\rd z}{B_z},
\label{solvea}  
\end{equation}
which may also be presented as
\begin{equation}
\eps^{ijk} \rd {r^j} B^k = 0.
\label{solveb}  
\end{equation}
\end{subequations}
Solutions of these relations define two surfaces, called ``magnetic surfaces'', that
are given by equations of the form
\begin{equation}
S_n({\bf r}) = \mathrm{constant}\qquad (n=1,2).
\label{magsurfs}
\end{equation}
It follows from \refeq{solve} that these also satisfy
\numeq{follows}{
{\bf B} \cdot \grad S_n = 0 \qquad (n=1,2),
}
that is, the normals to $S_n$ are orthogonal to $\bf B$, or $\bf B$ is
parallel to the tangent of~$S_n$.  The intersection of the two surfaces forms the
so-called ``magnetic lines'', that is, loci that solve the dynamical system
\numeq{loci}{
\frac{\dtr(\tau)}{\rd \tau} = {\bf B}\bigl(\bf r(\tau)\bigr),
}
where $\tau$ is an evolution parameter. Finally, the Gaussian potentials $\alpha$
and~$\beta$ are constructed as functions of~$\bf r$ only through a dependence
on the magnetic surfaces,
\begin{align}
\alpha({\bf r}) &= \alpha \bigl( S_n ({\bf r})\bigr),\nonumber\\
\beta({\bf r}) &= \beta \bigl( S_n ({\bf r})\bigr),
\label{depmags}
\end{align}
so that
\numeq{depmags2}{
\grad \alpha \times \grad\beta = (\grad S_1 \times \grad S_2) \eps^{mn}
\frac{\partial \alpha}{\partial S_m}\frac{\partial \beta}{\partial S_n}.
}
Evidently as a consequence of \refeq{follows}, $\grad
\alpha\times\grad\beta$ in~\refeq{depmags2} is parallel to~$\bf B$, and
because
$\bf B$ is divergence-free $\alpha$ and
$\beta$ can be adjusted so that the  norm of $\grad \alpha \times \grad\beta$ 
coincides with $|{\bf B}|$. Once $\alpha$ and
$\beta$ have been fixed in this way, $\theta$ can easily be computed from $\vec
A-\alpha\grad\beta$. 

Neither the individual magnetic surfaces nor the Gauss potentials are unique.
[By viewing $A$ as a canonical 1-form, it is clear that the expression
\refeq{1-form} retains its form after a canonical transformation of
$\alpha,\beta$.] One may therefore require that the Gaussian potentials $\alpha$
and $\beta$ simply coincide with the two magnetic surfaces: $\alpha=S_1$,
$\beta=S_2$. Nevertheless, for a given
$\bf A$ and $\bf B$ it may not be possible to solve \refeq{solve} explicitly.

The \CS\ integrand ${\bf A}\cdot {\bf B}$ becomes in the \Cpr\
\numeq{CSi}{
{\bf A}\cdot {\bf B} = \grad\theta\cdot (\grad \alpha\times\grad\beta)
= \grad\cdot(\theta\ {\bf B}) = {\bf B} \cdot \grad\theta.
}
Thus having identified the Gauss potentials $\alpha$ and $\beta$ with the two
magnetic surfaces, we deduce from \refeq{follows} and \refeq{CSi} three
equations for the three functions ($\theta$, $\alpha$, $\beta$) that comprise
the \Cpr.
\begin{align}
{\bf B} \cdot\grad \alpha &= {\bf B} \cdot\grad\beta = 0 \nonumber\\
{\bf B} \cdot\grad \theta &= \mbox{\CS\ density } {\bf A}\cdot {\bf B} 
\label{follows2}
\end{align}

Eq.~\refeq{CSi} also shows that in the \Cpr\ the \CS\ density becomes a total
derivative.
\numeq{totderiv}{
{\bf A} \cdot {\bf B} = \grad\cdot(\theta {\bf B})
}
This does \emph{not} mean that the \Cpr\ is unavailable when the \CS\ integral 
over all space is nonzero. Rather for a nonvanishing integral and well-behaved
${\bf B}$ field, one must conclude that the Clebsch function~$\theta$ is singular
either in the finite volume of the integration region or on the surface at infinity
bounding the integration domain. Then the \CS\ volume integral over ($\Omega$)
becomes a surface integral on the surfaces ($\partial\Omega$) bounding the
singularities.
\numeq{boundsing}{
\int_\Omega \mathrm{d}^3 r {\bf A} \cdot {\bf B} = \int_{\partial\Omega} \rd {\bf
S}\cdot\, (\theta {\bf B}) 
}
Eq.~\refeq{boundsing} shows that the \CS\ integral measures the magnetic flux,
modulated by~$\theta$ and passing through the surfaces that surround the
singularities of~$\theta$. 

The following  explicit example  illustrates the above points. 

 Consider the vector potential whose spherical components are
given by 
\begin{align}
A_r &= (\cos\Theta) a'(r),\nonumber\\
A_\Theta &= -(\sin\Theta)\frac1r \sin a(r),\nonumber\\
A_\Phi &= -(\sin\Theta)\frac1r \bigl(1-\cos a(r)\bigr).
\label{vecpot}
\end{align}
($r$, and $\Theta$, $\Phi$ denote the conventional radial coordinate and the
polar, azimuthal angles.)  The function $a(r)$ is taken to vanish at the origin, and
to behave as
$2\pi\nu$ at infinity ($\nu$~integer or half-integer). The corresponding
magnetic field reads 
\begin{align}
B_r &= -2(\cos\Theta)\frac1{r^2} \bigl(1-\cos a(r)\bigr),\nonumber\\
B_\Theta &=  (\sin\Theta)\frac1r a'(r)\sin a(r),\nonumber\\
B_\Phi &= (\sin\Theta)\frac1r a'(r) \bigl(1-\cos a(r)\bigr),
\label{corrmag}
\end{align}
and the \CS\ integral -- also called the ``magnetic helicity'' in the
electrodynamical context --  is quantized (in multiples of
$16\pi^2$) by the behavior of
$a(r)$ at infinity.
\begin{align}
\int \mathrm{d}^3 r {\bf A} \cdot {\bf B} &= 
-8\pi\int_0^\infty \rd r \frac {\rd\null}{\rd r} \bigl(a(r) -\sin a(r)\bigr) 
\nonumber\\
&= -16\pi^2\nu.
\label{atinf}
\end{align}

In spite of the nonvanishing magnetic helicity, a \Cpr\ for \refeq{vecpot} is readily
constructed. In form notation, it reads 
\numeq{constr}{
A=\rd{(-2\Phi)} +2\Bigl(1-\bigl(\sin^2 \frac a2\bigr) \sin^2\Theta\Bigr) 
\rd{\!\left(\Phi + \tan^{-1}\Bigl[ \bigl(\tan\frac a2\bigr)\cos\Theta\Bigr]\right).}
}
The
magnetic surfaces can be taken from  formula~\refeq{constr} to coincide with
the Gauss potentials. 
\begin{align}
S_1 &= 2\Bigl(1- \bigl(\sin^2 \frac a2\bigr)\sin^2\Theta\Bigr) =
\mbox{constant}\nonumber\\
S_2 &= \Phi + \tan^{-1} \Bigl[\bigl(\tan \frac a2\bigr)\cos\Theta\Bigr]=
\mbox{constant}
\label{gausspot}
\end{align}
The magnetic lines are determined by the intersection of $S_1$ and $S_2$. 
\begin{align}
\cos \frac a2 &= \eps \cos(\Phi - \phi_0)\nonumber\\
\sin\Theta &= \sqrt{\frac{1-\eps^2}{1-\eps^2\cos^2 (\Phi-\phi_0)}}
\label{maglines}
\end{align}
where $\eps$ and $\phi_0$ are constants. The potential $\theta= -2\Phi$ is
multivalued. Consequently the ``surface'' integral determining the \CS\ term
reads
\numeq{surfaceint}{
\int \mathrm{d}^3 r {\bf A} \cdot \vec B = \int \mathrm{d}^3 r \grad\cdot(-2\Phi {\bf B}) 
= -4 \pi \int_0^\infty r\rd r \int_0^\pi \rd \Theta B_\Phi\Bigr|_{\Phi=2\pi} \ .
}
That is, the magnetic helicity is the flux of the toroidal magnetic field through the
positive-$x$ ($x,z$) half-plane. 

Finally we remark on a subtle property of the Clebsch decomposition when used in variational calculations
\cite{DesJacPol}. Consider an ``action" of the form
\begin{equation}
I = I_0({\bf B}) + \frac{\mu}{2} \int\limits_\Omega d^3 r \bf A \cdot B. 
\label{b22}
\end{equation}
Variation of $\bf A$ gives
\begin{equation}
\delta I = \int\limits_\Omega d^3 r (\grad \times \mathcal{B} + \mu {\bf B}) \cdot
\delta
\vec A,
\label{b23}
\end{equation}
where $\mathcal{B} ({\bf r}) \equiv \frac{\delta I_0 ({\bf B})}{\delta {\bf B} ({\bf
r})}$. Demanding that I be stationary against variations of ${\bf A}$ requires the
vanishing of the term in parentheses, which is transverse, since the transverse part
of the variation $ \delta \bf A$ is arbitrary.
\begin{equation}
\vec \nabla  \times \mathcal{B} + \mu {\bf B} = 0,
\label{b24}
\end{equation}
Now let us examine the same problem in the Clebsch parameterization. The
Chern-Simons contribution to (\ref{b22}) reads, according to (\ref{CSi})
\begin{equation}
\frac{\mu}{2} \int\limits_\Omega \mathrm{d}^3  r \vec \nabla \cdot (\theta {\bf B}) =
\frac{\mu}{2} \int\limits_{\partial \Omega} d \ {\bf S} \cdot \theta \bf B.
\label{b25}
\end{equation}
In the gauge $\theta = 0$ (\ref{b25}) vanishes, and in any gauge it has no bulk
contribution, so its variation will never produce the second left-hand term in
(\ref{b24}). So how is (\ref{b24}) regained?

Returning to (\ref{b22}), we accept the fact that the variation of the last term vanishes, while
the variation of the first leaves
\begin{equation}
\delta I = \int\limits_\Omega d^3 r (\vec \nabla \times \mathcal{B}) \cdot ( \vec \nabla \beta \
\delta
\alpha - \vec \nabla \alpha \ \delta \beta).
\label{b26}
\end{equation}
Since $\delta \alpha$ and $\delta \beta$ are arbitrary, $I$ is stationary provided
\begin{subequations}
\begin{eqnarray}
(\vec \nabla \times \mathcal{B}) \cdot \vec \nabla \beta = 0, \label{b27a}\\
(\vec \nabla \times \mathcal{B}) \cdot \vec \nabla \alpha = 0.
\label{b27b}
\end{eqnarray}
\end{subequations}
We obtain two equations, which imply by (\ref{follows2}),
\begin{equation}
\vec  \nabla \times {\bf \mathcal{B}} + \mu ({\bf r}) \vec B = 0.
\label{b28}
\end{equation}
Transversality of $ \nabla \times \vec B$ and $\vec B $ further requires that
$\vec B \cdot \vec \nabla \mu ({\bf r}) = 0$, {\it i.e.}, a non-vanishing $\vec \nabla
\mu ({\bf r})$ ($=$ non-constant $\mu$) must lie in the ($\vec \nabla \alpha, \vec
\nabla \beta$) plane.

Since (\ref{b28}) is weaker than the parameteriztion-independent (\ref{b24}), we conclude
that the\\ Clebsch paramterization is somewhat incomplete, when used in variational
calculations that ignore surface terms. This is similar to the fact that in the \Cpr \
gauge potentials, which carry a non-vanishing Chern-Simons term ($=$ velocities
with non-vanishing  vortex helicity), encode the non-vanishing value in a surface
term.  

In Section \ref{Section7.4}, we shall present a different method, based on group theory, for
obtaining the Clebsch parameterization. This approach is then generalized to  non-Abelian
vector fields. 
\addtocontents{toc}{\protect\hrulefill\par}
}
\vspace{15pt}
\hrule 
\setcounter{equation}{0}
\subsection{Irrotational fluids}
The simplification found for 1-dimensional fluids, (\ref{three21}), presenting the velocity as a
derivative of a velocity potential
$\theta$, can be extended to two-and three-dimensional fluids that are irrotational: the vorticity vanishes.
\begin{eqnarray}
\vec \omega = \vec \nabla \times {\bf v} = 0 \label{one41} \\
{\bf v} = \vec \nabla \theta \label{one42}
\end{eqnarray}

The Clebsch parameterization (\ref{Cpar}) holds trivially; the potentials $\alpha, \beta$, the Casimirs $C_{(d)}$ vanish, and $N$
obeys (\ref{ntheta70}). This removes the obstruction to a canonical formalism. The Lagrangian (\ref{takenas}) becomes
\begin{equation}
L= -\int d r\ \rho \dot{\theta} - H |_{{\bf v} = \vec \nabla \theta},  \label{one43}
\end{equation}
and this can be derived by the Eckart procedure as in (\ref{three6}) - (\ref{three13}), where now only the continuity
equation is enforced by the Lagrange multiplier $\theta$, and the Gaussian potentials $\alpha$ and $\beta$  are omitted.

The Euler equation (\ref{eqone17}), with the pressure $P$ expressed in terms of the enthalpy $V'(\rho)$ as in
(\ref{eqone34}) may be integrated once to give the Bernoulli equation.
\begin{equation}
\dot{\theta} + \frac{1}{2} (\nabla \theta)^2 = - V'(\rho) \label{one44}
\end{equation}
This also follows by bracketing $\theta$ with $H$ in (\ref{one43}), since now there is only one non-vanishing and canonical
bracket.
\begin{equation}
\{\theta ({\bf r}), \rho ({\bf r'})\} = \delta({\bf r} - {\bf r'}) \label{one45}
\end{equation}
Observe that the Bernoulli equation (\ref{one43}) allows for the possibility of expressing  $\rho$ in terms of $\dot{\theta} +
\frac{1}{2} (\nabla \theta)^2$, through the inverse to $V'$. One can then eliminate $\rho$ in the continuity equation, leaving
a single (non-linear) equation for $\theta$.
\subsection{Relativistic fluids}
\setcounter{equation}{0}
We discuss the formalism for describing relativistic fluids; only the Euler approach is
treated. Usually the dynamics of a relativistic fluid is presented in terms of the
energy-momentum tensor, $\theta^{\mu\nu}$, and the equations of motion are just
the conservation equations $\partial_\mu \theta^{\mu\nu}=0$. This is analogous to the nonrelativistic situation
mentioned previously, where the nonrelativistic energy momentum  complex $T^{\mu
\nu}$ encapsulates the equations of motion for a nonrelativistic fluid.  [We denote the
relativistic energy-momentum tensor by $\theta^{\mu\nu}$, to distinguish it from
the nonrelativistic $T^{\mu\nu}$ introduced in \refeq{eqone44}-\refeq{eqone48}.
The limiting relation between the two is given below.] But we shall begin with a
Lagrange density. 

Inspired by the suggestive formula (\ref{eule88}), we consider
\numeq{Lagdens}{
{\cal L} = -j^\mu a_\mu - f(\sqrt{j^\mu j_\mu}).
}
Here $j^\mu$ is the current Lorentz vector $j^\mu= (c\rho, {\bf j})$ \cite{gravcos}. The $a_\mu$
comprise a set of auxiliary variables; in the relativistic analog of irrotational
fluids we take $a_\mu=\partial_\mu\theta$, more generally 
\numeq{relanalirr}{
a_\mu = \partial_\mu\theta + \alpha \partial_\mu\beta, 
}
so that the \CS\ density of $a_i$ is a total derivative 
[compare (\ref{cpars}), (\ref{three5})]. The function~$f$ depends on the
Lorentz invariant $j^\mu j_\mu = c^2 \rho^2 -{\bf j}^2$ and encodes the specific
dynamics (equation of state).

The energy momentum tensor for  
$\cal L$ is 
\numeq{enmomten}{
\theta_{\mu\nu} = -g_{\mu\nu} {\cal L} + \frac{j_\mu j_\nu}{\sqrt{j^\alpha
j_\alpha}} f' (\sqrt{j^\alpha j_\alpha}). 
}
 [One way to derive \refeq{enmomten} from \refeq{Lagdens} is to
embed that expression in an external metric tensor $ g_{\mu\nu}$, which is then
varied; in the variation $j^\mu$ and $a_\mu$ are taken to be metric-independent
and
$j_\mu= g_{\mu\nu} j^\nu$.] 
Furthermore,  varying $j^\mu$ in \refeq{Lagdens} shows that 
\numeq{varyingjmu}{
a_\mu = -\frac{j_\mu}{\sqrt{j^\alpha j_\alpha}} f'(\sqrt{j^\alpha j_\alpha}),
}
so that \refeq{enmomten}  becomes  
\numeq{enmombec}{
\theta_{\mu\nu} = -g_{\mu\nu} [nf'(n) - f(n)] + u_\mu u_\nu n f'(n). 
}
We have introduced the proper velocity $u_\mu$ by factoring $j_\mu$, as suggested
by Eckart.
\numeq{intrumu}{
j_\mu= nu_\mu\qquad u^\mu u_\mu=1
}
One sees that $n$ is proportional to the proper density and $1/n$  is proportional to the
specific volume.  Eq.~\refeq{enmombec} identifies the proper energy density~$e$
and the pressure~$P$ (which coincides with $\cal L$) through the conventional
formula \cite{Wein} 
\numeq{convform}{
\theta_{\mu\nu} = -g_{\mu\nu} P + u_\mu u_\nu (P + e). 
}
Therefore, in our case
\begin{align}
e &= f(n) \label{efn}\\
P &= n f'(n) - f(n)\ . 
\label{pnfn}
\end{align}

The thermodynamic relation involving entropy~$S$ reads
\numeq{thermts}{
P \rd{\Bigl(\frac1n\Bigr)} + \rd{\Bigl(\frac en\Bigr)} \propto \rd S, 
}
where the proportionality constant is determined by the temperature.
With \refeq{efn} and \refeq{pnfn}  the left side of \refeq{thermts} vanishes
and we verify that entropy is constant, that is, we are dealing with an
isentropic system, as has been stated in the very beginning.

For the free system, the pressure vanishes, so we choose $f(n) = cn$.
\numeq{pressvan}{
{\cal L}_0 = -j^\mu a_\mu - c\sqrt{j^\mu j_\mu}
} 
Other forms for $f$ give rise to relativistic fluid mechanics with other equations of state.

Taking the divergence or $\theta_{\mu \nu}$ in (\ref{enmombec}) leaves
\begin{eqnarray}
\partial^\mu \theta_{\mu \nu} = - n f'' (n) \partial_\nu n + n u^\mu \partial_\mu (
u_\nu f' (n))  + \partial_\mu (nu^\mu) u_\nu f' (n). \label{five12}
\end{eqnarray}
The first two terms on the right are orthogonal to $u^\nu$, the last one is parallel. So
the vanishing of the divergence of $\theta_{\mu \nu}$ implies the continuity equation.
\begin{equation}
0 = \partial_\mu (nu^\mu) = \partial_\mu j^\mu \label{five13}
\end{equation}
The vanishing of the remaining components is equivalent to
\begin{equation}
u^\mu (\partial_\nu u_\mu - \partial_\mu u_\nu) f' (n) + (g_{\nu \mu} -u_\nu u_\mu) \partial^\mu
n f'' (n) = 0.
\label{five14}
\end{equation}
This is the relativistic Euler equation.

The same two equations follow from the Lagrange density (\ref{Lagdens}). Variation
of $\theta$ and the Gauss potentials $\alpha$ and $\beta$ gives the current continuity
equation (\ref{five13}) and equations satisfied by $\alpha$ and $\beta$
\begin{equation}
u^\mu \ \partial_\mu \alpha = 0 = u^\mu \ \partial_\mu \beta \label{five15}
\end{equation}
Variation of $j^\mu$ gives (\ref{varyingjmu}), whose curl reads
\begin{eqnarray}
(\partial_\nu\ u_\mu - \partial_\mu \ u_\nu) f' (n) 
+ (u_\mu \ \partial_\nu \ n - u_\nu \ \partial_{\mu}\ n) f''
(n) = \partial_\mu\ a_\nu - \partial_\nu \ a_\mu = \partial_\mu \alpha \partial_\nu \beta - \partial_\nu \alpha \partial_\mu
\beta.
\label{five16}
\end{eqnarray}
Contracting this with $u^\mu$ makes the right side vanish by virtue of (\ref{five15})
and the left side coincides with (\ref{five14}).

It is especially intriguing to notice that
$\theta^{\mu\nu}$ is symmetric but $T^{\mu\nu}$ is not. To make the connection 
we recall that $u^\mu = 1/\sqrt{1-{\bf v}^2/c^2} (1, {\bf v}/c)$, we observe that
$n=\sqrt{\rho^2 c^2 -{\bf j}^2}$, set ${\bf j} = {\bf v} \rho$ and
conclude that $n=\rho c \sqrt{1-{\bf v}^2/c^2} \sim \rho c - (\rho {\bf v}^2/2c)$.
Also $f(n)$ is chosen to be
$cn + V(n/c)$, and thus $P=n f'(n) - f(n) = (n/c) V' (n/c) - V(n/c)$. It follows that 
\begin{align}
\theta^{oo} &= \frac{nc - ({{\bf v}^2 n}/{c^3})V'}{1-{\bf v}^2/c^2} + V \approx 
\frac{\rho c^2 -   \rho {\bf v}^2/2}{1-{\bf v}^2/c^2} + V(\rho) \nonumber\\
&\approx \rho c^2 + \frac{\rho {\bf v}^2}2 + V(\rho) = \rho c^2 + T^{oo}\ .
\label{itfollowsthat}
\end{align}
Thus, apart from the relativistic ``rest energy'' $\rho c^2$, $\theta^{oo}$ passes to
$T^{oo}$. The relativistic energy flux is $c\theta^{jo}$ (because
$\frac\partial{\partial x^\mu} \theta^{\mu o} = \frac1c \dot \theta^{oo} + 
\partial_j\theta^{jo}$).
\begin{align}
c \theta^{jo} &= \frac{v^j}{1-{\bf v}^2/c^2} \Bigl(nc + \frac nc V'\Bigr) \approx 
v^j \frac{\rho c^2 -   \rho {\bf v}^2/2 + \rho V'(\rho)}{1-{\bf v}^2/c^2}  \nonumber\\
&\approx j^j c^2 +  \rho v^j\bigl({\bf v}^2/2 +V'(\rho)\bigr) = j^j c   + T^{jo}\ 
\label{relenflux}
\end{align}
Again, apart from the $O(c^2)$ current, associated with the $O(c^2)$ rest energy in
$\theta^{oo}$, $T^{jo}$ is obtained in the limit. The momentum density is
$\theta^{oi}/c$ (because $\theta^{\mu\nu}$ has dimension of energy density).  
\numeq{dimenerdens}{
\theta^{oi}/c = \frac{v^i/c^2}{1-{\bf v}^2/c^2} \Bigl(nc + \frac nc V'\Bigr) \approx \rho v^i
= {\cal P}^i\ 
 }
Finally, the momentum flux is obtained directly from $\theta^{ij}$. 
\begin{align}
\theta^{ij} &= \delta^{ij} \Bigl(\frac nc V' - V\Bigr) + \frac{v^i v^j}{c^2 - {\bf v}^2} \Bigl(nc
+ \frac nc V'\Bigr)\nonumber\\
&\approx \delta^{ij} \bigl( \rho V'(\rho) - V(\rho)\bigr) + v^i v^j \rho = T^{ij}  
\label{finmomflux} 
\end{align}
\newpage
\section{ SPECIFIC MODELS ($d \ne 1$)}

We now examine irrotational models, both relativistic and nonrelativistic, for
which we shall specify an explicit force law and discuss further properties. 

\subsection{Two models, mostly in spatial dimensions $d \ne 1$.}
{\bf (i) Galileo-invariant nonrelativistic model: Chaplygin gas}\label{2.2i} 
\addcontentsline{toc}{subsubsection}{(i) Galileo-invariant nonrelativistic model: Chaplygin
gas}\\

\noindent Recall that the nonrelativistic Lagrangian for irrotational motion reads
\begin{equation} L^{\mathrm{Galileo}}=\int \rd r \Bigl(\theta\dot\rho - \rho
\frac{(\grad\theta)^2}{2}-V(\rho)\Bigr),
\label{Galil}
\end{equation}
 where $\grad\theta=\bf v$.  The Hamiltonian density $\cal H$ is
composed of the last two terms beyond the canonical $1$-form $\int  \rd r
\theta\dot\rho$
\begin{equation}
 H=\int \rd r \Bigl(\rho\frac{(\grad\theta)^2}{2}+V(\rho)\Bigr)=\int\rd r
{\cal H}.
\end{equation}
Varying (\ref{Galil}) with respect to $\rho$ produces the Bernoulli equation (1.3.4 ).
 Various expressions for $V$ appear in the literature.  $V(\rho)\propto
\rho^n$  is a popular choice, appropriate for the adiabatic equation of state. 
We shall be specifically interested in the Chaplygin gas \cite{Chp}. 
\numeq{ChaGas}{
V(\rho)=\lambda/ \rho, \qquad \lambda>0 
}
 According to what we said before, the Chaplygin gas has enthalpy
$V'=-{\lambda}/{\rho^2}$, negative pressure $P=-{2\lambda}/{\rho}$, 
and speed of sound $s=\sqrt{2\lambda}/\rho$ (hence $\lambda>0$).  

 Chaplygin introduced his equation of state as a mathematical approximation to
the physically relevant adiabatic expressions with $n>0$. (Constants are arranged
so that the Chaplygin formula is tangent at one point to the adiabatic
profile.)  Also it was realized that certain deformable solids can be
described by the Chaplygin equation of state. These days negative
pressure is recognized as a possible physical effect: exchange forces in atoms give
rise to negative pressure; stripe states in the quantum Hall effect may be a
consequence of negative pressure; the recently discovered cosmological constant
may be exerting negative pressure on the cosmos, thereby accelerating expansion. 

For any
form of $V$, the model possesses the Galileo symmetry, discussed previously as appropriate to
nonrelativistic dynamics. 
There are a total of $\fract{1}{2}(d+1)(d+2)$
Galileo generators in $d$ space plus one time dimensions.  Together with the
central term, $N$,  we have a total of 
$\fract{1}{2}(d+1)(d+2)+1$ generators.

A useful consequence of symmetry transformations is that they
map solutions of the equations of motion into new solutions.  Of course, ``new''
solutions produced by Galileo transformations are trivially related to the old
ones: they are simply shifted, boosted or rotated.  

But we shall now turn to the
specific Chaplygin gas model, with
$V(\rho)= {\lambda}/{\rho}$, which possesses additional and unexpected
symmetries.

The Chaplygin gas action and consequent Bernoulli equation for the Chaplygin
gas in
$(d,1)$ space-time read
\begin{gather} I^{\mathrm{Chaplygin}}_\lambda= \int \rd t \int \rd r
\Bigl(\theta\dot\rho -
\rho\frac{(\grad \theta)^2}{2} -
\frac{\lambda}{\rho}\Bigr)\label{chapgasact}\\
\dot\theta + \frac{(\grad \theta)^2}{2} =
\frac{\lambda}{\rho^2}\label{conseqBern}
\end{gather}
  This model possesses further space-time symmetries beyond those of
the Galileo group \cite{BazJac}.  First, there is a one-parameter~($\omega$, dimensionless)
time rescaling transformation
\begin{equation}\label{trt}
 t\rightarrow T=e^{\omega} t,
\end{equation}
 under which the fields transform as 
 \begin{align}
 \theta(t,{\bf r}) &\rightarrow \theta_\omega (t,{\bf r}) = e^\omega
 \theta(T,{\bf r}),\label{trt2}\\
 \rho(t,{\bf r}) &\rightarrow \rho_\omega (t,{\bf r}) = e^{-\omega} \rho(T,{\bf r}).\label{trt3}
 \end{align}
 Second, in $d$ spatial dimensions, there is a
vectorial, $d$-parameter~($\vec\omega$, dimension inverse velocity) space-time mixing transformation.
\begin{align} t&\rightarrow T(t,{\bf r}) = t+\vec\omega \cdot {\bf r} +
\fract{1}{2}\vec \omega^2
\theta(T,{\bf R})\label{eq:100}\\
{\bf r} &\rightarrow {\bf R}(t,{\bf r}) = {\bf r} + \vec \omega \,\theta(T,{\bf R})
\end{align}
 Note that the transformation law for the coordinates involves the
$\theta$ field itself.  Under this transformation, the fields transform according
to
\begin{align}
\theta(t,{\bf r}) &\rightarrow \theta_{\vec\omega}(t,{\bf r}) = \theta(T,{\bf R}),\label{eq:102}\\
\rho(t,{\bf r}) &\rightarrow \rho_{\vec\omega}(t,{\bf r}) = \rho(T,{\bf R})\frac{1}{|J|},
\end{align}
 with $J$ the Jacobian of the transformation linking $(T,{\bf R})\rightarrow
(t,{\bf r})$. 
\numeq{matrlink}{
J = \det \begin{pmatrix} 
\displaystyle\frac{\partial T}{\partial t} &\displaystyle  \frac{\partial
T}{\partial{r^j}}\\[2ex]
\displaystyle\frac{\partial R^i}{\partial t} &\displaystyle \frac{\partial
R^i}{\partial r^j}
\end{pmatrix} =
\Bigl(1-\vec\omega \cdot\grad \theta(T,{\bf R}) - \frac{\vec \omega^2}2\,  \dot\theta
(T,{\bf R})\Bigr)^{-1} }
(The time and space derivatives in the last element are with respect to~$t$
and~${\bf r}$.)
 One can tell the complete story for these transformations:
The action (\ref{chapgasact}) is invariant; Noether's theorem gives the conserved quantities,
which for the time rescaling is
\begin{equation} 
S=tH-\int \rd r \rho\theta \qquad\mbox{(time rescaling)},
\label{eq:105}
\end{equation}
 while for the space-time mixing one finds
\begin{equation}
{\bf G} = \int \rd r \left({\bf r} {\cal H} - \theta \vec{\cal P}\right)
\qquad\mbox{(space-time mixing).}
\label{eq:106}
\end{equation}
 The time independence of $S$ and $\bf G$  can be verified with the help of the
equations of motion (continuity and Bernoulli) \cite{BorHop}.  Poisson bracketing the
fields~$\theta$ and~$\rho$ with 
$S$ and $\bf G$ generates the appropriate infinitesimal transformation on the
fields. Note that unlike the Galileo constants of motion, the new constants of motion cannot be locally
expressed in terms of $\bf v$: their integrands depends locally on $\theta = \frac{\bf \nabla}{\nabla^2} \cdot {\bf v}$.

So now the total number of generators is the sum of the previous
$\fract12(d+1)(d+2) +1$ with $1+d$ additional ones.
\begin{equation}
\fract{1}{2}(d+1)(d+2) + 1 + 1 +d = \fract{1}{2}(d+2)(d+3)
\label{two16}
\end{equation}
 When one  computes the Poisson brackets of all  these with each other one finds 
the Poincar\'e Lie algebra in one higher spatial dimension, that is,
in $(d+1,1)$-dimensional space-time, where the Poincar\'e group possesses
$\fract12 (d+2)(d+3)$ generators.  Moreover, one verifies that
$(t,\theta,{\bf r})$ transform  linearly as a $(d+2)$ Lorentz vector in light-cone
components, with $t$ being the ``$+$'' component and $\theta$ the ``$-$''
component. \cite{Baz}

Presently, we shall use these additional symmetries to generate new solutions
from old ones, but, in contrast to the Galileo transformations, the new solutions will
be nontrivially linked to the former ones.  Note that the additional
symmetry holds even in the free theory. 

Before proceeding, let us observe that $\rho$ may be eliminated by using the
Bernoulli equation to express it in terms of $\theta$.  In this way, one is led to
the following
$\rho$-independent action for $\theta$ in the Chaplygin gas problem:
\begin{equation} 
I^{\mathrm{Chaplygin}}_\lambda=-2\sqrt\lambda \int dt \int \rd r
\sqrt{\dot\theta+\frac{(\grad
\theta)^2}{2}}.\label{rindependent}
\end{equation}
 Although this operation is possible only in the interacting case, the
interaction strength disappears from the equations of motion.
\numeq{interstren}{
\frac\partial{\partial t} \frac1{\sqrt{\dot\theta+\frac{(\grad
\theta)^2}{2}}} +\grad \cdot \frac{\grad\theta}{\sqrt{\dot\theta+\frac{(\grad
\theta)^2}{2}}} = 0 
}
$\lambda$ merely
serves as an overall factor in the action.

The action (\ref{rindependent}) looks unfamiliar; yet it is Galileo invariant.
[The combination $\dot\theta+\frac12 (\grad\theta)^2$ responds to Galileo
transformations without a 1-cocycle; see (1.2.71).]
 Also
(\ref{rindependent}) possesses the additional symmetries described
above, with $\theta$ transforming according to the previously recorded equations.

Let us discuss some solutions.  For example, the free theory is solved by
\begin{equation}
\theta(t,{\bf r}) = \frac{{\bf r}^2}{2t} 
\label{freesol}
\end{equation}
 which corresponds to the velocity
\begin{equation}
{\bf v}(t,{\bf r}) = \frac{{\bf r}}{t}.
\label{two20}
\end{equation}
Galileo transformations generalize this in an obvious manner into a set of  displaced, rotated and
boosted solutions.  Performing on the above formula for~$\theta$ the new transformations of
time-rescaling and space-time mixing (\ref{trt}), (\ref{trt2}), (\ref{eq:100})-(\ref{eq:102}), we find
that the  solution is invariant.

We can find a   solution similar to (\ref{freesol}) in the interacting case, for
$d>1$, which we henceforth assume. (The $d=1$ case will be separately
discussed in Section 3.) One verifies that a solution is
\begin{equation}
\theta(t,{\bf r}) = -\frac{{\bf r}^2}{2(d-1)t} \qquad \rho(t,{\bf r}) =
\sqrt{\frac{2\lambda}{d}}(d-1)\frac{|t|}{r} = \sqrt{\frac{2 \lambda}{d}} \ \frac{1}{v}
\label{disclat}
\end{equation}
\begin{equation}
{\bf v}(t,{\bf r}) =  -\frac{{\bf r}}{(d-1)t} \qquad {\bf j}(t,{\bf r}) = -\epsilon(t)
\sqrt{\frac{2\lambda}{d}} \bf \hat r.
\end{equation}
Note that the speed of sound, $s = \sqrt{2 \lambda} /\rho = \sqrt{d} v$,
exceeds~$v$.  Again this solution can be translated, rotated, and boosted. 
Moreover, the solution is time-rescaling--invariant.  However, the
space-time mixing transformation (\ref{eq:100})-(\ref{matrlink}) produces a wholly different solution. 
This is best shown graphically, where the $d=2$ case is exhibited (\emph{see
Figure}) \cite{AnDet}. 
\begin{figure}[hbtp]
\begin{gather*}
\BoxedEPSF{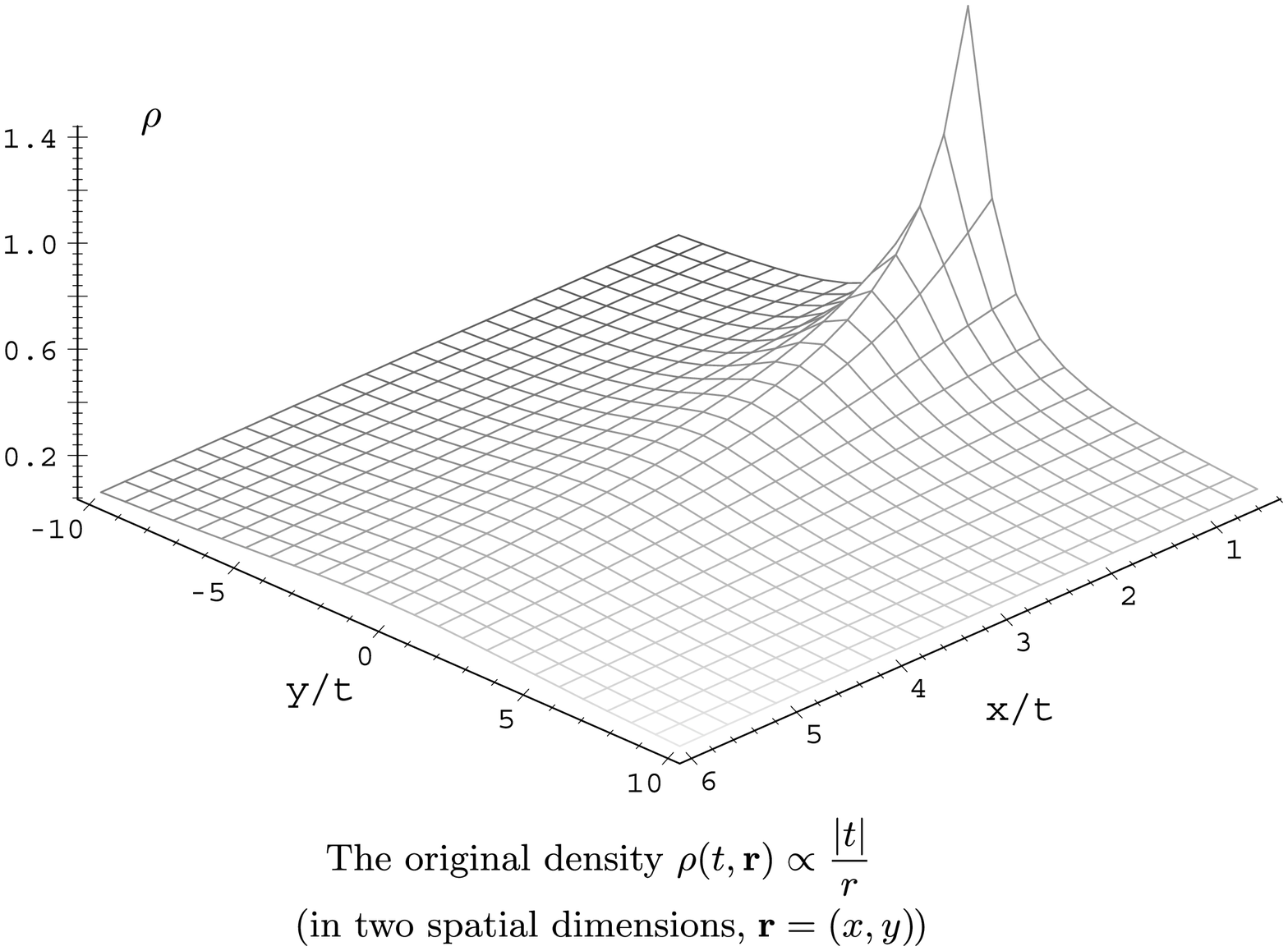 scaled 700}\\[1ex]
\BoxedEPSF{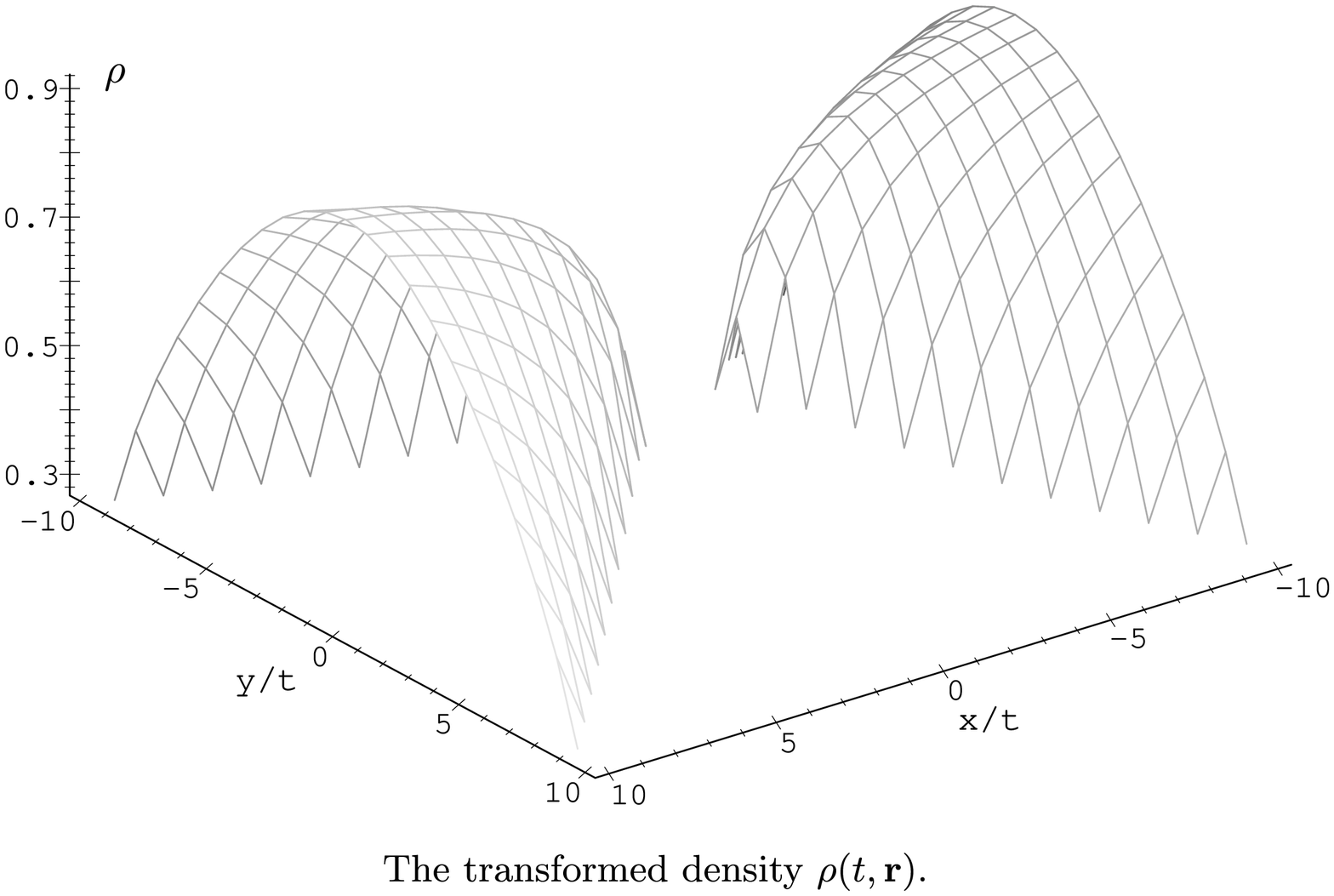 scaled 700}
\end{gather*}
\vspace*{-1pc}
\end{figure}

Another interesting solution, which is essentially one-dimensional (lineal), even
though it exists in arbitrary spatial dimension, is given by
\begin{eqnarray}
\theta(t,{\bf r}) &=& \Theta ({\bf \hat n}\cdot{\bf r}) + {\bf u} \cdot {\bf r} - \fract12
t\bigl({\bf u}^2 - ({\bf \hat n}\cdot {\bf u})^2\bigr). \nonumber \\
{\bf v} (t, {\bf r}) &=& {\bf u} + {\bf \hat n} \Theta' ({\bf \hat n} \cdot {\bf r})
\end{eqnarray}
Here ${\bf \hat n}$ is a spatial unit vector,  $\bf u$ is an arbitrary vector with
dimension of velocity, while $\Theta$ is an arbitrary function with static
argument. The
corresponding charge density is time-independent.
 \numeq{timeindep}{
\rho(t,{\bf r}) = \frac{\sqrt{2\lambda}}{{|\bf \hat n}\cdot{\bf u} + \Theta'({\bf \hat n}\cdot{\bf r})|} = \frac{2 \lambda}{|{\bf \hat
n} \cdot {\bf v}|}
 }
The current is static and divergenceless.
\numeq{curstat}{
{\bf j}(t,{\bf r}) = \sqrt{2\lambda} \Bigl(\frac{\bf v}{|{\bf \hat n} \cdot {\bf v}|}\Bigr)
}
The sound speed $s=\sqrt{2\lambda}/\rho = |{\bf \hat n} \cdot {\bf v}|$ is smaller than $v$.

Finally, we record a planar static solution to (\ref{interstren}), which depends on two orthogonal
unit vectors
${\bf \hat n}_1$, and ${\bf
\hat n}_2$ \cite{Oga}.
\begin{equation}
\theta(t, {\bf r}) = \Theta({\bf \hat n}_1\cdot{\bf r} / {\bf \hat n}_2\cdot{\bf r}). \nonumber
\label{solinterstren}
\end{equation}
\begin{equation}
{\bf v} (t, {\bf r}) = \frac{({\bf \hat n}_1 ({\bf \hat n}_2 {\bf r}) - {\bf \hat n}_2 ({\bf \hat n}_1\cdot {\bf r}))}{({\bf \hat n}_2 {\bf
r})^2} \ \Theta' ({\bf \hat n}_1 \cdot {\bf r}/{\bf \hat n}_2 \cdot {\bf r})
\end{equation}
This gives the density
\begin{equation}
\rho = \frac{\sqrt{2 \lambda}}{v}
\end{equation}
Now the sound speed coincides with $v$. \\
%\smallbreak

%\centerline{\hfill\hbox to 2in{\hrulefill} \hfill}
%\vspace*{-\smallskipamount}

%\begin{problem}\label{prob:6}
%\emph{
%Show that the solution for $\theta$ given in \refeq{freesol} is invariant under the
%time rescaling transformation \refeq{trt}, \refeq{trt2}, and under the
%space-time mixing transformation \refeq{eq:100}--\refeq{eq:102}.  }

%\centerline{\hfill\hbox to 2in{\hrulefill} \hfill}
%\end{problem}

\noindent {\bf (ii) Lorentz-invariant relativistic model: Born-Infeld model}
\addcontentsline{toc}{subsubsection}{(ii) Lorentz-invariant relativistic model: Born-Infeld model}\\

We now turn to a Lorentz-invariant generalization of our Galileo-invariant
Chaplygin model in
$(d,1)$-dimensional space-time.  We already know from
(1.2.73)--(1.2.78) how to construct the free Lagrangian with a relativistic kinetic energy.
\begin{equation} T({\bf v}) = -c^2\sqrt{1- {{\bf v}^2}/{c^2}}
\end{equation}
Mass has been scaled to unity, and we retain the velocity
of light $c$ to keep track of the nonrelativistic $c\rightarrow\infty$ limit. 
Evidently, the momentum is
\begin{equation}
{\bf p} = \frac{\partial T({\bf v})}{\partial {\bf v}}=\frac{{\bf v}}{\sqrt{1-{{\bf v}^2}/{c^2}}}.
\end{equation}
 Thus the free relativistic Lagrangian, with current conservation
enforced by the Lagrange multiplier~$\theta$, reads [compare~ (1.2.72), with Gaussian potentials, $\alpha$ and $\beta$
omitted]
\begin{equation} \bar L_0^{\mathrm{Lorentz}} = \int \rd r \Bigl(-c^2\rho
\sqrt{1- {{\bf v}^2}/{c^2}} +
\theta\bigl(\dot\rho+\grad\cdot({\bf v} \rho)\bigr)\Bigr).\label{laglor}
\end{equation}
 This may be presented in a Lorentz-covariant form in terms of a current
four-vector
$j^{\mu}=(c\rho, {\bf v}\rho)$.  $\bar L_0^{\mathrm{Lorentz}}$ of equation
(\ref{laglor}) is thus equivalent to [compare (\ref{pressvan})]
\begin{equation}
\bar L_0^{\mathrm{Lorentz}} = \int \rd r \Bigl(
  - j^\mu\partial_\mu \theta -c\sqrt{j^\mu j_\mu}\Bigr).
\label{Lorcov}
\end{equation}
 Eliminating ${\bf v}$ in (\ref{laglor}), we find that $\bf p$ is irrotational,
\begin{equation}
{\bf p} = \frac{\partial T}{\partial {\bf v}} = \frac{{\bf v}}{\sqrt{1-{\bf v}^2}/c^2}
= \grad \theta,
\qquad
{\bf v} =
\frac{\grad\theta}{\sqrt{1+ {(\grad\theta)^2}/{c^2}}},
\end{equation}
 and the free Lorentz-invariant Lagrangian reads 
\begin{equation}
 L_0^{\mathrm{Lorentz}}=\int\rd r \Bigl(\theta\dot\rho -\rho
c^2\sqrt{1+ {(\grad\theta)^2}/{c^2}}\, \Bigr).
\label{LiL}
\end{equation}

To find $L_0^{\mathrm{Galileo}}$ [(2.1.1) with $V =0$] as the nonrelativistic limit of
$L_0^{\mathrm{Lorentz}}$ in~\refeq{LiL}, a nonrelativistic~$\theta$ variable 
must be extracted from its relativistic counterpart. Calling the former
$\theta_{\mathrm{NR}}$ and the latter, which occurs in \refeq{LiL},
$\theta_{\mathrm{R}}$, we define
\numeq{thetaRNR}{
\theta_{\mathrm{R}} \equiv -c^2 t  + \theta_{\mathrm{NR}}.
}
It then follows that apart from a total time derivative
\numeq{tottimdder}{
 L_0^{\mathrm{Lorentz}} \mathop{\hbox to 3em{\rightarrowfill}}_{c\to\infty}
L_0^{\mathrm{Galileo}}.
}

 Next, one wants to include interactions.  While there are many ways to build
Lorentz-invariant interactions, we seek an expression that reduces to the
Chaplygin gas in the nonrelativistic limit.  Thus, we choose
\begin{equation}
 L_a^{\BI}=\int\rd r
\left(\theta\dot\rho - 
\sqrt{\rho^2 c^2+a^2}\sqrt{c^2+(\grad\theta)^2}\, \right),
\label{laglorint}
\end{equation}
 where $a$ is the interaction strength \cite{JacPol}.   (The reason for the nomenclature will
emerge presently.) We see from (2.1.4) that, as
$c\to\infty$,
\begin{equation} L_a^{\mathrm{\BI}}  \mathop{\hbox to
3em{\rightarrowfill}}_{c\to\infty} L_{\lambda= {a^2}/{2}}^{\mathrm{Chaplygin}}.
\end{equation}
[Again $\theta_{\mathrm{NR}}$ is extracted from $\theta_{\mathrm{R}}$ as in
\refeq{thetaRNR} and a total time derivative is ignored.]

Although it perhaps is not obvious, (\ref{laglorint}) defines a
Poincar\'e-invariant theory, and this will be explicitly demonstrated below. 
Therefore, $L_a^{\mathrm{\BI}}$ possesses Poincar\'e
symmetries in
$(d,1)$ space-time, with a total of $\fract{1}{2}(d+1)(d+2)+1$ generators, where
the last ``${}+1$''  refers to the total number~$N = \int \rd r \rho$.  

When $a=0$, the
model is free and elementary. It was demonstrated previously
[eqs. (1.2.73)--(1.2.81)] that the free equations of motion are
precisely the same as in the nonrelativistic free model, so the complete solution
(1.2.84)--(1.2.86)
works here as well.  For
$a\ne 0$, in the presence of interactions, one can eliminate
$\rho$ as before, and one is left with a Lagrangian just for the $\theta$ field.  It
reads
\begin{equation} L_a^{\mbox{\scriptsize Born-Infeld}}=-a\int\rd r
\sqrt{c^2-(\partial_\mu
\theta)^2}.
\label{B-I}
\end{equation}
 This is a Born-Infeld-type theory for a scalar field $\theta$; its Poincar\'e
invariance is manifest, and again, the elimination of $\rho$ is only possible
with nonvanishing $a$, which however disappears from the dynamics,
serving merely to normalize the Lagrangian. 

Although manifestly Lorentz covariant, the Lagrangian (\ref{B-I}) is not of the form (1.4.1). To achieve that
expression, we choose $f (n) = c \sqrt{a^2 + n^2}$, so that the pressure  $P$ becomes
\begin{equation}
P = -\frac{a^2 c}{\sqrt{a^2 + n^2}}. 
\label{pressl}
\end{equation}
This leads to the Chaplygin pressure as $c \to \infty$ with $\lambda = a^2 /2$. Now the Lagrange
density $\mathcal{L} = P$ coincides with (1.4.1 ) when $j_\mu$ is written as 
\begin{eqnarray}
j_\mu &=& \frac{a \partial_\mu \theta}{\sqrt{c^2 - (\partial_\mu \theta)^2}}, \nonumber\\
n &=& a\sqrt{\frac{(\partial_\mu \theta)^2}{c^2- (\partial_\mu \theta)^2}}.
\end{eqnarray}

The equations of motion that follow from~\refeq{laglorint} read
\begin{align}  
\dot\rho + \grad \cdot \Biggl(
 \grad\theta  \sqrt{\frac{\rho^2
c^2+a^2}{c^2+(\grad\theta)^2}}\,\Biggr) &= 0,
\label{followsfrom1}\\[1ex]
\dot\theta + \rho c^2 \sqrt{\frac{c^2+(\grad\theta)^2}{\rho^2
c^2+a^2}} &=0\ . \label{followsfrom2}\\[1ex]
\intertext{The density $\rho$ can be evaluated in terms of $\theta$ from
\refeq{followsfrom2}; then \refeq{followsfrom1} becomes}
\partial^\alpha \Bigl( \frac1{\sqrt{c^2-(\partial_\mu \theta)^2}}\, 
\partial_\alpha \theta\Bigr) &= 0, \label{followsfrom}
\end{align}
 which also follows from
\refeq{B-I}.  After $\theta_{\mathrm{NR}}$ is extracted from
$\theta_{\mathrm{R}}$ as in
\refeq{thetaRNR}, we see that in the
nonrelativistic limit $L_a^{\mbox{\scriptsize Born-Infeld}}$ \refeq{laglorint} or
\refeq{B-I} becomes $L_\lambda^{\mathrm{Chaplygin}}$
of~\refeq{chapgasact} or \refeq{rindependent},
\numeq{B-ItoC}{
L_a^{\mbox{\scriptsize Born-Infeld}}  \mathop{\hbox to
3em{\rightarrowfill}}_{c\to\infty} L_{\lambda=a^2/2}^{\mathrm{Chaplygin}},
 }
and the equations of motion \refeq{followsfrom1}--\refeq{followsfrom} reduce  to \refeq{conseqBern} and
\refeq{interstren}.

In view of all the similarities to the nonrelativistic Chaplygin gas, it comes as
no surprise that the relativistic Born-Infeld  theory possesses additional
symmetries. These additional symmetry transformations, which leave
(\ref{laglorint}) or (\ref{B-I}) invariant, involve a one-parameter ($\omega$, dimesionless) 
 reparameterization of time, and a $d$-parameter ($\vec
\omega$, dimension velocity) vectorial reparameterization of space. Both transformations are
field dependent \cite{ref20}.

The time transformation is given by an implicit formula involving also the
field~$\theta$, 
\begin{equation}
 t \to T(t, {\bf r}) = \frac{t}{\cosh \omega} -
\frac{\theta(T, {\bf r})}{c^2} \tanh
\omega,
\end{equation}  
while the field transforms according to
\begin{equation}
\theta(t, {\bf r}) \to \theta_\omega (t, {\bf r})= \frac{\theta(T,
{\bf r})}{\cosh \omega} + c^2 t \tanh  \omega\ .
\label{eq:135}
\end{equation}  
[We record here only the transformation on~$\theta$; how~$\rho$ transforms can
be determined from the (relativistic) Bernoulli equation, (\ref{followsfrom2}), which expresses~$\rho$ in terms
of~$\theta$. Moreover, \refeq{eq:135} is sufficient for discussing the invariance of
\refeq{B-I}.] The infinitesimal generator, which is time independent by virtue of
the equations of motion, is \cite{BorHop94}
\begin{align}
 S &= \int \rd r \Big(c^4 t \rho + \theta
\sqrt{\rho^2c^2+a^2} \sqrt{c^2 + (\grad \theta)^2}
\Big),\nonumber\\
&= \int \rd r (c^4 t \rho + \theta {\cal H}) \qquad\mbox{(time
reparameterization).}\label{infgener}
\end{align} 

A second class of transformations involving a reparameterization
of the spatial variables is implicitly defined by
\begin{equation}
{\bf r} \to {\bf R} (t, {\bf r}) = {\bf r} + \frac{\vec \omega}{c^2} \theta (t, {\bf R}) + \hat{\boldsymbol \omega} (\hat{\boldsymbol
\omega} \cdot {\bf r}) \Bigg(\sqrt{1 + \frac{{\boldsymbol \omega}^2}{c^2}} \
-1\Bigg),
\end{equation} 
\begin{equation}
\theta (t, {\bf r})  \to \theta_{\vec\omega} (t, {\bf r}) =
\sqrt{1 +\frac{{\boldsymbol \omega}^2}{c^2}} \ \theta (t, {\bf R}) + {\vec
\omega} \cdot {\bf r}.
\end{equation}  
The time-independent generator of the
infinitesimal transformation reads \cite{BorHop94}
\begin{align}
{\bf G} &=\int \rd r (c^2  {\bf r} \rho + \theta \rho
\grad\theta),\nonumber\\
&= 
\int
\rd r (c^2
 {\bf r} \rho +\theta {\bf \mathcal{P}})\qquad\mbox{(space re\pr).}
\label{eq:88}
\end{align} 
Of course the Born-Infeld action \refeq{laglorint} or \refeq{B-I} is invariant
against these transformations, whose infinitesimal form is generated by the
constants.

With the addition of $S$ and $\vec G$ to the previous
generators, the Poincar\'e algebra in $(d+1,1)$ dimension is
reconstructed, and $(t, {\bf r},\theta)$ transforms linearly as a
$(d+2)$-dimensional Lorentz vector (in Cartesian
components).  Note that this symmetry also
holds in the free, $a=0$, theory.

It is easy to exhibit solutions of the relativistic equation \refeq{followsfrom},
which reduce to solutions of the nonrelativistic, Chaplygin gas equation
\refeq{interstren} [after
$-c^2t$ has been removed, as in \refeq{thetaRNR}].   For example 
\numeq{nonrelChap}{
\theta(t,{\bf r}) = -c \sqrt{c^2t^2 + \frac{{\bf r}^2}{d-1}}
}
solves \refeq{followsfrom} and reduces to \refeq{disclat}.  The relativistic analog
of the lineal solution (2.1.23) is 
\numeq{relanal}{
\theta(t,{\bf r}) = \Theta({\bf \hat n}\cdot{\bf r}) + {\bf u}\cdot{\bf r} - ct
\sqrt{c^2 + {\bf u}^2 -({\bf \hat n} \cdot {\bf u})^2} .
}
 [Note that the above profiles continue to solve \refeq{followsfrom} even when
the sign of the square root is reversed, but then they no longer possess a 
nonrelativistic limit.] 

Additionally there exists an essentially relativistic solution, describing massless
propagation in one direction: according to \refeq{followsfrom}, $\theta$ can satisfy
the wave equation
$\sqcap\llap{$\sqcup$} \theta = 0$, provided
$(\partial_\mu \theta)^2 = $~constant, as for example with plane waves,
\numeq{withplanew}{
\theta(t,{\bf r}) = f({\bf \hat n}\cdot{\bf r} \pm ct),
}
 where $(\partial_\mu \theta)^2$ vanishes.  Then $\rho$ reads, from
\refeq{followsfrom2},
\numeq{eq:143}{
\rho = \mp \frac a{c^2} f' \ . 
}

\newpage
\subsection{Common ancestry: the Nambu-Goto action}
\setcounter{equation}{0}
The ``hidden'' symmetries and the associated transformation laws for the
Chaplygin and Born-Infeld models may be given a coherent setting by considering
the Nambu-Goto action for a p-brane in $(p +1)$ spatial dimensions, moving on
$(p +1,1)$-dimensional space-time.  In our context, a~p-brane is simply a
$p$-dimensional extended object: a $1$-brane is a string, a $2$-brane is a
membrane and so on.  A~p-brane in $(p +1)$ space divides that space in two.

The Nambu-Goto action reads
\begin{align}
I_{\rm NG} &= \int\rd{ \phi^0} \rd\phi {\cal L}_{\rm NG} = 
\int \rd{ \phi^0} \rd{
\phi^1}
\cdots \rd{
\phi^p}
\sqrt{G},\label{N-Gr1}\\
 G &= (-1)^p \det \frac{\partial X^\mu}{\partial \phi^\alpha}
\frac{\partial X_\mu}{\partial \phi^\beta}.
\end{align}
Here $X^\mu$ is a $(p+1,1)$ ``target space-time" (p-brane) variable, with $\mu$
extending over the range $\mu=0,1,\dots,p,p+1$.  The $\phi^\alpha$ are
``world-volume'' variables describing the extended object with $\alpha$ ranging
$\alpha=0,1,\dots,p$; $\phi^r$,
$r=1, \dots, p$, parameterizes the $p$-dimensional p-brane that evolves
in $\phi^0$.

The Nambu-Goto action is parameterization invariant, and we shall show that
two different choices of parameterization (``light-cone'' and ``Cartesian'') lead to
the Chaplygin gas and Born-Infeld actions, respectively.  For both
parameterizations we choose
$(X^1, \dots, X^p)$ to coincide with $(\phi^1,
\dots, \phi^p)$, renaming them as ${\bf r}$ (a $p$-dimensional vector). This is
usually called the ``static \pr''. (The ability to carry out this parameterization
globally  presupposes that the extended object is topologically trivial; in the
contrary situation, singularities will appear, which are spurious in the sense that
they disappear in different parameterizations, and parameterization-invariant
quantities are singularity-free.)\\

\noindent {\bf (i) Light-cone parameterization}\\ [-18pt]
\addcontentsline{toc}{subsubsection}{(i) Light-cone parameterization}

For the light-cone parameterization we define $X^\pm$ as $\frac{1}{\sqrt{2}}
(X^0 \pm X^{p+1})$.  $X^+$ is renamed $t$ and identified with $\sqrt{2\lambda}\,
\phi^0$.  This completes the fixing of the parameterization and the remaining variable
is $X^-$, which is a function of $\phi^0$ and $\vec\phi$, or after redefinitions, of $t$ and
${\bf r}$.  $X^-$ is renamed as
 $\theta(t,{\bf r})$ and then the Nambu-Goto action in this
parameterization coincides with the Chaplygin gas action 
$I^{\mathrm{Chaplygin}}_\lambda$ in \refeq{rindependent} \cite{Gold}. \vspace{12pt}

\noindent {\bf (ii) Cartesian parameterization}\\[-18pt]
\addcontentsline{toc}{subsubsection}{(ii) Cartesian parameterization}

For the second, Cartesian parameterization $X^0$ is
renamed $ct$ and identified with $c \phi^0$.  The remaining target space
variable $X^{p+1}$, a function of $\phi^0$ and $\vec\phi$, equivalently of $t$
and ${\bf r}$, is renamed
$\theta(t,{\bf r})/c$.  Then the Nambu-Goto action reduces
to the Born-Infeld action $\int \rd t L_a^{\BI}$, (\ref{B-I}) \cite{Gold}.\\

\noindent{\bf (iii) Hodographic transformation}
\addcontentsline{toc}{subsubsection}{(iii) Hodographic transformation}\\

There is another derivation of the Chaplygin gas from the Nambu-Goto action
that makes use of a hodographic transformation, in which independent and
dependent variables are interchanged.  Although the derivation is more
involved than the light-cone/static \pr\ used in Section 2.2(i) above,
the hodographic approach is instructive in that it gives a natural definition for
the density~$\rho$, which in the above static \pr\ approach is determined
from~$\theta$ by the Bernoulli equation~\refeq{conseqBern}.

We again use light-cone combinations: $\frac{1}{\sqrt{2}}
(X^0  + X^{p+1})$ is called $\tau$ and is identified with $\phi^0$, while 
$\frac{1}{\sqrt{2}} (X^0 - X^{p+1})$ is renamed~$\theta$. At this stage the
dependent, target-space variables are~$\theta$ and the transverse coordinates
${\bf X}\colon X^i$, $(i=1,\ldots,p)$, and all are functions of the
world-volume parameters $\phi^0=\tau$ and
$\vec\phi\colon \phi^r$, $(r=1,\ldots,p)$; $\partial_\tau$ indicates differentiation
with respect to $\tau=\phi^0$, while $\partial_r$ denotes derivatives with
respect to $\phi^r$. The induced metric
$G_{\alpha\beta} = \frac{\partial X^\mu}{\partial\phi^\alpha} \frac{\partial
X_\mu}{\partial\phi^\beta}$ takes the form
\numeq{inducmetr}{
G_{\alpha\beta} = \begin{pmatrix}
G_{oo} & G_{os}\\
G_{ro} & -g_{rs}
\end{pmatrix} = 
\begin{pmatrix}
2\partial_\tau\theta - (\partial_\tau {\bf X})^2& \partial_s\theta -
\partial_\tau {\bf X} \cdot \partial_s {\bf X}\\ 
\partial_r\theta -
\partial_r {\bf X} \cdot \partial_\tau {\bf X}&
-\partial_r {\bf X} \cdot \partial_s {\bf X}
\end{pmatrix}.
}
The Nambu-Goto Lagrangian now leads to the canonical momenta
\begin{align}
\frac{\partial {\cal L}_{\mathrm NG}}{\partial {\partial_\tau} {\bf X}} &= {\bf p},\label{NGLa}\\[1ex]
\frac{\partial {\cal L}_{\mathrm NG}}{\partial {\partial_\tau}  \theta} &=
\Pi,\label{NGLb}
\end{align}
and can be presented in first-order form as 
\numeq{foform}{
 {\cal L}_{\mathrm NG} = {\bf p}\cdot \partial_\tau {\bf X} +
\Pi\partial_\tau\theta + \frac1{2\Pi} ({\bf p}^2+g) +
u^r ({\bf p} \cdot \partial_r {\bf X} + \Pi\partial_r \theta),
}
where $g=\det g_{rs}$ and 
\numeq{Lagmult}{
u_r\equiv \partial_\tau {\bf X} \cdot \partial_r {\bf X} - \partial_r \theta 
}
acts as a Lagrange multiplier. Evidently the equations of motion are
\begin{subequations}\label{eveqmot}
\begin{align}
\partial_\tau {\bf X} &= - \frac1\Pi {\bf p} - u^r \partial_r 
{\bf X},\label{eveqmota}\\
\partial_\tau\theta &= \frac1{2\Pi^2} ({\bf p}^2+g) - u^r
\partial_r\theta,\label{eveqmotb}\\
\partial_\tau{\bf p} &= -\partial_r \Bigl(\frac1\Pi g g^{rs}\partial_s{\bf X}\Bigr)
- \partial_r(u^r{\bf p}),\label{eveqmotc}\\
\partial_\tau \Pi &= -\partial_r ( u^r \Pi).\label{eveqmotd}
\end{align}
\end{subequations}
Also there is the constraint
\numeq{alsoconstr}{
{\bf p}\cdot\partial_r {\bf X}  + \Pi \partial_r \theta = 0.
}
[That $u^r$ is still given by \refeq{Lagmult} is a consequence of
\refeq{eveqmota} and \refeq{alsoconstr}.] Here $g^{rs}$ is inverse to~$g_{rs}$, and
the two metrics are used to move the $(r,s)$ indices.  The theory still possesses an
invariance against redefining the spatial parameters with a $\tau$-dependent
function of the parameters; infinitesimally: $\delta\phi^r= -f^r(\tau, \vec\phi)$,
$\delta\theta = f^r\partial_r \theta$, $\delta X^i = f^r \partial_r X^i$. This freedom
may be used to set $u^r$ to zero and $\Pi$ to~$-1$. 

Next the hodographic
transformation is performed: Rather than viewing the dependent variables
${\bf p}$, $\theta$, and ${\bf X}$ as functions of $\tau$ and $\vec \phi$,
${\bf X}(\tau,\vec \phi)$ is inverted so that $\vec \phi$ becomes a function of
$\tau$ and ${\bf X}$ (renamed $t$ and ${\bf r}$, respectively), and ${\bf p}$ and
$\theta$ also become functions of~$t$  and~${\bf r}$. It then
follows from the chain rule that the constraint~\refeq{alsoconstr} (at $\Pi=-1$)
becomes
\numeq{constbec}{
0=\frac{\partial X^i}{\partial \phi^r} \Bigl(p^i -\frac\partial{\partial
X^i}\theta\Bigr),
 }
and is solved by 
\numeq{constbecsolv}{
{\bf p} = \grad \theta\ .
}
Moreover, according to the chain rule and the implicit function theorem, the partial
derivative with respect to~$\tau$ at fixed $\vec\phi$ [this derivative is present in
\refeq{foform}] is related to the partial derivative with respect to~$\tau$ at
fixed ${\bf X}={\bf r}$ by
\numeq{relpartder}{
\partial_\tau =\partial_t + \grad\theta\cdot\grad,
}
where we have used the new name ``$t$'' on the right. Thus the Nambu-Goto
Lagrangian -- the $\phi$ integral of the Lagrange density \refeq{foform}
(at $u^r=0$, $\Pi=-1$) -- reads
\begin{subequations}\label{NGLII}
\numeq{NGLIIa}
{
L_{\mathrm NG} = \int \rd\phi \Bigl(
{\bf p}\cdot\grad\theta - \dot\theta - \grad\theta\cdot\grad \theta -
\fract12 ({\bf p}^2 + g)
\Bigr).
}
But use of \refeq{constbecsolv} and of the Jacobian relation $\rd\phi =\rd r
\det \frac{\partial\phi^s}{\partial X^i} = \frac{\rd r}{\sqrt g}$ shows that
\numeq{NGLIIb}{
L_{\mathrm NG} = \int \rd r \Bigl(
-\frac1{\sqrt g}\dot\theta - \frac1{2\sqrt g}(\grad \theta)^2 - \fract12
\sqrt g
\Bigr).
}
With the definition
\numeq{withdef}{
\sqrt g = \sqrt{2\lambda}/\rho,
}
$L_{\mathrm NG}$ becomes, apart from a total time derivative
\numeq{TotTimder}{
L_{\mathrm NG} =\fract1{\sqrt{2\lambda}} \int \rd r \Bigl(
\theta \dot\rho -  \rho\frac{(\grad \theta)^2}{2}  - \frac\lambda\rho
\Bigr).
}
Up to an overall factor, this is just the Chaplygin gas Lagrangian in 
\refeq{chapgasact}.

The present derivation has the advantage of relating the density~$\rho$ to the
Jacobian of the ${\bf X} \to \vec \phi$ transformation: $\rho = \sqrt{2\lambda}
\det \frac{\partial\phi^s}{\partial X^i}$. (This in turn shows that the
hodographic transformation is just exactly the passage from Lagrangian to
Eulerian fluid variables.)
\end{subequations}

\subsection{Interrelations}\label{interrel}
\setcounter{equation}{0}
The relation to the Nambu-Goto action explains the origin of the
hidden $(p+1,1)$ Poincar\'e group in our two nonlinear models on
$(p,1)$ space-time: the ($p+1, 1$)Poincar\'e invariance is what remains of the
reparameterization invariance of the Nambu-Goto action after choosing
either the light-cone or Cartesian parameterizations.  (In this context, recall that the light-cone
subgroup of the Poinca\'{r}e group is isomorphic to the Galileo group in one lower dimension
\cite{sussk}) Also the nonlinear, field dependent form of the transformation laws leading to the
additional symmetries is understood: it arises from the identification of some of the dependent
variables ($X^\mu$) with the independent variables $(\phi^\alpha)$.

The complete integrability of the $d=1$ Chaplygin gas and Born-Infeld
model is a consequence of the fact that both descend from a string in
2-space; the associated Nambu-Goto theory being completely
integrable. We shall discuss this in Section 3.  

We observe that in addition to the nonrelativistic descent from the
Born-Infeld theory to the Chaplygin gas, there exists a mapping of one
system on another, and between solutions of one system and the other,
because both have the same p-brane ancestor.  The mapping is achieved
by passing from the light-cone parameterization to the Cartesian, or
vice-versa.  Specifically this is accomplished as follows.\\

\noindent{\bf (i) Chaplygin gas \boldmath$\to$ Born-Infeld:}
\addcontentsline{toc}{subsubsection}{(i) Chaplygin gas \boldmath$\to$ Born-Infeld:}\\
  
Given $\theta_{NR} (t,{\bf r})$, a
nonrelativistic solution, determine $T(t,{\bf r})$ from the equation
\numeq{eq:172}{
T+\frac{1}{c^2} \theta_{NR} (T,{\bf r}) = \sqrt{2} \, t.
} 
Then the relativistic solution is
\numeq{relatsol}{
\theta_R(t,{\bf r})=\frac{1}{\sqrt{2}} c^2 T - \frac{1}{\sqrt{2}}
\theta_{NR} (T,{\bf r}) = c^2 (\sqrt2T - t).
} \\

\noindent{\bf (ii) Born-Infeld \boldmath$\to$ Chaplygin gas:}\\
\addcontentsline{toc}{subsubsection}{(ii) Born-Infeld \boldmath$\to$ Chaplygin gas:}
Given $\theta_R(t,{\bf r})$, a relativistic solution, find $T(t,{\bf r})$ from
\begin{equation}
T+\frac{1}{c^2} \theta_{R} (T,{\bf r}) = \sqrt{2} \, t.
\end{equation} 
Then the nonrelativistic solution is
\numeq{nonrelsol}{
\theta_{NR}(t,{\bf r})=\frac{1}{\sqrt{2}} c^2 T - \frac{1}{\sqrt{2}}
\theta_{R} (T,{\bf r}) = c^2 (\sqrt2T - t).
} 

\begin{figure}
\begin{center}
% \leavevmode
% \epsfysize=150pt{\epsffile{jackiwfig.eps}}
%\fbox{Some figure goes here}
\BoxedEPSF{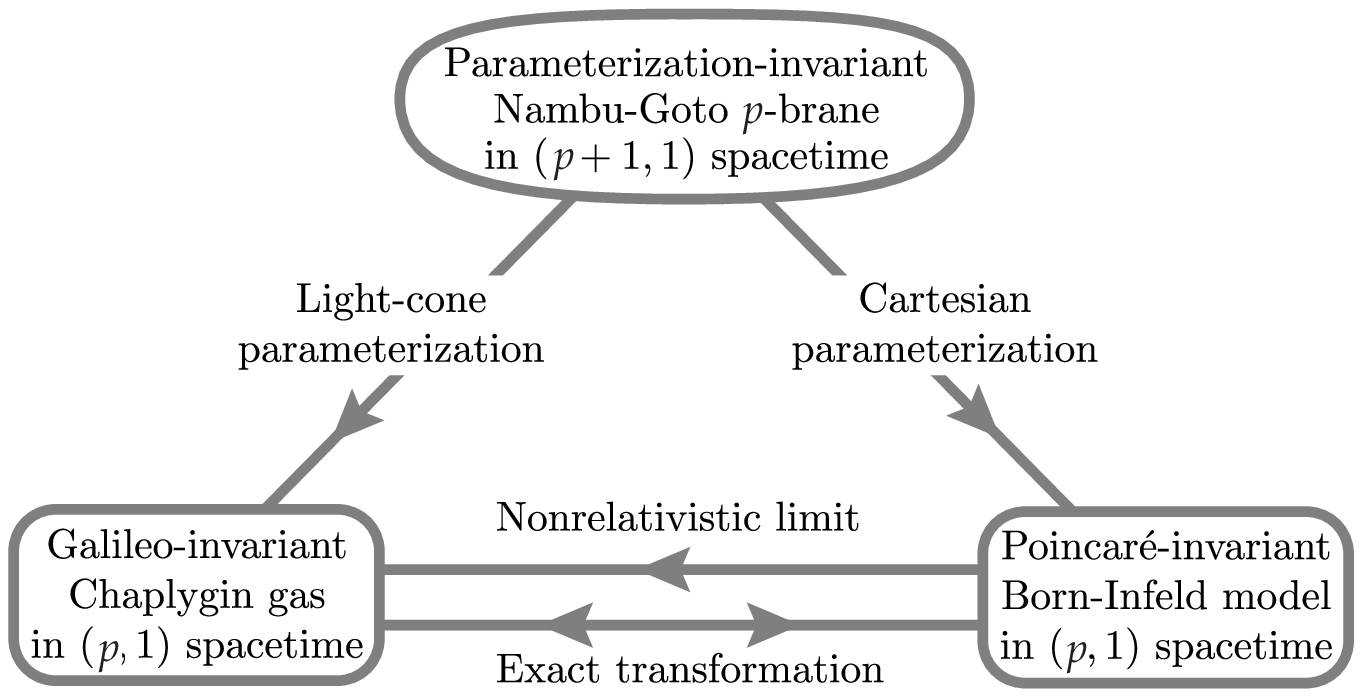}
 \end{center}
\centerline{Dualities and other relations between nonlinear
equations.}
\end{figure}

The relation between the different models is depicted in the {\it Figure} below. 

A final comment:  Recall that the elimination of~$\rho$, both in the
nonrelativistic (Chaplygin) and relativistic (Born-Infeld) models is possible only in
the presence of interactions. Nevertheless, the $\theta$-dependent
($\rho$-independent) resultant Lagrangians contain the interaction strengths
only as overall factors; see \refeq{rindependent} and \refeq{B-I}. It is these
$\theta$-valued Lagrangians that correspond to the Nambu-Goto action in
various parameterizations. Let us further recall the the Nambu-Goto action also
carries an overall multiplicative factor: the p-brane ``tension'', which has been
suppressed in~\refeq{N-Gr1}. Correspondingly, for a ``tensionless'' p-brane, the
Nambu-Goto expression vanishes, and cannot generate dynamics. This suggests
that an action for ``tensionless'' p-branes could be the noninteracting fluid
mechanical expressions \refeq{chapgasact}, \refeq{laglorint}, with vanishing
coupling strengths
$\lambda$ and $a$, respectively. Furthermore, we recall that the noninteracting
models retain the higher, dynamical symmetries, appropriate to a p-brane in
one higher dimension.

\newpage

\section{SPECIFIC MODELS ($d=1$)}\label{sec:6}

In this Section, we shall discuss nonrelativistic/relativistic models in one
spatial dimension. Complete integrability has  been established for both the
Chaplygin gas \cite{Nutku} and the Born-Infeld theory \cite{BarChe}. We can now
understand this to be a consequence of the complete integrability of the
Nambu-Goto 1-brane (string) moving on 2-space (plane), which is the antecedent
of both models. [Therefore, it suffices to discuss only the Chaplygin gas, since
solutions of the Born-Infeld model can then be obtained by the mapping
\refeq{eq:172}--\refeq{relatsol}.] 

As remarked previously, in one dimension there is no vorticity, and the
nonrelativistic velocity~$v$ can be presented as a derivative with respect to the
single spatial variable of a potential~$\theta$. 
Similarly, the relativistic momentum $p=v/\sqrt{1-v^2/c^2}$ is a derivative of
a potential~$\theta$.   In both cases the potential is canonically conjugate to the
density
$\rho $ governed by the canonical 1-form  $\int \rd x \theta\dot\rho $.
Moreover, it is evident that at the expense of a spatial nonlocality, one may
replace $\theta$ by its antiderivative, which is $p$ both nonrelativistically and
relativistically (nonrelativistically $p = v $), so that in both cases the Lagrangian
reads 
\numeq{repltheta}{
L = -\fract12 \int \rd x \rd y \rho(x) \eps(x-y)\dot{p}(y) - H\ .
} 
For the Chaplygin gas and the Born-Infeld models, $H$ is given respectively by 
\begin{align}
H^{\mathrm{Chaplygin}} &= \int \rd x \Bigl( \fract12 \rho {p}^2 +
\frac\lambda\rho\Bigr) \label{ChapH}\\
H^{\BI} &= \int \rd x \bigl( \sqrt{\rho^2c^2+ a^2} \sqrt{c^2+p^2}\bigr)\ .
\label{ChapBI}
\end{align}
The equations of motion are, respectively
\begin{align}
&\mbox{Chaplygin gas:} & \dot\rho + \frac\partial{\partial x}(p\rho)
&=0,\label{eqmot234}\\
& & \dot p + \frac\partial{\partial x}\Bigl( \frac{p^2}2 -
\frac\lambda{\rho^2}\Bigr) &=0,\label{eqmot235}\\
& &\mbox{or}\quad  \frac\partial{\partial t} \frac1{\sqrt{\dot\theta +
\fract{p^2}2}} +
\frac\partial{\partial x} \frac p{\sqrt{\dot\theta +
\fract{p^2}2}} &=0. \label{eqmot236} \\[3ex]
&\mbox{Born-Infeld model:} &  \dot\rho + \frac\partial{\partial
x}\biggl(p \sqrt{\frac{\rho^2c^2 + a^2}{c^2+p^2}}\biggr) &=0,\label{eqmot237}\\
& & \dot p + \frac\partial{\partial x}\biggl(\rho c^2 \sqrt{\frac{c^2 + p^2}{\rho^2
c^2 + a^2}}\biggr) &=0,\label{eqmot238}\\
& &\llap{$\displaystyle\mbox{or}\quad  \frac\partial{c^2\partial t} \biggl(
\frac{\dot\theta}{\sqrt{c^2-\fract1{c^2} \dot\theta^2 + p^2}}\biggr)$} 
- \frac\partial{\partial x} \biggl(\frac{p}{\sqrt{c^2-\fract1{c^2}
\dot\theta^2 + p^2}}\biggr) &=0. \label{eqmot239}
\end{align}
In the above, eqs.~\refeq{eqmot236} and \refeq{eqmot239} result by determining
$\rho$ in terms of $\theta$ ($p=\theta^\prime$; dash indicates differentation with respect to spatial
argument) from
\refeq{eqmot235} and \refeq{eqmot238}, and using that expression for $\rho$ in
\refeq{eqmot234} and \refeq{eqmot237}.

\subsection{Chaplygin gas on a line}

\setcounter{equation}{0}
{\bf (i) Specific solutions}
\addcontentsline{toc}{subsubsection}{(i) Specific solutions}\\

Classes of solutions for a Chaplygin gas in one dimension can be given in closed
form. For example, to obtain general, time-rescaling--invariant solutions, we make
the \emph{Ansatz} that $\theta\propto 1/t$. Then \refeq{interstren} or
\refeq{eqmot236} leads to a second-order nonlinear differential equation for the
$x$-dependence of~$\theta$. Therefore solutions involve two arbitrary constants,
one of which fixes the origin of~$x$ (we suppress it);  the other we call $k$, and
take it to be real. The solutions then read
\numeq{solthenread}{
\theta(t,x) = - \frac1{2k^2 t} \cosh^2 kx\ .
}
[Other solutions can be obtained by relaxing the reality condition on $k$ and/or
shifting the argument $kx$ by a complex number. In this way one  finds that
$\theta$ can also be $\frac1{2k^2 t} \sinh^2 kx$, $\frac1{2k^2 t} \sin^2 kx$,
$\frac1{2k^2 t} \cos^2 kx$; but these lead to singular or unphysical forms
for~$\rho$.] The density corresponding to \refeq{solthenread} is found from
\refeq{conseqBern} or \refeq{eqmot235} to be 
\numeq{densecorresp}{
\rho(t,x) = \sqrt{2\lambda} \frac{k\left|t\right|}{\cosh^2 kx}\ . 
}
The velocity/momentum $v=p=\theta^\prime$ is 
\numeq{velmom}{
v(t,x) = p(t,x) = -\frac1{kt} \sinh kx \cosh kx,
}
while the sound speed
\numeq{sounspee}{
s(t,x) = \frac{\cosh^2 k x}{k\left|t\right|},
} 
is always larger than $\left| v\right|$. Finally, the current $j=  \rho
\frac{\partial\theta}{\partial x}$ exhibits a kink profile, 
\numeq{kinkprof}{
 j(t,x) = -\eps(t) \sqrt{2\lambda} \tanh kx,
}
which is suggestive of complete integrability. 

Another particular  solution is the Galileo boost of the static profiles 
(2.1.24), (2.1.25):
\begin{align}
p(t,x) &= p(x-ut)\label{galboost1},\\
\rho(t,x) &= \frac{\sqrt{2\lambda}}{\left|p-u\right|}\ .
\end{align}
Here $u$ is the boosting velocity and $p(x-ut)$ is an arbitrary function of its
argument (provided $p\neq u$). Clearly this is a constant profile
solution, in linear motion with velocity~$u$ \cite{grndland}. 

Further evidence  for complete integrability is found by identifying an infinite
number of constants of motion. One verifies that the following quantities
\numeq{infnumconsmo}{
I_n^{\pm} = \int \rd x \rho \Bigl(p \pm \frac{\sqrt{2\lambda}}\rho \Bigr)^n\ ,
\quad n=0, \pm1, \ldots
 }
are conserved. 

The combinations $p \pm \frac{\sqrt{2\lambda}}\rho $ are just the velocity
$(\pm)$ the sound speed, and they are known as Riemann coordinates.
\numeq{riemcoor}{
R_{\pm} = p \pm \frac{\sqrt{2\lambda}}\rho  
}
The equations of motion for this system [continuity \refeq{eqmot234} and Euler 
 \refeq{eqmot235}] can be succinctly presented in terms
of $R_{\pm}$: 
\numeq{succpres}{
\dot R_{\pm} = - R_{\mp} R^\prime_{\pm}\ .
}\\

\noindent {\bf (ii) General solution for the Chaplygin gas on a line}
\addcontentsline{toc}{subsubsection}{(ii) General solution for the Chaplygin gas on a line}\\

The general solution to the Chaplygin gas can be found by linearizing the
governing equations (continuity and Euler) with the help of a Legendre
transform, which also effects a hodographic transformation that exchanges the
independent variables $(t, x)$ with the dependent ones $(\rho, \theta)$;
actually instead of
$\rho$ we use the sound speed $s = \sqrt{2\lambda}/\rho$ and instead of
$\theta$ we use the momentum $p = \theta^\prime$.  

Define
\numeq{eq:245}{
\psi(p,s) = \theta(t,x) - t\dot\theta(t,x) - x \theta^\prime(t,x)\ . 
}
From the Bernoulli equation we know that 
\numeq{eq:246}{
\dot\theta = -\fract12 p^2 + \fract12 s^2\ .
}
Thus 
\numeq{eq:247}{
\psi(p,s) = \theta(t,x) + \frac t2 (p^2-s^2) - xp,
}
and the usual Legendre transform rules govern the derivatives.
\begin{subequations}\label{eq:248}
\begin{align}
\frac{\partial\psi}{\partial p} &= tp -x \label{eq:248a}\\
\frac{\partial\psi}{\partial s} &= -ts \label{eq:248b}
\end{align}
\end{subequations}
 It remains to incorporate the continuity equation
\refeq{eqmot234} whose content must be recast by the
hodographic transformation. This is achieved by rewriting  equation
\refeq{eqmot234} in terms of $s = \sqrt{2\lambda}/\rho$
\numeq{eq:249}{
\frac{\partial s}{\partial t} + p \frac{\partial s}{\partial x} - s
\frac{\partial p}{\partial x} = 0\ 
}
Next \refeq{eq:249} is presented as a relation between Jacobians
\begin{subequations}\label{eq:250}
\numeq{eq:250a}{
\frac{\partial (s,x)}{\partial (t,x)} + p \frac{\partial (t,s)}{\partial (t,x)} - s
\frac{\partial (t,p)}{\partial (t,x)} = 0,
}
which is true because here $\partial x/\partial t= \partial t/\partial x=0$.
Eq.~\refeq{eq:250a} implies, after multiplication by $\partial (t,x)/\partial
(s,p)$
\begin{align}
0 &= \frac{\partial (s,x)}{\partial (s,p)} + p \frac{\partial (t,s)}{\partial (s,p)} - s
\frac{\partial (t,p)}{\partial (s,p)}\nonumber\\
&= \frac{\partial x}{\partial p}  - p \frac{\partial t}{\partial p}  -
s\frac{\partial t}{\partial s}\ .\label{eq:250b}
\end{align}
The second equality holds because now we take $\partial s/\partial p= \partial
p/\partial s=0$. Finally, from \refeq{eq:247}, \refeq{eq:248} it follows that
\refeq{eq:250b} is equivalent to 
\numeq{eq:250c}{
\frac{\partial^2 \psi}{\partial p^2}  - \frac{\partial^2 \psi}{\partial s^2}  +
 \frac2s \frac{\partial \psi}{\partial s} = 0\ .
}
\end{subequations}
This linear equation is solved by two arbitrary functions of $p\pm s$ ($p\pm
s$ being just the Riemann coordinates)
\numeq{eq:251}{
\psi(p,s) = F(p+s) - sF'(p+s) + G(p-s) + sG'(p-s)\  
}

In summary, to solve the Chaplygin gas equations, we choose two functions $F$
and $G$, construct $\psi$ as in \refeq{eq:251}, and regain $s$
($=\sqrt{2\lambda}/\rho$), $p$ ($= \theta^\prime$), and $\theta$
from
\refeq{eq:247},
\refeq{eq:248}. In particular, the solution \refeq{solthenread},
\refeq{densecorresp} corresponds to
\numeq{eq:252}{
F(z) = G(-z) = \pm \frac z{2k} \ln z,
}
where the sign is correlated with the sign of~$t$. \vspace{15pt}
\hrule 
\addtocontents{toc}{\protect\hrulefill\par\vspace{-10pt}}
\subsection*{C. Sidebar on the integrability of the cubic potential in d=1}
\addcontentsline{toc}{section}{\quad \ C. Sidebar on the integrability of the cubic potential in one dimension}

Although it does not belong to the models that we have discussed, the cubic
potential for 1-dimensional motion, $V(\rho) = \ell \rho^3/3$, is especially
interesting because it is secretly free [see (\ref{eqone54b})] -- a fact that is exposed when Riemann
coordinates are employed. For this problem these read $R_{\pm} =
p\pm\sqrt{2\ell} \rho$ and again they are just the velocity $(\pm)$ the sound
speed. In contrast to \refeq{succpres} the Euler and continuity equations for this
system decouple: $\dot R_{\pm} = - R_{\pm} R^\prime_{\pm}$.
Indeed, it is seen that
$ R_{\pm}$ satisfy essentially the free Euler equation [compare with
\refeq{three15b} at $V''=0$ and identify $R_{\pm}$ with~$v$]. Consequently, the solution
\refeq{eule82}--\refeq{rhoray2} works here as well.

Recall the previous remark in Section 1.2 (iv)  on the Schr\"odinger group
[Galileo
and SO(2,1)]: in one dimension the cubic potential is invariant against this
group of transformations, and in all dimensions the free theory is
invariant. Therefore a natural speculation is that the
secretly noninteracting nature of the cubic potential in one dimension is a
consequence of Schr\"odinger group invariance. 

Another interesting fact about a one-dimensional nonrelativistic fluid with cubic
potential is that it also arises in a collective, semiclassical description of
nonrelativistic free fermions in one dimension, where the cubic potential
reproduces fermion repulsion \cite{JevSak}. In spite of the nonlinearity of the
fluid model's equations of motion, there is no interaction in the underlying
fermion dynamics. Thus,  the presence of the Schr\"odinger group  and the 
equivalence to free  equations for this fluid system is an understandable
consequence.
\addtocontents{toc}{\protect\hrulefill\par}
 \vspace{15pt}
\hrule 

\subsection{Born-Infeld model on a line}
\setcounter{equation}{0}

Since the Born-Infeld system is related to the Chaplygin gas by the transformation
described in Section~\ref{interrel}, there is no need to discuss
separately Born-Infeld solutions. Nevertheless, the formulation in terms of
Riemann coordinates is especially succinct and gives another view on the
Chaplygin/Born-Infeld relation. 

The Riemann coordinates $R_{\pm}$ for the Born-Infeld model are contructed
by first defining
\begin{align}
\frac1c\, \theta^\prime  = p/c &= \tan \phi_p,\nonumber\\
a/\rho c &= \tan \phi_\rho, \label{eq:253}\\
\intertext{and}
R_{\pm}  &= \phi_p \pm \phi_\rho\ . \label{eq:254}
\end{align}
The 1-dimensional version of the equations of motion
\refeq{followsfrom1}, \refeq{followsfrom2}, that is, \refeq{eqmot237},
\refeq{eqmot238} can be presented as 
\numeq{eq:255}{
\dot R_{\pm} = -c(\sin R_{\mp}) R^\prime_{\pm}\ . 
}

The relation to the Riemann description of the Chaplygin gas can now be seen 
in two ways:  a nonrelativistic limit and an exact transformation. For the former,
we note that at large~$c$, $\phi_p \approx p/c$, $\phi_\rho\approx a/\rho c$ so
that
\numeq{eq:256}{
R_{\pm}^{\BI} \approx \frac1c \Bigl(p \pm \frac a\rho\Bigr) = \frac1c
R_{\pm}^{\prime \mathrm{Chaplygin}} \Bigr|_{\lambda=a^2/2}\ .
}
Moreover, the equation \refeq{eq:255} becomes, in view of \refeq{eq:256},
\numeq{eq:257}{
\frac1c \dot R_{\pm}^{\mathrm{Chaplygin}} = - R_{\mp}^{\mathrm{Chaplygin}}
\frac1c\frac\partial{\partial x} R_{\pm}^{\mathrm{Chaplygin}},
 }
so that  \refeq{succpres} is regained. On the other hand, for the exact
transformation we define new Riemann coordinates in the relativistic, Born-Infeld 
case by
\numeq{eq:258}{
{\cal R}_{\pm} = c \sin R_{\pm}\ .
}
Evidently \refeq{eq:255} implies that ${\cal R}_{\pm}$ satisfies the
nonrelativistic equations \refeq{succpres}, \refeq{eq:257} when
$R_{\pm}$ solves the relativistic equation \refeq{eq:255}. Expressing
${\cal R}_{\pm}$ and  $R_{\pm}$ in terms of the corresponding nonrelativistic
and relativistic variables produces a mapping between the two sets. Calling
$p_{\mathrm{NR}}$, $\rho_{\mathrm{NR}}$ and $p_{\mathrm{R}}$,
$\rho_{\mathrm{R}}$ the  momentum and density of the nonrelativistic
and of the relativistic theory, respectively, the mapping implied by
\refeq{eq:258} is 
\begin{align}
p_{\mathrm{NR}} &=
\frac{c^2 \rho_{\mathrm{R}}
p_{\mathrm{R}}}{\sqrt{(p^2_{\mathrm{R}} +
c^2)(\rho^2_{\mathrm{R}}c^2 + a^2)}},\nonumber\\[2ex]
\rho_{\mathrm{NR}} &= \frac1{c^2} 
\sqrt{(p^2_{\mathrm{R}} + c^2)(\rho^2_{\mathrm{R}}c^2 + a^2)}\ .
\label{eq:259}
\end{align}
As can be checked, this maps the Chaplygin equations into the Born-Infeld
equations. But the mapping is not canonical.

We record the infinite number of constants of motion, which put into evidence
the (by now obvious) complete integrability of the Born-Infeld equations on a
line. The following quantities are time-independent:
\numeq{eq:260}{
I_n^{\pm} = ac^{n-1} \int \rd x \frac{(\phi_p \pm
\phi_\rho)^n}{\sin\phi_\rho\cos \phi_p}\ , \quad
n=0,\pm1,\ldots
 }
The nonrelativistic limit takes the above into \refeq{infnumconsmo}, while
expressing $I_n^{\pm}$ in terms of ${\cal R}_{\pm}$ according to
\refeq{eq:258} shows that the integrals in \refeq{eq:260} gives rise to a
series of the integrals in \refeq{infnumconsmo}.

In the relativistic model $\rho$ need not be constrained to be positive
(negative $\rho$ could be interpreted as antiparticle density). The
transformation $p \to -p$, $\rho \to -\rho$ is a symmetry and can be
interpreted as charge conjuguation. 

Further, $p$ and $\rho$ appear in 
an equivalent way. As a result, this theory enjoys a 
duality transformation.
\numeq{eq:261}{
\rho \to \pm\frac{a}{c^2} p \qquad  p \to \pm\frac{c^2}{a}
\rho\ 
}
Under the above, both the canonical structure and the
Hamiltonian remain invariant. Solutions are mapped in
general to new solutions. Note that the nonrelativistic limit
is mapped to the ultra-relativistic one under the above duality.
Self-dual solutions, with $\rho = \pm\frac{a}{c^2} p$, satisfy
\numeq{eq:262}{
\dot{\rho} = \mp c \frac\partial{\partial x} \rho^\prime, 
}
and are, therefore, the chiral relativistic solutions that
were presented at the end of Section 2.1(ii).  In the self-dual
case, when $p$ is eliminated from the canonical 1-form and from
the Hamiltonian with the help of \refeq{eq:261}, one arrives at an action for
$\rho$, which coincides (apart from irrelevant constants) with the
self-dual action, constructed some time ago \cite{FloJac}.
\begin{align}
\biggl\{
\fract12 &\displaystyle\int \rd t \rd x \rd y  \dot\rho(x)\epsilon (x-y)
p(y)  - \int  \rd t \rd x \sqrt{\rho^2c^2+a^2} \sqrt{c^2+p^2}  \rd t
\bigg\}\biggr|_{p=\frac{c^2}a \rho} \nonumber\\
= \frac{2c^2}a \biggl\{ \fract14  &\displaystyle\int \rd t \rd x \rd y
\dot\rho(x)\epsilon (x-y)
\rho(y)  
- \frac c2 \int \rd t \rd x\Bigl( \rho^2(x) + \frac{a^2}{c^2}\Bigr)
\biggr\}       \label{eq:263}
\end{align}

\subsection{General solution of the Nambu-Goto  theory
 for a
(p=1)-brane (string) in two spatial dimensions  (on a plane)}
\setcounter{equation}{0}

The complete integrability of the 1-dimensional Chaplygin gas and  Born-Infeld theory,
as well as the relationships between the two, derives from the fact that the
different models descend  by
fixing in different ways the parameterization invariance of the Nambu-Goto 
theory a for string on a plane.  At the same time, the equations governing  the
planar motion of a string can be solved completely. Therefore it is instructive
to see how the string solution produces the Chaplygin solution \cite{Baz}. 

We follow the development in Section 2.2(iii). The Nambu-Goto action
reads 
\begin{subequations}\label{eq:264}
\begin{align}
I_{\textrm{NG}} &= \int \rd{\phi^0} L_{\textrm{NG}},\label{eq:264a}\\
 L_{\textrm{NG}} &= \int \rd {\phi^1} {\cal L}_{\textrm{NG}},\label{eq:264b}\\
{\cal L}_{\textrm{NG}}&= \Bigl[ -\det \frac{\partial X^\mu}{\partial\phi^\alpha}
\frac{\partial X_\mu}{\partial\phi^\beta}\Bigr]^{1/2}\ .\label{eq:264c}
\end{align}
\end{subequations}
Here $X^\mu$, $\mu=0,1,2$, are string variables and $(\phi^0, \phi^1)$ are its
parameters. As in Section 2.2(iii), we define light-cone combinations
$X^{\pm} = \frac1{\sqrt2}(X^0\pm X^2)$, rename $X^-$ as $\theta$, and choose
the parameterization $X^+=\phi^0\equiv \tau$. After suppressing the
superscripts on $\phi^1$ and $X^1$, we construct the Nambu-Goto Lagrange
density as
\begin{align}
{\cal L}_{\textrm{NG}}&= \det\nolimits^{1/2} \begin{pmatrix}
2\partial_\tau \theta -(\partial_\tau X)^2 & u \\
u & -(\partial_\phi X)^2
\end{pmatrix},\label{eq:265}\\
u &= \partial_\phi \theta - \partial_\tau X \partial_\phi X.\label{eq:266}
\end{align}
Equations of motion are presented in Hamiltonian form:
\numeq{eq:267}{
p = \frac{\partial{\cal L}_{\textrm{NG}}}{\partial\partial_\tau X}, \qquad
\Pi= \frac{\partial{\cal L}_{\textrm{NG}}}{\partial\partial_\tau \theta},
}
\begin{subequations}\label{eq:268}
\begin{align}
\partial_\tau X &= -\frac1\Pi p - u \partial_\phi X,\label{eq:268a}\\
\partial_\tau \theta &= \frac1{2\Pi^2} \bigl(
          p^2 + (\partial_\phi X)^2  \bigr) - u \partial_\phi \theta,\label{eq:268b}\\
\partial_\tau p &= -\partial_\phi \Bigl(\frac1\Pi \partial_\phi
X \Bigr) - \partial_\phi (up),\label{eq:268c}\\
\partial_\tau \Pi &= - \partial_\phi (u\Pi),\label{eq:268d}
\end{align}
\end{subequations}
and there is the constraint
\numeq{eq:269}{
p\partial_\phi X + \Pi \partial_\phi\theta = 0\ .
}
There still remains the reparameterization freedom of replacing $\phi$ by an
arbitrary function of $\tau$ and $\phi$; this freedom may be used to set
$u=0$, $\Pi=-1$. Consequently, in the fully parameterized equations of
motion  eq.~\refeq{eq:268d} disappears; instead of \refeq{eq:268a} and
\refeq{eq:268c}, we have $\partial_\tau X = p$, $\partial_\tau p =
\partial^2_\phi X$, which imply
\begin{subequations}\label{eq:270}
\numeq{eq:270a}{
(\partial_\tau^2 -\partial_\phi^2)X = 0, 
}
\refeq{eq:268b} reduces to
\numeq{eq:270b}{
\partial_\tau \theta = \fract12 \bigl[ (\partial_\tau X)^2 + (\partial_\phi
X)^2\bigr], 
 }
and the constraint \refeq{eq:269} requires
\numeq{eq:270c}{
\partial_\phi \theta = \partial_\tau X \partial_\phi X\ .
}
\end{subequations}
Solution to \refeq{eq:270a} is immediate in terms of two functions $F_{\pm}$, 
\numeq{eq:271}{
x(\tau,\phi) = F_+ (\tau +\phi) + F_- (\tau-\phi)
}
and then \refeq{eq:270b}, \refeq{eq:270c} fix $\theta$.
\numeq{eq:272}{
\theta(\tau,\phi) = \int^{\tau+\phi} \rd z \bigl[F'_+(z)\bigr]^2 + 
 \int^{\tau-\phi} \rd z \bigl[F'_- (z)\bigr]^2\ 
}
This completes the description of a string moving on a plane.

But we need to convert this information into a solution of the Chaplygin gas,
and we know from Section 2.2(iii) that this can be accomplished by a
hodographic transformation: instead of $X$ and $\theta$ as a function of
$\tau$ and $\phi$, we seek $\phi$ as a function of $\tau$ and $X$, and this
renders $\theta$ to be a function of $\tau$ and $X$ as well. The density $\rho$
is determined by the Jacobian $\left|\partial X/\partial \phi\right|$. 

Replace $\tau$ by $t$ and $X$ by $x$ and define $\phi$ to be $f(t,x)$. Then from
\refeq{eq:271} it follows that
\numeq{eq:273}{
x = F_+ \bigl(t + f(t,x)\bigr) + F_- \bigl(t - f(t,x)\bigr)\ .
}
This equation may be differentiated with respect to $t$ and $x$, whereupon
one finds
\begin{subequations}\label{eq:274}
\begin{align}
\dot{f} &= -\frac{F'_+(t+f) + F'_-(t-f)}{F'_+(t+f) - F'_-(t-f)},
      \label{eq:274a}\\
f^\prime &=  \frac1{F'_+(t+f) - F'_-(t-f)}\ .
      \label{eq:274b}
\end{align}
\end{subequations}

Thus the procedure for constructing a Chaplygin gas solution is to choose two
functions $F_{\pm}$, solve the differential equations \refeq{eq:274} for $f$, and
then the fluid variables are
\begin{align}
\theta(t,x) &= \int^{t+f(t,x)} \bigl[F'_+(z)\bigr]^2 \rd z 
                       + \int^{t-f(t,x)} \bigl[F'_-(z)\bigr]^2 \rd z, \label{eq:275}\\
\frac{\sqrt{2\lambda}}\rho &= \left| F'_+(t+\phi) - F'_-(t-\phi)\right|\ .
                              \label{eq:276}
\end{align}

One may verify directly that \refeq{eq:275} and \refeq{eq:276} satisfy the
required equations: Upon differentiating \refeq{eq:275} with respect to $t$ and
$x$, we find
\begin{subequations}\label{eq:277}
\begin{align}
\dot{\theta}&= (F'_+)^2 (1+
 \dot{f})  + (F'_-)^2 (1-
\dot{f})\nonumber\\
        &= -2F'_+ F'_-    \label{eq:277a}\\
\theta^\prime&= (F'_+)^2 
f^\prime  - (F'_-)^2  
 f^\prime\nonumber\\
        &= F'_+ + F'_-.    \label{eq:277b}
\end{align}
\end{subequations}
The second equalities follow with the help of \refeq{eq:274}. From
\refeq{eq:277} one sees that
\numeq{eq:278}{
\dot{\theta} + \fract12 (\theta^\prime)^2 = 
\fract12 ( F'_+ - F'_- )^2 = \frac\lambda{\rho^2} 
 }
the last equality being the definition \refeq{eq:276}. Thus the Bernoulli (Euler)
equation holds. For the continuity equation, we first find from \refeq{eq:276}
and \refeq{eq:277}
\begin{subequations}\label{eq:279}
\begin{align}
\dot{\rho}&= 
\pm \frac{\partial}{\partial t} \frac{\sqrt{2\lambda}}{F'_+ - F'_-} \nonumber\\
     &= \mp \frac{\sqrt{2\lambda}}{(F'_+ - F'_-)^2} \Bigl[
F''_+ (1 +\dot{f}) -
F''_- (1 -\dot{f})\Bigr] \nonumber\\
        &= \pm \frac{2\sqrt{2\lambda}}{(F'_+ - F'_-)^3}  
              \bigl( F''_+ F'_- + F''_- F'_+  \bigr), \label{eq:279a}\\[2ex]
\intertext{\pagebreak[1]}
\frac{\partial }{\partial x}(\rho \theta^\prime)&=
     \frac{\partial }{\partial x}\Bigl(\pm \sqrt{2\lambda}
\frac{F'_+  + F'_- }{ F'_+ - F'_-}\Bigr)  \nonumber\\
 &= \mp \frac{\sqrt{2\lambda}}{(F'_+ - F'_-)^2}  
              \bigl(F''_+ F'_- + F''_- F'_+  \bigr) f^\prime\nonumber\\
&= \mp \frac{2\sqrt{2\lambda}}{(F'_+ - F'_-)^3}  
              \bigl( F''_+ F'_- + F''_- F'_+  \bigr). 
\label{eq:279b}
\end{align}
\end{subequations}
The last equalities follow from \refeq{eq:274};  since \refeq{eq:279a} and 
\refeq{eq:279b} sum to zero, the continuity equation holds.

We observe that the differentiated functions $F'_{\pm}$ are just the Riemann
coordinates: from \refeq{eq:277b} and \refeq{eq:276} [with the absolute value
ignored] we have 
\numeq{eq:280}{
p \pm \frac{\sqrt{2\lambda}}\rho \equiv R_{\pm} = 2F'_{\pm}\ .
}
Also it is seen with the help of \refeq{eq:274} that the Riemann formulation
\refeq{succpres} of the Chaplygin equations is satisfied by $2F'_{\pm}$. 

The constants of motion \refeq{infnumconsmo} become proportional to 
\begin{align}
I^{\pm}_n &\propto \int \rd x \frac1{F'_+ - F'_-} \bigl[  F'_{\pm}\bigr]^n
\nonumber\\
      &= \int \rd x \frac{\partial f}{\partial x} \bigl[  F'_{\pm}(t\pm f)\bigr]^n
\nonumber\\ 
      &\propto \int \rd z  \bigl[  F'_{\pm}(z) \bigr]^n\ .\label{eq:281}
\end{align}
Finally we remark that the solution 
\refeq{solthenread}, \refeq{densecorresp} corresponds to 
\numeq{eq:282}{
F_+(z) = - F_- (z) = \pm  \frac{\ln z}{2k}\ .
}
There exists a relation between the two functions~$F$ and~$G$  in \refeq{eq:251},
which encode the Chaplygin gas solution in the linearization approach of
Section 3.1(ii), and the above two functions $F_{\pm}$, which do the same
job in the Nambu-Goto approach. The relation is that $2F'_+$ is inverse to $2F''$
and $2F'_-$ is inverse to $2G''$, that is,
\begin{align}
2F'' [2F'_+ (z)] &=z, \nonumber\\
2G'' [2F'_- (z)] &=z. \label{eq:282b}
\end{align}

\newpage
\section{SUPERSYMMETRIC FLUID MECHANICS} 

%\subsection{Introduction}

As explained in Section
1, classical fluids describe particles moving collectively. The fluid inherits its mechanical properties, 
such as energy, nomentum and
angular momentum from the corresponding underlying particle properties.

One consequence of this is that classical fluids cannot carry intrinsic spin.
To be precise, the angular momentum with respect to the center of mass of a small volume $V$
scales as its mass (which scales as $V$) times the residual velocity of the fluid about the center
of mass (which scales like $\ell$- the linear dimension of $V$) times the distance to the center of
mass (which also scales like $\ell$). Therefore, the self-angular momentum density scales lie
$\ell^2$, and goes to zero with $\ell$.

The inclusion of a spin density in fluids can be achieved in an essentially 
quantum mechanical formulation by introducing Grassmann (anticommuting) 
variables in the description, the spin density being represented as a bilinear
in the Grassmann variables. This description reveals the possibility of
implementing within fluid mechanics supersymmetry transformations, which
effectively mix spin and kinematical degrees of freedom. Particular forms
of the Hamiltonian, generalizing the classical Chaplygin gas, admit these
supersymmetry transformations as an invariance and generate conserved
quantities.

Supersymmetry poses severe restrictions in the dynamics. In general,
supersymmetric extended objects cannot be formulated in arbitrary dimensions,
and this holds true for supersymmetric fluids. It is natural, therefore,
that the supersymmetric Chaplygin models are essentially related to, and derive 
from, higher-dimensional supersymmetric membrane actions in a way similar 
to the one already exposed in Section 2.2. As such, they also enjoy nontrivial
higher-dimensional relativistic symmetries which are not apparent from 
their action. 

In the following we shall analyze the case of planar \cite{JacPol2} and
lineal \cite{BergJac} fluids, which devolve from the motion of membranes or strings in
$(3+1)$ or $(2+1)$ dimensional spacetimes, respectively.

Lineal and planar supersymmetric fluid models seem to exhaust
the possibilities for the supersymmetric Nambu-Goto/fluid connection.  
For a higher
dimensional generalization, the reduction program would begin with a
$p$-brane in $p+2$ dimensional space-time, giving rise to a fluid in
$p+1$ dimensional space-time.  While there are no constraints on $p$ in the
purely bosonic case, supersymmetric extensions are greatly
constrained: the list of possible ``fundamental'' 
super $p$-branes (i.e. with only scalar supermultiplets in the
world-volume) contains only the above two cases:
$p=2$ in four dimensions and $p=1$ in three dimensions \cite{DuffFiel}.

\subsection{Supersymmetric fluid in $(2+1)$ dimensions}
We begin by  positing the fluid model. The Chaplygin gas Lagrangian 
is supplemented
by Grassmann variables $\psi_a$ that are Majorana 
spinors [real, two-component: $\psi_a^* = \psi_a$, $a=1,2$, $(\psi_1\psi_2)^* =
\psi_1^*\psi_2^*$].

The associated Lagrange density reads
\begin{equation}
{\cal L} = -\rho (\dot\theta - \fract12 \psi \dot\psi) - \fract12 \rho
(\grad\theta - \fract12 \psi \grad \psi)^2 -
\frac{\lambda}{\rho}  -\frac{\sqrt{2\lambda}}{2} \, \psi
\vec\alpha \cdot \grad \psi\ .\label{lagsusy}
\end{equation} 
Here $\alpha^i$ are two ($i=1,2$), $2 \times 2$, real
symmetric Dirac ``alpha'' matrices; in terms of Pauli matrices
we can take $\alpha^1=\sigma^1$, $\alpha^2=\sigma^3$. 
Note that the matrices satisfy the following relations, which
are needed to verify subsequent formulas.
\begin{align}
\epsilon_{ab} \alpha^i_{bc} &= \epsilon^{ij} \alpha^j_{ac}
\nonumber \\
\alpha^i_{ab} \alpha^j_{bc} &= \delta^{ij} \delta_{ac}
-\epsilon^{ij} \epsilon_{ac}
\nonumber \\
\alpha^i_{ab} \alpha^i_{cd} &= \delta_{ac} \delta_{bd}
- \delta_{ab} \delta_{cd} + \delta_{ad} \delta_{bc}
\end{align}
$\epsilon_{ab}$ is the $2 \times 2$ antisymmetric matrix
$\epsilon \equiv i \sigma^2$.  In equation (\ref{lagsusy}) $\lambda$ is
a coupling strength which is assumed to be positive.  
  The Grassmann term enters with coupling
$\sqrt{2\lambda}$, which is correlated with the strength of the Chaplygin
potential $V(\rho)={\lambda}/{\rho}$  in order to ensure supersymmetry, as we
shall show below.  It is evident that the velocity should be defined as 
\begin{equation}
{\bf v} = \grad \theta - \fract12 \psi \grad \psi\ .\label{clebfer}
\end{equation} 

The Grassmann variables directly give rise to a Clebsch
formula for ${\bf v}$, and provide the Gauss potentials.  The
two-dimensional vorticity reads
$\omega=\epsilon^{ij}\partial_i v^j = -\fract12
\epsilon^{ij}\partial_i \psi \partial_j \psi = -\fract12
\grad \psi \times \grad \psi$. The variables 
$\{\theta, \rho\}$ remain a canonical pair, while the
canonical 1-form in (\ref{lagsusy}) indicates that the
canonically independent Grassmann variables are
$\sqrt{\rho} \, \psi$ so that the antibracket of the $\psi$'s is
\begin{equation}
\{ \psi_a {(\bf r)}, \psi_b ({\bf r}')\} = -\frac{\delta_{ab}}{\rho({\bf r})}
\delta({\bf r}- {\bf r}')\ .\label{eq:2.5}
\end{equation} 
One verifies that the algebra (\ref{eqone36}) - (\ref{eqone41})
is satisfied, and further, one has
\begin{align}
\{ \theta ({\bf r}), \psi({\bf r})\} &= -\frac{1}{2\rho({\bf r})}
\psi ({\bf r}) \delta ({\bf r}-{\bf r}') \label{eq:2.5a}, \\[1ex]
\{ {\bf v} ({\bf r}), \psi ({\bf r}')\}
&= 
-\frac{\grad \psi({\bf r})}{\rho({\bf r})}
 \delta({\bf r}-{\bf r}') \label{eq:2.5b}, \\[2ex]
\{ \vec{\cal P} ({\bf r}), \psi({\bf r}') \} &= -\grad \psi({\bf r})
\delta({\bf r}-{\bf r}')\ .
\label{eq:2.5c}
\end{align}
The momentum density $\vec{\cal P}$ is given by the bosonic formula $\vec{\cal
P} ={\bf v}\rho$, but the Grassmann variables are hidden in $\bf v$, by virtue
of~\refeq{clebfer}.

The equations of motion read
\begin{align}
\dot\rho + \grad \cdot ({\bf v}\rho) &= 0, \label{eq:2.6a}
\\
\dot\theta + {\bf v} \cdot \grad \theta
&= \fract12 {\bf v}^2 + \frac{\lambda}{\rho^2}+
\frac{\sqrt{2\lambda}}{2 \rho}\, \psi \vec\alpha \cdot
\grad \psi, \label{eq:2.6b} \\[1ex] 
\dot\psi + {\bf v} \cdot \grad \psi 
&= \frac{\sqrt{2\lambda}}{\rho} \, \vec\alpha
\cdot \grad \psi, \label{eq:2.6c}
\end{align}
and together with (\ref{lagsusy}) they imply
\begin{equation}
\dot{\bf v} + {\bf v} \cdot \grad {\bf v} = 
\grad \frac{\lambda}{\rho^2} +
\frac{\sqrt{2\lambda}}{\rho}\, (\grad \psi) \vec\alpha \cdot
\grad \psi\ .
\label{eq:2.6d}
\end{equation} 

All these equations may be obtained by bracketing with the
Hamiltonian
\begin{equation}
H=\int \rd{\null^2 r}   \Bigl(\fract12 \rho {\bf v}^2 + \frac{\lambda}{\rho}
+\frac{\sqrt{2\lambda}}{2}\, \psi \vec\alpha \cdot \grad
\psi \Bigr) = \int \rd{\null^2 r} {\cal H}
\label{eq:2.7}
\end{equation} 
when (\ref{eqone36}), (\ref{eqone40}) as well as  \refeq{eq:2.5}--\refeq{eq:2.5b}
are used.

We record the components of the energy-momentum ``tensor'',
and the continuity equations they satisfy.  The energy
density ${\cal E}=T^{oo}$, given by
\begin{equation}
{\cal E} = \fract12 \rho {\bf v}^2 +
\frac{\lambda}{\rho}+\frac{\sqrt{2\lambda}}{2}\, \psi
\vec\alpha \cdot  \grad \psi  = T^{oo},
\label{eq:2.8}
\end{equation} 
satisfies a continuity equation with the energy flux $T^{jo}$.
\begin{equation}
T^{jo}
=   \rho  v^j   \bigl(\fract12  {\bf v}^2 - \frac\lambda{\rho^2}\bigr) + 
\frac{\sqrt{2\lambda}}{2}\, \psi \alpha^j {\bf v} \cdot 
\grad \psi - \frac{\lambda}{\rho} \psi \partial_j \psi
+  \frac{\lambda}{\rho} \epsilon^{jk} \psi \epsilon
\partial_k \psi
\label{eq:2.9b} 
\end{equation} 
\begin{equation}
\dot T^{oo} + \partial_j T^{jo}= 0
\label{eq:2.9a} 
\end{equation} 
This ensures that the total energy, that is, the
Hamiltonian, is time-indepen\-dent.  Conservation of the total
momentum
\begin{equation}
{\bf P} = \int \rd{\null^2 r}   \vec{\cal P}
\label{eq:2.10}
\end{equation} 
follows from the continuity equation satisfied by the
momentum density ${\cal P}^i=T^{oi}$ and the momentum flux, that is, the stress
tensor $T^{ij}$.
\begin{gather}
T^{ji}= \rho v^i v^j - \delta^{ij} \Bigl(\frac{2\lambda}{\rho} 
+ \frac{\sqrt{2\lambda}}{2}\,  \psi \vec\alpha \cdot \grad
\psi \Bigr) +  \frac{\sqrt{2\lambda}}{2}\, \psi \alpha^j
\partial_i \psi 
\\
\dot T^{oi} + \partial_j T^{ji}= 0 
\end{gather} 
But $T^{ij}$  is not
symmetric in its spatial indices, owing to the presence of spin
in the problem.  However, rotational symmetry makes it
possible to effect an ``improvement'', which modifies the
momentum density by a total derivative term, leaving the
integrated total momentum unchanged (provided surface
terms can be ignored) and rendering the stress tensor
symmetric.  The improved quantities are
\begin{gather}
\hskip-1.5in {\cal P}^i_I = T^{oi}_I  = \rho v^i + \fract18
\epsilon^{ij}
\partial_j (\rho \psi \epsilon \psi),
\label{imprquant}\\
 T^{ij}_I
= \rho v^i v^j -\delta^{ij} \Bigl(\frac{2 \lambda}{\rho} + 
\frac{\sqrt{2\lambda}}{2} \, \psi \vec\alpha \cdot \grad
\psi \Bigr) + \frac{\sqrt{2\lambda}}{4} \, \Big( \psi 
\alpha^i \partial_j \psi + \psi \alpha^j \partial_i \psi  \Big)
\nonumber \\ 
{}-
\fract18 \partial_k \Big[ (\epsilon^{ki} v^j + \epsilon^{kj}
v^i) \rho \psi \epsilon\psi\Big], \qquad\qquad\qquad\\ 
\dot T^{oi}_I + \partial_j T^{ij}_I = 0\ .
\end{gather}
It immediately follows from the symmetry of $T^{ij}_I$ that the angular
momentum
\begin{equation}
M=\int \rd{\null^2 r}   \epsilon^{ij} r^i {\cal P}^j_I
= \int \rd{\null^2 r}  \rho \epsilon^{ij} r^i v^j + \fract14\int \rd{\null^2 r}
\rho\psi\epsilon\psi 
\end{equation} 
is conserved. The first term is clearly the orbital part
(which still receives a Grassmann contribution through $\bf v$),
whereas the second, coming from the improvement, is the spin
part. Indeed, since
$\frac{i}{2}\epsilon=\fract12 \sigma^2 \equiv \Sigma$, we
recognize this as the spin matrix in (2+1)~dimensions. The
extra term in the improved momentum density~\refeq{imprquant},  $\fract18
\epsilon^{ij}\partial_j (\rho\psi\epsilon\psi)$, can then be
readily interpreted as an additional localized momentum
density, generated by the nonhomogeneity of the spin density.
This is analogous to the magnetostatics formula giving the
localized current density~${\bf j}_m$ in a magnet in terms of its
magnetization ${\bf m}$: ${\bf j}_m = \grad\times\bf m$. All in
all, we are describing a fluid with spin.

Also the total number 
\begin{equation}
N=\int \rd{\null^2 r}  \rho
\label{eq:2.15new} 
\end{equation} 
is conserved
by virtue of the continuity equation (\ref{eq:2.6a}) satisfied
by $\rho$.  Finally, the theory is Galileo invariant, as is seen
from the conservation of the Galileo boost,
\begin{equation}
{\bf B} = t {\bf P}-\int \rd{\null^2 r}   {\bf r}\rho
\label{eq:2.16} 
\end{equation} 
which follows from (\ref{eq:2.6a}) and (\ref{eq:2.10}).  The
generators $H, {\bf P}, M, {\bf B}$ and $N$ close on the (extended)
Galileo group.  [The theory is not Lorentz invariant in
$(2+1)$-dimensional space-time, hence the energy flux
$T^{jo}$ does not coincide with the momentum density,
improved or not.]

We observe that $\rho$ can be eliminated from
(\ref{lagsusy}) so that ${\cal L}$ involves only $\theta$ and
$\psi$.  From (\ref{eq:2.6b}) and (\ref{eq:2.6c}) it follows
that
\begin{equation}
\rho=\sqrt\lambda \bigl(\dot\theta - \fract12 \psi \dot\psi + \fract12
{\bf v}^2\bigr)^{- 1/2}\ .
\label{eq:2.17} 
\end{equation} 
Substituting into (\ref{lagsusy}) produces the supersymmetric generalization of
the Chaplygin gas Lagrange density in \refeq{rindependent}.
\begin{equation}
{\cal L} = - 2\sqrt\lambda \, \Big\{\sqrt{ 2\dot\theta -  \psi
\dot\psi +  (\grad \theta - \fract12 \psi \grad \psi)^2}
+ \fract12 \psi \vec\alpha \cdot \grad \psi \Big\} 
\label{eq:2.18} 
\end{equation} 
Note that the coupling strength has disappeared from the
dynamical equations, remaining only as a normalization factor
for the Lagrangian.  Consequently the above elimination of
$\rho$ cannot be carried out in the free case, $\lambda=0$.

\subsection{Supersymmetry}
\setcounter{equation}{0}
As we said earlier, this theory possesses 
supersymmetry.  This
can be established, first of all, by verifying that the
following two-component supercharges are time-independent 
Grassmann quantities.
\begin{equation}
Q_a =  \int \rd{\null^2 r}   \Big[ \rho {\bf v} \cdot  (\vec\alpha_{ab}
\psi_b) +  \sqrt{2\lambda} \psi_a \Big]\ 
\label{eq:2.19} 
\end{equation} 
Taking a time derivative and using the evolution equations
(\ref{eq:2.6a})--(\ref{eq:2.6d}) establishes that $\dot{Q}_a=0$.

Next, the supersymmetric transformation rule for the dynamical variables is
found by constructing a bosonic symmetry generator~$Q$, obtained by contracting 
the Grassmann charge with a constant Grassmann parameter $\eta^a$,
 $Q=\eta^aQ_a$, and commuting with the dynamical variables.  Using the canonical
brackets one verifies the following field transformation rules.
\begin{align}
\delta \rho = \{ Q, \rho\} &= -\grad \cdot
\rho( \eta\vec\alpha\psi)
\label{eq:2.20a} \\[1ex]
 \delta \theta = \{ Q, \theta\} &=
-\fract12( \eta\vec\alpha \psi) \cdot \grad \theta - \fract14
 ( \eta\vec\alpha\psi) \cdot \psi \grad\psi +
\frac{\sqrt{2\lambda}}{2\rho}\,  \eta  \psi
\label{eq:2.20b} \\[1ex] 
 \delta \psi = \{ Q, \psi\} &=
-(\eta\vec\alpha \psi) \cdot \grad \psi -
{\bf v} \cdot \vec\alpha  \eta - \frac{\sqrt{2\lambda}}{\rho}\, \eta
\label{eq:2.20c} \\[1ex] 
 \delta \vec v = \{ Q, \vec v\} &=
-(\eta\vec\alpha \psi) \cdot \grad {\bf v} +
\frac{\sqrt{2\lambda}}{\rho}\,
 \eta \grad \psi\ 
\end{align}
Supersymmetry is
reestablished by determining the variation of the action $\int
\rd t \rd{\null^2 r}  {\cal L}$  consequent to the above field variations:  the
action is invariant.  One then reconstructs the supercharges
(\ref{eq:2.19}) by Noether's theorem. 
Finally, upon computing the bracket of two supercharges,
one finds
\begin{equation}
\{ \eta^a_1 Q_a, \eta^b_2 Q_b\} = 2  (\eta_1 \eta_2) H,
\end{equation} 
which again confirms that the charges are time-independent.
\begin{equation}
\{ H, Q_a\} = 0\
\end{equation} 

Additionally a further, kinematical, supersymmetry can be
identified.  According to the equations of motion the following
two supercharges are also time-independent.
\begin{equation}
\bar{Q}_a = \int \rd{\null^2 r}  \rho \psi_a\label{susyq}
\end{equation} 
$\bar{Q}=\bar{\eta}^a \bar{Q}_a$ effects a shift of the
Grassmann field.
\begin{align}
\bar{\delta} \rho = \{ \bar{Q}, \rho\} &= 0
\\
\bar{\delta} \theta = \{ \bar{Q}, \theta\} &=
-\fract12(  \bar{\eta}\psi) \\ 
\bar{\delta} \psi = \{ \bar{Q}, \psi\} &=
- \bar{\eta}\\
\bar{\delta} \vec v = \{ \bar{Q}, \vec v\} &= 0\ 
\end{align}
This transformation leaves the Lagrangian invariant, and
Noether's theorem reproduces (\ref{susyq}).  The algebra
of these charges closes on the total number $N$,
\begin{equation}
\{ \bar{\eta}_1^a \bar{Q}_a, \bar{\eta}_2^b \bar{Q}_b \} 
=  (\bar{\eta}_1 \bar{\eta}_2) N,
\label{eq:2.25} 
\end{equation} 
while the algebra with the generators (\ref{eq:2.19}), closes
on the total momentum, together with a central extension,
proportional to volume of space $\Omega = \int \rd{\null^2 r}$.
\begin{equation}
\{ \bar{\eta}^a \bar{Q}_a, \eta^b Q_b \} 
=    (\bar{\eta}\vec\alpha \eta) \cdot {\bf P} +   \sqrt{2\lambda}\,
(\bar{\eta}\epsilon \eta) \Omega\ 
\label{eq:2.26} 
\end{equation} 
The supercharges $Q_a, \bar Q_a$, together with the Galileo
generators ($H, {\bf P}, M, {\bf B}$),  and with
$N$ form a superextended Galileo algebra. The
additional, nonvanishing brackets are
\begin{align}
 \{ M, Q_a\} &= \fract12 \epsilon^{ab}Q_b,\\
 \{ M, \bar Q_a\} &= \fract12 \epsilon^{ab}\bar Q_b,\\
 \{{\bf B},  Q_a\} &= \vec\alpha_{ab}\bar Q_b\ .
\end{align}

\subsection{Supermembrane connection}
\setcounter{equation}{0}

The equations for the supersymmetric Chaplygin fluid devolve
from a supermembrane Lagrangian, $L_M$.  We shall
give two different derivations of this result, which make use
of two different parameterizations for the
parameter\-ization-invariant membrane action and give rise,
respectively, to \refeq{lagsusy} and \refeq{eq:2.18}. The two
derivations follow what has been done in the bosonic case in
Sections~2.2 (i) and 2.2 (iii).

We work in a light-cone gauge-fixed theory:  The supermembrane
in 4-dimensional space-time is described by coordinates
$X^\mu$ $(\mu=0,1,2,3)$, which are decomposed into
light-cone components $X^\pm=\frac{1}{\sqrt{2}} (X^0 \pm
X^3)$ and transverse components $X^i$ $\{i=1,2\}$.  These
depend on an evolution parameter $\phi^0\equiv\tau$ and two
space-like parameters $\phi^r$ $\{r=1,2\}$.  Additionally
there are two-component, real Grassmann spinors $\psi$,
which also depend on $\tau$ and $\phi^r$.  In the light-cone
gauge, $X^+$ is identified with $\tau$, $X^-$ is renamed
$\theta$, and the supermembrane Lagrangian is \cite{ref:3}
\begin{equation}
L_M=\int \rd{\null^2 \phi}  {\cal L}_M = -\int \rd{\null^2 \phi}  \, 
\{\sqrt{G} -
\fract12
\epsilon^{rs}
\partial_r \psi \balpha \partial_s \psi \cdot {\bf X} \},
\label{eq:3.1} 
\end{equation}
where $G=\det G_{\alpha\beta}$.
\begin{align}
G_{\alpha\beta} &=  
\begin{pmatrix}
G_{oo} &\quad  G_{os} \\ 
G_{ro} & -g_{rs}
\end{pmatrix}
=  
\begin{pmatrix}
2 \partial_\tau \theta-(\partial_\tau {\bf X})^2
- \psi \partial_\tau \psi & \quad u_s \\ 
u_r & -g_{rs}
\end{pmatrix}\label{eq:3.2}
\\[2ex]
%\end{align}
%\begin{align}
G &= g\Gamma
\nonumber \\[1ex]
\Gamma &\equiv  2\partial_\tau \theta-(\partial_\tau {\bf X})^2
 -  \psi \partial_\tau \psi
+g^{rs} u_r u_s \nonumber  \\[1ex] 
g_{rs} &\equiv \partial_r {\bf X} \cdot \partial_s {\bf X}
\ ,  \quad g=\det g_{rs} \nonumber  \\[1ex] 
u_s &\equiv \partial_s \theta - \fract12 \psi
\partial_s\psi - \partial_\tau {\bf X} \cdot \partial_s {\bf X} 
\label{eq:3.3}
\end{align}
Here $\partial_\tau$ signifies differentiation with respect to
the evolution parameter $\tau$, while $\partial_r$ differentiates with respect to
the space-like parameters $\phi^r$; $g^{rs}$  is the inverse of
$g_{rs}$, and the two are used to move the $(r,s)$ indices.  Note that
the dimensionality of the transverse coordinates $X^i$ is the
same as of the parameters $\phi^r$, namely two.

\subsubsection*{(i) Hodographic transformation}
\addcontentsline{toc}{subsubsection}{(i) Hodographic transformation}

To give our first derivation following the procedure in Section~2.2(iii), we
rewrite the Lagrangian in canonical, first-order form, with the help of bosonic
canonical momenta defined by
\begin{subequations}\label{eq:3.4}
\begin{align}
\frac{\partial {\cal L}_M}{\partial \partial_\tau {\bf X}} &=
{\bf p} = -\Pi  \partial_\tau {\bf X} - \Pi u^r \partial_r {\bf X},
\label{eq:3.4a} \\[1ex]
\frac{\partial {\cal L}_M}{\partial \partial_\tau \theta} &=
\Pi = \sqrt{{g}/{\Gamma}}\ .
\label{eq:3.4b}
\end{align}
\end{subequations}%
(The Grassmann variables already enter with first-order derivatives.) The
supersymmetric extension of \refeq{foform} then takes the form
\begin{align}
{\cal L}_M&={\bf p} \cdot \partial_\tau {\bf X} + \Pi \partial_\tau
\theta - \fract12\Pi \psi \partial_\tau\psi + \frac{1}{2\Pi}
({\bf p}^2+g) +
\fract12 \epsilon^{rs}
\partial_r \psi \balpha \partial_s \psi \cdot {\bf X} 
\nonumber\\[1ex] 
&\quad {}+ u^r \Big({\bf p} \cdot\partial_r {\bf X}  + \Pi \partial_r
\theta -
\fract12 \Pi \psi \partial_r \psi \Big)\ .
\label{eq:3.5} 
\end{align}
In (\ref{eq:3.5}) $u^r$ serves as a Lagrange multiplier
enforcing a subsidiary condition on the canonical variables, and $g=\det g_{rs}$. 
The equations that follow from (\ref{eq:3.5}) coincide with
the Euler-Lagrange equations for
\refeq{eq:3.1}.  The theory still possesses an
invariance against redefining the spatial parameters with a
$\tau$-dependent function of the parameters.  This freedom
may be used to set $u_\tau$ to zero and fix $\Pi$ at $-1$.  Next
we introduce the hodographic transformation, as in Section~2.2(iii), 
whereby independent-dependent variables are
interchanged, namely we view the $\phi^r$ to be
functions of $X^i$.  It then follows that the constraint on
(\ref{eq:3.5}), which with
$\Pi=-1$ reads
%\begin{subequations}
\begin{equation}
 {\bf p} \cdot\partial_r {\bf X} - \partial_r \theta + \fract12 \psi
\partial_r \psi =0,
\label{eq:3.6a}%
\end{equation}
becomes
\begin{equation}
\partial_r {\bf X} \cdot \Big({\bf p} -\grad \theta +
\fract12 \psi\grad \psi \Big) =0\ .
\label{eq:3.6b}%
\end{equation}
Here ${\bf p}$, $\theta$ and $\psi$ are viewed as functions of
${\bf X}$, renamed ${\bf r}$, with respect to which acts the gradient
$\grad$.  Also we rename ${\bf p}$ as ${\bf v}$, which
according to (\ref{eq:3.6b}) is
\begin{equation}
{\bf v} = \grad \theta -
\fract12 \psi\grad \psi\ .
\label{eq:3.6c}%
\end{equation}
%\end{subequations}%

As in Section~2.2(iii), from the chain rule and the implicit function theorem 
it follows that
\begin{equation}
\partial_\tau = \partial_t + \partial_\tau {\bf X} \cdot
\grad,
\label{eq:3.7}
\end{equation}
and according to (\ref{eq:3.4a}) (at $\Pi=-1$, $u^r=0$)
$\partial_\tau {\bf X} = {\bf p} = {\bf v}$.  Finally, the measure 
transforms according to $\rd{\null^2 \phi}  \to
\rd{\null^2 r}  \frac{1}{\sqrt{g}}$.  Thus the Lagrangian
for (\ref{eq:3.5}) becomes, after setting $u^r$ to zero and
$\Pi$ to $-1$,
\begin{subequations}
\begin{gather}
L_M=\!\int \!\frac{\rd{\null^2 r}}{\sqrt{g}} \Bigl( {\bf v}^2 -\dot\theta - {\bf v}
\!\cdot\!
\grad \theta + \fract12 \psi(\dot\psi + {\bf v} \cdot \grad
\psi) -\fract12({\bf v}^2+g)\nonumber\\
\hbox to 2in{\hfill} -\fract12 \epsilon^{rs}\, \psi
\alpha^i\, \partial_j \psi \, \partial_s X^j\,
\partial_r X^i
\Bigr)\ .
\label{eq:3.8}
\end{gather}
But $\epsilon^{rs} \partial_s X^j \partial_r X^i = \epsilon^{ij}
\det \partial_r X^i = \epsilon^{ij} \sqrt{g}$.  After
$\sqrt{g}$ is renamed $\sqrt{2\lambda}/\rho$,
(\ref{eq:3.8}) finally reads
\begin{equation}
L_M= \frac1{\sqrt{2\lambda}}  \int \rd{\null^2 r}\,  \Big(
{-\rho} (\dot\theta -
\fract12 \psi\dot\psi) -
\fract12 \rho (\grad \theta - \fract12 \psi \grad
\psi)^2 - \frac{\lambda}{\rho} - \frac{\sqrt{2\lambda}}{2} 
\psi \balpha \times \grad \psi \Big)\ .
\label{eq:3.9}
\end{equation}
\end{subequations}
Upon replacing $\psi$ by $\frac{1}{\sqrt{2}} (1-\epsilon)
\psi$, this is seen to reproduce the Lagrange density
\refeq{lagsusy}, apart from an overall factor.

\subsubsection*{(ii) Light-cone parameterization}
\addcontentsline{toc}{subsubsection}{(ii) Light-cone parameterization}
For our second derivation, we return to
(\ref{eq:3.1})--(\ref{eq:3.6b}) and use the remaining
reparameterization freedom to equate the two $X^i$
variables with the two $\phi^r$ variables, renaming both as
$r^i$.  Also $\tau$ is renamed as $t$. This parallels the method in
Section~2.2(i).  Now in (\ref{eq:3.1})--(\ref{eq:3.3}) $g_{rs} = \delta_{rs}$,
and
$\partial_\tau {\bf X}=0$, so that (\ref{eq:3.3}) becomes simply
\begin{align}
G=\Gamma &= 2\dot\theta -  \psi \dot\psi + u^2
\label{eq:3.10} \\[1ex]
 {\bf u} &= \grad \theta - \fract12 \psi \grad \psi\ .
\label{eq:3.11}
\end{align}
Therefore the supermembrane Lagrangian (\ref{eq:3.1}) reads
\begin{equation}
L_M=-\int \rd{\null^2 r} \biggl\{ \sqrt{2 \dot\theta -  \psi \dot\psi +
\bigl(\grad \theta -\fract12 \psi \grad \psi\bigr)^2}
+ \fract12 \psi \balpha \times \grad \psi \biggr\}\ .
\label{eq:3.12}
\end{equation}
Again a replacement of $\psi$ by $\frac{1}{\sqrt{2}}
(1-\epsilon) \psi$ demonstrates that the integrand
coincides with the Lagrange density in (\ref{eq:2.18}) (apart
from a normalization factor).

\subsubsection*{ (iii) Further consequences of the supermembrane
connection}
\addcontentsline{toc}{subsubsection}{(iii) Further consequences of the supermembrane
connection}

Supermembrane dynamics is Poincar\'e invariant in
(3+1)-dimensional space-time.  This invariance is hidden
by the choice of light-cone parameterization: only the
light-cone subgroup of the Poincar\'e group is left as a
manifest invariance.  This is just the $(2+1)$ Galileo group
generated by $H$, ${\bf P}$, $M$, ${\bf B}$, and $N$.  (The
light-cone subgroup of the Poincar\'e group is isomorphic to
the Galileo group in one lower dimension. \cite{sussk})  The Poincar\'e
generators not included in the above list correspond to
Lorentz transformations in the ``$-$'' direction.  We expect
therefore that these generators are ``dynamical'', that is,
hidden and unexpected conserved quantities of our
supersymmetric Chaplygin gas, similar to the situation with
the purely bosonic model.

One verifies that the following quantities
\begin{align}
S &= tH-\int \rd{\null^2 r}\,  \rho \theta 
\label{eq:3.14} \\[1ex]
{\bf G} &= \int \rd{\null^2 r}  ({\bf r} {\cal H} - \theta \bcP_I - \fract18
\psi \balpha \balpha \cdot \bcP_I \psi) \nonumber \\
&= \int \rd{\null^2 r} ({\bf r} {\cal H}  - \theta \bcP - \fract14
\psi \balpha \balpha \cdot \bcP \psi) 
\label{eq:3.15}
\end{align}
are time-independent by virtue of the equations of motion
(\ref{eq:2.6a})--(\ref{eq:2.6d}), and they supplement the Galileo
generators to form the full $(3+1)$ Poincar\'e algebra, which becomes
the super-Poincar\'e algebra once the supersymmetry is taken
into account. Evidently \refeq{eq:3.14}, \refeq{eq:3.15} are the supersymmetric
generalizations of \refeq{eq:105}, \refeq{eq:106}.

We see that planar fluid dynamics can be extended to
include Grassmann variables, which also enter in a
supersymmetry-preserving interaction.  Since our
construction is based on a supermembrane in
(3+1)-dimensional space-time, the fluid model is
necessarily a planar Chaplygin gas.  
In the next section we shall derive a lineal version of the
supersymmetric Chaplygin gas starting from a superstring
in (2+1)-dimensional space-time.

\subsection{Supersymmetric fluid in $(1+1)$ dimensions}

The one-dimensional case is in principle simpler since, in one
spatial dimension, the canonical structure can be straightforwardly
realized. The physical implications of adding Grassmann variables,
however, are somewhat limited 
since there is no vorticity and no spin in one space dimension.

Nevertheless, a supersymmetric version of the lineal Chaplygin gas
can be constructed. This is achieved along the same lines as in
2+1 dimensions, by considering a
superstring moving on a plane and again fixing the parametrization
invariance. The construction is analogous to what has already been
done in one higher dimension: the Nambu-Goto action for a
supermembrane in (3+1)-dimensions gives rise, in a specific
parametrization, to a supersymmetric planar Chaplygin gas. %\cite{RJAPPRD}  

As we shall demonstrate, the supersymmetric extension enjoys the same
integrability properties as the purely bosonic, lineal Chaplygin gas,
as a consequence of the complete integrability for the dynamics of the
superstring on the plane.

\subsubsection*{(i) Superstring formulation}
\setcounter{equation}{0}
\addcontentsline{toc}{subsubsection}{(i) Superstring formulation}

We begin with the Nambu-Goto superstring in 3-dimensional space-time, 
%\cite{Gauntlett}

\beq
\label{stringaction}  I = -\int \rmd \tau \rmd \sigma \,(\sqrt{g} - i
\epsilon^{ij} \partial_i X^{\mu} \bar{\psi} \gamma_{\mu} \partial_j
\psi),
\eeq
where
\beqar
g &=& -\mbox{det} \{\Pi^{\mu}_{i}\Pi^{\nu}_{j}\eta_{\mu\nu} \},\\
\Pi^{\mu}_{i} &=& \partial_{i}X^{\mu} - i \bar{\psi} \gamma^{\mu}
\partial_i \psi \ .
\eeqar
In these expressions $\mu,\nu$ are spacetime indices running over $0,
1, 2$ and $i,j$ are world-sheet indices denoting $\tau$ and $\sigma$.  We now go to the light-cone
gauge where we define $X^{\pm}=\frac{1}{\sqrt{2}}(X^0 \pm X^2)$.
$X^{+}$ is identified with the timelike parameter $\tau$, $X^{-}$ is
renamed $\theta$, and the remaining transverse component $X^1$ is renamed $x$.  We
can choose a 2-dimensional Majorana representation for the
$\gamma$-matrices: 
$$\gamma^0=\sigma^2, \ \ \gamma^1= -i\sigma^3, \ \ \gamma^2= i \sigma^1,$$
 such that $\psi$ is a real, two-component spinor.  A remaining
fermionic gauge choice sets  
$$\gamma^{+} \psi=0,$$
where $\gamma^{\pm}=\frac{1}{\sqrt{2}}(\gamma^0 \pm \gamma^2)$.  Thus $\psi$
is further reduced to a real, one-component Grassmann field.  Finally we define
the complex conjugation of a product of Grassmann fields $(\psi_1
\psi_2)^\star = \psi^{\star}_{1}\psi^{\star}_2$ so as to eliminate $i$
from Grassmann bilinears in our final expression.  The light-cone
gauge-fixed Lagrange density becomes:
\beq
{\cal L} = -\sqrt{g\Gamma} + \sqrt{2} \psi \partial_{\sigma}\psi,
\eeq
where
\beqar
g&=&(\partial_{\sigma}x)^2, \\
\Gamma &=& 2 \partial_{\tau}\theta - (\partial_{\tau}x)^2 - 2\sqrt{2} \psi \partial_{\tau}\psi +
\frac{u^2}{g},\\
u&=& \partial_{\sigma}\theta - \partial_{\tau}x \partial_{\sigma}x - \sqrt{2}\psi\partial_{\tau}\psi \ .
\eeqar
In the above equations, $\partial_{\sigma}$ and $\partial_{\tau}$
denote partial derivatives with respect to the spacelike and timelike
world-sheet coordinates.  The canonical momenta
\beqar
\label{pdef} p&=&\frac{\partial {\cal L}}{\partial (\partial_{\tau}x)} =
\sqrt{\frac{g}{\Gamma}}( \partial_{\tau}x + \frac{u}{g} \partial_{\sigma} x), \\ 
\Pi&=&\frac{\partial {\cal L}}{\partial (\partial_{\tau}\theta)}= -\sqrt{\frac{g}{\Gamma}},
\eeqar
satisfy the constraint equation
\beq \label{constraint}
p \partial_{\sigma} x + \Pi \partial_{\sigma} \theta - \sqrt{2} \Pi
\psi \partial_{\sigma} \psi = 0
\eeq
and can be used to recast ${\cal L}$ into the form
\beq
{\cal L} = p\partial_{\tau}x + \Pi \partial_{\tau}\theta +
\frac{1}{2\Pi}(p^2+g)+\sqrt{2}\psi\partial_{\sigma}\psi-\sqrt{2}\Pi\psi\partial_{\tau}\psi
+ u(p \partial_{\sigma}x + \Pi \partial_{\sigma}\theta - \sqrt{2} \Pi
\psi \partial_{\sigma} \psi),
\eeq
where $u$ is now a Lagrange multiplier enforcing the constraint.
We use the remaining parameterization freedom to fix $u=0$ and $\Pi=-1$
and perform a hodographic transformation, interchanging independent 
with dependent variables.  The partial derivatives
transform by the chain rule:
\beqar
\partial_{\sigma} &=& (\partial_{\sigma}x) \partial_x = \sqrt{g} \partial_x\ , \\
\partial_{\tau} &=& \partial_t + (\partial_{\tau}x) \partial_x =
\partial_t + v\partial_x \ ,
\eeqar
and the measure transforms with a factor of $1/\sqrt{g}$.  Finally,
after renaming $\sqrt{g}$ as $\sqrt{2\lambda}/\rho$, we obtain the
Lagrangian for the Chaplygin ``super'' gas in (1+1)-dimensions, 
\beq
\label{lagr}
L = \frac{1}{\sqrt{2\lambda}} \int {\rm d\null}x \, { \{
 -\rho(\dotth-\frac12\psi\dotpsi)-\frac12\rho v^2 -
\frac{\lambda}{\rho} + \frac{\sqrt{2\lambda}}{2}\psi\, \psi^\prime \} }\ ,  
\eeq
where according to (\ref{pdef}) and (\ref{constraint}) (at $u=0$ and $\Pi=-1$)
\beq
\label{vdef} v=p= \theta^\prime -\frac12 \psi \psi^\prime \ .
\eeq
We have used $\rho$ and $v$ in anticipation of their role as the
fluid density and velocity, and we demonstrate below that they indeed satisfy
appropriate equations of motion.  For convenience we have also
rescaled $\psi$ everywhere by a factor of $2^{-3/4}$. The Lagrangian
(\ref{lagr}) agrees with the limiting case of the planar fluid in (\ref{lagsusy}). 
We note that as for the planar case, a more straightforward derivation
leads to the fluid Lagrangian of (\ref{lagr}) with $\rho$ integrated
out.  Specifically, if the parameterization freedom is used directly to
equate the spacelike and timelike coordinates $\sigma$ and $\tau$ with
$x$ and $t$, we obtain
\beq
L' = - \int {\rm d\null}x \,{\Bigl( \sqrt{2 \dotth - \psi \dotpsi +
v^2} - \frac12 \psi\, \psi^\prime \Bigr)}\ ,
\eeq
where $v$ is defined as in (\ref{vdef}).  This form of the Lagrangian
can be obtained from (\ref{lagr}) after $\rho$ is eliminated using the
equations of motion for $\theta$ and $\psi$, shown below. 

\subsubsection*{(ii) Supersymmetric Chaplygin gas}
\addcontentsline{toc}{subsubsection}{(ii) The supersymmetric Chaplygin gas}
\subsubsection*{(a) Equations of motion}
\addcontentsline{toc}{subsubsection}{(a) Equations of motion}

The following equations of motion are obtained by variation of
the Lagrangian (\ref{lagr}).
\beqar
\label{rhoeq} \dotrho + \partial_x(\rho v) &=& 0 \\
\label{psieq} \dotpsi + \Bigl( v+\frac{\sql}{\rho}\Bigr)\psi^\prime &=& 0\\
\label{theq} \dotth + v \theta^\prime &=&\frac12v^2+\frac{\lambda}{\rho^2}-\frac{\sql}{2\rho}
\psi \psi^\prime
\\
\label{veq} \dotv + v v^\prime &=& \partial_x\Bigl(\frac{\lambda}{\rho^2}\Bigr) \ 
\eeqar
Naturally, there are only three independent equations of motion as
(\ref{veq}) is obtained from (\ref{psieq}), (\ref{theq}) and
(\ref{vdef}). Equations (\ref{rhoeq}) and (\ref{veq}) are seen to be just the continuity
and Euler equations for the Chaplygin gas.  Note that these do not see
the Grassmann variables directly. 

We now pass to the Riemann coordinates, which for this system are
(velocity $\pm$ sound speed $\sql/\rho$): 
\beq
R_{\pm} = \Bigl(v \pm \frac{\sql}{\rho}\Bigr) \ .
\eeq
In terms of the Riemann coordinates, the equations of motion obtain
the form
\beqar
\label{Req} \dot{R_{\pm}} &=& - R_{\mp} R^\prime_{\pm}, \\
\label{psiReq}\dotpsi &=& -R_{+} \psi^\prime, \\
\label{thReq} \dotth &=& -\frac12 R_{+}R_{-}-\frac12 R_{+} \psi \psi^\prime \ .
\eeqar
The equations in (\ref{Req}) contain the continuity and Euler equations and
are known to be integrable.  It is readily verified that
equation (\ref{psiReq}) for $\psi$ is solved by any function of
$R_{-}$, 
\beq
\label{sol}\psi = \Psi(R_{-}),
\eeq
and hence the fluid model is completely integrable.  That this is the case
should come as no surprise considering that we began with an
integrable world-sheet theory.  

At this
point it may seem curiously asymmetric that equation (\ref{psiReq})
for the Grassmann field should contain the $R_{+}$ Riemann coordinate
and not the $R_{-}$ companion coordinate.  In fact, the reverse would have
been the case if the sign of the $\sql$ term in (\ref{lagr}) had
been opposite.  The entire model is consistent with this substitution,
which is just the choice of identifying $\sqrt{g}$ with plus or
minus the sound speed $\sql/\rho$.

The energy-momentum tensor is constructed from (\ref{lagr}), and
its components are
\beqar
\label{hamil} T^{00} &=& {\cal H} = \frac12 \rho v^2 + \frac{\lambda}{\rho} -
\frac{\sql}{2}\psi \psi^\prime, \\
T^{01} &=& {\cal P} = \rho v, \label{fortwesev} \\
T^{10} &=& \frac{\rho v}{2} R_{+}R_{-} - \frac{\sql}{2}R_{+}
\psi \psi^\prime, \\
T^{11} &=& \rho  R_{+}R_{-} \ .
\eeqar
The expected conserved quantities of the system, the generators of the
Galileo group, are verified to be time-independent using the equations
of motion.  We have
\beqar
\label{Neq}N&=&\int \rmd x \, \rho, \\
P&=&\int \rmd x \, \rho v, \\
H&=&\int \rmd x \, \Bigl(\frac12 \rho v^2 + \frac{\lambda}{\rho} -
\frac{\sql}{2}\psi \psi^\prime \Bigr), \\
%&=& H_{0} - \frac{\sql}{2} \int{\rmd x \, \ppp} \equiv H_0 - \frac{\sql}{2}A_0, \\
\label{Beq}B&=&\int \rmd x \, \rho (x-vt) = \int{\rmd x \, x\rho} - t P,
\eeqar
%where it should be noted that $H_{0}$, the Hamiltonian without the
%last term, is 
%still separately time independent. 
Although  some generators look purely bosonic, there are still
Grassmann fields hidden 
in $v$ according to its definition (\ref{vdef}).  

In going to Riemann
coordinates, we can observe a ladder of conserved charges of the form
\beq
\label{bocharges}I^{\pm}_{n}=\int \rmd x \, \rho R_{\pm}^{n}  \ .
\eeq
The first few values of $n$ above give
\beqar
I^{\pm}_{0} &=& N, \\
I^{\pm}_{1} &=& P \pm \sql \Omega, \\
I^{+}_{2} &=& 2 H \ , 
%= 2 \Bigl(H_0 \mp \frac{\sql}{2} A_0 \Bigr),
\eeqar
where $\Omega$ is used to denote the length of space $\int \rmd x$.  (Note
that $I^{-}_{2},$ would correspond to the Hamiltonian of the theory
with $\sql$ replaced by its negative).  
%the  plus or minus choice has been absorbed
%into the freedom of defining $H$ to be $H_0 \mp \frac{\sql}{2}A_0$ as
%described above.  

In Section (4.2) we identified two different supersymmetry generators,
which correspond in one space dimension to the time independent
quantities
\beqar
\label{Qtildeeq} \tilde Q &=& \int \rmd x \, \rho \psi, \\
\label{Qeq}Q &=& \int \rmd x \, \rho \Bigl(v-\frac{\sql}{\rho}\Bigr) \psi \ .
\eeqar
These are again special cases ($n=0$ and $n=1$) of a ladder of
conserved supercharges described by
\beq
\label{supercharges}Q_{n} = \int \rmd x \, \rho R_{-}^n \psi \ .
\eeq
We see that the supercharges evaluated on the solution (\ref{sol})
reproduce the form of the bosonic charges \refeq{infnumconsmo}. 
 
Let us observe that there exist further bosonic and fermionic
conserved charges.  For example, one may verify that the bosonic charges
%\beqar\label{Acharges}
%I^{\pm}_{na} = \int \rmd x \, \rho R_{\pm}^n \Bigl(\frac{\partial R_{\pm}}{\rho}\Bigr)^a \ , \\
%K_{na} = \int \rmd x \, \rho R_{-}^n \Bigl(\frac{\partial R_{-}}{\rho}\Bigr)^a \ppp \\
%Q_{na} = \int \rmd x \, \rho R_{-}^n \Bigl(\frac{\partial R_{-}}{\rho}\Bigr)^a \psi \\
%S_{na} = \int \rmd x \, R_{-}^n \Bigl(\frac{\partial R_{-}}{\rho}\Bigr)^a \partial_x \psi \\
%\eeqar
%are all conserved for positive-integer values of the indices $n$ and
%$a$.  
\beqar
\label{Ibeq}\int \rmd x \, \rho R_{\pm}^n \Bigl(\frac{
R^\prime_{\pm}}{\rho}\Bigr)^m  \\
\label{Aeq} \int \rmd x \, \rho R_{-}^n \Bigl(\frac{\psi \psi^\prime}{\rho}\Bigr)
\eeqar 
are conserved, as are the fermionic charges
\beq\label{Keq} 
\int \rmd x \, \rho R_{-}^n \Bigl(\frac{\psi^\prime}{\rho}\Bigr).
\eeq
Conserved expressions involving higher derivatives may also be constructed.
The conservation of these quantities is easily understood when
the string world-sheet variables are used.  Then the above are written
as $\int \rmd \sigma R_{\pm}^n (\partial_\sigma R_{\pm})^m$, $\int
\rmd \sigma R_{-}^n (\psi \partial_\sigma \psi)$, and $\int
\rmd \sigma R_{-}^n (\partial_\sigma \psi)$, respectively.  Furthermore when
$R_{\pm}$ are evaluated on solutions, they become functions of $\tau
\pm \sigma$ [see Section 3.3], so that integration over $\sigma$ extinguishes the
$\tau$ dependence, leaving constant quantities.

\subsubsection*{(b) Canonical structure}
\addcontentsline{toc}{subsubsection}{(b) Canonical structure}

The equations of motion (\ref{rhoeq})-(\ref{theq}) can also be obtained
by Poisson bracketing with the Hamiltonian (\ref{hamil}) if the following
canonical brackets are postulated.
\beqar
\lbrace \theta(x),\rho(y) \rbrace &=& \delta(x-y)\\
\lbrace \theta(x),\psi(y)\rbrace &=&-\frac{\psi}{2\rho}\delta(x-y)\\
\lbrace \psi(x),\psi(y) \rbrace &=&-\frac{1}{\rho}\delta(x-y) 
\eeqar
where the last bracket, containing Grassmann arguments on both sides is
understood to be the anti-bracket. 
With these one verifies that the conserved charges in
(\ref{Neq})-(\ref{Beq}) generate the appropriate Galileo symmetry
transformations on the dynamical variables $\rho$, $\theta$, and
$\psi$.  Correspondingly the supercharges (\ref{Qtildeeq}),(\ref{Qeq})
generate super transformations.
\beqar
\tilde\delta\rho&=&0 \qquad\qquad\quad \, \delta\rho = -\eta \partial_x(\rho\psi) \\
\tilde\delta\theta &=& -\frac12 \eta \psi \qquad\quad \delta\theta =
-\frac12 \eta R_{+} \psi  \\
\tilde\delta\psi&=&-\eta \qquad\qquad \, \delta\psi = -\eta \psi \psi^\prime - \eta R_{-}
%\delta\rho &=& -\eta (\rho\psi)' \\
%\delta\theta &=& -\frac12 \eta R_{+} \psi \\
%\delta\psi &=& -\eta \psi \psi' - \eta R_+
\eeqar
which leave the Lagrangian (\ref{lagr}) invariant.
The algebra of the bosonic generators reproduces the algebra of the
(extended) Galileo group. The algebra of the supercharges is
\beqar
\{\bar{\eta} Q, \eta Q\} &=& 2 \bar{\eta}\eta   H,\\
\{\bar{\eta} \tilde Q, \eta \tilde Q\} &=&  \bar{\eta} \eta N, \\
\{\bar{\eta} \tilde Q, \eta Q\} &=& \bar{\eta} \eta(P - \sqrt{2\lambda}  \Omega) ,\\
\{B, Q\} &=& \tilde Q \ .
\eeqar

\subsubsection*{(c) Additional symmetries of the fluid model}
\addcontentsline{toc}{subsubsection}{(c) Additional symmetries of the fluid model}

As mentioned above, since the fluid model descends from the
superstring, it should possess an enhanced symmetry beyond the Galileo
symmetry in (1+1)-dimensions.  In fact, the following conserved charges
effecting time rescaling and space-time mixing are
also verified:
\beqar
D &=& t H- \int \rmd x \, \rho\theta \ , \\
G &=& \int \rmd x \, (x{\cal H} - \theta{\cal P}) \ ,
\eeqar

The
Galileo generators supplemented by $D$ and $G$ together satisfy the
Lie algebra of the (2+1)-dimensional Poincar\'{e} group, with $N$, $P$, and $H$ corresponding to the three translations and with
$B$, $D$ and $G$ forming the (2+1)-dimensional Lorentz group $SO(2,1)$:
\begin{equation}
\begin{array}{ccc}
\big\{B,D\big\}= B,\hfill
&\big\{G,B\big\}=D,\hfill
&\big\{D,G\big\}=G \ ,
\end{array}
\end{equation}
with Casimir
\beqar
C = B \circ G + G \circ B + D \circ D \ .
\eeqar
Adjoining the supercharges results in the super-Poincar\'{e} algebra
of (2+1)-dimensions. The Lorentz charges do not belong to the infinite towers of constants of
motion mentioned earlier.  Rather, they act as raising and lowering
operators.  One verifies for the $Q_n$ and $I_n^{+}$:
\begin{equation}
\begin{array}{ccc}
\big\{B,I_n^{+}\big\}=-n I_{n-1}^{+},\hfill
&\big\{D,I_n^{+}\big\}=(n-1) I_{n}^{+},\hfill
&\big\{G,I_n^{+}\big\}=(\frac{n}{2}-1) I_{n+1}^{+},\hfill
\\
\big\{B,Q_n\big\}=-n Q_{n-1},\hfill
&\big\{D,Q_n\big\}=(n-\frac12) Q_{n},\hfill
&\big\{G,Q_n\big\}=(\frac{n}{2}-\frac12) Q_{n+1}.\hfill
\end{array}
\end{equation}
(Note that the
$\{B,I_2^{+}\}$ bracket coincides with $\{B,2H\}$, which should equal
$-2P$ according to the Galileo algebra. But the above result,
viz. $-2I_1^{+}$, gives $-2(P-\sql\Omega)$.  This central addition arises from a term of
the form $\int\rmd x \rmd y \sql \,  x \delta^\prime(x-y),$
whose value is ambiguous, depending on the order of integration.)
The brackets with the $I_n^{-}$ do not close, but the
$I_n^{-}$ can be modified by the addition of another tower of constant
quantities, namely those of (\ref{Aeq}):
\beq
\label{Itildeeq}
\tilde{I}_n^{-} = I_n^{-} - \sql n(n-1) \int \rmd x
R_{-}^{n-2}\psi \psi^\prime \ .
\eeq
The modified constants obey the same algebra as $I_n^{+}$
\begin{equation}
\begin{array}{ccc}
\big\{B,\tilde{I}_n^{-}\big\}=-n \tilde{I}_{n-1}^{-},\hfill
&\big\{D,\tilde{I}_n^{-}\big\}=(n-1) \tilde{I}_{n}^{-},\hfill
&\big\{G,\tilde{I}_n^{-}\big\}=(\frac{n}{2}-1) \tilde{I}_{n+1}^{-}.
\end{array}
\end{equation}
Evidently $I_n^{+}$, $\tilde{I}_n^{-}$, and $Q_n$ provide irreducible, infinite dimensional representations for $SO(2,1)$, with the Casimir, in adjoint action, taking the form $l(l+1)$, and $l=1$ for $I_n^{+}$, $\tilde{I}_n^{-}$, and $l=1/2$ for $Q_n$.

We inquire about the algebra of the towers of extended charges
$I_n^{+}$, $\tilde{I}_n^{-}$, and $Q_n$.  While some (bosonic) brackets vanish,
others provide new constants of motion like those in
(\ref{Ibeq})-(\ref{Keq}) and their generalizations with more
derivatives.  Thus it appears that one is dealing with an open (super) algebra.

A final comment:
We have presented supersymmetric versions of fluid dynamical models in two
and one space dimension. These models enjoy supersymmetry as well as
extra ``dynamical'' symmetries tracing back to their origin in higher
dimensional supermembrane models. Other investigations of supersymmetric fluids are
reported in Ref. \cite{groonib}.

There remain some obvious open questions.
One is what other fluid interactions can be
obtained from the rich factory of (super)branes. For example, string theory
$D$-branes have gauge fields living on them.  Such gauge fields would
presumably remain in passing to a fluid model and may thus provide a
model of (super)magnetohydrodynamics from $D$-branes. 

Another question is the construction and properties of fluid models with
Grassmann variables in arbitrary dimensions. Such models would not
descend from supermembrane models and, consequently, would not enjoy
supersymmetry or other extended symmetries.
Note, nevertheless, that Grassmann Gauss potentials~$\psi$ can be used even
in the absence of
supersymmetry. For example, the above models with the last explicitly
fermion-dependent term omitted, possess a conventional, bosonic
Hamiltonian without supersymmetry, while the Grassmann
variables are hidden in ${\bf v}$ and occur only in the canonical
1-form.  These models would describe fluids with fundamental spin degrees
of freedom and it would be worthwhile to explore their physical properties
in this description.

\newpage
\section{NON-ABELIAN FLUIDS}

\subsection{Introduction}

In this Section we generalize fluid dynamics to systems
with non-Abelian charges.
We begin with some comments on the physical contexts in which such a
generalization might be useful and the scope and limitations of possible
approaches to the problem.

The quintessential example of a physical system with non-Abelian charges
is the quark-gluon plasma.
High energy collisions of heavy nuclei can produce a plasma state of quarks and gluons.
This new state of matter has recently been of great interest both theoretically
and in experiments at the RHIC facility and at CERN.
In fact, there is growing evidence that such a state has already been
achieved at
the RHIC facility \cite{mclerran}.
In attempting a theoretical description, there are basically two
approaches that we can use.
Since the plasma is at high temperatures, one can argue that the average
energy per particle is high enough to justify the use of
perturbative Quantum
Chromodynamics by virtue of asymptotic freedom.
However, it is known that because of the infrared divergences
various resummations, such as
summing hard thermal loop contributions, have to be done
before a perturbative expansion with control of the infrared
degres of freedom can be set up \cite{pisarski}. One has to address also
the question of chromomagnetic screening, because unlike the Abelian
plasma, there can be spatial screening of magnetic type interactions \cite{GPY}.
The expected end result is then a good description valid
at high
temperatures and for plasma states that are not too far from equilibrium,
since one is perturbing around the equilibrium state.
An alternative
approach, which may be more suitable for nondilute plasmas or for
situations far from equilibrium, would be to use a fluid mechanical
description.

We begin by observing that many of the general comments
given in the introduction, on deriving fluid mechanics from an underlying particle theory by statistical averages, will apply 
in the non-Abelian context as well.
Specifically for the  quark-gluon plasma,
some work along these lines was done many years ago using single
particle kinetic equations \cite{heinz}. The one-particle kinetic
equation takes the form
\beq
P^\mu \left[ {\partial \over \partial X^\mu} + g {\cal Q}_a F^a_{\mu\nu}
{\partial \over \partial P_\nu} + g f_{abc}A^b_\mu {\cal Q}_c
{\partial \over \partial {\cal Q}_a}\right] f(X,P,{\cal Q})= {\cal C} (f).
\label{na1}
\eeq
Here $f(X,P,{\cal Q})$ is the one-particle distribution function and ${\cal C}$
is the collision integral term taking account of scatterings of the
particles. $A^a_\mu$ and $F^a_{\mu\nu}$ are the potential and field for a
non-Abelian theory based on a gauge group with structure constants
$f_{abc}$. Here ${\cal Q}_a$ represents the classical color charge of the particle.

It may be interesting to note that, for a collisonless plasma, {\it i.e.}, with
${\cal C}=0$, the
Boltzmann equation (\ref{na1}) is the
equation for the distribution function for single particles obeying the
standard classical equations of motion
for non-Abelian particles --
the so-called Wong equations \cite{wong} -- which are 
\beqar
m{dX^\mu \over d\tau}&=&
P^\mu, \nonumber\\
m{dP_\mu \over d\tau}&=& g{\cal Q}_a F^a_{\mu\nu}P^\nu, \label{na2}\\
m{d{\cal Q}_a\over d\tau}&=& -g f_{abc} (P^\mu A^b_\mu ){\cal Q}_c.
\nonumber
\eeqar
As we shall see shortly, the motion of the color degrees of freedom can also be
described in a phase  space way; the appropriate space is the Lie group
modulo the maximal torus.

The Boltzmann equation (\ref{na1})
is invariant under gauge transformations
in the sense that if
$f(X,P,{\cal Q})$ solves (\ref{na1}), then so does
$f(X,P, U^{-1}{\cal Q} U)$.
As in the Abelian case, one may, in a dilute system,
seek a perturbative
solution of the form $f= f^{(0)}+g f^{(1)}+...$, where
$f^{(0)}=n_p$ is the equilibrium distribution;
the perturbative corrections can then give
the transport coefficients and fluid equations of motion \cite{heinz}.
Such an approach suffers from the same limitations mentioned in the introduction, namely, that
it can only be justified for dilute systems near equilibrium and in a semiclassical
approximation.
However, the fact that the equations for the Abelian fluid have a fairly large regime of validity,
despite the fact that it can be derived from an underlying particle theory
in a limited context, then prompts us to ask for
an \textit{a
priori} derivation of a non-Abelian fluid mechanics, which
incorporates the non-Abelian degrees of freedom, coupling to a
non-Abelian gauge field, \textit{etc.} This theory may be valid for
dense, nonperturbative and nondilute systems.  
This is the goal of this section.
In proceeding with the development of such a theory, it is useful to keep in mind some 
guidelines or desirable features.
First of all, a canonical
or symplectic formulation (at least in the conservative limit)
is important for quantization, so we should aim for this.
 At the same time, the analysis
based on the kinetic equations still  remains useful to us as a guide
for arriving at the equations of interest.
Our analysis is based on the paper in Ref \cite{bistro}.

\subsection{Non-Abelian Euler variables}\label{abeEul}
\setcounter{equation}{0}  % reset counter

A possible form for the non-Abelian Euler fluid variables may be inferred
from the single-particle equations of motion (\ref{na2}) by a procedure analogous to
what was one in Section 1.1 for the Abelian case.

The single-particle non-Abelian current is defined in terms of
the Lagrange variables $X^\mu$ and ${\cal Q}_a$, 
\beq
J^\mu_a (t, {\bf r} ) = \int d\tau ~ {\cal Q}_a(\tau ) ~
{d X^\mu (\tau ) \over d\tau} ~\delta ( X^0(\tau ) - ct)
~\delta ({{\bf X}} (\tau ) - {\bf r} ).
\label{na4a}
\eeq
With the parametrization $X^0 (\tau ) = c \tau$,
\beqar
\rho_a (t, {\bf r} )&=& {\cal Q}_a (t) ~\delta ( {{\bf X}}(t) - {\bf r} ),\label{na4b}\\
{{\bf J}}_a (t, {\bf r} )&=&  {\cal Q}_a (t) ~ {\dot {{\bf X}}}(t) ~\delta ({{\bf X}}(t)
- {\bf r} ).\label{na4c}
\eeqar
This generalizes (\ref{eqone2}) and (\ref{eqone6}) by inclusion of the dynamical
non-Abelian charge ${\cal Q}_a$ which satisfies (\ref{na2}), so that
$J^\mu_a$ is covariantly conserved.
\beq
(D_\mu \, J^\mu)_a \equiv \del_\mu J^\mu_a ~+~ f_{abc} A^b_\mu J^\mu_c =0
\label{na4d}
\eeq
(In the present case, the current is defined without the mass factor and it is
normalized by ${\cal Q}_a$.)
Passage to the fluid description is achieved as in Section 1.1. In the
many-particle case, dynamical quantities are decorated with
the particle label $n$, as in (1.1.8)-(\ref{eqone10}), which is summed in the
definition of the current. Then in the continuum fluid limit, $n$ is replaced by
the continuous variable ${\bf x}$ and the charge and current densities read
\beqar
\rho_a (t, {\bf r} ) &=& \int d^3x ~{\cal Q}_a (t, {\bf x} ) ~\delta ({{\bf X}}(t, {\bf x} ) - {\bf r}
),\label{na4e}\\
{{\bf J}}_a(t,{\bf r} ) &=&\int d^3x ~ {\cal Q}_a (t,{\bf x} ) ~{\dot {{\bf X}}}(t, {\bf x} )
~\delta ({{\bf X}}(t, {\bf x} )- {\bf r} ), \label{na4f}
\eeqar
with ${\cal Q}_a (t, {\bf x} )$ satisfying
\beq
{\dot {\cal Q}_a}(t, {\bf x} ) + f_{abc} \left[
c A^b_0 (t, {{\bf X}}(t,{\bf x} )) + {\dot {{\bf X}}}(t, {\bf x} )
\cdot {\vec A}^b (t, {{\bf X}}(t,{\bf x} ))\right] {\cal Q}_c (t, {\bf x} )
=0.
\label{na4g}
\eeq
[Notice that replacing a discrete sum by an integral over $\bf x$ forces ${\cal Q}_a (t,
{\bf x})$ to be a charge density.] 

Observe that, just as in the Abelian case discussed in
Section 1.1, the ${\bf x}$-integration evaluates ${\bf x}$ at $ \vec \chi (t,{\bf r} )$, the
inverse of
${{\bf X}}$, and the Jacobian factor $|\det \del X^i / \del x^j|_{{\bf x}= \boldsymbol \chi}$
is just the Abelian charge density $\rho$ [see (\ref{eighteen}), (\ref{eqone19})].
Thus the non-Abelian quantities factorize.
\beqar
\rho_a (t, {\bf r} )&=& Q_a (t, {\bf r} ) ~\rho (t, {\bf r} )\label{na4h}\\
{{\bf J}}_a (t, {\bf r} ) &=& Q_a (t, {\bf r} ) ~\rho (t, {\bf r} )
~{{\bf v}}(t, {\bf r} ) \label{na4i}
\eeqar
Equivalently
\beq
J^\mu_a (t, {\bf r} ) = Q_a (t, {\bf r} ) ~ j^\mu (t, {\bf r} ),
\label{na4j}
\eeq
where
\beqar
Q_a (t, {\bf r} ) &=& {\cal Q}_a (t, {\bf x} ) \vert_{{\bf x} = \chi},\label{na4k}\\
\rho (t, {\bf r} )~ Q_a (t, {\bf r} ) &=& \int dx {\cal Q}_a (t, {\bf x} ) ~ \delta ({{\bf X}}
(t, {\bf x} ) - {\bf r} ).\label{na4l}
\eeqar
As a consequence of its definition, the Abelian current factor $j^\mu = (c\rho, {\bf v} \rho)$
satisfies its own continuity equation (\ref{eqone16}).  Moreover, differentiating (\ref{na4l}) with
respect to time and using (\ref{eqone16}) and (\ref{na4g}) results in an equation for
${\dot Q}_a$,
\beq
{\dot Q}_a (t, {\bf r} ) ~+~ {{\bf v}}(t,{\bf r} )\cdot \nabla Q_a(t,{\bf r} ) =
-f_{abc} \left[
c A^b_0 (t, {\bf r} ) + {{\bf v}} (t, {\bf r} )
\cdot {\vec A}^b (t,{\bf r} )\right] Q_c (t, {\bf r} )
\label{na4m}
\eeq
which can also be written as
\beq
j^\mu (D_\mu Q)_a =0. \label{na4n}
\eeq
This is analogous to the Abelian equations (1.2.74). Equations
(\ref{na4m}) and (\ref{na4n}) can be understood from the fact that the covariantly
conserved current (\ref{na4a}), (\ref{na4d}) factorizes according to (\ref{na4j})
into a group variable $Q_a$ and a conserved Abelian current
$j^\mu$. Consistency of (\ref{eqone16}), (\ref{na4d}) and (\ref{na4j})
then enforces (\ref{na4n}).
Evidently this is a generalization of the particle Wong equation
(\ref{na2}); thereofre we shall refer to it as the fluid Wong equation.

We recognize that the formulas
(\ref{na4e}) and (\ref{na4f}) are the non-Abelian version of
the Lagrange variable-Euler variable
correspondence [see (\ref{eqone3}) and (\ref{eqone4})].
Also, (\ref{na4g}) is
the field generalization
of the particle Wong equation, presented in the Lagrange formalism, and
(\ref{na4m}) and (\ref{na4n}) are the Euler version
of the same.
The decomposition of the non-Abelian current
in (\ref{na4j}) is the non-Abelian version of the Eckart decomposition
(\ref{intrumu}).
Indeed, (\ref{na4j}) may be further factored as
in (\ref{intrumu}).
\beq
J^\mu_a (t, {\bf r} )= Q_a (t, {\bf r} ) ~n (t, {\bf r} )~ u^\mu (t, {\bf r} )
\label{na4p}
\eeq

In the remainder of Section 5, we are guided in our construction of
a dynamical model for non-Abelian fluid mechanics and ``color" hydrodynamics by
the above properties of the non-Abelian current, which follow from the very general
arguments, based on a particle picture for the substratum of
a fluid. In Sidebar G, at the end of this Section, we present a different model,
based on a field theoretic fluid substratum.

\subsection{Constructing the action}
\setcounter{equation}{0}  % reset counter

The equations of motion for the non-Abelian fluid in the Euler formulation include
the kinematical equations:  continuity (\ref{na4d}) and Wong (\ref{na4n}) that are
satisfied by the non-Abelian current, which is factorized as in (\ref{na4j})-(\ref{na4p}). Still
needed is the Euler force equation, analogous to (\ref{five14}), which specifies the dynamics.
We present this by constructing an action whose variation reproduces the kinematical
equations and gives a model for the dynamical equation. 

The algebra underlying the
non-Abelian theory is realized with anti-Hermitian generators $T_a$ satisfying
\beq
[T^a, ~T^b ] = f_{abc}~T^c,
\label{na5a}
\eeq
and normalized by
\beq
{\rm tr }~ T^a T^b = - {1\over 2} \, \delta^{ab}.
\label{na5b}
\eeq
In a canonical particle theory we expect that the algebra
(\ref{na5a}) is reproduced by Poisson
brackets for corresponding symmetry generators.
In a field theory, we expect to find a copy of (\ref{na5a}) at each point in space,
leading to the Poisson brackets
\beq
\{ \rho_a({\bf r}),~ \rho_b({\bf r}') \}
	= f_{abc}  ~\rho_c({\bf r} ) ~\delta({\bf r}-{\bf r}'),
\label{na5}
\eeq
which generalize (\ref{eqone36}). [A common time argument is suppressed.]
Upon quantization the brackets become commutators and 
acquire the factor $i /\hbar$.
(Schwinger terms do not spoil the quantum algebra
unless there are anomalies in the gauge symmetry \cite{refbib44}.
Of course, we assume that the theory is anomaly-free.)

%%The action which leads to the commutation rules
%%(\ref{na5}) is the Kirillov-Kostant form
%%\beq
%%{\cal S} = 2~ \mu~ \int dt~ \Tr (T_3 g^{-1} {\dot g}).
%%\label{na6}
%%\eeq
%%We have taken the group $SU(2)$ as an example;
%%hence $g$ is an element of
%%$SU(2)$. It may be specified by three parameters
%%$\theta^a$ and taken to be of the form
%%$\exp ( T_a \theta^a )$ where $T_a = - \sigma_a/2 i$ and $\sigma_a$  are the Pauli
%%matrices.
%%The action (\ref{na6}) is invariant under
%%$g\rightarrow g \exp (T_3 \vf )$ modulo surface terms.
%%Therefore, the theory is defined on the
%%two-dimensional sphere $SU(2)/U(1) = S^2$.
%%The observables are given by
%%$ S_a= -2\Tr (g T_3 g^{-1} T_a)$;
%%they can be shown to obey
%%the commutation rules
%%$[ S_a, S_b ] = i \ep_{abc} S_c$.
The action which leads to the commutation rules (\ref{na5}) is the field theoretic version of the
Kirillov-Kostant form, which for a particle (not a field) reads
\beq
I_{K K} = 2 n \int dt~ tr T^3 g^{-1} {\dot g}.
\label{na6}
\eeq
where $n$ is a normalization constant and we have taken the group $SU(2)$ as a concrete example: $g
\in SU(2), g= exp (T^a \varphi_a), T^a = \sigma^a/2i$ and
$\sigma^a$ are Pauli matrices. The action (\ref{na6}) is
invariant under $g \to g \, exp \, T^3 \, \varphi$, modulo
surface terms. Therefore, the theory governed by (\ref{na6}) is
defined on the 2-dimensional sphere $SU(2)/U(1) = S^2$. The
observables are given by $q_a = -2 \, tr \, (g
\, T^3\, g^{-1}\, T^a)$, one can show that they obey commutation rules
$\{q_a, q_b\} = \varepsilon_{abc} q_c$, which is the
single-particle version of (\ref{na5}). Upon quantization, the
quantum Hilbert space will consist of one unitary irreducible
representation of the group $SU(2)$ with the highest weight or
$j$-value given by $j = \half n$. The symplectic 2-form associated with (\ref{na6}) is the field of a
magnetic monopole on the two-sphere
$S^2 = SU(2)/U(1)$. The Dirac quantization rule
requires that $n$ be an integer, consistent with
$j$ being half-integral.
Interpreting the single irreducible representation
of the group as representing the charge degrees of freedom,
the action (\ref{na6}) describes a single particle with $SU(2)$ non-Abelian
charges. One can in fact use (\ref{na6})
as part of an action for the Wong equations of motion \cite{bacjac}, \cite{bal}.

More generally, consider a Lie group $G$ with $H$ denoting its Cartan subgroup (or
maximal torus).
The required generalized action is given by
\beq
I_0 =  \sum^r_{s=1}~{n_s} \int dt~ tr K_{(s)} g^{-1} {\dot g}
\label{na9}
\eeq
where $g\in G$ and $n_s$ are the highest weights defining a unitary irreducible
respresentation of $G$, $K_{(s)}$ are the diagonal generators of the
Cartan subalgebra $H$ of
$G$. The summation in (\ref{na9}) extends up to the rank $r$ of the
algebra, but some of the $n$'s could vanish.
The action
(\ref{na9}) is invariant under
$g
\rightarrow gh$,
$h\in H$ and time independent, so that the phase space is $G/H$, which is known to be a K\"ahler
(and symplectic) space. Quantization leads to a finite dimensional Hilbert space which
carries a unitary irreducible representation of $G$ labelled by the highest weights
$n_s$.

%%Given this structure, we see that we can generate
%%(\ref{na5}) by an action of the form
%%\beq
%%{\cal S} =  \sum_{{\bf r}} ~ \sum^r_{s=1}~{w_s} \int dt~ \Tr (K_{(s)} g^{-1} {\dot g})
%%\label{na10}
%%\eeq
%%We have approximated space by a lattice of points
%%to give a clear meaning to the idea of having a copy of the Lie algebra
%%at each point.
%%This action (\ref{na9}) is not yet what we want, there
%%is one other point we must take account of.
%%The matrix charge density $\rho = \rho_a T_a$
%%can be diagonalized at each
%%point in space by a transformation $g({\bf r} )$ so that we may
%%write $\rho = g^{-1} \rho_{diag} g$.
%%The diagonal density $\rho_{diag}$ is of the form
%%$\sum_s n_s K_{(s)}$, where $n_s$ are a set of scalar
%%densities. In general, $n_s$ can depend on ${\bf r}$, so we see that
%%the action must be summed over ${\bf r}$ with appropriate density factors.
%%Passing to the continuum limit keeping this in mind, we
%%see that the Lagrange density we are seeking must be of the form
Given this structure, we see that field theoretic generalization, which would give rise to (\ref{na5}), appears
as  $\int dt\, dr\, \sum ^r_{s=1} n_{(s)} \, tr \, (K_{(s)} \, g^{-1} \, \dot{g})$,
where now $n_s$ and $g$ depend on $\bf r$, while the $K_{(s)}$
remain as constant elements of the Cartan subalgebra of the
group. Thus we take as the Lagrange density for our
non-Abelian fluid dynamics the formula
\beq
{\cal L}= \sum_{s=1}^{r} j^\mu_{(s)}
2 ~\tr  K_{(s)} g^{-1}D_\mu g
-f(n_{(1)},n_{(2)},\ldots, n_{(r)}) + {\cal L}_{gauge}.
\label{na11}
\eeq
Here $j^\mu_{(s)}$ are a set of Abelian currents; they may be
taken to be in the Eckart form
$j^\mu_{(s)} = n_{(s)} u^\mu_{(s)}$,
where $u^\mu$'s are four-velocity vectors, with
$	u^\mu_{(s)} u_{(s) \mu} = 1 $.
As far as the variational problem of this action is concerned, we regard
$n_{(s)}$ as given by $j^\mu_{(s)}$ via $n_{(s)} = \sqrt{ j^\mu_{(s)} j_{\mu (s)}}$.
The space and time components of the currents are given by
$j^\mu_{(s)}= (c\rho_{(s)} , \rho_{(s)} \vec{v}_{(s)} )$, $\rho_{(s)} = n_{(s)}
u^0_{(s)}$.
In equation (\ref{na11})
$n_{(s)}$ are the invariant densities for the diagonal directions
of the Lie algebra.

Comparison with the usual form of the action for an Abelian fluid shows that what we
have obtained is the non-Abelian analogue of the irrotational part of the
flow. In the Abelian case, and without the gauge field coupling,
equation (\ref{na11}) entails a single contribution, $s=1$, and
$2\tr (K_1 g^{-1} \del_\mu g) = - \del_\mu \theta$, with vanishing vorticity.
In the non-Abelian
case, the vorticity is still nonvanishing.
One can easily generalize (\ref{na11}) to
include the other Gaussian components of the Clebsch parametrized vector which
couples to
$j^\mu_{(s)}$. This gives the Lagrangian
\begin{equation}
\La = \sum\limits_{s=1}^{r} j^\mu_{(s)} \left\{
2 \tr  K_{(s)}
g^{-1}D_\mu g +a_{\mu (s)}\right\}
    -f(n_{(1)},n_{(2)},\ldots, n_{(r)}) + \La_{gauge}
\label{na13a}
\end{equation}
where $a_{\mu (s)}$ is given by
\beq
a_{\mu (s)} = \alpha_{(s)}\partial_\mu \beta_{(s)}.\label{na13b}
\eeq

For the rank - one group $SU(2)$, with its single Cartan
element, the $s$-sum in (\ref{na11}) is exhausted by a single
element. We see that the $SU(2)$ fluid has one component.
For higher rank groups, the non-Abelian fluid can have up to
$r$ components, but fewer - indeed even just a single flow -
are possible when some of the densities
$n_{(s)}$ vanish. Mathematically, single-component fluids are the simplest, but
physically it is unclear what kinematical regimes of a quark-gluon plasma, for
example, would admit or even require such a reduction in flows.

The covariant derivative in (\ref{na11}), namely,
\beq
D_\mu g=\d_\mu g + A_\mu g, \label{na12}
\eeq
involves a dynamical non-Abelian gauge potential $A_\mu=A_\mu^a
T^a$ whose dynamics is provided by $\La_{gauge}$.
The first term in $\La$
contains the canonical 1-form for our theory and determines the
symplectic structure
ad the canonical brackets. We have added the Hamiltonian
density part, the function $f (n_{(1)},n_{(2)},\ldots, n_{(r)})$
which
describes the fluid dynamics. The theory is invariant under gauge transformations
with group element $U$
\beq
\begin{array}{rcl}
g &\to& U^{-1} g \\
A_\mu &\to& U^{-1}\left(A_\mu + \d_\mu\right) U\;.
\end{array}
\label{na13}
\eeq
where $U\in G$. In Side bar D we show that the canonical structure of our theory leads to the
charge-density algebra (\ref{na5}).

\vskip .1in
\hrule
\vskip .1in
{\addtocontents{toc}{\protect\hrulefill\par\vspace{-10pt}}
\noindent{\bf  D. Sidebar on the charge density algebra}
\addcontentsline{toc}{section}{\quad \ D. Sidebar on the charge density algebra}
\vskip .1in
\renewcommand{\theequation}{D.\arabic{equation}}
\setcounter{equation}{0}  % reset counter

The portion of the Lagrange density (\ref{na14}) that determines
the Poisson bracket is
\beq\label{eqb1}
\La_{canonical}=\rho \,2\tr K g^{-1}\dot{g} = \rho \,2\tr Q \dot{g} g^{-1}\;.
\eeq
where $Q \equiv g K g^{-1}$. The phase space is parameterized 
by the scalar density $\rho$ and parameteris $\varphi_a$ specifying the elements $g$ of the group
$G$. Thus the phase space is identified as as
\[
\{\mbox{set of all maps}\, \rho \, ({\bf r}) = {\bf R}^3 \to {\bf
R}_+, g({\bf r}) = {\bf R}^3 \to G \}
\]
With a parametrization of the group element, \textit{e.g.}
$g(\varphi) = e^{T^a \varphi_a}$, one sees that $\dot{g} g^{-1}$ has the form $ \dot{\varphi}_a
C^{a}_{\ b}(\varphi) T^b$, where the non-singular matrix $C^{a}_{\;
b}(\varphi)$ is defined by
\beq\label{eqb2}
C^{a}_{\ b}(\varphi) T^b = \frac{\d g(\varphi)}{\d \varphi_a}
g^{-1}(\varphi)\;.
\eeq
Thus
\beq\label{eqb3}
\La_{canonical}=-\rho \; \dot{\varphi}_a C^{a}_{\ b}Q_b
	= -\dot{\varphi}_a C^{a}_{\ b} \rho_b
\eeq
We give a new name to the combination
$C^a_{\ b} \rho_b$,
\beq
\Pi^a \equiv - C^a_{\, b} \rho_b
\label{eqb3b}
\eeq
The (\ref{eqb3}) reads
\beq
\La_{canonical}= \Pi^a \dot{\vf}_a
\label{eqb3a}
\eeq
Consequently, applying the formalism explained in
Sidebar A(a), we conclude immediately that $\Pi^a$ and $\vf_a$ are canonically
conjugate.
Moreover, the charge density
can be expressed in terms of $\Pi^a$ and $c^b_{\ a}$, the inverse
to $C^a_{\ b}$,
\beq\label{eqb4}
\rho_a= - c^{a}_{\ b} \Pi^b\;.
\eeq
\par
The non-Abelian charge density $\rho_a$ is a function of
$(t,{\bf r})$ and for (\ref{na5}) we need the bracket with
another density evaluated at $(t,{\bf r}')$. (The common $t$-dependence is
suppressed.) Since the dependence of $ c^{a}_{\ b}$ on $\varphi$ involves no spatial
derivatives of $\varphi$, it is clear that the brackets will be local in
${\bf r}-{\bf r}'$, just as is the bracket between $\varphi$
and $\Pi$.
\beqar\label{eqb5}
\{\rho_a({\bf r}),\rho_b({\bf r'})\}&=& \left(
	c^{a}_{\; a'}\frac{\d  c^{b}_{\; b'}}{\d \varphi_{a'}} \Pi^{b'}
	- \;a \leftrightarrow b\right)\delta \left({\bf r}-{\bf r}'\right)
\nonumber\\
&=& \left(
      -c^{a}_{\; a'}c^{b}_{\; c'}\frac{\d  C^{c'}_{\; c''}}
	{\d \varphi_{a'}}c^{c''}_{\; b'} \Pi^{b'}
      - \;a \leftrightarrow b\right)\delta \left({\bf r}-{\bf r}'\right)
\nonumber\\
&=& \left(c^{a}_{\; a'}c^{b}_{\; c'}\frac{\d  C^{c'}_{\; c''}}{\d
\varphi_{a'}}\rho_{c''}
      - \;a \leftrightarrow b\right)\delta \left({\bf r}-{\bf r}'\right)
\eeqar
To evaluate the derivative with respect to $\varphi$, return to
(\ref{eqb2}) and observe
\beqar\label{eqb6}
\frac{\d  C^{c'}_{\; c''}}{\d \varphi_{a'}} &=& -\frac{\d}{\d \varphi_{a'}}
	2\tr \frac{\d g}{\d \varphi_{c'}} g^{-1} T^{c''} 
\nonumber \\
&=& - 2 \tr\left( \frac{\d^2 g}{\d \varphi_{a'}
	\d \varphi_{c'}} g^{-1} - C^{c'}_{\; d'} T^{d'}
	C^{a'}_{\; d''} T^{d''}\right) T^{c''}
\eeqar
The first term in the parentheses is symmetric in $(a',c')$; when
inserted in (\ref{eqb5}) it produces a symmetric contribution in
$(a,b)$ and does not contribute when antisymmetrization in $(a,b)$
is effected. What is left establishes (\ref{na5}).
\beqar\label{eqb7}
\{\rho_a({\bf r}),\rho_b({\bf r}')\}&=&
\left(
      c^{a}_{\; a'} c^{b}_{\; c'} C^{c'}_{\; d'} C^{a'}_{\; d''}
	2 \tr T^{d'} T^{d''} T^{c''} \rho_{c''}
      - \;a \leftrightarrow b\right)
	\delta \left({\bf r}-{\bf r}'\right)
\nonumber\\
&=&
-\left( 2\tr T^{a}T^{b}T^{c''}\rho_{c''} - \;a \leftrightarrow
b\right)\delta \left({\bf r}-{\bf r}'\right)
\nonumber\\
&=& - 2 \tr f_{abd} T^{d} T^{c''}\rho_{c''}
	\delta \left({\bf r}-{\bf r}'\right)
\nonumber\\
&=&  f_{abc} \rho_{c}({\bf r})
	\delta \left({\bf r}-{\bf r}'\right)
\eeqar
For simplicity here we examined the single-channel case. The multi-channel case can
be treated similarly.

It is instructive to rederive the above result by using the general 
theory of canonical transformations, described in
Sidebar A (a), (d).

From (\ref{eqb3}) we see that the canonical 1-form has components
\begin{eqnarray}
a_\rho &=& 0, \nonumber\\
a_{\varphi_a} &=& - \rho C^a_{\ b} Q_b = - C^a _{\ b} \,\rho_b
\label{lthree}
\end{eqnarray}
[These generalize  $a_i$ in (\ref{lag}).] The symplectic 2-form [generalizing $f_{ij}$ in (A.3)]
reads
\begin{eqnarray}
f_{\rho \rho} ({\bf r}, {\bf r'}) &=&0, \nonumber\\
f_{\rho \varphi_a} ({\bf r}, {\bf r'}) &=& - \delta ({\bf r} - {\bf r'})
C^a_{\ b} \, Q_b, \nonumber\\ 
f_{\varphi_a \varphi_b}\, ({\bf r},
{\bf r'}) &=& \delta ({\bf r} - {\bf r^\prime}) \, \rho\, C^a_{\ c}\, C^b_{\ d} \,
Q_e \, f_{cde}.
\label{Leleven}
\end{eqnarray}
(The delta-function arises because the 1-form is a function of the coordinates and involves no coordinated derivatives.) It follows form (A.12) that the generator of an infinitesimal canonical transformation
\begin{eqnarray}
\delta \rho &=& - v^\rho \nonumber\\
\delta \varphi_a &=& - v^{\varphi_a}
\label{ltwelve}
\end{eqnarray}
obeys
\begin{subequations}\label{6.13}
\begin{equation}
\int d r' [v^{\varphi_a} f_{\varphi_a \rho}] = v^{\varphi_a} \, C^a_{\ b} \, Q_b =
\frac{\delta G}{\delta\rho}
\label{lthirteena}
\end{equation}
\begin{eqnarray}
\int d r' [v^\rho \, f_{\rho \varphi_a} + v^{\varphi_b} \, f _{\varphi_b \, \varphi_a}] 
= - v^\rho \, C^a_{\ b} \, Q_b - v^{\varphi_b}\, \rho \, C^a_{\ c} \, C^b_{\ d} \, Q_e \, f_{cde}
 =\frac{\delta G}{\delta \varphi_a}
\label{lthirteenb}
\end{eqnarray}
\end{subequations}

Let us now consider the left translation of $g$:
$\delta g = \epsilon_a \, T^a \, g,$ or equivalently $\delta \varphi_a = \epsilon_b \,
c^b_{\, a},\\ \delta Q_a = f_{abc} \, \epsilon_b \, Q_c,$ $\delta \rho_a = f_{abc} \,
\epsilon_b \, \rho_c$. Here $\epsilon_a$ is a function of ${\bf r}$. Thus we have $v^\rho = 0,
v^{\varphi_a}=\epsilon_b \, c^b_{\ a}$ and (\ref{6.13}) are solved by
\begin{equation}
G= -\int d r \, \rho \, Q_a \, \epsilon_a = -\int d r \, \epsilon_a \, \rho^a\, .
\label{lfourteen}
\end{equation}
From the general theory (A.13), we learn that by Poisson bracketing $G$ generates the above
transformation on any function of the phase space variables. Thus
\begin{eqnarray}
\bigg\{\int d r \, \epsilon^a ({\bf r}) \, \rho_a ({\bf r}), \rho_b ({\bf r'}) \bigg\} &=& 
-\delta \rho_b ({\bf r'})
\nonumber\\ &=& f_{abc} \, \epsilon_b \, ({\bf r'}) \, \rho_c \, ({\bf r'})\, .
\label{lfifteen}
\end{eqnarray}
Stripping away the arbitrary function $\epsilon_a({\bf r})$ reproduces (\ref{eqb7}).

Next we consider a right translation of $g$ by the Lie algebra element $K$, present in the canonical
1-form:
$\delta g = g\, K \lambda$, or equivalently $\delta \varphi_a = \lambda \, Q_b \, c^b_{\ a}$, leading to
$v^\rho=0, \quad v^{\varphi_a} = - \lambda \, Q_b \, c^b_{\ a}$. Here $\lambda$ is a function of $\bf
r$. Eqs. (\ref{6.13}) are now solved by
\begin{equation}
G= - \int d r \, \lambda \, \rho\, ,
\label{lsixteen}
\end{equation}
when $K$ is normalized to $tr \, K^2 = - \frac{1}{2}$. It then follows from (A.13), that once
$\lambda$ is stripped away,
\begin{equation}
\{\rho({\bf r}), g({\bf r}^\prime)\} = - \delta({\bf r}-{\bf r'})\, g\, K\, .
\label{lseventeen}
\end{equation}
and
\begin{equation}
\{\rho ({\bf r}), \ \rho({\bf r'})\}=0\, .
\end{equation}
\vspace{6pt}
\hrule 
\addtocontents{toc}{\protect\hrulefill\par}
}
\vspace{12pt}

\subsection{Equations of motion}\label{eqmo}
\setcounter{equation}{0}  % reset counter

In working out the equations of motion and other consequences of this theory, it is
instructive to consider first the simpler case of a single flow.
This is obtained for $G= SU(2)$, but could also occur in higher rank groups.

\subsubsection*{(i) Single component flow}
\addcontentsline{toc}{subsubsection}{(i) Single component flow}

The single flow Lagrangian takes the form
\beq
\La= j^\mu 2 \tr K g^{-1}D_\mu g - f(n) + \La_{gauge}\, ,
\label{na14}
\eeq
where $K =\sigma_3/2i$.
Now we have a single Abelian current $j^\mu$ which can be
decomposed  as
\beq
j^\mu=(c\rho, {\bf v} \rho)=n \,u^\mu\;,\qquad u^\mu u_\mu= 1\, .
\label{na15}
\eeq

The current $J^\mu_a$ to which
$A_\mu^a$ couples is easily worked out from (\ref{na14}).
Upon defining
\beq
g K g^{-1} \equiv Q = Q_a T^a \, ,
\label{na15a}
\eeq
we find
\beqar
J^\mu_a &=& (c\rho_a , {{\bf J}}_a )= Q_a j^\mu = Q_a ~n~u^\mu\, ,
\label{na16}\\
\rho_a&=& \rho Q_a.\nonumber
\eeqar
$Q_a$ may be thought of as the charge of a single particle, with
$\rho_a= \rho Q_a$ as the non-Abelian charge density at the point ${\bf r}$.
Notice that this is of the
Eckart form where the currents are given by charge densities multiplied by the
velocity vector. This agrees with the general discussion in Section (\ref{abeEul}), where we
showed that a particle based model for the fluid leads to the Eckart forms
(\ref{na4h}), (\ref{na4i}).

The gauge invariance of the Lagrangian (\ref{na14}) shows that
the current $J^\mu\equiv
J_a^\mu T^a $ must be covariantly conserved, {\it i.e.},
\beq
(D_\mu J^\mu )_a = \del_\mu J^\mu_a ~+~ f_{abc} A^b_\mu J^\mu_c
=0\label{na17}
\eeq
Further the current $j^\mu$ satisfies an ordinary conservation law,
\beq
\del_\mu j^\mu =0\label{na17a}
\eeq
In Sidebar E, we show that both conservation
laws are a consequence of invariance of the action with respect to
variations of the group element $g$: \negthickspace arbitrary left variations of $g$
lead to covariant conservation of $ J^\mu_a$ (\ref{na17}) while
the particular variation $\delta g= g K \lambda$, with $\lambda$ an arbitrary
function of space-time, ensures that
$j^\mu$ is conserved as in (\ref{na17a}).
We shall see later that these can be interpreted in terms of the Wong
equations. Notice that these conservation laws
lead to the fluid Wong equation
\beq
j^\mu (D_\mu Q)_a =0
\label{na18}
\eeq
as has already been noted in (\ref{na4n}).
\vskip .1in
\hrule
\vskip .1in
{\addtocontents{toc}{\protect\hrulefill\par\vspace{-6pt}}
\noindent{\bf E. Sidebar on varying the group element $g$}
\addcontentsline{toc}{section}{\quad \ E. Sidebar on varying the group element $g$}
\vskip .1in
\renewcommand{\theequation}{E.\arabic{equation}}
\setcounter{equation}{0}  % reset counter

We determine the variation of
\beq\label{eqa1}
I_0=\int\dd t \dd r \sum\limits_{s=1}^{r} j^\mu_{(s)}\,
	2\tr K_{(s)} g^{-1}D_\mu g\, ,
\eeq
when $g$ is varied either arbitrarily or in the specific manner
\beq\label{eqa2}
g^{-1}\delta g = K_{(s')} \lambda\, ,
\eeq
where $\lambda$ is an arbitrary function on space-time. This
will provide the needed results (\ref{na17}) and (\ref{na17a})
for the single channel situation, as well as  for many channels in (\ref{na33}) and
(\ref{na34}), below.
\par
Recall the definitions $Q_{(s)}=g K_{(s)}g^{-1}$ and
$D_\mu g = \d_\mu g + A_\mu g$, which implies
$D_\mu g^{-1} = \d_\mu g^{-1} - g^{-1} A_\mu$.
First, the variation of $g^{-1}D_\mu g$ is established.
\beq\label{eqa3}
\delta\left(g^{-1}D_\mu g\right)
	= -g^{-1}\delta g g^{-1} D_\mu g
	+ g^{-1}D_\mu \delta g
\eeq
To evaluate the last term, note that $D_\mu \delta g =
D_\mu\left(g g^{-1} \delta g\right)= (D_\mu g) g^{-1} \delta g +
g D_\mu\left(g^{-1} \delta g\right)$.
Thus
\beq\label{eqa4}
\delta\left(g^{-1}D_\mu g\right) = \d_\mu \left(g^{-1} \delta g\right) +
	[ g^{-1} D_\mu g,g^{-1} \delta g]\;.
\eeq
Inserting (\ref{eqa4}) into the variation of $I_0$ in (\ref{eqa1}),
integrating by parts, and rearanging the trace with $K_{(s)}$ gives
\beq\label{eqa5}
\delta I_0 = -\int \dd t \dd r \sum\limits_{s=1}^{r} \left(\d_\mu
j^\mu_{(s)} \, 2 \tr K_{(s)} g^{-1} \delta g + j^\mu_{(s)} 2 \tr \left[
g^{-1} D_\mu g , K_{(s)}\right] g^{-1}\delta g\right)\;.
\eeq
\par
Considering first arbitrary variations: the vanishing of $\delta
I_0$ requires
\beq\label{eqa6}
\sum\limits_{s=1}^{r}\left(\d_\mu j^\mu_{(s)} K_{(s)}
	+ j^\mu_{(s)} \left[ g^{-1} D_\mu g , K_{(s)}\right]\right)=0
\eeq
or, after sandwiching the above between $g\ldots g^{-1}$,
\beq\label{eqa7}
\sum\limits_{s=1}^{r}\left(\d_\mu j^\mu_{(s)} Q_{(s)}
	+ j^\mu_{(s)} \left[ D_\mu g
	g^{-1}, Q_{(s)}\right]\right)=0\;.
\eeq
Finally we verify that
\beq\label{eqa8}
\left[ D_\mu g g^{-1}, Q_{(s)}\right]= D_\mu Q_{(s)}\;,
\eeq
so that the desired results (\ref{na17})  and (\ref{na33}) follow.
\beq\label{eqa9}
\sum\limits_{s=1}^{r}\left(\d_\mu j^\mu_{(s)} Q_{(s)}
	+ j^\mu_{(s)}  D_\mu Q_{(s)} \right)
	=D_\mu\left(\sum\limits_{s=1}^{r} j^\mu_{(s)}
	Q_{(s)}\right)= D_\mu J^\mu =0  
\eeq
\par
Next we consider the specific variation (\ref{eqa2}) and separate
the sum (\ref{eqa5}) into the term $s=s'$ and $s\neq s'$. After
a rearrangement of the last term in (\ref{eqa5}), we get
\beqar\label{eqa10}
\delta I_0 &=& -\int\dd t \dd r \left( \d_\mu j^\mu_{(s')} \, 2\tr
K_{(s')}K_{(s')}\lambda +  j^\mu_{(s')} \, 2\tr g^{-1}D_\mu g
 [K_{(s')},K_{(s')}]\lambda\right)
\nonumber \\ &&
+\sum\limits_{s\neq s'} \left( \d_\mu j^\mu_{(s)} \, 2\tr
K_{(s)}K_{(s')}\lambda +  j^\mu_{(s)} \, 2\tr g^{-1}D_\mu g
 [K_{(s)},K_{(s')}]\lambda\right)\, .
\eeqar
The first commutator vanishes; so does the second when $K_{(s)}$
and $K_{(s')}$ commute, \textit{i.e.} when they belong to the
Cartan subalgebra. Also $2\tr K_{(s)}K_{(s')}= - K_{(s)}^a
K_{(s')}^a $; for $s'=s$ this is constant, while for $s'\neq s$
it vanishes when it is arranged that distinct elements of the
Cartan algebra are selected.  Thus for stationary variations
$j^\mu_{(s)}$ must be conserved, and (\ref{na17a}) as well as
(\ref{na34}) are validated.
\vskip .1in
\hrule
\vskip .1in
}
\setcounter{equation}{7}
It remains to derive the Euler equation. This is accomplished by
varying $j^\mu$; stationary variation requires
\beq
2 \tr \left[ Q (D_\mu g) g^{-1}\right] =
2 \tr (K g^{-1} D_\mu g ) =  \frac{u_\mu}{c^2} f'(n)\, ,
\label{na19}
\eeq
which we call the non-Abelian Bernoulli equation. The Euler
equation then follows, as in the Abelian case, by taking the curl\, .
\beq
\d_\mu \left(2\tr Q (D_\nu g) g^{-1}\right) -
	\d_\nu \left(2\tr Q (D_\mu g) g^{-1}\right)
= \d_\mu \left(\frac{u_\nu}{c^2} f'(n)\right)
	- \d_\nu \left(\frac{u_\mu}{c^2} f'(n)\right)
\label{na20}
\eeq
In Sidebar F, we show that manipulating the left
side allows rewriting (\ref{na20}) as
\beq
2 \tr (D_\mu Q) (D_\nu g) g^{-1}
	+ 2 \tr  Q F_{\mu\nu} 
=  \d_\mu \left(\frac{u_\nu}{c^2} f'(n)\right)
      - \d_\nu \left(\frac{u_\mu}{c^2} f'(n)\right)\;.
\label{na21}
\eeq
Finally, contracting with $j^\mu=n u^\mu$ and using (\ref{na17})
produces the relativistic, non-Abelian Euler equation.
\beq
	\frac{n u^\mu}{c^2}\d_\mu \left(u_\nu f'(n)\right)
	- n \d_\nu  f'(n)=2 \tr  J^\mu F_{\mu\nu}
\label{na22}
\eeq
The left side is of the form of the usual Abelian Euler equation;
the
right side describes the non-Abelian Lorentz force acting on the charged fluid.

\vskip .1in
\hrule
\vskip .1in
{%\addtocontents{toc}{\protect\hrulefill\par}
\noindent{\bf F. Sidebar on manipulating equation (\ref{na20})}
\addcontentsline{toc}{section}{\quad \ F. Sidebar on manipulating equation (5.4.9)}
\vskip .1in
\renewcommand{\theequation}{F.\arabic{equation}}
\setcounter{equation}{0}  % reset counter

Observe that the first term in  (\ref{na20}) equals
\beq\label{eqc1}
\d_\mu 2 \tr  Q (D_\nu g) g^{-1}=
2 \tr \bigg( (D_\mu Q) (D_\nu g) g^{-1}+
Q (D_\mu D_\nu g ) g^{-1} - Q  (D_\nu g) g^{-1}  (D_\mu g) g^{-1}\bigg).
\eeq
The first term on the right side is rewritten with the help of
(\ref{eqa8}) and combined with the last term, leaving
\begin{displaymath}
2\tr \left( Q (D_\mu D_\nu g) g^{-1} - Q (D_\mu g) g^{-1} (D_\nu g) g^{-1}
\right).
\end{displaymath}
After antisymmetrization in $(\mu,\nu)$, the left side of
(\ref{eqc1}) reads
\beq\label{eqc2}
2\tr Q \left(([D_\mu, D_\nu]g) g^{-1}
	-[ (D_\mu g) g^{-1} ,(D_\nu g) g^{-1}]\right)=
	2\tr\left( Q F_{\mu\nu}
	- [Q,(D_\mu g) g^{-1}] (D_\nu g) g^{-1}\right).
\eeq
When (\ref{eqa8}) is used again, (\ref{eqc2}) becomes the left
side of (\ref{na21}).
\vskip .1in
\hrule
\vskip .1in
\addtocontents{toc}{\protect\hrulefill\par}}
\setcounter{equation}{11}
The curved space generalization of the action for the Lagrangian (\ref{na13})
is given by
\beq
I = \int \sqrt{-\eta}~ d^4x~ \bigg( j^\mu 2 \tr  K g^{-1}D_\mu g
 - f(n) \bigg) ~+~I_{gauge} \, ,
\label{na23}
\eeq
where $n = \sqrt{ j^\mu j^\nu \eta_{\mu\nu}}$. (We use $\eta$ for the metric even in
curved space to avoid confusion with $g$, the group element.)
The variation of this action with respect to the metric,
$\delta I = - \half \int \sqrt{-\eta} ~\delta\eta_{\mu\nu}
~{\Theta}^{\mu\nu}$ identifying the total energy-momentum tensor
${\Theta}^{\mu\nu}$ as
\beqar
{\Theta}^{\mu\nu}&=& \theta^{\mu\nu} ~+~ \theta^{\mu\nu}_{gauge}\, ,\nonumber\\
\theta^{\mu\nu}&=&-\eta^{\mu\nu} [n
f'(n) - f(n)]
	+\frac{u^\mu u^\nu}{c^2} n f'(n)\, .
\label{na24}
\eeqar
We have used equation (\ref{na18}) to eliminate $\tr K g^{-1}
D_\mu g$ in the matter part of the energy-momentum tensor
$\theta^{\mu\nu}$. This behaves as the corresponding Abelian
expression (\ref{enmombec}), except that now there is an interaction
with a non-Abelian gauge field.

The divergence of $\theta^{\mu\nu}$
entails two independent parts: one proportional to $u^\nu$ and
the other orthogonal to it, compare (\ref{five12}).
\beq
\d_\mu \theta^{\mu\nu}= \d_\mu (n u^\mu) \frac{u^\nu f'(n)}{c^2}
	+ n\left[\frac{u^\mu}{c^2} \d_\mu (u^\nu f'(n))
	-\d^\nu f'(n) \right]
\label{na25}
\eeq
The first vanishes by the virtue of (\ref{na17a}) and the rest is
evaluated from the Euler equation (\ref{na22}), leaving
\beq
%\label{eq327b}
\d_\mu \theta^{\mu\nu}=  2 \tr J_\mu F^{\mu\nu} \, ,
\label{na26}
\eeq
which is canceled by the divergence of the gauge-field
energy-momentum tensor.
\beq
\d_\mu \theta^{\mu\nu}_{gauge}= - 2 \tr
	 J_\mu F^{\mu\nu} \label{na27}
\eeq
Thus we have conservation of the total energy-momentum tensor.

We record the nonrelativistic limit of the Euler equation
(\ref{na22}).
For small velocities, we may write, as in Section 1.4.
\beqar
n &\approx& \rho -\frac{1}{2 c^2}\rho {\bf v}^2 \nonumber\\
u^\mu &\approx& (c , {\bf v})\label{na28}
\eeqar
Further, we take $f(n)$ to be of the form
\beq
f=n c^2 +V(n)\, .
\label{na29}
\eeq
With these simplifications, we find that the
nonrelativistic limit for the spatial component of (\ref{na22})
gives the Euler equation with a non-Abelian Lorentz force
\beq
\dot{\bf v} + {\bf v}\cdot\nabla {\bf v} = \vec{force}
	+ Q_a \vec{E}^a + \frac{{\bf v}}{c} \times Q_a \vec{B}^a\, ,
\label{na30}
\eeq
where $\vec{force}$ is the pressure force coming from the potential
$V$ (and is therefore Abelian in nature), while non-Abelian force
terms involve the chromoelectric and chromomagnetic fields.
\beq
E^i_a = c F_{0i}^a \;, \qquad
B^i_a = -\frac{1}{2} \epsilon^{ijk} F^a_{jk}\label{na31}
\eeq

It is seen that the non-Abelian fluid moves effectively in a
single direction specified by $\vec{j}=\rho\vec{v}$. Nevertheless,
it experiences a non-Abelian Lorentz force.

Eventhough we wrote down the curved space
action (\ref{na23}) primarily to obtain the energy-momentum tensor,
we note that it may
be useful for relativistic astrophysics with the
quark-gluon plasma as in early universe or perhaps in the interior of neutron stars.

\subsubsection*{(ii) Multi-component flow}
\addcontentsline{toc}{subsubsection}{(ii) Multi-component flow}
\setcounter{equation}{20}

We now return to the more general multi-component action (\ref{na11}).
The current which couples to the gauge
potential is now
\beq
J^\mu=\sum\limits_{s=1}^{r} Q_{(s)} j^\mu_{(s)} \;,
\qquad \mbox{with}\qquad
Q_{(s)} = g K_{(s)} g^{-1}\;.
\label{na32}
\eeq
Arbitrary variation of $g$ ensures that (\ref{na32}) is
covariantly conserved, $D_\mu J^\mu =0$,
but we also need the conservation of
individual $j^\mu_{(s)}$.  This is achieved by considering special
variations of $g$ of the form
$\delta_{(s)} g = g K_{(s)} \lambda_{(s)}$.
These variations of $g$ lead to
\beq
\d_\mu j^\mu_{(s)} = \dot{\rho}_{(s)} + \nabla\cdot ({\bf v}_{(s)} \rho_{(s)}
)=0
\label{na33}
\eeq
The fluid Wong equation which follows from the conservation of the
Abelian and non-Abelian currents now reads
\beq
\sum\limits_{s=1}^{r} j^\mu_{(s)}D_\mu Q_{(s)}=0 \;.
\label{na34}
\eeq
Varying the individual $j^\mu_{(s)}$ in (\ref{na13a})
produces the Bernoulli
equations
\beq
2 \tr  Q_{(s)} (D_\mu g) g^{-1} = \frac{u_\mu}{c^2}\,
f^{(s)},
\label{na35}
\eeq
where
\beq
 f^{(s)} \equiv
	\frac{\d}{\d n_{(s)} } f(n_{(1)},n_{(2)},\ldots,n_{(r)}).
\label{na35a}
\eeq
As in the single channel case,
the curl of equation (\ref{na35}) can be cast in the
form
\beq
\frac{1}{c^2} \left\{ \d^\mu \left(u^\nu_{(s)} f^{(s)}\right)
      - \d^\nu \left(u^\mu_{(s)} f^{(s)}\right)\right\}
=2 \tr\bigg((D^\mu Q_{(s)}) (D^\nu g) g^{-1}
      +  Q_{(s)} F^{\mu\nu} \bigg) 
\;.
\label{na36}
\eeq
When contracted with $j^\mu_{(s)}=n_{(s)} u^\mu_{(s)}$, this leaves
\beq
\frac{n_{(s)}u^{\mu}_{(s)} }{c^2}
	\d_\mu \left(u^\nu_{(s)} f^{(s)}\right)
      - n_{(s)} \d^\nu  f^{(s)}
=j_{\mu (s)} 2 \tr \bigg( (D^\mu Q_{(s)}) (D^\nu g) g^{-1}
      + Q_{(s)} F^{\mu\nu} \bigg)
\;.
\label{na37}
\eeq

However, unlike in the single channel case, the right side does
not simplify since $j_{\mu (s)} Q_{(s)} $ cannot be replaced by
$J_\mu$ because the latter requires summing over $s$. Also the
first right-hand term in (\ref{na37}) does not vanish since
(\ref{na34}) requires summation over $s$. Equations (\ref{na37}) are the Euler
equations for the multicomponent non-Abelian fluid.

The matter part of the energy-momentum tensor is now given as
\beq
\theta^{\mu\nu}=-g^{\mu\nu} \left(
	\sum\limits_{s=1}^{r} n_{(s)} f^{(s)} - f\right)
	+ \sum\limits_{s=1}^{r}
	\frac{ u^\mu_{(s)} u^\nu_{(s)} }{c^2} n_{(s)} f^{(s)}\;.
\label{na38}
\eeq
Its divergence is of the form (\ref{na25}).
\beq
\d_\mu \theta^{\mu\nu} =  \sum\limits_{s=1}^{r}\left\{
	\left(\d_\mu (n_{(s)}u^\mu_{(s)})\right)
	\frac{u^\nu_{(s)} f^{(s)}}{c^2}
	+  n_{(s)}\left[ \frac{u^\mu_{(s)} }{c^2}
	\d_\mu\left( u^\nu_{(s)} f^{(s)}\right)
	-  \d^\nu  f^{(s)}\right]
\right\}
\label{na39}
\eeq
Using (\ref{na33}) and (\ref{na37}), the right side of this
equation is evaluated as
\begin{displaymath}
\sum\limits_{s=1}^{r} j_{\mu (s)} 2 \tr 
	\bigg( ( D^\mu Q_{(s)}) (D^\nu g) g^{-1}
	+  Q_{(s)} F^{\mu\nu}\bigg) 
.
\end{displaymath}
Since now we are summing over all channels, it follows from
(\ref{na32}) and (\ref{na34}) that, as before,
\beq
\d_\mu \theta^{\mu\nu} =   2 \tr J_\mu F^{\mu\nu} \, .
\label{na40}
\eeq

Some simplifications which lead to a more transparent physical
picture occur if
the dynamical potential separates
\beqar
f(n_{(1)},\ldots,n_{(r)}) &=& \sum\limits_{s=1}^{r} f_{(s)}(n_{(s)})\nonumber\\
f^{(s)}&=& f'_{(s)}
\label{na41}
\eeqar
Then the left side of (\ref{na37}) refers only to variables
labeled $s$, while the right side may be rewritten with the help of
the generalized Wong equation (\ref{na34}) to give
\beq
\frac{n_{(s)} u^{\mu}_{(s)}}{c^2} \d_\mu \left(
	u^\nu_{(s)} f'_{(s)}\right) - n_{(s)} \d^\nu  f'_{(s)}
=2 \tr \bigg(J_{\mu} F^{\mu\nu} 
 - \sum\limits_{s'\neq s}^{r} j_{\mu (s')}
      \left( Q_{(s')} F^{\mu\nu} + (D^\mu Q_{(s')}) (D^\nu g)
g^{-1}\right) \bigg) \;.
\label{na42}
\eeq
Thus in the addition to the Lorentz force, there are forces
arising from the other channels $s'\neq s$.

Note that for separated dynamics (\ref{na41}), the energy-momentum
tensor also separates,
\beq
\theta^{\mu\nu}=\sum\limits_{s=1}^{r} \theta^{\mu\nu}_{(s)}
=\sum\limits_{s=1}^{r}\left\{ -g^{\mu\nu}
	\left( n_{(s)} f'_{(s)} - f_{(s)}(n_{(s)}) \right)
	+  \frac{ u^\mu_{(s)} u^\nu_{(s)} }{c^2}
	n_{(s)} f'_{(s)}\right\}\;.
\label{na43}
\eeq
but the divergence of each individual $T^{\mu\nu}_{(s)}$ does not
vanish. This is just as expected since energy can now be
exchanged
between the different channels and with the gauge field;
this is also evident
from the equation of motion (\ref{na42}). It is clear that this
fluid moves with $r$ different velocities ${\bf v}_{(s)}$.

The single-channel Euler equation (\ref{na22}) is expressed in
terms of physically relevant quantities (currents, chromomagnetic
fields); the many-channel equation (\ref{na37}) involves,
additionally, the gauge group element $g$.  One may simplify
that equation by going to special gauge, for example $g=I$,
so that the right side of (\ref{na37}) reduces to
\beq\
j^{\mu}_{(s)} 2 \tr \bigg( (D_\mu Q_{(s)}) (D_\nu g) g^{-1}
      +  Q_{(s)} F_{\mu\nu} \bigg)
= j^\mu_{(s)} 2 \tr  K_{(s)}  (\d_\mu A_\nu-\d_\nu A_\mu)
\label{na44}
\eeq
while the Wong equation (\ref{na34}) becomes
\beq
\sum\limits_{s=1}^{r} j^\mu_{(s)} [A_\mu,K_{(s)}]=0.
\label{na45}
\eeq
It is interesting that in this gauge the nonlinear terms in
$F_{\mu\nu}$ disappear.

Observe that the inclusion of the $\alpha, ~\beta$-components of the
current, or the use of the Lagrangian (\ref{na13a}), does not change the form of the
equations of motion for fluids when expressed in terms of the velocities
and densities. The expressions for these quantities in terms of
the group parameters  and $\alpha$, $\beta$ will, of course, be altered.

\vskip .1in
\hrule
\vskip .1in
\addtocontents{toc}{\protect\vspace{29pt}}
{
\noindent{\bf G. Sidebar on field-based fluid mechanics}
\addtocontents{toc}{\protect\hrulefill\par\vspace{-10pt}}
\addcontentsline{toc}{section}{\quad \ G. Sidebar on field-based fluid mechanics}
\vskip .1in
\renewcommand{\theequation}{G.\arabic{equation}}
\setcounter{equation}{0}  % reset counter
In the body of our review, we presented Eulerian variables, their connection to
Lagrangian variables and their equations of motion in a picture derived from an
underlying physical reality composed of point particles, whose discrete
distribution is approximated by a continuum. For the Abelian case this was developed
in Section 1.1. The hallmark of this approach is that the fluid current factorizes
in an Eckart form. For the Abelian current we have
\beq
j^\mu = nu^\mu\, ,
\label{P1}
\eeq
which is no restriction at all, but for the non-Abelian current
the further factorization is nontrivial,
\beq
J^\mu_a = Q_a j^\mu = Q_a ~nu^\mu\, .
\label{P2}
\eeq

Here we present an alternative view of fluid mechanics, which is field-based
as opposed to particle-based. In the Abelian case, it results in fluid equations
that coincide with those of the particle-based derivation. This is in
keeping with
the fact that the Abelian Eckart form is not restrictive. However for the
non-Abelian situation, the field-based picture results in equations which are
different and much less elegant than the particle-based ones discussed in the body
of the text.

A field-based realization of the Euler equations for an Abelian fluid is provided by
the Madelung ``hydrodynamical" rewriting of the Schr\"odinger equation \cite{Mad}.
\beq
i\hbar \, \dot{\psi}
= -{\hbar^2 \over 2m} \vec \nabla^2 \psi
\label{P3}
\eeq
(We consider only the free equation.)
Upon presenting the wave function as
\beq
\psi (t, {\bf r} ) = \sqrt{\rho (t, {\bf r} )}~ e^{im \theta (t,{\bf r})/\hbar}
\label{P4}
\eeq
we find that the imaginary part of (\ref{P3}) results in the continuity equation
for
$j^\mu = (c\rho , {\bf j})$, where the spatial current ${\bf j}$
is also the quantum current
$(\hbar /m)~{\rm Im} \psi^* \vec \nabla \psi$.
When ${\bf j}$ is written as $ {{\bf v}} \rho$, ${{\bf v}}$ is identified as
$\nabla \theta$; the velocity is irrotational.
The real part of (\ref{P3}) gives the Bernoulli equation, with a quantum force
derived from
\beq
V = {\hbar^2 \over 2m^2} (\vec \nabla \sqrt{\rho} )^2
= {\hbar^2 \over 8m^2} {(\vec \nabla \rho )^2 \over \rho}
\label{P5}
\eeq
The Euler equation follows by taking the gradient of the Bernoulli equation.
In this way, we arrive again at the conventional irrotational fluid.

The story changes if we start from a non-Abelian Schr\"odinger equation.
Again we consider the free case with nonrelativistic kinematics.
Thus the equation involves a multi-component wave function
$\Psi$, with
\beq
i\hbar \dot{\Psi}
= -{\hbar^2 \over 2m} \vec \nabla^2 \Psi
\label{P6}
\eeq
The color degrees of freedom lead to the conserved non-Abelian current.
\beqar
J^\mu_a &=& ( c \rho_a , {{\bf J}}_a ) \nonumber\\
\rho_a = i \Psi^\dagger T^a \Psi,  && ~~~~
{{\bf J}}_a = {\hbar \over m} {\rm Re} \Psi^\dagger T^a \vec \nabla \Psi
\label{P7}
\eeqar
The singlet current $j^\mu = ( c\rho , {\bf j})$ is also conserved.

For definiteness and simplicity, we shall henceforth
assume that the group is $SU(2)$ and that the representation is the
fundamental one: $T^a={\sigma^a}/{2i}$, $\{T^a,T^b\}=-\delta_{ab}/2$.
We shall also set the mass $m$ and Planck's constant $\hbar$ to unity \cite{Bohm55}. The
non-Abelian analogue of the Madelung decomposition
(\ref{P4}) is
\beq\label{P8}
\Psi=\sqrt{\rho} ~g u\, ,
\eeq
where $\rho$ is the scalar $\Psi^\dagger\Psi$, $g$ is a
group element, and $u$ is a constant vector that points in
a fixed direction [{\it e.g.}, for $SU(2)$ the two-component
spinor $u$ could be taken as $u_1 =1,~u_2 =0$, then $i
u^\dagger T^a u= \delta^a_{3}/2$].  The singlet density is $\rho$,
while the singlet current ${\bf j} = I m \Psi^\ast \vec \nabla \Psi$ is
\beq\label{P9}
{\bf j}= {\bf v}\rho\;, \qquad {\bf v}\equiv
	-i u^\dagger  \:g^{-1} \vec\nabla g\:u
\;.
\eeq
With the decomposition (\ref{P8}), the color density (\ref{P7})
becomes
\beq\label{P10}
\rho_a=Q_a \rho\;,\qquad Q_a= i u^\dagger\: g^{-1}T^a g \: u = i R_{ab}~
u^\dagger T^b u =\ R_{ab}~ t^b/2,
\eeq
where $R_{ab}$ is in the adjoint representation of the group and
the unit vector $t^a$ is defined as $t^a/2= i u^\dagger T^a u$. On
the other hand, the color current reads
\beq\label{P11}
{\bf J}_a =\frac{1}{2} \rho~ R_{ab} ~u^\dagger \:(T^b
	\: g^{-1}\vec \nabla g + g^{-1}\vec \nabla g \: T^b) \: u \;,
\eeq
which with the introduction of
\beqar
g^{-1}\nabla g &\equiv& -2{\bf v}^a T^a\, ,\label{P12}\\
{\bf v}&=&{\bf v}^a t^a\, ,\label{P13}
\eeqar
may be presented as
\beq
{\bf J_a}=\frac{\rho}{2} R_{ab} ~{\bf v}^b\;.
\label{P14}
\eeq
Unlike the Abelian model, the vorticity is nonvanishing.
\beq
\vec \nabla\times{\bf v}^a = \epsilon^{abc} {\bf v}^b\times {\bf v}^c
\label{P15}
\eeq
\par
A difference between the Madelung approach and the previous
particle based one is that the color current is not proportional
to the singlet current. Equation (\ref{P14}) may be decomposed as
\beq
{\bf J}_a =  Q_{a}\rho {\bf v}
	+\frac{\rho}{2} R_{ab} {\bf v}^b_\perp
\label{P16}
\eeq
where the ``orthogonal'' velocity ${\bf v}_\perp^a$ is defined as
\beq
{\bf v}^a_\perp=\left(\delta^{ab}-t^a t^b\right) {\bf v}^b\;.
\label{P17}
\eeq
Equation (\ref{P15}) shows that color current possesses
components that are orthogonal to the singlet current.
\par
In a Postscript at the end of this Sidebar,
 we derive for the $SU(2)$ case
the decomposition of the Schr\"odinger equation with the
parametrization (\ref{P8}). Two equations emerge: one regains
the conservation of the Abelian current and the
other is the ``Bernoulli'' equation.
\beq
\left(  g^{-1} \dot{g}\right)^a = \left[ {\bf v}^b\cdot {\bf v}^b
	- \frac{\vec \nabla^2 \sqrt{\rho}}{\sqrt{\rho}}\right] t^a
	+ \frac{1}{\rho} \vec \nabla\cdot \left(\rho \epsilon_{abc}
	{\bf v}^b t^{c}\right)
\label{P18}
\eeq
It is further verified that the covariant conservation
of the color current is a consequence of the Abelian continuity equation
 and
(\ref{P18}). However, there is no Wong equation because the
color current is not proportional to the conserved singlet
current. Finally, using the identity, which follows from
the definition (\ref{P12}),
\beq
\dot{\bf v}^a = -\frac{1}{2}\vec \nabla \left( g^{-1} \dot{g}
\right)^a + \epsilon_{abc} {\bf v}^b \left(g^{-1} \dot{g}
\right)^c\, , \label{P19}
\eeq
one can deduce an Euler equation for $\dot{\bf v}^a$ from
(\ref{P18}).
\par
We record the energy and momentum density
\beqar
\mathcal{E}&=&\frac{1}{2}\vec \nabla{\Psi^\dagger} \cdot \vec \nabla \Psi
= \frac{1}{2}\rho {\bf v}^a\cdot {\bf v}^a
	+ \frac{\vec \nabla\rho \cdot \vec \nabla\rho}{8 \rho}\label{P20}
\\
\vec{\mathcal{P}}&=&\frac{i}{2}\left(
	\vec \nabla \Psi^\dagger \Psi
	-\Psi^\dagger \vec \nabla \Psi \right)= {\bf v}\rho \label{P21}
\eeqar
Both parallel and orthogonal components of the velocitiy contribute
to the energy density but only the parallel component ${\bf v}$
contributes to the momentum density. It is clear that within
the present approach the fluid color flows in every direction in
the group space, but the mass density is carried by the unique
velocity ${\bf v}$.  This is in contrast to our previous approach
where all motion is in a single direction
or at most in the directions of the Cartan elements of the Lie
algebra (Section \ref{eqmo}).
\par
The difference between the two approaches is best seen from a
comparison of Lagrangians. For the color Schr\"odinger theory in
the Madelung representation
\beqar
\La_{Schrodinger}&=&\frac{i}{2}\left(\Psi^\dagger\dot{\Psi}
	- \dot{\Psi}^\dagger \:\Psi \right)
	-\frac{1}{2} \vec \nabla\Psi^\dagger\cdot \vec \nabla \Psi \label{P22}
\\
&=& i\rho \:u^\dagger \:g^{-1}\dot{g}\: u
	-\frac{1}{2}\rho {\bf v}^a\cdot {\bf v}^a
	-\frac{(\vec \nabla\rho)^2}{8 \rho}.\label{P23}
\eeqar
With $u\otimes u^\dagger \equiv I/2 - 2 i K$, the free part of
the above reads
\beq
\La^0_{Schrodinger} = \rho \: 2\tr K g^{-1}\dot{g}
-\frac{1}{2} \rho {\bf v}^a\cdot {\bf v}^a\, .
\label{P24}
\eeq
On the other hand, the free part of the Lagrange density
(\ref{na14})
in the nonrelativistic limit, with $f(n)$ given by (\ref{na29}), is
\beqar
\La^0&=&\rho \:2\tr\bigg( K g^{-1}\dot{g} 
	+ K {\bf v} \cdot g^{-1}\vec \nabla g\bigg)
	- \sqrt{\rho^2 (c^2-{\bf v}^2)}\, ,
\nonumber \\
&\approx& \rho \:2\tr\, \bigg( K g^{-1}\dot{g} 
	+ K {\bf v} \cdot g^{-1}\vec \nabla g \bigg)
	- \rho c^2+\frac{1}{2}\rho {\bf v}^2\, ,
\label{P25}\\
&=& \rho \:2\tr\,  K g^{-1}\dot{g} 
	-\frac{1}{2}\rho {\bf v}^2 - \rho c^2\, , \nonumber
\eeqar
where we have used ${\bf v}=-2 \tr\, K g^{-1} \nabla g$,
which follows upon the variation of ${\bf v}$, in the next-to-last
equality above. Thus the canonical $1$-form is the same for both
models while the difference resides in the velocity dependence
of their respective Hamiltonians.  Only the singlet ${\bf v}$
enters (\ref{P25}) while the Madelung construction uses the
group vector ${\bf v}^a$.
\par
Finally, note that while the Euler equation, which emerges when
(\ref{P18}) and (\ref{P19}) are combined, intricately couples
all directions of the fluid velocity ${\bf v}^a$, it does admit
the simple solution ${\bf v}^a = {\bf v}t^a$, with ${\bf v}$ obeying
the Abelian equations that arise from (\ref{P3})-(\ref{P4}).
\vskip .1in
\noindent{\it Postscript}:\\
When (\ref{P8}) is inserted into (\ref{P6}), and use is made
of the definition (\ref{P12}), we find in the $SU(2)$ case
\beq
\frac{1}{2}i\dot{\rho} \,u + i\rho (g^{-1}\dot{g})^a T^a \,u =
	-\frac{1}{2}\sqrt{\rho} \vec \nabla^2\sqrt\rho \,u
	+\vec \nabla (\rho {\bf v}^a) T^a \,u
	+\frac{1}{2}\rho {\bf v}^a\cdot {\bf v}^a \, u\, .
\label{P26}
\eeq
Next (\ref{P26}) is premultiplied by $u^\dagger$, where it implies
\beq
i\dot{\rho}  + \rho (g^{-1}\dot{g})^a {t^a}  =
	-\sqrt{\rho} \vec \nabla^2\sqrt\rho
	-i \vec \nabla \left(\rho {\bf v}^a {t^a}\right)
 	+\rho {\bf v}^a\cdot {\bf v}^a \;.
\label{P27}
\eeq
The imaginary part reproduces the continuity equation for the
singlet current, while the real part gives
\beq
 (g^{-1}\dot{g})^a T^a  =
-\frac{1}{\sqrt{\rho}} \vec \nabla^2\sqrt\rho
 +{\bf v}^a\cdot {\bf v}^a \;.
\label{P28}
\eeq
To obtain further information, we premultiply (\ref{P26}) with
$u^\dagger T^{b}$. This gives
\beq\label{eqd4}
\dot{\rho} t^b - i\rho (g^{-1}\dot{g})^b
	+\rho (g^{-1}\dot{g})^a \epsilon_{bac} {T^c}
= i\sqrt{\rho} \vec \nabla^2\sqrt\rho T^b
	- \vec \nabla \cdot\left(\rho {\bf v}^b \right)
	-i \vec \nabla\cdot\left( \rho {\bf v}^a\right) \epsilon_{bac} {T^c}
 	-i\rho {\bf v}^a\cdot {\bf v}^a T^b\;.
\eeq
The imaginary part gives (\ref{P18}) while the real part is
identically satisfied by virtue of the Abelian continuity equation
 and (\ref{P18}).
\addtocontents{toc}{\protect\hrulefill\par}}
\vskip .1in
\hrule
\vskip .1in
\newpage
\section{NONCOMMUTATIVE FLUIDS}
Most of the  fluids examined so far arise out of underlying particle
systems. [The sole exception is the Madelung-like construction based
on an underlying field -- the Schr\"{o}dinger field (both Abelian and
non-Abelian) -- discussed in Sidebar G].  This becames manifest in the
Lagrange description, in which the coordinates of the underlying
particle substratum are explicitly involved. The transition to the
Euler description, then, allows us to express the system in terms of
purely fluid quantities, namely density and velocity.

It is possible to consider fluids whose Euler description does
not derive from an explicit underlying Lagrangian description. Although
such fluids would look conventional in terms of density and velocity,
they would nevertheless be effective descriptions of exotic underlying
theories.

A concrete and interesting realization of this is noncommutative
fluids. In these, the fundamental degrees of freedom represent ``particles''
on a noncommutative space, in which coordinates become non commuting
operators and the notion of points breaks down. 

Noncommutative spaces were introduced by Heisenberg to
ameliorate ultraviolet infinites in quantum field theory. They have
been considered in mathematics and they arise as particular
projections of quantum systems in an intense magnetic field (see Section \ref{Sec7.2}) and
also as special limits of string theory. Gauge theory on these spaces
becomes particularly attractive, since it fuses spatial and internal
degrees of freedom into one coherent formalism. As we shall
demonstrate, such theories can be viewed as noncommutative fluids
and the transitition to the Euler description will emerge as the
transformation mapping them to equivalent commutative effective
theories, known to the string literature as the Seiberg-Witten map. 
The presentation follows Ref. \cite{JackiwPi}.

\subsection{Review of noncommutative spaces}\label{s1}
Noncommutative spaces are described in terms of coordinates that are 
noncommuting operators. In the simplest realization, `flat'
noncommuting coordinates are characterized by a constant, antisymmetric 
tensor $\tn ij$.

\numeq{1.1}{
[x^i, x^j] = i\tn ij\ 
}

We must specify whether $\vth$ possesses an
inverse~$\vom$.

\numeq{1.6n}{
\tn ij \on jk = \delta^i_k
}
An inverse can exist in even dimensions, provided $\vth$ is nonsingular, but $\vom$ will
not exist in odd dimensions, where the antisymmetric $\vth$ always possesses a zero
mode. We shall assume the generic situation: nondegenerate $\vth$ with no zero modes
in even dimensions, where we give a Euclidean treatment: all
coordinates are spatial; while in odd dimensions there is one zero
mode, chosen in the time direction. Therefore, in all dimensions only spatial variables are noncommuting.

Fields $\mathfrak{f}$ are defined as arbitrary functions
of the operator coordinates $x^i$ (and $t$ in odd dimensions). They can be
expressed in a Taylor expansion as sums of monomials involving (ordered)
products of the coordinates. They are, effectively,
themselves operators on the same footing as the $x^i$ (explicitly
depending on the commuting ccordinate $t$ in odd dimensions). The 
derivative of a function $\partial_i \mathfrak{f}$ is defined, as usual, by eliminating
$x^i$ once from each place in monomials in which it appears. Equivalently, this
can be achieved through the adjoint action of the $x^j$ themselves:

\numeq{44.44}{
\partial_i \mathfrak{f} = -i \omega_{ij} [ x^j , \mathfrak{f} ] \, ,
}
which clearly has the desired effect on monomials. 
The volume integral over the full noncommutative space, on the other
hand, can be expressed as the trace of the corresponding operator over
a Heisenberg-like Hilbert space on which the $x^i$ act.

\numeq{66.66}{
\int \mathfrak{f} = \sqrt{ \det (2\pi \vth) } \Tr \mathfrak{f}
}
Products of fields in the above space
are defined as the usual ordered products of the corresponding operators.
We may trade the above noncommutative description for one involving
usual, commuting functions by introducing a new, nonlocal,
noncommutative product between functions, called star-product. 
First,
order all monomials in a fully symmetric way in the $x^i$ (``Weyl ordering''),
which can always be achieved by using\rfq{1.1}. This creates a one-to-one
correspondence between noncommutative operators and commutative
functions. 
The derivative and volume integral of the noncommutative function as defined
 in\rfq{44.44}  and \rfq{66.66} map,
in fact, to the usual derivative and integral of the commutative function $\mathfrak{f}$.
The product of two operators $\mathfrak{f}$ and $\gk$, on the other hand,
upon Weyl-reordering, 
corresponds to a new function, called the star-product of the corresponding
commutative functions $\mathfrak{f}$ and $\gk$ and denoted by $\mathfrak{f} \star \gk$, is
defined by
\begin{equation}
(f \ast g) (x) = exp \frac{i}{2} \, \theta^{ij} \, \frac{\partial}{\partial x^i}
\frac{\partial}{\partial y^i}\, f(x) \, g(y) \mid_{x=y}\, .
\label{6one5}
\end{equation}
We may further define the $\ast$-commutator of two functions as
their antisymmetrized $\ast$-product:

\numeq{55.55}{
[ \mathfrak{f} , \gk ]_\star = \mathfrak{f} \star \gk - \gk \star \mathfrak{f} 
}
The noncommuting coordinates $x^i$, in particular, map to the usual
commuting coordinates and their $\ast$-commutator reproduces the
noncommutative space relations\rfq{1.1}.
\begin{equation}
[x^i, x^j]_\ast = i \, \theta^{ij}
\label{6one7}
\end{equation}
The two formulations, operator and star, are obviously equivalent and
will be used interchangeably.

For future use let us record here the formulas of noncommuting,
electrodynamics in 4-dimensional space. The vector potential $\hat{A}_\mu$
is a function of the noncommuting coordinates $x^\mu$, and it undergoes
gauge transformations, which infinitesimally read
\begin{equation}
\delta\hat{A}_\mu = \partial_\mu \hat{\lambda} - i \, [\hat{A}_\mu, \hat{\lambda}]_\ast
\equiv D_\mu \, \hat{\lambda}.
\label{6one8}
\end{equation}
Here $\hat{\lambda}$ is the gauge function, also depending on the non commuting
coordinates. A field strength $\hat{F}_{\mu \nu}$ is defined so that it
transform covariantly: under gauge transformations
\begin{equation}
\delta \hat{F}_{\mu \nu} = -i [\hat{F}_{\mu \nu}, \hat{\lambda}]_\ast\, .
\label{6one9}
\end{equation}
The formula for $\hat{F}_{\mu \nu}$ therefore reads
\begin{equation}
\hat{F}_{\mu \nu} = \partial_\mu \hat{A}_\nu \, -\partial_\nu \, \hat{A}_\mu
-i [\hat{A}_\mu, \hat{A}_\nu]_\ast\, ,
\label{6one10}
\end{equation}
and the equations satisfied by the $\hat{F}_{\mu \nu}$ are
\begin{eqnarray}
\frac{1}{2} \, \varepsilon^{\alpha \beta \mu \nu} \, D_\beta \, \hat{F}_{\mu
\nu} &=&0\, , \label{6one11}\\
D_\mu \, \hat{F}^{\mu \nu} &=& \hat{J}^\nu\, ,
\label{6one12}
\end{eqnarray}
where $\hat{J}^\nu$ is a noncommuting source current. Note that even though
we are dealing with electrodynamics, the noncommutativity requires a
non-Abelian, Yang-Mills-like structure for the relevanat expressions.

The noncommutative space structure\rfq{1.1} remains invariant under
a group of transformations.
Subjecting the spatial \co s to an infinitesimal \co\ transformation
\numeq{1.2}{
\delta \vx = - {\bf f}(\vx)\, ,
}
for some operator functions $f^i$,
and requiring that\rfq{1.1} remain unchanged results in the condition

\numeq{1.3}{
-[\fvx i, x^j] -[x^i, \fvx j] = 0\, ,
}
which in turn implies  by\rfq{1.1} that 

\numeq{1.4}{
-\pas k \fvx i \tn kj - \pas k  \fvx j \tn ik  = 0\ .
}
The left side is recognized as the Lie derivative of a contravariant tensor,

\numeq{1.5}{
L_f \tn ij = f^k \pas k \tn ij - \pas k  f^i \tn kj - \pas k f^j \tn ik\, ,
}
with the first term on the right vanishing since $\vth$ is constant. So the
\nc\ algebra\rfq{1.1} is preserved by those coordinate transformations that 
leave $\vth$ invariant: $L_f \vth = 0$.

To solve for ${\bf f}$ in even dimensions, we define

\numeq{1.7n}{
f^i = \tn ij g_j\quad (i,j=1,\ldots,2n)\ .
}
This entails no loss of generality, because $\vth$ is nonsingular (by hypothesis).
Then\rfq{1.4} becomes
\begin{subequations}\label{3.1}
\numeq{3.1a}{
\tn i\ell \pas k g_\ell  \tn kj + \tn j\ell \pas k  g_\ell \tn ik  = 0\ .
}
Because $\vth$ is nonsingular and antisymmetric, this implies

\numeq{3.1b}{
 \pas k g_\ell  - \pas \ell  g_k  = 0 \, ,
}

\end{subequations}
\noindent or

\numeq{3.2}{
  g_\ell  = \pas\ell\phi\ .
}
Thus we have 

\numeq{3.3}{
f^i = \tn ij \pas j  \phi 
}
for the \ct s (in even dimensions) that leave $\vth$ invariant. Since

\numeq{1.11}{
\grad\cdot {\bf f} = 0\ 
}
the transformations are volume preserving; the  Jacobian of the finite 
\df\ is unity.  However, except in two dimensions, these are not the most 
general volume-preserving transformations.  Nevertheless, they form a group: 
the Lie bracket of two transformations like\rfq{1.4}, 
$f_1^i = \tn ij \paj\phi_1$ and $f_2^i = \tn ij \paj\phi_2$,
takes the same form, $\tn ij\paj(\tn k\ell \partial_k\phi_1 \partial_\ell \phi_2)$.  
The group is the symplectic subgroup of volume-preserving \df s that also 
preserve~$\tn ij$. 
In two dimensions, where we can set $\tn ij=\theta \en ij$, the above 
transformations exhaust all the area-preserving transformations. 

In odd dimensions, where (by assumption) $\vth$ possesses 
a single zero mode, for definiteness we orient the
coordinates so that the zero mode lies in the first direction (labeled 
$0\to$~time) and
$\vth$, confined to the remaining (spatial) dimensions, is nonsingular.

\numeq{1.12}{\begin{aligned}
\theta^\mn &= \biggl(\begin{matrix} 
0 & 0\\
0 &\tn ij
\end{matrix} \biggr) \quad  (i,j = 1,\ldots,2n) \\
\tn ij \on  jk &= \delta_k^i
\end{aligned}
}
The infinitesimal \df s that preserve $\vth$ are
\numeq{1.13}{
f^\mu = \Biggl\{ \begin{array}{l}
f(t) \\
\displaystyle 
\tn ij \pa{x^j} \phi\tvx\, .
\end{array} 
}
These still form a group. Two transformations, 
$(f_1,\phi_1)$ and $(f_2,\phi_2)$,
possess a  Lie bracket of the same form\rfq{1.13}, with 
$(f_2\partial_t f_1 - f_1\partial_t f_2, f_2\partial_t\phi_1 - f_1\partial_t\phi_2 
+ \tn k\ell \partial_k \phi_1\partial_\ell\phi_2)$. 
But the space-time volume is not preserved: $\partial_\mu f^\mu \neq
0$. (Of course, at fixed time, the spatial volume is preserved.) 
%\subsection{Fluid formulations revisited}

Unit-Jacobian diffeomorphisms also leave invariant the equations 
for an ideal fluid, in the Lagrange formulation of fluid mechanics, 
and in particular a planar (two dimensional) fluid
supports area-preserving diffeomorphisms.; see Section \ref{Sec1.2} (ii). This coincidence of invariance 
suggests that other aspects of noncommutativity possess analogs in the 
theory of fluids,
whose familiar features can therefore clarify some obscurities of
noncommutativity. (A similar point of view concerning the quantum Hall effect was taken in Ref. \cite{sussarv})
We shall explore this connection
and shall demonstrate that the \sw\ map \cite{SiebWit} between
\ncg\ and commuting gauge fields corresponds to the mapping between 
the Lagrange and Euler formulations of fluid mechanics. We shall obtain
a simple derivation of the explicit ``solution'' to the \sw\ map in even
dimensions \cite{OkaOo} and will extend it to odd dimensions.

The two formulations of fluid dynamics (Lagrange and Euler) can be put 
in the proper context for the noncommutative space setting.
The natural Poisson
(commutator) structure, present in the Lagrange description of a
fluid, and the possibility of introducing a vector potential to describe
the evolution of comoving coordinates, will be recognized as classical
precursors of analogous
noncommuting entities. Within this framework, we shall show how noncommuting
gauge fields respond to coordinate transformations, generalizing previously
established results. \cite{PiJac}

As explained in Section 1.1, the Lagrange description uses the coordinates of the
particles comprising the fluid: ${\bf X}(t, {\bf x})$.  These are labeled by the comoving ccordinates~$ \vx$, which
are the coordinates of some initial reference configuration, {\it e.g.}, ${\bf X}(0,{\bf x}) = {\bf x}$, see
(1.1.11). We may parameterize the evolution of $\bf X$ by defining

\numeq{1.14}{
X^i \tvx = x^i + \tn ij \han j (t, {\bf x})
}
which looses no generality provided $\vth$ is nonsingular. As will be seen below, $\bf\hat{A}$
behaves as a noncommuting, Abelian vector potential. 

\subsection{\ Noncommutative Gauge theory}\label{s2}
\subsubsection*{(i) Commuting theory with Poisson structure}\label{s2ssA}
\addcontentsline{toc}{subsubsection}{(i) Commuting theory with Poisson structure}
\setcounter{equation}{0}
We introduce into the Lagrange fluid description the (nonsingular) antisymmetric tensor
$\vth$. This allows for a natural definition of a  Poisson bracket, which may be viewed as a
classical precursor of the noncommutativity of coordinates. 
We define the bracket by 
%\begin{subequations}\label{2.2}

\numeq{2.2a}{
\{\mathcal O_1,\mathcal O_2\} = \tn ij \frac{\partial \mathcal O_1}{\partial x^i} \frac{\partial \mathcal O_2}{\partial x^j} }
so that

\numeq{2.2b}{
 \{ x^i, x^j\} = \tn ij\ .
}

%\end{subequations} 
It follows from the definition\rfq{1.14} that

\numeq{2.4a}{ 
\{X^i,X^j\}  = \tn ij + \tn ik \tn j\ell \hat F_{k\ell} 
}
with

\numeq{2.5}{
\hat F_{ij} = \pa{x^i} \ha_j - \pa{x^j} \ha_i + \{\ha_i, \ha_j\}\ .
} 

It is seen that the structure of the gauge field $\tenshf$ is as in a noncommuting theory,
with the Poisson bracket replacing the $\ast$ commutator of two potentials~$\vha$,
compare (6.1.10). Also, in the limit that the deviation of $\bf X$  from the reference
configuration~$\bf x$ is small, that is, for small~$\vha$, we recover a conventional
Abelian gauge field. 

The above formulas are understood to hold either in even dimensions for a purely spatial
Euclidean formulation (there is no time variable) or in odd-dimensional space-time for
 spatial components. ($\bf X$ and $\bf x$ are spatial vectors, without time
components.)

\subsubsection*{(ii) Coordinate transformations in the commuting theory (even dimensions)}\label{s2ssB}
\addcontentsline{toc}{subsubsection}{(ii) Coordinate transformations in the commuting theory (even
dimensions)}

In even dimensions, the $\vth$-preserving transverse \df, which also implements the
reparameterization symmetry of the Lagrange fluid, acts on $\vX$  through the bracket
according to (1.2.24) and \rfq{3.3} as
\begin{equation}
\delta_\phi \, \vX = {\bf f} \cdot \vec\nabla \, \vX = \theta^{ij} \frac{\partial \vX}{\partial
x^i} \frac{\partial \phi}{\partial x^j}.
\label{6two5}
\end{equation}
This may be presented with help of the bracket defined in (\ref{2.2a}).

\numeq{2.7}{
\delta_\phi \vX {(\vx)} = \tn ij \frac{\partial \vX {(\vx)}}{\partial x^i}  \frac{\partial\phi(\vx)}{\partial x^j} 
= \{ \vX {(\vx)}, \phi(\vx)\} \
}
Because $\delta \vX$ compares the transformed and untransformed $\vX$
at the same argument,
$\delta \han i = \on ij \delta X^j$ and the volume-preserving
diffeomorphism (\ref{6two5}),\rfq{2.7} induces a gauge transformation on $\vha$:
[Compare (\ref{6one9}), (\ref{6one10})].

\begin{subequations}\label{2.8}

\begin{align}
\delta_\phi \vha(\vx)  &= \grad\phi(\vx) + \{\vha(\vx), \phi(\vx)\} 
\equiv \vD \phi \label{2.8a}\\
\delta_\phi \hat F_{ij}(\vx) &= \{\hat F_{ij} (\vx), \phi(\vx)\} \label{2.8b}
\end{align}
\end{subequations} 
We see that the dynamically sterile relabeling diffeomorphism of the parameters in the
Lagrange fluid leads to an equally sterile gauge transformation, under which $\vX$ and
$\tenshf$ transform covariantly, as in\rfq{2.7}, \rfq{2.8b}. 

Next we consider a \df\ of the target space.

\numeq{2.9}{
\delta_\fkf \vX  = - \fkf (\vX) 
}
In contrast to the previous relabelings, this transformation is dynamical,
deforming the fluid configuration. Quantities

\numeq{2.8n}{
C_n (\vX) = \frac1{2^n n!}  
\eps_{i_1j_1\cdots
i_n j_n} \{X^{i_1}, X^{j_1}\}\cdots \{X^{i_n}, X^{j_n}\}\, ,
}
which are defined in $d=2n$~dimensions, respond to the transformation\rfq{2.9} in a
noteworthy fashion. One verifies that

\numeq{2.9n}{
\delta_\fkf C_n (\vX) =  -\grad\cdot\fkf(\vX)C_n(\vX)\, ,
}
so that transverse (volume-preserving) target-space \df s leave $C_n$ invariant.
Eq.\rfq{2.9n} is most easily established by recognizing that

\numeq{2.10n}{
C_n (\vX) = \text{Pfaff} \{X^i, X^j\} 
= \det\nolimits^{1/2} \{X^i, X^j\}  
=\det\nolimits^{1/2} \vth \det \frac{\partial X^i}{\partial x^j}\ .}

The significance of these transformations is evident from (1.1.19), which shows that
$1/\rho(\vcr) = C_n(\vX)\bigr|_{\vx=\vchi(\vcr)}$ when $\det\nolimits^{1/2} \vth$
is identified with
$1/\rho_0$. 
 The transformation law for~$\rho$ under
transverse target space
\df s becomes

\numeq{2.11}{
\delta_\fkf \rho(\vcr) =  \fkf \cdot \grad \rho(\vcr)\ .  
}
It follows that 
this transformation
leaves  invariant 
 all terms in the Lagrangian that depend only
on~$\rho$ [like the potential in (1.2.30)]. 

When we restrict the transverse, target-space \df s to those that also leave $\vth$
invariant, {\it i.e.},\rfq{3.3} (of course in two dimensions this is not a restriction), 
further quantities are left invariant. These are constructed as in\rfq{2.8n}, but with any
number of brackets $\{ X^i , X^j \}$ replaced by~$\tn ij$. 

It is interesting to combine the \df\ of the parameter space with that of the target space,
for a simultaneous transformation on both spaces. To this end we chose the form of the
target space transformation to coincide with that of the  reparameterization/relabeling
transformation.

\numeq{2.12}{
\mathfrak{f}^i(\vX) =  \tn ij\frac{\partial \phi(\vX)}{\partial X^j} \ . 
} 
As we shall show below,
this results in a gauge-covariant coordinate transformation on the vector
potential~$\vha$, once a further gauge transformation is carried out. Thus we consider 
$\Delta \equiv \delta_\phi + \delta_\mathfrak{f}$, 

\numeq{2.13}{
\Delta X^i = \{X^i, \phi(\vx)\} - \tn ij\frac{\partial \phi(\vX)}{\partial X^j} \ .  
}
[Note that any deviation of $\fkf^i (\vX)$ from $\tn ij\, {\partial\phi(\vX)}/{\partial X^j} $ may be
attributed to~$\phi$, and can be removed by a further gauge transformation.]
 However, covariance is not preserved in\rfq{2.13}:
$\vX$ on the left is covariant, but on the right in the Poisson bracket there occurs
$\phi(\vx)$, which is not covariant. The defect may be remedied by combining
$\Delta\vX$ with a further gauge transformation,

\numeq{2.14}{
\delta_{\textrm{gauge}} \vX =   \{\vX, \phi(\vX) - \phi(\vx)\}\, ,
}
so that in $\Delta + \delta_{\textrm{gauge}} \equiv \hat\delta$ we have a covariant
transformation rule: 

\numeq{2.15}{
\hat\delta X^i =   \{X^i, \phi(\vX)\} - \tn ij \frac{\partial\phi(\vX)}{\partial X^j}\, ,
}
which in turn implies that $\vha$ transforms as 

\numeq{2.16}{
\hat\delta \ha_i =  \on ij \{X^j, \phi(\vX)\} - \frac{\partial\phi(\vX)}{\partial X^i}\ .
}

To recognize this transformation more clearly, we present it as
\begin{subequations}\label{2.17}
\numeq{2.17a}{
\hat\delta \ha_i =  \on ij \{X^j, X^k\} \frac{\partial\phi(\vX)}{\partial X^k} - \frac{\partial\phi(\vX)}{\partial
X^i}\, , }
and use\rfq{2.4a} to find
\numeq{2.17b}{
\hat\delta \ha_i =  \tn k\ell \hat F_{i\ell} \frac{\partial\phi(\vX)}{\partial X^k} 
  = f^k (\vX)\hat F_{ki}\ .}
\end{subequations}

Note that in the final expression\rfq{2.17b} the response of $\vha$ is entirely covariant:
it involves the covariant curvature $\tenshf$ and the diffeomorphism function~${\bf f}$
evaluated on the covariant argument~$\vX$. This expression is precisely the
gauge-covariant \ct. (This was derived, by a somewhat
different method \cite{PiJac}.)

\subsubsection*{\bf (iii) Coordinate transformations in the noncommuting theory
with $\ast$-products\linebreak
 (even dimensions)}\label{s2ssC} 
\addcontentsline{toc}{subsubsection}{(iii) Coordinate transformations in the noncommuting theory
with $\ast$-products  (even dimensions)}

The above development may be taken over directly into  a
noncommutative field theory by replacing Poisson brackets by $-i$~times~$\ast$-
commutators, so that\rfq{2.2b} goes over into\rfq{1.1}. Eq.\rfq{1.14} remains
and\rfq{2.4a},\rfq{2.5} become

%\begin{align}
%[X^i,X^j]_\star &= i\tn ij + i\tn ik \tn j\ell \hat F_{k\ell}\label{2.18}\\
%\hat F_{ij} &= \pa{x^i} \ha_j - \pa{x^j} \ha_i - i[\ha_i, \ha_j]_\star\ .\label{2.19}
%\end{align}
\begin{equation}
[X^i,X^j]_\star = i\tn ij + i\tn ik \tn j\ell \hat F_{k\ell}\label{2.18}
\end{equation}
with $\hat F$ given by (\ref{6one10}).
The covariant transformation rules\rfq{2.15} and\rfq{2.16} may be used in the
noncommutative context, provided a sensible ordering prescription is set for~$\phi(\vX)$.
This we do as follows. Define
\begin{subequations}\label{2.20}
\numeq{2.20a}{
\Phi = \int \rd x \phi(\vX)
}
where $\phi(\vX)$ is a series of (star) powers of~$\vX$:
\numeq{2.20b}{
\phi(\vX) = c + c_i X^i + \fract12 c_{ij}X^i\star X^j + \fract13 c_{ijk} X^i\star X^j \star X^k +
\cdots\cdot
 }
\end{subequations}
[We are not concerned about convergence of the integral\rfq{2.20a}, 
since we are interested in local
quantities like \rfq{2.20b} or \rfq{2.23} below.]
The integration over $\vx$ (the argument of~$\vX$) ensures that $\Phi$ is invariant (in an
operator formalism the integral becomes the trace of the operators). The $c$-coefficients
in\rfq{2.20b} are required to be invariant agaisnt cyclic index shuffling (so that $\Phi$ and $\phi$ possess the same
number of free parameters).  Also we require $\phi$ to be Hermitian. [This ensures, e.g., that
$c_{ij}$ is real symmetric; that $\Re  c_{ijk}$ is entirely symmetric and that $\Im  c_{ijk}$ is
entirely antisymmetric (which is impossible in two dimensions).] Then\rfq{2.15}
and\rfq{2.16} become
 \begin{align}
\hat\delta X^i &= -i[X^i, \phi(\vX)]_\star - \tn ij \frac{\delta\Phi}{\delta
X^j}\label{2.21}\\[0.5ex]
\hat\delta \ha_i &= -i\on ij[X^j, \phi(\vX)]_\star  -  \frac{\delta\Phi}{\delta X^i}\label{2.22}
\end{align}
where now the last entries employ a functional derivative:
\numeq{2.23}{
\frac{\delta\Phi}{\delta X^i} = c_i +  c_{ij} X^j + c_{ijk}  X^j \star X^k +
\cdots\cdot
 }

In two dimensions, 
the ordering prescription\rfq{2.20} and its consequence\rfq{2.23}
preserve the invariance of the $[X^i, X^j]_\star$ commutator against the target space \df\
[last term in\rfq{2.21}]. Thereby a property of the classical Poisson bracket
[{\it c.f.}\negthinspace\rfq{2.9n} at $n=1$] is maintained in the noncommuting theory.

With $\phi(\vX)$ at most quadratic in~$\vX$ (${\bf f}$ at most linear), one
readily verifies the result inRef. \cite{PiJac}.
\numeq{2.24}{
\delta \ha_i = \fract12 \bigl\{ f^j (\vX)\star \hat F_{ji}  +   \hat F_{ji} \star f^j
(\vX)
\bigr\}\ 
 }
But with more general $\phi$ (${\bf f}$ containing quadratic and higher powers) there
arise further reordering terms. 

\subsubsection*{(iv) Coordinate transformations in commuting and noncommuting
theories \linebreak \indent \indent (odd dimensions)}\label{s2ssE}
\addcontentsline{toc}{subsubsection}{(iv) Coordinate transformations in commuting and noncommuting theories
(odd dimensions)}

In odd dimensions, with the $\vth$-preserving transformation function given
by\rfq{1.13}, the relabeling transformation on 
the base space is
\numeq{3.13}{\begin{aligned}
\delta_\phi \vX\tvx &=  \tn ij \pa{x^j} \phi\tvx \pa{x^i} \vX\tvx  
    + f(t)  t \, \dot{\vX} \tvx \\[1ex]
  &= \bigl\{\vX\tvx, \phi\tvx  \bigr\} + f(t)\dot{\vX}\tvx   \ .
\end{aligned}
}
The fluid coordinate $\vX$ has components only in the spatial directions. Here the Poisson
bracket is defined with the nonsingular~$\tn ij$. 

For the target space \df\ we again take the formula\rfq{2.12}, so that the combined,
noncovariant transformation $\Delta\equiv\delta_\phi+\delta_\mathfrak{f}$ reads 
\numeq{3.14}{
\Delta  X^i = \bigl\{ X^i , \phi\tvx\bigr\}
 +   f(t)\dot X^i\tvx - \tn ij \frac{\partial\phi\tvX}{\partial X^j}\ . 
}
This is modified by the gauge transformation
\numeq{3.15}{
\delta_{\textrm{gauge}} \vX = \bigl\{ \vX, \phi\tvX - \phi\tvx \bigr\}
 -   \bigl\{ \vX, f(t)\han 0 \tvx \bigr\} 
}
resulting in the covariant transformation $\Delta + \delta_{\textrm{gauge}} 
\equiv \hat\delta$.

\numeq{3.16}{
\hat\delta  X^i = \bigl\{ X^i , \phi\tvX\bigr\}
  - \tn ij \frac{\partial\phi\tvX}{\partial X^j} + f(t) D{X^i} 
}
Here $D{X^i} = \dot X^i + \{ \ha_0, X^i \}$, where $\ha_0$ is a connection introduced to
render the time derivative covariant against time-dependent gauge transformations,
generated by $\phi$. This is achieved when the gauge transformation law for $\ha_0$ is 

\numeq{3.17}{
\delta_\phi \ha_0=  \dot\phi +   \bigl\{ \han0,  \phi \bigr\} \ .
}
The spatial components of the vector potential are introduced as before in\rfq{1.14}, and $D X^i$ becomes

\numeq{3.20}{
  D{X^i} = \tn ij \Bigl( \skew5\dot{\hat A}_j  -\partial_j \han 0 + \{\han 0, \han j\} \Bigr)
 = \tn ij \hat F_{0j}\ .
}
The covariant transformation law of $\vha$ follows from\rfq{2.18},\rfq{3.16},
and\rfq{3.20}: 

\numeq{3.21}{\begin{aligned}
\hat\delta \han i &= \on ij \bigl\{ X^j , \phi\tvX\bigr\}
  -  \frac{\partial\phi\tvX}{\partial X^i} +  \on ij f(t) D{X^j} \\[1ex]
  &= f^j \tvX \hat F_{ji} + f(t) \hat F_{0i} = f^\mu \tvX \hat F_{\mu i}\ .
\end{aligned}
}
It remains to fix the transformation law of $\han 0$. This requires specifying
$\delta_\mathfrak{f} \han 0$. Since 

\numeq{2.35}{
\delta_\mathfrak{f} \han i =  -  \frac{\partial\phi\tvX}{\partial X^i}
} it is natural
to take 

\numeq{2.36}{
\delta_\mathfrak{f} \han 0 =  -  \dot{\phi}\, \tvX
}
(The time derivative acts on the first argument only.)
Thus we have from\rfq{3.17} and\rfq{2.36}

\numeq{2.37}{
\Delta \han 0 =  \dot{\phi}\tvx + \{\han 0, \phi\tvx\} -  
\dot{\phi}\tvX\ . 
}
After adding to this a gauge transformation generated by
$\phi\tvX - \phi\tvx$ we are left with

\numeq{3.22}{
\begin{aligned}
  \hat\delta \han0  &= \frac{\partial\phi\tvX}{\partial X^i} \dot{X}^i + \{\han 0,
 \phi\, \tvX\}\\
 &= \frac{\partial\phi}{\partial X^i} \rD{X^i} =  f^i \tvX \hat F_{i0} =  f^\mu \tvX \hat F_{\mu 0}\ .
\end{aligned}
}
Eqs.\rfq{3.21} and\rfq{3.22} coincide with the formula obtained in a conventional
commuting gauge theory \cite{jacKIW}. 

Similar results follow within the noncommuting formalism, once the now familiar ordering
prescription is given for $\phi\tvX$ and $\Phi = \int \rd x \phi\tvX$. In the \nc\
theory \rfq{3.21} and\rfq{3.22} are regained, up to reordering terms. 

\subsection{Seiberg-Witten map = Euler fluid-Lagrange fluid map}
\setcounter{equation}{0}
By considering various limits within string theory, Seiberg and Witten found that
noncommuting fields can be mapped onto non-local functions of commuting fields \cite{SiebWit}.
The principle behind the mapping can be stated as a requirement of stability agaisnt
gauge transformations in the following sense. Consider the noncommuting gauge
potential $\hat A_\mu$ to be a functional of the commuting gauge potential $A_\mu$
and of $\theta$. It is then required that a commuting gauge transformation performed
on the commuting gauge potential  with commuting gauge function $\lambda:
A_\mu \to A^\lambda_\mu \equiv A_\mu + \partial_\mu \, \lambda$, can be equivalently
achieved by performing noncommuting gauge transformation on the noncommuting gauge
potential with noncommuting gauge function $\hat{\lambda}: \, \hat A_\mu \to
\hat A^\lambda_\mu \equiv (e^{i\hat{\lambda}}) \ast [\hat A_\mu + i \,
\partial_\mu] \ast  (e^{i\hat{\lambda}})^{-1}$. Thus
\begin{subequations}
\begin{equation}
\hat A^{\hat{\lambda}}_\mu (A_\mu) = \hat A_\mu (A_\mu + \partial_\mu \lambda)
\label{6three1a}
\end{equation}
This in turn implies a differential equation in $\theta$ for $\hat A_\mu$.
\begin{equation}
\frac{\partial \hat A_\mu}{\partial \theta^{\alpha \beta}} =-\frac{1}{8} \{\hat A_\alpha,
\partial_\beta \, \hat A_\mu + \hat F_{\beta \mu}\}^+_\ast - (\alpha \gets \beta)
\label{6three1b}
\end{equation}
\end{subequations}
with ``initial" condition $\hat A_\mu \mid_{\theta =0} = A_\mu$. $\{ \ ,\ \}^+_\ast$ denotes the
$\ast$-anticommutator. One can solve (\ref{6three1b}) order-by-order in $\theta$, with the lowest order solution
being
\begin{equation}
\hat A_\mu = A_\mu -\frac{1}{2} \theta^{\alpha \beta} \, A_\alpha (\partial_\beta \,
A_\mu + F_{\beta \mu}) + \cdot \cdot \cdot
\label{6three2}
\end{equation}
but in general, the equations for different $\alpha$ and $\beta$ contained in
(\ref{6three1b}) are not integrable, except in two dimensions where $\theta^{\alpha
\beta}$ involves a single quantity $\theta^{\alpha \beta} = \varepsilon^{\alpha \beta}
\theta$.

Aside from its intrinsic mathematical interest as providing a connection between
commuting and noncommuting fields, the Seiberg-Witten map also serves a
practical purpose. The noncommuting field strength is not gauge invariant, rather,
just as in Yang-Mills theory, it is gauge covariant: $\hat F_{\mu \nu} \to (e^{i \hat
\lambda}) \ast \hat F_{\mu \nu} \ast (e^{i \hat
\lambda})^{-1}$. But unlike in Yang-Mills theory, there are no local gauge invariant
quantities. To obtain a gauge invariant result one must integrate over $x^\mu$, or
(equivalently) consider the trace over the Hilbert space on which the operators $x^\mu$
act. However, if one wishes to compare predictions of noncommuting
electrodynamics with those of ordinary commuting electrodynamics (to set limits on
the amount of noncommutativity in Nature) non local quantities are not useful,
because the physical content of ordinary electromagnetism is expressed by local
quantities (waves, energy and momentum densities, etc.). Here the Seiberg-Witten
map provides a resolution: map the noncommuting theory onto a commuting one,
from which local gauge invariant quantities may be extracted \cite{sp11}. 

We now show that the inverse Seiberg-Witten map is equivalent to the map between
Lagrange and Euler descriptions for fluids. The argument is first presented in two
Euclidean dimensions and (2+1)-dimensional space-time, where (\ref{6three1b}) is
integrable (because it involves a single derivative variable). Then the argument is
taken to higher dimensions.

\subsubsection*{(i) \sw\ map in (2) and (2+1) dimensions}\label{s2ssD}
\addcontentsline{toc}{subsubsection}{(i) \sw\ map in (2) and (2+1) dimensions}

To construct the \sw\ map in two Euclidean dimensions, we (temporarily) introduce a
time dependence in the fluid variables (but not into the \df\ functions -- only spatial
variables are transformed) and 
observe that $(c \rho, \vv \rho)$ form a conserved 3-vector~$j^\alpha$ 
[also true  in the noncommuting theory when an ordered definition for
$\delta(\vX\tvx - \vcr)$ is given -- this will be provided below].
Therefore, the dual of~$j^\alpha$, $\eps_{\mn\alpha} j^\alpha$,  
satisfies a Bianchi identity and can be presented as the curl of a potential, apart from
additive and multiplicative constants.

\begin{gather}
 \eps_{\mn\alpha} j^\alpha \propto F_\mn + \text{constant}\label{2.25}\\
F_\mn = \partial_\mu A_\nu - \partial_\nu A_\mu \label{2.26}
\end{gather}
Note $j^\alpha$, $F_\mn$, $A_\mu$ are ordinary functions, even in the noncommuting
setting, since the noncommuting variables~$\vX$ are integrands (in an operator
formalism, their trace is involved). In particular, the spatial tensor is determined
by~$\rho$.

\numeq{2.27}{
\pa{r^i} A_j (\vcr)  - \pa{r^j} A_i (\vcr) = F_{ij}(\vcr) 
   = - \eps_{ij} (\rho -\rho_0)
 =  - \eps_{ij} \rho_0 \Bigl(\int \rd x \delta\bigl( \vX(\vx) - \vcr\bigr) - 1\Bigr)
 }
(The time dependence is now suppressed.)
$\vX$ contains $\vha$, as in\rfq{1.14}. Since $\bf X$ is (noncommuting) gauge covariant, the
integral in\rfq{2.27} is (noncommuting) gauge invariant. Therefore,\rfq{2.27} serves to
define an (inverse) \sw\ map between the noncommuting (hatted) and commuting
(unhatted) variables.
The additive ($\eps_{ij} \rho_0 $) and multiplicative ($-1$) constants are fixed by
requiring agreement at small $\vha$.
 It still remains to give a proper ordering to the $\delta$-function
containing~$\vX$. This we do by a Fourier transform prescription

\numeq{2.28}{
\int \rd r e^{i\vk\cdot\vcr} F_{ij}(\vcr)
   = -\eps_{ij}\rho_0 \int \rd x \bigl( e^{i\vk\cdot\vX(\vx)}_\star - e^{i\vk\cdot\vx}\bigr)\, ,
 }
and the ordering (Weyl ordering) is defined by the expansion of the exponential in ($\ast$
product) powers: $e^{i\vk\cdot\vX}_\star\equiv 
1+ i \vk \cdot \vX - \frac12  \vk \cdot \vX \star \vk \cdot \vX + \cdots$.

When the exponential $ e^{i\vk\cdot\vX}_\star$
is written explicitly in terms of~$\vha$:
$\exp_\star i( k_i x^i + \theta k_i \en ij \ha_j)$, factoring the exponential 
into $e^{i\vk\cdot\vx}$~times another factor involves the Baker-Hausdorff 
lemma, (because the individual terms in the exponent do not $\ast$ commute). This leads
to an open Wilson line integral \cite{unkn}. In that form\rfq{2.28} is seen to 
coincide with the known solution to the
\sw\ map \cite{OkaOo}, which is now also recognized as nothing but an instance of the
Lagrange$\,\to\,$Euler map of fluid mechanics.

To construct the \sw\ map in (2+1)-dimensional space-time we consider
the conserved current, defined in (\ref{eqone13}) and (\ref{eqone14}), except that now the time
dependence is retained  throughout and the 
derivative is gauged with $\han0$:

\numeq{3.23}{
  \vj\tvr = \int\rd x \bigl(\dot \vX + \{\han0, \vX\}\bigr) \delta (\vX-\vcr)
}
The operator ordering is prescribed in momentum space with the exponential (Weyl)
ordering and\rfq{3.23} in the noncommuting theory becomes

\numeq{3.24}{
\vj \tvk  \equiv
\int \rd r e^{i\vk\cdot\vcr} \vj \tvr = \int\rd x e^{i\vk\cdot\vX}_\star \bigl(\dot{\vX} - 
i[\han 0,
\vX]_\star\bigr)
\ . }
Note that the commutator does not contribute to current conservation because it is
separately transverse.

\numeq{3.25}{
 \int\rd x e^{i\vk\cdot\vX}_\star \bigl[\han 0, \vk\cdot\vX\bigr]_\star =
\int\rd x \han 0  \bigl[\vk\cdot\vX, e^{i\vk\cdot\vX}_\star\bigr]_\star = 0 
}
Therefore the 3-current is conserved as before. Its dual, $\eps_{\mn\alpha} j^\alpha$
satisfies the Bianchi identity, so the \sw\ mapping reads

\numeq{3.26}{\begin{aligned}
\int\rd r e^{i\vk\cdot\vcr}  \bigl(1 - \fract12 \tn ij F_{ij}\bigr) &= 
     \int\rd x e^{i\vk\cdot\vX}_\star, \\[1ex]
\int\rd r e^{i\vk\cdot\vcr} F_{0i} &= 
   \on ij \int\rd x e^{i\vk\cdot\vX}_\star \bigl(\dot X^j - i[\han0, X^j]_*\bigr)\\
   &= \int\rd x e^{i\vk\cdot\vX}_\star \hat F_{0i} \ .
\end{aligned}
}
Formulas\rfq{2.28} and\rfq{3.26} may be verified by comparison with the explicit
$\mathcal O(\theta)$
\sw\ map (\ref{6three2}), which  for field strengths implies
\numeq{3.27}{
F_\mn = \hat F_\mn - \tn \alpha\beta (\hat F_{\alpha\mu} \hat F_{\beta\nu} -
        \han \alpha \partial_\beta \hat F_\mn)\ .
}
Upon setting $\tn \alpha0 = 0$, $\tn ij = \theta \en ij$ and 

\numeq{3.28}{
e^{i k_i (x^i +\tn ij\han j)}_\star = e^{i\vk\cdot\vx} \Bigl(1 + i\theta k_i \en ij \han j 
- \fract12 \theta^2 k_i k_m \en ij \en mn \han j \han n \Bigr)
}
it is recognized that\rfq{2.28} and\rfq{3.26} reproduce\rfq{3.27}.

\subsubsection*{(ii) \sw\  map in higher even dimensions}\label{s3}
\addcontentsline{toc}{subsubsection}{(ii) \sw\  map in higher even dimensions}

%\subsection{Even dimensions}\label{s3ssA}
In dimensions higher than three the correspondence between the Bianchi
identity and the conservation of particle current is lost. The derivation
of the \sw\ map calls for higher conserved currents, whose duals are
two-forms. 

The introduction of such currents can be motivated by starting again from
the commutative particle density $\rho$ as expressed in (\ref{eqone13}) and
its inverse $\rho^{-1}$ as expressed in (\ref{eqone19}). Their product

\numeq{aaa}{
1 =  \int \rd x  \delta \bigl(\vX - \vcr\bigr)
\det \frac{\partial X^i (\vx)}{\partial x^j} 
}
is independent of the fluid profile $\vX (\vx)$ and constitues a topological
invariant. The Jacobian determinant in the above can be expressed in
terms of the square-root determinant (Pfaffian) of the antisymmetric 
matrix $\{ X^j , X^k \}$:

\numeq{bbb}{
1 = \frac{\rho_0}{2^n n!} \int \rd x  \delta \bigl(\vX - \vcr\bigr)
\epsilon_{i_1 , j_1 , \dots ,i_n , j_n} \{ X^{i_1} , X^{j_1} \}
\cdots \{ X^{i_n} , X^{j_n} \} = \rho_0 \int \rd x 
\delta \bigl(\vX - \vcr\bigr)  C_n (\vX )
}
where, in analogy with the 2-dimensional case, we identified 
${\rm Pfaff} ({\vth})$ with $1/\rho_0$. Removing all
$n$ Poisson brackets from the above recovers the full density $\rho$.
The removal of a single Poisson bracket $\{ X^i , X^j \}$, then,
produces a sort of residual density $\rho_{ij}$ in the corresponding
dimensions, which becomes a candidate for the \sw\ commutative field
strength:

\numeq{bbc}{
\rho_{ij} = \frac{\rho_0}{2^{n-1} (n-1)!} \int \rd x  \delta 
\bigl(\vX - \vcr\bigr) \, \epsilon_{i,j,i_2,j_2, \dots ,i_n ,j_n} 
\{ X^{i_2} , X^{j_2} \} \cdots \{ X^{i_n} , X^{j_n} \}
}
The current dual to $\rho_{ij}$, in momentum space,

\numeq{ccc}{
J^{j_1 \dots j_{2n-2}} = \frac{\rho_0}{2^{n-1} (n-1)!}
\int \rd x  e^{i \vk \cdot \vX} \{ X^{[ j_1} , X^{j_2} \}
\cdots \{ X^{j_{2n-3}} , X^{j_{2n-2} ]} \}
}
(the indices are fully antisymmetrized) is gauge invariant and conserved, 
ensuring that $\rho_{ij}$ satisfies the Bianchi identity. 

The corresponding current in the noncommutative case can be written by 
turning products
into $\ast$-products and Poisson brackets into ($-i$ times) $\star$-commutators. 
The ordering
of the exponential and other factors above has to be fixed in a way
which ensures that the obtained current is conserved. Various such orderings
are possible. For definiteness, we pick the ordering corresponding to the
choice made in:

\begin{eqnarray}
J^{j_1 \dots j_{2n-2}} = \frac{\rho_0}{(2i)^{n-1}}
\int \rd x \negthickspace&& \negthickspace  \int_0^1 ds_1 \cdots \int_0^1 ds_{n-1}
\delta \left( 1- \sum_{i=1}^{n-1} s_i \right) \nonumber \\
\negthickspace&& \negthickspace e_*^{i s_1 \vk \cdot \vX } \bigl[ X^{[ j_1} , X^{j_2} \bigr]_*  * 
\cdots e_*^{i s_{n-1} \vk \cdot \vX } * \bigl[ X^{j_{2n-3}} , 
X^{j_{2n-2}] } \bigr]_*
\label{ddd}
\end{eqnarray}
This corresponds to Weyl-ordering the exponential and distributing it
in all possible ways between the different commutators. Note that the
volume of the $s_i$-integration space reproduces the factor $1/(n-1)!$
present in \rfq{ccc}.

To express compactly the above and to facilitate the upcoming derivations, 
we introduce antisymmetric tensor notation. We define the
basis 1-tensors ${\rm v}_j$ representing the derivative vector field
$\partial_j$, and corresponding one-forms $\rd x^j$. We consider the 
fundamental 1-tensor $\rX$ and the 1-form $\rk$. 

\numeq{eee}{
\rX = X^j {\rm v}_j ~,~~~~ \rk = k_j {\rd x}^j
}
All tensor products will be understood as antisymmetric.

\numeq{fff}{
{\rm v}_j {\rm v}_k \equiv \frac{1}{2} \left(
{\rm v}_j \wedge {\rm v}_k - {\rm v}_k \wedge {\rm v}_j \right), \mbox{\it etc.}
}
This amounts to considering ${\rm v}_j$ and $\rd x^k$ as anticommuting 
quantities. Scalar products are given by the standard contraction .

\numeq{ffg}{
{\rm v}_j \cdot {\rd x}^k = \delta_k^j
}
We also revert to operator notation, dispensing with $\ast$-products
and writing $\Tr$ for $\rho_0 \int \rd x$. 
Finally, we simply write $\int_{(n-1)}$
for the ($n-1$)-dimensional $s_i$-integration.

\numeq{fgg}{
\int_{(n-1)} \equiv \frac{1}{(2i)^{n-1}} \int_0^1 ds_1 \cdots
\int_0^1 ds_{n-1} \delta \left( 1- \sum_{i=1}^{n-1} s_i \right)
}

Overall, the current in \rfq{ddd} is written as the rank-($2n-2$)
antisymmetric tensor $\rJ$,

\numeq{ggg}{
\rJ = \tr  \int_{(n-1)}
e^{i s_1 \rk \cdot \rX } \rX \rX \cdots e^{i s_{n-1} \rk \cdot \rX } 
\rX \rX\, ,
}
and its conservation is expressed by the contraction $\rk \cdot \rJ =0$.
The contraction of $\rk$ acts on each $\rX$ in a graded fashion.
Using cyclicity of trace and invariance under relabeling the $s_i$, 
this becomes

\numeq{hhh}{
\rk \cdot \rJ = (n-1) \tr  \int_{(n-1)}
e^{i s_1 \rk \cdot \rX } \bigl[ \rk \cdot \rX , \rX \bigr]
e^{i s_2 \rk \cdot \rX } \rX \rX \cdots e^{i s_{n-1} \rk \cdot \rX } \rX \rX
}

Using the identity

\numeq{iii}{
\bigl[ e^{i s \rk \cdot \rX} , \rX \bigr] =
\int_0^s  ds_1 e^{i s_1 \rk \cdot \rX } \bigl[ i \rk \cdot \rX , \rX \bigr]
e^{i (s - s_1 ) \rk \cdot \rX }
}

we can absorb the $s_1$-integration in \rfq{hhh} and bring it to the form

\numeq{jjj}{
\rk \cdot \rJ = -\frac{1}{2} \tr  \int_{(n-2)}
\bigl[ e^{i s_2 \rk \cdot \rX} , \rX \bigr] 
\rX \rX \cdots e^{i s_{n-1} \rk \cdot \rX } \rX \rX
}
Finally, using once more the cyclicity of trace, we see that the above
contraction vanishes. This proves that the tensor $\rJ$ is conserved and,
as a consequence, its dual $\rho_{jk}$ satisfies the Bianchi identity.
As in the 2-dimensional case, we put

\numeq{kkk}{
F_{jk} (\vk ) = \rho_{jk} (\vk) - \omega_{jk} \delta (\vk)\, ,
}
and recover the commuting Abelian field strength, which can, in turn, be
expressed in terms of a (commutative) Abelian potential $A_j$.

In the above manipulations we freely used cyclicity of trace. In general
this is dangerous, since the commuted operators may not be trace class. 
Assuming, however, that $\vX$ becomes asymptotically $\vx$ for large distances,
the presence of the exponentials in the integrand
ensures that this operation is permissible.

As mentioned previously, the fully symmetric ordering is not the only
one that leads to an admissible $\rho_{jk}$. As an example, in the
lowest-dimensional nontrivial case $d=4$ we can alter the ordering
by splitting the commutator as

\numeq{lll}{
J^{jk} = \frac{1}{2i} \tr  e^{i \vk \cdot \vX } \bigl[ X^j , X^k \bigr]
 ~ \to ~
J_f^{jk} = -i \tr \int_0^1 ds f(s) e^{i s \vk \cdot \vX } X^j 
e^{i (1-s) \vk \cdot \vX } X^k\, . 
}
If $f(s) = - f(1-s)$ the above will be antisymmetric in ($j,k$) and 
conserved, as can explicitly be verified. Further, if $f(s)$ satisfies

\numeq{mmm}{
\int_0^1 ds (2s-1) f(s) = 1\, ,
}
then \rfq{lll} will also have the correct commutative limit. We obtain an
infinity of solutions depending on a function of one variable $f(s)$.
This arbitrariness reflects the fact that the \sw\ equations are not
integrable and therefore the solution for $\vth = 0$ depends on the
path in the $\vth$-space taken for integrating the equations. For
$d=4$ the parameter space is a plane and the
path from a given $\vth$ to $\vth =0$ on the plane
can be parametrized by a function of a single variable, just like
$J_f^{jk}$. The various solutions are related through field redefinitions.

\subsubsection*{(iii) \sw \ map in  higher odd  dimensions}\label{s3ssB}
\addcontentsline{toc}{subsubsection}{(iii) \sw \ map in  higher odd  dimensions}
%\vspace*{1in}

The situation in odd dimensions differs in that we need to specify
separately the components of the conserved current in the commutative
and noncommutative directions. For $d=2n+1$ the current is of rank
$2n-1$ and it can be constructed by a procedure analogous to the
even-dimensional case: We start from the expression for the total
particle current $j^\mu$ (\ref{eqone13}) and (\ref{eqone14}) and introduce 
$2n-2$ commutators, one less than the number which would fully saturate it 
to $(c, \vv )$. The temporal components $J^{0 j_1 \dots j_{2n-2}}$ can be
expressed as a rank-$(2n-2)$ antisymmetric spatial tensor $\rJ^0$,
while the spatial components $J^{j_0 j_1 \dots j_{2n-2}}$ can be expressed
as a rank-$(2n-1)$ antisymmetric tensor $\rJ$. Their fully
ordered expressions are

\numeq{nnn}{
\rJ^0 = \frac{1}{n-1} \tr  \int_{(n-1)}
e^{i s_1 \rk \cdot \rX } \rX \rX \cdots e^{i s_{n-1} \rk \cdot \rX } \rX \rX,
}

\numeq{ooo}{
\rJ = \tr  \int_{(n)} e^{i s_0 \rk \cdot \rX} D \rX
e^{i s_1 \rk \cdot \rX } \rX \rX \cdots e^{i s_{n-1} \rk \cdot \rX } \rX \rX\, .
}
The above expressions can be unified by introducing a temporal component
for the field $X^\mu$, namely $X^0 \equiv t$ (which is obviously commutative),
and extending the one-tensor $\rX$ also to include $X^0 {\rm v}_0$. Further,
we can Fourier transform in time and define $\rk = k_\mu {\rd x}^\mu$ to
include also the frequency $k_0$. Then the corresponding (space-time)
($2n-1$)-tensor $\rJ$ acquires the form

\numeq{oop}{
\rJ = \int dt ~\tr  \int_{(n)} e^{i s_1 \rk \cdot \rX} \, D \rX \,
e^{i s_2 \rk \cdot \rX } \rX \rX \dots e^{i s_n \rk \cdot \rX } \rX \rX\, .
}
$X^0$ is absent in $\rX \rX$ and, since $D X^0 = 1$, only $s_0 + s_1$ appears 
in the temporal component of $\rJ$; integrating over $s_1$ reproduces
the factor $1/(n-1)$ appearing in \rfq{nnn}.

The above current is obviously gauge invariant. We shall prove that it
is also conserved, that is, it satisfies $\rk \cdot \rJ = 0$.
The contraction is
\begin{eqnarray}
\rk \cdot \rJ = \int dt ~\tr  \int_{(n)} \negthickspace \negthickspace \negthickspace \negthickspace &&
\negthickspace\negthickspace
\negthickspace
\negthickspace
\Bigl\{ e^{i s_1 \rk \cdot \rX} \, \rk \cdot D\rX \,
e^{i s_2 \rk \cdot \rX } \rX \rX \cdots e^{i s_n \rk \cdot \rX } \rX \rX
\nonumber \\
&-& \sum_{m=2}^n 
e^{i s_1 \rk \cdot \rX} \, D\rX \, \rX \rX \cdots
e^{i s_m \rk \cdot \rX } \bigl[ \rk \cdot \rX , \rX \bigr] 
e^{i s_{m+1} \rk \cdot \rX } \cdots \rX \rX \Bigr\}
\end{eqnarray}
(with $s_{n+1} =0$). By formula \rfq{iii} and a similar one for the
covariant time derivative, the above can be rewritten as

\begin{eqnarray}
\rk \cdot \rJ = \int dt ~\tr  \int_{(n-1)} \negthickspace \negthickspace \negthickspace \negthickspace &&
\negthickspace\negthickspace
\negthickspace
\negthickspace \Bigl\{
De^{i s_1 \rk \cdot \rX} \,
e^{i s_2 \rk \cdot \rX } \rX \rX \cdots e^{i s_{n-1} \rk \cdot \rX } \rX \rX
 \nonumber \\
&-& \sum_{m=2}^{n-1} 
e^{i s_1 \rk \cdot \rX} \, D\rX \, \rX \rX \cdots
\bigl[ e^{i s_m \rk \cdot \rX } , \rX \bigr] 
\rX \rX \cdots \rX \rX \Bigr\}.
\end{eqnarray}
Due to the cyclicity of trace, the sum above telescopes and only the
first term of the $m=2$ commutator and the second term of the $m=n-1$
commutator survive. Altogether we obtain

\begin{eqnarray}
\rk \cdot \rJ &=& \int dt ~\tr  \int_{(n-1)} \left(
D e^{i s_1 \rk \cdot \rX} + D \rX \rX + \rX D \rX \right)
e^{i s_2 \rk \cdot \rX } \rX \rX \cdots e^{i s_{n-1} \rk \cdot \rX } \rX \rX
\nonumber \\
&=& \int dt ~\tr  \int_{(n-1)} D \left( e^{i s_1 \rk \cdot \rX} \rX \rX \right) 
\cdots e^{i s_{n-1} \rk \cdot \rX } \rX \rX \nonumber \\
&=& \int dt ~\tr  \int_{(n-1)} \frac{1}{n-1} D \left(
e^{i s_1 \rk \cdot \rX} \rX \rX
\cdots e^{i s_{n-1} \rk \cdot \rX } \rX \rX \right) \\
&=& \int dt ~\frac{d}{dt} \tr  \int_{(n-1)} \frac{1}{n-1}
e^{i s_1 \rk \cdot \rX} \rX \rX
\cdots e^{i s_{n-1} \rk \cdot \rX } \rX \rX \nonumber \\
&=& 0, \nonumber
\end{eqnarray}
which proves the conservation of $\rJ$. Its dual $\rho_\mn$ satisfies
the ($2n+1$)-dimensional Bianchi identity and can be used to define the
commutative Abelian field stength.

\begin{eqnarray}
F_{ij} (\vk ) &=& \rho_{ij} (\vk) - \omega_{ij} \delta (\vk) \\
F_{0i} (\vk ) &=& \rho_{0i} (\vk )
\end{eqnarray}

In the above we gave separate derivations of the \sw\ map for even
and odd dimensions. The two can be unified by demonstrating that each
case can be obtained as a dimensional reduction of the other in one
more dimension. This is treated next section.

\subsubsection*{ (iv) Dimensional reduction}\label{s3ssC}
\addcontentsline{toc}{subsubsection}{(iv) Dimensional reduction}

It is quite straightforward to see that the even dimensional \sw\ map
is obtained from the $d=2n+1$ map by dimensional reduction.
We assume a time-independent configuration in which $X^j$ 
($j=1,\dots ,2n$) do not depend on $t$ and $A_0$ vanishes.
In this case $D \rX$ vanishes and so does
$\rJ$ in\rfq{ooo}; only the component $\rJ^0$ 
in\rfq{nnn} survives, reproducing the $2n$-dimensional solution.

The reduction from a fully noncommutative $d=2n+2$ case to the $d=2n+1$
case is only slightly subtler. For concreteness, we shall take 
$t \equiv x^0$ to be canonically conjugate to the last dimension, 
call it $z \equiv x^{2n+1}$, which will be reduced; that is, 

\numeq{separ}{
[ t , z ] = i \theta_0 ~~(\theta_0 = \theta^{0,2n+1})~,~~~
[ t , x^i ] = [ z , x^i ] = 0 ~~(i=1,\dots ,2n).
}
This can always be achieved with an orthogonal rotation of the $x^\mu$.
The reduced configuration consists of taking all fluid coordinates other 
than $X^{2n+1}$ to be independent of $x^{2n+1}$ and, further, the gauge
potential corresponding to $z=x^{2n+1}$ to vanish. Specifically,
\begin{eqnarray}
X^i &=& X^i ( \vx , t), \\
X^0 &=& t, \\
X^{2n+1} &=& z + \theta_0 A_0 ( \vx , t) .
\end{eqnarray}
With this choice the corresponding field strengths become

\begin{align}
[X^i,X^j] &= i\tn ij + i\tn ik \tn j\ell \hat F_{k\ell}\label{fij},\\
[ X^i , X^0 ] &= 0, \\
[ X^i , X^{2n+1} ] &= i \theta_0 ( D_0 X^i - i [X^i , A_0 ] ) 
= i\theta^{ij} \theta^{2n+1,0} \hat F_{j0},
\end{align}
with $\hat F_{\mu \nu}$ ($\mu,\nu=0,\dots,2n$) the field strength of
a noncommutative $d=2n+1$ theory.

The corresponding $d=2n+2$ \sw\ map reduces to the $d=2n+1$ map.
Indeed, the current $\rJ$ in\rfq{ggg}, now, is a rank-$2n$ antisymetric 
tensor. When all its indices are spatial ($1,\dots,2n$) it becomes a fully
saturated topological invariant, that is, a constant; this reproduces
a constant $\rho_{0,2n+1}$. When one of its indices is $0$ and
the rest are spatial it vanishes, leading to $\rho_{i,2n+1} = 0$.
When one of its indices is $2n+1$ and the rest are spatial it reproduces
expression \rfq{ooo}. Finally, when two of its indices are $0,2n+1$
and the rest are spatial it reproduces\rfq{nnn}, recovering the full
commuting ($2n+1$)-dimensional Abelian field strength. 

We stress that the above reductions are not the most general ones.
Indeed, mere invariance of the fluid configuration with respect to
translations in the extra dimension does not require the vanishing of
the gauge field in the corresponding direction. This means that we
could choose $X^0 = t + H (\vx ,t)$ (instead of $X^0 = t$) in both
$d=2n+1$ and $d=2n+2$. The corresponding reduced theory contains an
extra Higgs scalar in the adjoint representation of the (noncomutative)
U(1) gauge group. Our \sw\ map in this situation reproduces, with
no extra effort, the space-time derivatives of a corresponding
commuting `Higgs' scalar.

The above complete reduction scheme 
($2n+2 \to 2n+1 \to 2n \to \dots$) is reminiscent of the topological 
descent equations relevant to gauge anomalies. It is possible to consider
the fluid analogs of noncommutative topological actions and the mapping
of topologically nontrivial configurations \cite{Polyc}, but we shall not
consider these issues here.

A final comment:
We demonstrated that gauge theory on noncommutative spaces has an effective
description as a classical fluid. Other nonclassical situations may
also be describable in a fluid dynamical language, thus revealing their
dynamics as pertaining to a spacial kind of fluid. In particular,
quantum mechanical many-body states could be effectively described
in this fashion. This will be explored in the  final Section.

\newpage
\section{MISCELLANEOUS TOPICS}
\subsection{Quantized fluid mechanics}
Eulerian fluid mechanics, even though it can be formulated as an independent
dynamical system in its own right, can also be understood as a good description for
the  physics of a distribution of particles at large length scales. At these scales, large
compared to the typical particle  separations, a continuum approximation can be
made and the underlying particulate nature is seldom apparent or needed. If this
fluid mechanics is quantized, the result is a quantum field theory where the
hydrodynamical variables of density and velocity become quantum operators. For
such an analysis, it is mathematically irrelevant whether the fluid theory emerged
as an approximation to the particle theory.

For quantization, the Poisson brackets (1.2.36), (1.2.39) and (1.2.40) are replaced by
$i/\hbar$ times a quantum commutator of the $\rho$ and $\bf v$ operators. There is
no ordering ambiguity in the quantal version of (1.2.40) because $\rho$ commutes
with
$\omega_{ij}$, according to (1.2.39) and (1.2.41). Ordering does have to be prescribed
when the Hamiltonian (1.2.33) is promoted to an Hermitian operator. The kinetic
term may be taken as $\frac{1}{2} \, v^i \, \rho \, v^i$ or as $\frac{1}{4} (\rho {\bf
v}^2 + {\bf v}^2 \rho)$; the two coincide because the $[\rho, {\bf v}]$ commutator is a
c-number. The Heisenberg equations of motion then imply that the current
contributing to the continuity equation (1.1.16) is the ordered Hermitian quantity
\numeq{eq:7.1.1}{
{\bf j} = \frac{1}{2} (\rho {\bf v} + {\bf v} \rho),
}
which also satisfies the quantum commutator analogs of (1.2.37) and (1.2.38). In the
Euler force equation (1.1.17) the term non-linear in {\bf v} emerges upon
commutation with the quantum Hamiltonian as $\frac{1}{2} \, (v^k \, \partial_k \, v^i
+ \,\partial_k \, v^i \, v^k)$.

%insert text
It is noteworthy that the nature of the particles which constitute the fluid, in
particular their statistics, does not seem to be important. For this reason, attempts to
 relate superfluidity to quantized hydrodynamics have been criticized \cite{Feyn}.
However, one can see that this is clearly not the whole story by the following
argument.

Consider the classical action for a (non-linear) Schr\"{o}dinger field on a
$d$-dimensional space ${\bf R}^d; \\ d = 2,3$ are the cases of particular interest to us.

\numeq{eq:7.1.2}{
I = \int d t  dr [i\hslash\,  \psi^\ast \, \dot{\psi} - \frac{\hslash^2}{2} \vec \nabla
\psi^\ast
\cdot
\vec \nabla \psi - \mathcal{V} (\psi^\ast \psi)]
}
 With the Madelung {\it Ansatz} (G.4) $(m=1)$, the action \refeq{eq:7.1.2} is brought
to a form appropriate for an irrotational fluid.
\numeq{eq:7.1.3}{I = \int dt  dr [\theta \dot{\rho} - \frac{1}{2} \rho (\vec \nabla
\theta)^2-V(\rho)],
}
where the classical potential $\cal{V}$ acquires an addition.
\numeq{eq:7.1.4}{
V(\rho) = \mathcal{V}(\rho) \frac{\hbar^2}{2} (\vec \nabla \sqrt{\rho})^2}
% [insert text-being typset, so check carefully]
We see that the Schr\"{o}dinger field theory (7.1.2) and fluid mechanics are
equivalent with a reinterpretation of certain quantities, the particular nature of the
fluid (such as its equation of state) will be characterized by the function $V(\rho)$.
The Schr\"{o}dinger theory (7.1.2) can be quantized as a fermion field or a boson
field. Therefore it must be possible to quantize the fluid theory (which is after all the
same thing, being a particular parametrization or choice of coordinates on the
classical phase space of the theory) as either a fermion theory or a boson theory.
Since the symplectic structure, given by the $\theta \dot{\rho}$-term, is what is
relevant to this consideration, the specific nature of $V(\rho)$ should not matter; it
should be possible to quantize any fluid, irrespective of its dynamics, with either
statistics.

This observation is certainly not new. The fluid theory is described in terms of the
density $\rho$ and the current ${\bf j} = \rho \vec \nabla \theta$. The algebra of
these observables, the current algebra, is given by (1.2.36)-(1.2.38).
This algebra is the same whether $\psi$, $\psi^\ast$ obey bosonic or fermionic
commutation rules. Various observables of interest, such as the Hamiltonian, momentum density, etc., can be constructed in terms of $\rho$, ${\bf j}$. Therefore, one can take the
algebra of $\rho$, ${\bf j}$ as the starting point and construct the quantum theory
in terms of a unitary irreducible representation of this algebra of
operators.
It is known that this algebra has many inequivalent realizations,
allowing for the freedom of choosing different statistics, bosons and fermions
corresponding to different representations \cite{Dashen}. The fluid theory can indeed
carry information about the statistics of the particles which compose it.

More generally, even for a one-component fluid, there are additional variables
needed. Considering the case of three spatial dimensions as an example, the free
action is
\begin{equation}
I_0 = - \int dt \ d r\, [\rho\, (\dot{\theta} + \alpha \dot{\beta}) + \frac{1}{2} \rho\,
(\vec
\nabla \theta + \alpha \vec \nabla \beta)^2].
\label{7.1.5}
\end{equation}
The fluid velocity is given in the Clebsch form ${\bf v} = \vec \nabla \theta + \alpha
\vec \nabla \beta$. The Clebsch parameterization of the velocities can be expressed as
[see Section 7.3 (i), below]
\begin{equation}
{\bf v} = i \, {\mathrm t} {\mathrm r} (\sigma_3 g^{-1} \vec \nabla g),
\label{7.1.6}
\end{equation}
where the group element $g$ is  an element of SU(2).
\begin{equation}
g = e^{\frac{\sigma^3}{2i}\beta} \, e^{\frac{\sigma^2 }{2i}\gamma} \,
e^{\frac{\sigma^3}{2i}\theta}\
\label{7.1.7}
\end{equation}
Using (\ref{7.1.7}) with (\ref{7.1.6}) gives the Clebsch parametrization for $\bf v$
with
$\alpha = \cos \gamma$.
Even though we have the additional variables $\alpha, \beta$ ,  the current algebra is
the same as before. Evidently the same current algebra is realized by fluids that
move with or without vortices.

Generally speaking, in two dimensions, the existence of inequivalent
representations is related to the nontrivial connectivity of the
space of fields.
 If the phase space is simply connected, it cannot support double-valued (or many-valued) wave functions and one 
does not have the possibility of  different statistics. The topology of the
phase space is therefore relevant to the question we are considering and
one can try to characterize the inequivalent representations for the fluid
current algebra along these lines. To see how this arises, it is instructive to
consider the configuration space for identical particles. For $N$ particles in
two dimensions, a point in the configuration space is given by $({\bf x}_1,
{\bf x}_2, ..., {\bf x}_N)$, where $ {\bf x}_i \in {\bf R}^2$. The identity of
particles tells us that we must make the identification
\beq
({\bf x}_1, {\bf x}_2, ...,{\bf x}_i , ..., {\bf x}_j, ..., {\bf x}_N) ~\sim ~
({\bf x}_1, {\bf x}_2, ...,{\bf x}_j, ..., {\bf x}_i, ..., {\bf x}_N) 
\label{7.1.7a}
\eeq
Further, we must impose the condition that the locations of the particles do not coincide,
i.e., ${\bf x}_i \neq {\bf x}_j$ if $i\neq j$. The resulting configuration space ${\cal C}_N$
has nontrivial connectivity; in fact, the first homotopy group is given by
$\Pi_1 ( {\cal C}_N)= B_N$, where $B_N$ is the braid group. 
The fact that $\Pi_1 ({\cal C}_N) \neq 0$ shows that the configuration space can support
many-valued wave functions. Under exchange of particles, the wave functions
get a phase; the various phase factors for different exchanges form a unitary
irreducible representation of $\Pi_1 ({\cal C}_N)$. The phase for a single exchange can be arbitrary and so we get the possibility of arbitrary statistics.
(An analogous construction can be done in three dimensions; the first
homotopy group is then the permutation group $S_N$ and we get the
possibility of fermions or bosons.)

Returning to the case of two dimensions, ${\bf X} = (x,y)$, notice that the existence of
different representations is also related to the existence of a flat potential, which
distinguishes between the representations. For the $N$-particle Heisenberg algebra,
we can write a representation in terms of complex variables $w = x-i y, {\bar w} = x +
iy$.
\beqar
{\hat {\bf X}_i} &=&  (w_i, {\bar w}_i) \nonumber\\
{\hat P}_{w_i} &=&  {\hat p}_{w_i}  + {i\over 2k} \sum_{j\neq i} {1\over w_i -w_j}
\nonumber\\
{\hat P}_{\bw_i} &=&  {\hat p}_{\bw_i}- {i\over 2k} \sum_{j\neq i} {1\over {\bar w}_i -{\bar
w}_j}\label{7.1.7b} \\ 
{\hat p}_{w_i} = -i {\del \over \del w_i }, &&
{\hat p}_{\bw_i} = -i {\del \over \del {\bar w}_i } \nonumber
\eeqar
$k$ is a parameter specifying the
representation, ${\hat p}$ gives the standard Schr\"odinger representation,
and the summed expressions comprise the potentials, which make the
representation (7.1.9) different from the standard one. It is easily verified
that the quantities in (7.1.9) obey the Heisenberg algebra. [The commutator
of the  momenta leads to $\delta ({\bf x}_i - {\bf x}_j)$, but since we do not
allow coincidence of  particle locations (or the wave functions vanish at
coincidences), this is zero.] We can write the representation (7.1.9) as
\beq
{\hat {\bf X}} = U^{-1} {\bf x} U, \hskip .5in {\hat P} = U^{-1} {\hat p} U
\label{7.1.7c}
\eeq
where $U$ is given by
\beq
U = \exp \left( -{i \over 2k} \sum [ \ln (w_i -w_j) - \ln ({\bar w}_i - {\bar w}_j ) ]
\right)
\label{7.1.7d}
\eeq
The representation (\ref{7.1.7b}) is however not equivalent to the Schr\"odinger representation
because $U$ is not single-valued and so does not give a unitary transformation.
Notice that the addition to the expression for the momentum operators is a ``flat potential'',
in the sense that its field strength, which is given by the commutators
$[{\hat P}_{w_i} , {\hat P}_{\bw_j}] $, is zero (with the condition of removing coincidences
of particle locations). This flat potential can be written as $U^{-1} [{\hat p}, U]$,
but it should be kept in mind that $U$ is not a genuine unitary transformation.

Similar results hold in general.
Let $\{\phi_a\}$ be a set of observables, which may include the identity,
obeying  a commutation algebra of the  form
\begin{equation}
[\phi_a, \phi_b] = C^c_{ab} \phi_c.
\label{7.1.9}
\end{equation}
Let $\phi^{(1)}_a$ and $\phi^{(2)}_b$ be two representations. We write $\phi^{(2)}_a =
\phi^{(1)}_a + \mathcal{A}_a$. The fact that $\phi^{(1)}_a$ and $\phi^{(2)}_a$ obey the
same algebra shows that $\mathcal{A}_a$ is a flat potential, {\it i.e.,}
\begin{equation}
[\phi^{(1)}_a, \mathcal{A}_b] - [\phi^{(1)}_b, \mathcal{A}_a] + [\mathcal{A}_a,
\mathcal{A}_b] - C^c_{ab} \mathcal{A}_c =0.
\label{7.1.10}
\end{equation}
If $\mathcal{A}_a = U^\dagger [\phi^{(1)}_a, U] $ for some unitary transformation U,
the two representations are unitarily equivalent. However, if $\mathcal{A}_a$ obeys
the zero-curvature condition (\ref{7.1.10}) but cannot be written as $\mathcal{A}_a
= U^\dagger [\phi^{(1)}_a, U]$ for some unitary $U$, the two
representations are inequivalent. 

Clearly, the
existence of inequivalent representations and the
existence of a flat potential which cannot be written as $\mathcal{A}_a = U^\dagger
[\phi^{(1)}_a, U] $ are related to the topology of the space of fields, or the phase space,
if we are thinking in terms of canonical quantization.
The $U$'s which we use form a representation of some nontrivial
$\Pi_1 ({\cal C})$; flat potentials
require in general that the first cohomology group
of
the phase space should be nontrivial. 
We have seen this explicitly at the level of particles.
When we generalize to a field theory one can ask whether statistics can be
obtained in such terms. In fact, a
description of the boson and fermion representations and an expression for the
corresponding flat potential have been obtained \cite{Dashen}. It is thus
possible to identify an operator, namely the flat potential $\mathcal{A}_a$,
which can distinguish the statistics of the underlying
particles. Nevertheless, the situation is not entirely satisfactory.
The flat potential has been obtained only for a subspace with fixed
value of
the particle number $N$ and it is in terms of
the phase of the fermionic ground state wave function for $N$ particles.
In our opinion, this answer
partially begs the question.
Also, the connection to the topology of the phase space in unclear.
A more direct method, in terms of the fields, would certainly be better. 

Turning to the
field variables $\rho$, ${\bf j}$, we
first note that, already, the use of  a group element to parameterize the velocities 
as in (\ref{7.1.7}) entails making certain additional
assumptions about the topology of the phase space. One has to determine on a
physical grounds whether such a parameterization is justified and if so, which group
should be used. For the parameterization (\ref{7.1.7}), the phase space is given by
\begin{equation}
\mathcal{P} = \{\mbox{set of all maps}\, \rho : {\bf R^3 \to R}_+, \, g \, : {\bf R^3} \to G
= SU (2) \}
\label{7.1.8}
\end{equation}
The topology of the phase space is specific to $SU (2)$: there is a compact $U(1)$
direction corresponding to $\theta$. Since $\rho$ and $\theta$ are canonically
conjugate, the operator $U = \mbox{exp}\, (i2 \pi \, \int \, d r \, \rho)$ shifts $\theta$
by $2\pi$. The compactness of $\theta$ requires that all observables be invariant
under the action of  $U$; equivalently, the spectrum of $ \int d r \, \rho$
must be an integer spaced. Thus the compactness of $U(1)$ leads to an
underlying particle description, with the integer portion of
$\int d r \rho$ being the number of particles $N$. Therefore, for
hydrodynamics with an underlying particle structure, the use of a compact
$\theta$-variable is appropriate. Notice also that $ \int d r \, \rho$ is a Casimir
operator for the algebra so that one can also restrict to representations with a fixed
value of $N$. If we have only
$(\rho,
\theta)$, this argument just recaptures the Schr\"{o}dinger field description.

One can now try to see how the flat potentials mentioned above can arise.
in the field description.
So far, there is no direct
construction of the potential $\mathcal{A}_a$  in terms of $\rho$ and $\bf j$.
In our formulation of the Clebsch parametrization, the
phase space given by (7.1.8) does not have nontrivial first homotopy group. 
One might argue that this is because we have to impose some further condition,
analogous to the condition of excluding coincidence of locations at the particle level.
It may be that a similar condition must be imposed, expressed in fluid variable
terms, on the space of (\ref{7.1.8}) to obtain the necessary structure for a quantum
hydrodynamics, which takes account of particle statistics. It is not yet clear what
such a condition would be. Also, representations of the algebra of observables are
characterized, in purely algebraic terms, by the values of the Casimir operators.
Thus, alternatively, one can ask the question: What are the Casimirs in terms of the
fluid observables which discriminate between the different statistics? How are these
Casimirs related to the topology of the phase space? As yet there is no completely
satisfactory answer to these questions.

\subsection{Fluids in  intense magnetic fields}\label{Sec7.2}
\setcounter{equation}{0}
Another interesting dynamical system is a charged fluid in an intense magnetic field.
This exhibits a noncommutativity, which is the fluid mechanical version of
noncommuting coordinates for a point particle in a magnetic field, so intense that it
effects reduction to the lowest Landau level.

\subsection*{(i) Particle noncommutativity in the lowest Landau level}
Before describing the motion of a charged fluid in an intense magnetic
field, we review the story for point particles on a plane, with an external and
constant magnetic field $\bf B$ perpendicular to the plane \cite{Dunne}.
The equation of motion for the 2-vector ${\bf r} = (x,y)$ is 
\begin{equation}
m \dot{v}^i = \frac{e}{c}\epsilon^{ij}v^j B+f^i({\bf r}),
\label{eq31}
\end{equation}
where $\bf v$ is the velocity $\dot{{\bf r}}$, and $\bf f$ represents
other forces, which we take to be derived from a potential $V$: 
${\bf f} = - {\bf\nabla}V$. The limit of large $B$ is equivalent to
small $m$. Setting the mass to zero in (\ref{eq31}) leaves a first
order equation.
\begin{equation}
\dot{r}^i = \frac{c}{eB}\epsilon^{ij}f^j({\bf r})
\label{eq32}
\end{equation}
This may be obtained by taking Poisson brackets of $\bf r$ with 
the Hamiltonian
\begin{equation}
H_0 = V,
\label{eq33}
\end{equation}
provided the fundamental brackets describe noncommuting coordinates,
\begin{equation}
\{ r^i,r^j \} = \frac{c}{eB}\epsilon^{ij},
\label{eq34}
\end{equation}
so that 
\begin{equation}
\dot{r}^i = \{ H_0, r^i \} = \{ r^j, r^i \}\partial_j V =
\frac{c}{eB}\epsilon^{ij}f^j({\bf r}).
\label{eq35}
\end{equation}

The noncommutative algebra (\ref{eq34}) and the associated dynamics
can be derived in the following manner.
The Lagrangian for the equation of motion (\ref{eq31}) is
\begin{equation}
L = \frac{1}{2}m {\bf v}^2 + \frac{e}{c}{\bf v} \cdot {\bf A} - V.
\label{eq36}
\end{equation}
When we choose the gauge ${\bf A} = (0,Bx)$ and set
$m$ to zero, (\ref{eq36}) leaves
\begin{equation}
L_0 = \frac{eB}{c} x \dot{y} - V(x,y),
\label{eq37}
\end{equation}
which is of the form $p\dot{q} - h(p,q)$, and one sees that
$(\frac{eB}{c}x,y)$ form a canonical pair.  This implies (\ref{eq34}),
and identifies $V$ as the Hamiltonian.  

Additionally, we give a canonical derivation of noncommutativity in
the $m \rightarrow 0$ limit, starting with the Hamiltonian
\begin{equation}
H = \frac{ {\vec \pi}^2 }{2m} + V.
\label{eq38}
\end{equation}
$H$ gives (\ref{eq31}) upon bracketing with $\bf r$ and ${\bf v}$, 
provided the following brackets hold.
\begin{eqnarray}
&&\{ r^i, r^j \} = 0 \label{eq39a} \\
&&\{ \pi^i, r^i \} = \delta^{ij} \label{eq39b} \\ 
&&\{ \pi^i,\pi^j \} = -\frac{eB}{c} \epsilon^{ij} \label{eq39c} 
\end{eqnarray}
Here $\boldsymbol \pi$ is the kinematical (non-canonical) momentum,
$m \dot{{\bf r}}$, related to the canonical momentum $\bf p$
by $\vec \pi = {\bf p} - \frac{e}{c}{\bf A}$.

We wish to set $m$ to zero in (\ref{eq38}).
This can only be done provided $\vec \pi$ vanishes, and
we impose $\vec \pi = 0$ as a constraint.  But according to
(\ref{eq39c}),  the bracket of the constraints 
$\{ \pi^i, \pi^j\} \equiv C^{ij} = -\frac{eB}{c}\epsilon^{ij}$ is non-zero.
Hence we must introduce Dirac brackets:
\begin{equation}
\{ O_1, O_2 \}_D = \{ O_1, O_2 \} 
-\{ O_1, \pi^k \}(C^{-1})^{kl} \{ \pi^l, O_2 \}. 
\label{eq310}
\end{equation}
With (\ref{eq310}), any Dirac bracket involving $\vec \pi$ vanishes,
so $\vec \pi$ may indeed be set to zero. But the Dirac bracket of two 
coordinates is now non-vanishing.
\begin{equation}
\{ r^i, r^j \}_D = -\{r^i, \pi^k\}\frac{c}{eB}
\epsilon^{kl} \{ \pi^l, r^j \}  = \frac{c}{eB}\epsilon^{ij}
\label{eq311}
\end{equation}
In this approach, noncommuting coordinates arise as 
Dirac brackets in a system constrained to lie in the lowest 
Landau level.

A quantum mechanical perspective on this result is  the following \cite{Magro}. Let us label the
degenerate quantum Landau states by ($N,n$), where $N$  labels the Landau level
and
$n$ the degeneracy of that level. Consider the computation of the matrix element in
the lowest $(N=0)$ Landau level of the commutator of $x$ as $y$.
\begin{eqnarray}
<0, n |[x, y]| 0, \Tilde{n}> = <0, n |x y| 0, \Tilde{n}> - <0, n \, |y x| 0, \Tilde{n}>
\nonumber\\ 
=\sum_{Nn'}(<0, n |x| N n'> <N n' |y| 0, \Tilde{n}> - <0, n |y| N n'> <N n' |x| 0,
\Tilde{n}>)
\label{7.2.14}
\end{eqnarray}
When the intermediate state sum is carried over all Landau levels, the two terms in
the sum cancel each other, and we find a vanishing result; coordinates commute.
Suppose however, we truncate the sum at the lowest level. A simple calculation gives
a non-vanishing result, consistent with (7.2.4). This shows that noncommuting
coordinates arise from a truncation of the Hilbert space.

[An amusing generalization determines the coordinate commutator when the
first $N$ Landau levels are retained, both in the external states and the
intermediate state sum. The result is that the only non-vanishing matrix element of
the $[x,y]$ commutator is in the highest Landau level.
\numeq{eq:7.2.15}{
<N,n |[x,y] |N, n'>= -\frac{i \hslash c}{e B} \, (N+1) \delta \, (n, n')
}
Here $\delta (n,n')$ is either the discrete or continuous  delta function, depending
whether the Landau degeneracy is exhibited in a discrete or continuous manner.
The result reduces to the previous, when only the lowest Landau level is kept, and
 also shows that as more and more states are included, the non commutativity is
pushed into ever higher states.]

%[Insert Z. Guralnik text]
\subsubsection*{(ii) Field noncommutativity in the lowest Landau level}
\addcontentsline{toc}{subsubsection}{(i) Particle noncommutativity in the lowest
Landau level}
\addcontentsline{toc}{subsubsection}{(i) Field noncommutativity in the
lowest Landau level}

We now turn to the equations of a charged fluid with density 
$\rho$ and mass parameter $m$ (introduced
for dimensional reasons) moving on a plane with velocity $\bf v$ in an
external magnetic field perpendicular to the plane. $\rho$ and
$\bf v$ are
functions of $t$ and $\bf r$ and give an Eulerian description
of the fluid.  The equations that are satisfied are the
continuity equation (1.2.42),  and the Euler equation (1.2.43), which acquires an
additional term, beyond the force team, due to the magnetic interaction [compare
(5.4.19) reduced to the Abelian case].
\begin{equation}
m \dot{v}^i + m{\bf v} \cdot {\vec \nabla} v^i = 
\frac{e}{c}\epsilon^{ij}v^jB + f^i
\label{eq313}
\end{equation}
Here $\bf f$ describes additional forces, e.g. 
$-\frac{1}{\rho} {\bf\nabla} P$ where $P$ is pressure.
We shall take the force to be derived from a potential in the 
form 
\begin{equation}
{\bf f}({\bf r}) = -{\vec\nabla} 
\frac{\delta}{\delta\rho({\bf r})} \int d^2 r V.
\label{eq314}
\end{equation}
[For isentropic systems, the pressure is only a function of
$\rho$.
Here we allow more general dependence of $V$ on $\rho$
({\it e.g.} nonlocality or dependence on derivatives of $\rho$).]

The relevant equations follow by bracketing 
$\rho$ and $\bf v$ with the Hamiltonian
\begin{equation}
H= \int d^2r \left( \rho \frac{ {\vec \pi}^2 }{2m} + V \right).
\label{eq315}
\end{equation}
provided that fundamental brackets are taken as
\begin{eqnarray}
&&\{ \rho({\bf r}), \rho({\bf r}^{\prime}) \} = 0 \label{eq316a} \\
&&\{ \vec \pi({\bf r}), \rho({\bf r}^{\prime}) \} = 
{\bf\nabla} \delta({\bf r} - {\bf r}^{\prime}) \label{eq316b} \\
&&\{ \pi^i({\bf r}), \pi^j({\bf r}^{\prime}) \} =
-\epsilon^{ij}\frac{1}{\rho}\left(m \omega({\bf r}) + 
\frac{eB}{c}\right) \delta({\bf r}-{\bf r}^{\prime}) \label{eq316c}
\end{eqnarray}
where $\epsilon^{ij}\omega({\bf r})$ is the vorticity 
$\partial_iv^j -\partial_jv^i$, and $\vec \pi = m {\bf v}$. These are just the rescaled
(by $m$) versions of (1.2.36), (1.2.39) and (1.2.40), except that the  last is modified
to include the constant magnetic field.

We now consider a strong magnetic field and take the limit 
$m\rightarrow 0$, which is equivalent to large $B$. 
Equations (\ref{eq313}) and (\ref{eq314})
reduce to
\begin{equation}
v^i = -\frac{c}{eB}\epsilon^{ij} \frac{\partial}{\partial r^j}
\frac{\delta}{\delta\rho({\bf r})} \int d^2r V.
\label{eq317}
\end{equation}
Combining this with the continuity equation
gives the equation for the density ``in the lowest Landau level.''
\begin{equation}
\dot{\rho}({\bf r}) = \frac {c}{eB}\frac{\partial}{\partial r^i}
\rho({\bf r}) \epsilon^{ij}\frac{\partial}{\partial r^j}
\frac{\delta}{\delta \rho({\bf r})}\int d^2r V
\label{eq318}
\end{equation}
(For the right hand side not to vanish, V must not be solely a 
function of $\rho$.)

The equation of motion (\ref{eq318}) can be obtained by 
bracketing with the Hamiltonian
\begin{equation}
H_0 = \int d^2r V,
\label{eq319}
\end{equation}
provided the charge density bracket is non-vanishing, showing 
non-commutativity of the $\rho$'s \cite{Ger}.
\begin{equation}
\{ \rho({\bf r}), \rho({\bf r}^{\prime}) \} = 
-\frac{c}{eB}\epsilon^{ij}\partial_i\rho({\bf r})
\partial_j\delta( {\bf r}-{\bf r}^{\prime} )
\label{eq320}
\end{equation}

$H_0$ and this bracket may be obtained from (\ref{eq315}) and
(\ref{eq316a}) -- (\ref{eq316c}) with the same Dirac procedure presented for
the particle case: We wish to set $m$ to zero in (\ref{eq315});
this is possible only if $\vec \pi$ is constrained to vanish.
But the bracket of the $\vec \pi$'s is non-vanishing, even at $m=0$,
because $B \ne 0$. Thus at $m=0$ we posit the Dirac brackets
\begin{eqnarray}
&&\{ O_1({\bf r}_1), O_2({\bf r}_2) \}_D = \nonumber \\ 
&&\{ O_1({\bf r}_1), O_2({\bf r}_2) \} 
-\int d^2{\bf r}_1^{\prime} d^2{\bf r}_2^{\prime}
\{ O_1({\bf r}_1), \pi^i({\bf r}_1^{\prime}) \}
(C^{-1})^{ij}({\bf r}_1^{\prime}, {\bf r}_2^{\prime}) 
\{ \pi^j({\bf r}_2^{\prime}), O_2({\bf r}_2) \},
\nonumber \\
\label{eq321}
\end{eqnarray}
where $C^{ij}$ is the bracket in (7.2.21), (at $m=0$) so that
\begin{equation}
(C^{-1})^{ij}({\bf r}_1, {\bf r}_2) = \frac{c}{eB}
\epsilon^{ij}\rho({\bf r}_1) \delta({\bf r}_1- {\bf r}_2). 
\label{eq322}
\end{equation}
Hence Dirac brackets with $\vec \pi$ vanish, and the Dirac 
bracket of densities is non-vanishing as in (\ref{eq320}).
\begin{eqnarray}
&&\{ \rho({\bf r}),\rho({\bf r}^{\prime}) \}_D = \nonumber \\
&&- \frac{c}{eB}\int d^2 r^{\prime\prime}
\{ \rho( {\bf r} ), \pi^i({\bf r}^{\prime\prime} \}
\rho({\bf r^{\prime\prime}}) \epsilon^{ij} 
\{ \pi^j({\bf r}^{\prime\prime}), \rho({\bf r}^{\prime}) \}= \nonumber \\
&&- \frac{c}{eB} \epsilon^{ij}\partial_i \rho({\bf r})
\partial_j\delta({\bf r} - {\bf r}^{\prime})
\label{eq323}
\end{eqnarray}

The $\rho$--bracket enjoys a more appealing expression in momentum space. 
Upon defining 
\begin{equation}
{\tilde{\rho}}({\bf p}) = \int d^2r e^{i{\bf p}\cdot{\bf r}}
\rho({\bf r})
\label{eq324}
\end{equation}
we find
\begin{equation}
\{ {\tilde{\rho}}({\bf p}), {\tilde{\rho}}({\bf q}) \}_D=
-\frac{c}{eB}\epsilon^{ij}p^iq^j
\tilde{\rho}({\bf p} + {\bf q}).
\label{eq325}
\end{equation}
The brackets (\ref{eq320}), (\ref{eq325}) give the algebra of area
preserving diffeomorphisms. Indeed (7.2.30) follows from the full diffeomorphism
algebra (1.2.38) with the identification that $\rho$ in (7.2.30) is related to $j^i$ in
(1.2.38) by $\rho = \varepsilon^{ij} \partial_i \, j^j$. 

The form of the charge density bracket (\ref{eq320}), (\ref{eq323}), 
(\ref{eq325}) can be understood by reference to the particle 
substructure for the fluid. Take as in (1.1.9)
\begin{equation}
\rho({\bf r}) = \sum_n \delta({\bf r} - {\bf r}_n), 
\label{eq326}
\end{equation}
where $n$ labels the individual particles.  The coordinates of 
each particle satisfy the non-vanishing bracket (\ref{eq34}).
Then the $\{ \rho({\bf r}), \rho({\bf r}^{\prime}) \}$ bracket
takes the form describe above.

\subsubsection*{(iii) Quantization}
\addcontentsline{toc}{subsubsection}{(iii) Quantization}
\label{quantize}

Quantization before the reduction to the lowest Landau level is
straightforward. For the particle case (\ref{eq39a}) -- (\ref{eq39c}) 
and for the 
fluid case (\ref{eq316a}) -- (\ref{eq316c})
we replace brackets with $i/\hbar$
times commutators.  After reduction to the lowest Landau level 
we do the same for the particle case thereby arriving at the 
``Peierls substitution,'' which states that the effect of 
an impurity [$V$ in (\ref{eq36})] on the lowest Landau energy
level can be evaluated to lowest order by viewing the $(x,y)$ arguments 
of $V$ as non-commuting variables.

However, for the fluid case quantization presents a choice.
On the one hand, we can simply promote the 
brackets (\ref{eq320}), (\ref{eq323}), (\ref{eq325}) to a 
commutator by multiplying by $i/\hbar$.
\begin{eqnarray}
&&[ \rho({\bf r}),\rho({\bf r}^{\prime}) ] = 
i\hbar\frac{c}{eB} \epsilon^{ij}\partial_i \rho({\bf r}^{\prime})
\partial_j\delta({\bf r} - {\bf r}^{\prime}) \label{eq327a} \\
&&\left[ \tilde{\rho}({\bf p}), \tilde\rho({\bf q}) \right] =
i \hbar \frac{c}{eB} \epsilon^{ij} p^i q^j \tilde\rho({\bf p} + {\bf q})
\label{eq327b}
\end{eqnarray}

Alternatively we can adopt the expression (\ref{eq326}), for the operator 
$\rho({\bf r})$, where the ${\bf r}_n$ now satisfy the
non-commutative algebra,
\begin{equation}
\left[  r_n^i,  r_{n^{\prime}}^j \right] = -i\hbar \frac{c}{eB} 
\epsilon^{ij}\delta_{nn^{\prime}}
\label{eq328}
\end{equation}
and calculate the $ \rho$ commutator as a derived
quantity.

However, once ${\bf r}_n$ is a non-commuting operator, functions
of $ {\bf r}_n$, even $\delta-$functions,
have to be ordered. We choose the Weyl ordering,  which is equivalent to 
defining the Fourier transform as 
\begin{equation}
\tilde{\rho}({\bf p}) = \sum_n e^{i{\bf p} \cdot {\bf r}_n}.
\label{eq329}
\end{equation}
With the help of (\ref{eq328}) and the Baker-Hausdorff lemma, we
arrive at the ``trigonometric algebra''. \cite{Fairlie} 
\begin{equation}
[\tilde{\rho}({\bf p}), \tilde{\rho}({\bf q})]=
2i \sin \left( \frac{\hbar c}{2eb} \epsilon^{ij}p^iq^j \right)
\tilde{\rho}({\bf p}+ {\bf q})
\label{eq330}
\end{equation}
This reduces to (\ref{eq327b}) for small $\hbar$.

This form for the  commutator, (\ref{eq330}), 
is connected to a Moyal $\ast$ product  in
the following fashion.  For an arbitrary c-number function $f({\bf r})$
define
\begin{equation}
<f> = \int d^2r \rho({\bf r})f({\bf r}) = 
\frac{1}{(2\pi)^2}\int d^2 p \tilde{\rho}({\bf p}) \tilde{f}(-{\bf p}).
\label{eq331}
\end{equation}
Multiplying (\ref{eq330}) by 
$\tilde{f}(-{\bf p})\tilde{g}(-{\bf q})$ and integrating
gives
\begin{equation}
[<f>, <g>] = <h>,
\label{eq332}
\end{equation}
with
\begin{equation}
h({\bf r}) = (f*g)({\bf r}) - (g*f)({\bf r})
\label{eq333}
\end{equation}
where the $*$ product is defined in (6.1.5). 
%\begin{equation}
%(f*g)({\bf r}) = e^{\frac{i}{2}\frac{\hbar c}{eb} 
%\epsilon^{ij} \partial_i \partial_j^{\prime}}
%f({\bf r})g({\bf r}^{\prime})|_{ {\bf r}^{\prime} = {\bf r}}.
%\label{eq334}
%\end{equation}
Note however that only the commutator is mapped into the $\ast$ 
commutator.  The product $<f><g>$ is not equal to $<f*g>$.

The lack of consilience between (\ref{eq327b}) and (\ref{eq330})
is an instance of the Groenwald-VanHove theorem which establishes 
the impossibility of taking over into quantum mechanics all classical
brackets. Equations (\ref{eq330}) -- (7.2.39) explicitly
exhibit the physical occurrence of the $\ast$ product for fields in
a strong magnetic background. 
 
\newpage
\subsection{Fluids and ferromagnets}
\setcounter{equation}{0}
A quantum system that has some similarity in its description to fluid mechanics is the ferromagnet \cite{VPN}.
 The requirements that we put on the ferromagnetic theory is that it give rise to a local spin algebra, [compare
(5.3.3)]
\numeq{eq:7.3.1}{
[S^a ({\bf r}), S^b ({\bf r^\prime})] = i \hslash\, \varepsilon_{abc} \, S^c ({\bf r}) \delta^3({\bf r -
r^\prime}) }
and that some Hamiltonian $H$ encode the specific nature of the ferromagnet. Thus we expect the action to be 
\numeq{7.3.2}{
I = I_o - \int dt\,  H.
}
Here $I_0$ involves the canonical 1-form that leads to \refeq{eq:7.3.1}, where the $S^a$
are definite functions of the canonical variables.

\subsubsection*{(i) Spin algebra}
\addcontentsline{toc}{subsubsection}{(i) Spin algebra}

In our treatment of non Abelian fluids we obtained the Poisson bracket version of \refeq{eq:7.3.1}, {\it i.e} (5.3.3), 
 by starting from (D.1)
specialized to $SU(2)$: $g \epsilon SU(2), K = \sigma^3 /2i$,
\numeq{7.3.3}{
S_a T^a = g \, K \, g^{-1}.
}
However, for the ferromagnet, there is no place for  a density variable $\rho$; 
rather we should set $\rho$ in (D.1) to a constant and posit for the canonical portion of the
Lagrange density
\numeq{7.3.4}{
\mathcal{L}_0 = -i t r \, \sigma^3 \, g^{-1} \, \dot{g}.
}
The canonical variables are three in number, corresponding to the three parameters specifying 
the $SU(2)$ group element. Consequently the 2-form involves a singular matrix
($3\negthinspace\times\negthinspace3$ and anti-symmetric) with no inverse. Nevertheless, one can
overcome this obstacle and conclude that
\refeq{eq:7.3.1} continues to hold with $S_\alpha$ defined by
\refeq{7.3.2}.

This is achieved with the following steps. With the notation of Sidebar D, 
we determine that the $\mathcal{L}_0$ reads [compare (D.2 and (D.3)]
\numeq{7.3.5}{
\mathcal{L}_0 = \dot{\varphi} \, C^a_{\ b} \, S_b.
}
The canonical $1-$ form has components $a_a = C^a_{\ b} \, S_b$ [compare (D.10)] and the
2-form reads
\begin{eqnarray}
\frac{\delta}{\delta \varphi^a} \, a_b \, - \frac{\delta}{\delta \varphi^b}\, a_a \equiv
f_{a b}, \nonumber\\ f_{ab} \, ({\bf r}, {\bf r^\prime}) = - \delta ({\bf r} - {\bf r^\prime})
C^c_{\ a} C^d_{\ b} S^e \, \varepsilon_{cde}.
\label{7.3.6}
\end{eqnarray}

Consider now the left translation of $g$: $\delta g = \epsilon^a \, \sigma^a g/2i$, or 
equivalently $\delta \varphi_a = -\epsilon^b c^b_{\ a}$, which leads to
$\delta S^a = \varepsilon_{abc} \, \epsilon^b \, S^c$. The generator of this 
transformation satisfies, according to (A.2), \\ 
\begin{equation}
\frac{\delta G}{\delta \varphi_a}
= - \epsilon^f \, c^f_{\ b} \, C^c_{\ b} \, C^d_{\ a} \, S^e \, \varepsilon_{cde} 
= - \epsilon^c \, C^d_{\ a} S^e \, \varepsilon_{cde}. 
\end{equation}
This equation is solved
by
\numeq{7.3.7}{
G = -\int d r \, S^a \epsilon^a.
}
From the general theory, (A.18), we knew that Poisson bracketing with $G$ generates
the above transformation on any function of the phase space variables, in particular on
$S^b$.
\numeq{7.3.8}{
\varepsilon_{abc}\, \epsilon^b ({\bf r^\prime}) S^c ({\bf r^\prime}) = 
\{G, \, S^b ({\bf r^\prime}) \} = \{ \int d r \, \epsilon^a {(\bf r)} \, S^a ({\bf r}), S^b
({\bf r^\prime}) \} }
Stripping away the arbitrary parameter $\epsilon^a ({\bf r})$ leads to a Poisson bracket,
which gives \refeq{eq:7.3.1} upon quantization.

It is instructive to obtain this result by applying the procedures in Sidebar A(b), 
relevant to singular 2-forms. Note that $S^a$ satisfies
\numeq{7.3.9}{
\frac{\delta}{\delta \varphi_b} \, S^a = c^a_{\, m}  \, f_{m b}.
}
Consequently it follows from (A.10) that the admissibility criterion (A.13) is obeyed.
The bracket of local spins reads
\numeq{7.3.10}{
\{S^a ({\bf r}), S^b \, ({\bf r^\prime})\} = \int d r^{\prime\prime}  d r^{\prime\prime\prime} \ \frac{\delta}{\delta
\varphi_c ({\bf r^\prime})} S^a ({\bf r})
\, f^{cd} ({\bf r^{\prime} }, {\bf r^{\prime \prime}}) \frac{\delta}{\delta \varphi_d ({\bf r^{\prime\prime}})} S^b ({\bf
r^\prime}).}
It follows from \refeq{7.3.5}, (A.10), (A.11) and \refeq{7.3.8} that \refeq{7.3.9} reduces to the Poisson
bracket version of \refeq{eq:7.3.1}.

\subsubsection*{(ii) Momentum density algebra}
\addcontentsline{toc}{subsubsection}{(ii) Momentum density algebra}

The above shows that the ferromagnet in the continuum approximation, may be considered, as far as the canonical structure is concerned, as fluid mechanics with a constant density
$\rho$. It has been known for some time that there is difficulty in defining a momentum density for the
ferromagnet \cite{FHaldane}. The momentum density must generate the coordinate transformation 
$\delta {\bf r} = - \vec \delta ({\bf r}), \delta \varphi_a  = \vec \delta\cdot \vec
\nabla \varphi_a$,
\numeq{7.3.11}{
v^{\varphi_a} = -\vec \delta \cdot \vec \nabla \, \varphi_a
}
and the generator $G$ should solve [see (A.17), (D.12), (D.13b)]
\begin{eqnarray}
\frac{\delta G}{\delta \varphi_a} &=& \int d r\, \vec \delta \cdot \vec \nabla \, \varphi_b \, C^c_{\ b} \, C^d_{\ a} \, S^e \,
\varepsilon_{cde}\nonumber \\ &=& \frac{\delta}{8 \varphi_a} \int d r \, \vec \delta \cdot t r \vec \nabla \, g \, g^{-1} \,
S
 + i \int d r \, \vec \nabla \cdot  \vec \delta  t r \frac{\delta g}{\delta \varphi_a} \, g^{-1} \, S
\label{7.3.12}
\end{eqnarray}
The second equality follows from the first with (D.2) and \refeq{7.3.3}. The last term prevents equation \refeq{7.3.11} from being
integrable. Therefore, a generator of local translations, {\it i.e.} a momentum density, cannot be defined, except in the case
that
$\vec
\nabla \cdot
\vec
\delta = 0$. Translations with transverse $\boldsymbol \delta$ correspond to volume-preserving transformations; and are canonically
implemented with the generator
\numeq{7.3.12}{
G (\vec \delta) = \int d r  \vec \delta \cdot t r \vec \nabla \, g \, g^{-1} \, S.
}
They include total momentum and orbital angular momentum.

It is instructive to see why the procedures in Section A(b) fail to a produce proper generator, which by Noether's theorem for the Lagrange density
\refeq{7.3.4} is
\numeq{eq:7.3.13}{
\mathcal{P} = a_a \, (\varphi) \, \vec \nabla \, \varphi_a = 2 t r \vec \nabla \, g \, g^{-1} \, S.
}
The bracket of the quantities
\begin{subequations} \label{7.14}
\begin{equation}
\{\P^i ({\bf r}), \P^j ({\bf \Tilde{r}}\} = \int d r^\prime \, d r^{\prime\prime} \bigg(\frac{\delta}{\delta \varphi_a ({\bf r^\prime})} \,
\P^i ({\bf r})\bigg)
\, f^{ab} ({\bf r^\prime}, {\bf r}^{\prime\prime}) \bigg(\frac{\delta}{\delta \varphi_b ({\bf r^{\prime\prime}})} \P^j ({\bf \Tilde{r}})\bigg)
\label{7.3.15a}
\end{equation}
is evaluated with the help of the projected inverse to $f^{a b}$, which satisfies (A.11). The result is 
\begin{eqnarray}
\{\P^i ({\bf r}), \P^j ({\bf r^\prime})\} &=& 
\P^i ({\bf r^\prime}) \partial_i \, \delta ({\bf r} - {\bf r^\prime}) +  
\P^j ({\bf r}) \partial_j \, \delta ({\bf r} - {\bf r^\prime})\\
&-& \vec \triangle^j \, ({\bf r^\prime}) \, \partial_i \, \delta ({\bf r}-
{\bf r^\prime})\nonumber  - \vec \triangle^i  ({\bf r}) \partial_j \delta ({\bf r} - {\bf r^\prime}),
\label{7.3.15b}
\end{eqnarray}
\end{subequations}
where
\numeq{7.3.15}{
\vec \triangle \equiv a_m \, P^{\ m}_a \,  \vec \nabla \, \varphi_a,
}
$P^m_a$ being the projector on the zero-modes of $f_{ab}$.
The usual momentum density algebra should not contain the $\vec \triangle$ modification(7.3.16),  \refeq{7.3.15}, [compare
(1.2.38)]. This modification also prevents the quantities $\int d r \vec \delta ({\bf r}) \vec {\bf P} ({\bf r})$ from realizing the full
diffeomorphism algebra (1.2.62)-(1.2.64), except in the restricted case of volume preserving diffeomorphisms $\vec \nabla \cdot \vec
\delta =0$. Finally we remark that with non vanishing $\vec \triangle$ the acceptability condition (A.13) is not met, and the bracket
\refeq{7.3.15} does not satisfy the Jacobi identity.

A solution to all these problems with implementing arbitrary diffeomorphisms is found through 
the connection to fluid mechanics, where the ferromagnetic phase space is enlarged to include $\rho$ as in (D1) \cite{VPN}. Then the
2-form is non-singular, the symplectic structure and conventional Poisson brackets exist [see (D.5)]. Either the equation (A.17) or
Noether's theorem give the generator arbitrary diffeomorphism, {\it viz.} as the momentum density, as
\numeq{7.3.16}{
\P = 2 \rho \, t r \, \vec \nabla \, g \, g^{-1} \, S,
}
which satisfies the proper algebra (1.2.38), or (\ref{7.3.15b}) without the $\vec \triangle$ spoiler.
The $\rho$-extended phase space can be reduced, in the Dirac sense, by imposing the first class constraint $\rho$=constant. This will
achieve the ferromagnet's phase space.
\newpage
\setcounter{equation}{0}
\subsection{Non-Abelian Clebsch parameterization \\
\small (or,  non-Abelian Chern-Simons term as a surface integral---a holographic
presentation)}\label{Section7.4} In Sidebar B we described the Clebsch parameterization
(\ref{threescalar}) for a 3-dimensional vector potential, which casts the Abelian Chern-Simons density
(\ref{CSi}) into a total derivative form (\ref{CSi}), so that its 3-dimensional volume integral obtains its
value from the 2-dimensional surface that bounds the integration volume. One may pose an analogous
question about a non-Abelian gauge potential: how should it be parameterized so that the non-Abelian
Chern-Simons term
\numeq{7.4.1}{
CS(A) = \varepsilon^{ijk} (A^a_i \, \partial_j \, A^a_j + \frac{1}{3} \, f_{abc} \, A^a_i \, A^b_j \, A^c_k) = - \varepsilon^{ijk} t r (\frac{1}{2} \, A_i \,
\partial_j \, A_k \, + \frac{1}{3} \, A_i \, A_j \, A_k)
}
becomes a total derivative, and its integral becomes a surface term? (In the second equality we use a matrix representation of the Lie
algebra: ${\bf A} = {\bf A}^a \, T^a ,  t r \, T^a \, T^b = - \delta^{u b}/2.)$ This question has been answered in different ways by physicists
and by mathematicians. We shall here present both constructions, but first we explain the reason for the difference.

Consider $CS(A)$ as a 3-form on an arbitrary manifold
\begin{eqnarray}
CS(A)&=& A^a \,  d \, A^a \, + \frac{1}{3} \, f_{abc} \, A^a \, A^b \, A^c \nonumber \\
A^a&\equiv& A^a \, d x^i
\label{7.4.2}
\end{eqnarray}
(In our form notation, we suppress the wedge product.) It follows that $d \, CS(A)$ is a 4-form.
\numeq{7.4.3}{
d\, CS(A) = F^a \, F^a\\
\quad F^a = \frac{1}{2} \, F^a_{ij} \, d x^i \, d x^j
}
If the Chern-Simons term is a total derivative, $CS(A) = d \, \Theta$, the $d\, CS(A)=0$.
So the possibility of expressing the Chern-Simons as a total derivative requires that $F^a \, F^a$ vanish. The ``physics" approach to the problem
achieves vanishing of the 4-form $d\, CS(A) = F^a \, F^a$ by working on a 3-dimensional manifold, which does not support 4-forms. [The Abelian
Clebsch parameterization (\ref{threescalar}) is given in 3-space!] The ``mathematics" approach remains with the 4-dimensional space, viewed as a
K\"{a}hler manifold, but requires that certain components of $F$ vanish, see below.

\subsubsection*{(i) Total-derivative form for the Chern-Simons term in 3-space}
\addcontentsline{toc}{subsubsection}{(i) Total-derivative form for the Chern-Simons term in 3-space}
In Sidebar B, eqs. (B.5) - (B.10), we gave an analytic/geometric construction of the Abelian Clebsch parameterization for a vector potential.
However, there is another approach to this Abelian problem, which contains clues for the non-Abelian generalization. So we re-analyze the Abelian
case.  

The alternate method for constructing the Clebsch parameterization for an Abelian potential relies on projecting
the potential from a non-Abelian one, specifically, one for
$SU(2)$. We consider an SU(2) group element~$g$ and a pure gauge SU(2) gauge field, whose matrix-valued 1-form is 
\numeq{gfield}{
g^{-1} \rd g = V^a \frac{\sigma^a}{2i}.
}
where $\sigma^a$ are Pauli matrices. It is known that 
\numeq{knowntotderiv}{
\tr (g^{-1} \rd g)^3 = -\fract14 \eps_{abc} V^a V^b V^c = -\fract32 V^1 V^2 V^3
}
is a total derivative; indeed its spatial integral measures the winding number of
the gauge function~$g$ \cite{TrJaZuWi}. Since $V^a$ is a pure gauge, we have
\numeq{Va}{
\rd{V^a} = -\fract12 \eps_{abc} V^b V^c,}
so that if we define an Abelian gauge potential $A$ by selecting one SU(2)
component of~\refeq{gfield} (say the third) $A=V^3$, the Abelian \CS\ density
for~$A$ is a total derivative, as is seen from the chain of equations that relies
on~\refeq{knowntotderiv} and~\refeq{Va},
\numeq{chaineq}{
A\rd A = V^3 \rd{V^3} = -V^1V^2V^3 = \fract23 \tr (g^{-1}\rd g)^3,
}
and concludes with an expression known to be a total derivative.
Of course $A=V^3$ is not an Abelian pure gauge. 

Note that $g$ depends on three arbitrary functions, the three SU(2) local gauge
functions. Hence $V^3$ enjoys sufficient generality to represent the
3-dimensional vector $A$. Moreover, since  $A$'s Abelian \CS\ density is
given by $\tr(g^{-1}\rd g)^3$, which is a total derivative, a \Cpr\ for $A$ is
easily constructed. We also observe that when the SU(2) group element~$g$ has
nonvanishing winding number, the resultant Abelian vector possesses a
nonvanishing
\CS\ integral, that is, nonzero magnetic helicity. Specifically, the example of the
Clebsch-parameterized gauge potential in (\ref{vecpot}), \refeq{constr} is gotten by
projecting onto the third direction of a pure gauge $SU(2)$ potential constructed from the
group element $g = exp (\sigma^a/2 i) \hat{r}^a a (r) 2 A= i\, t r \, \sigma^3 \,
g^{-1} \, d g$. An even more direct example is given by the $SU(2)$ group element $g=
e^{\frac{\sigma^3}{2i}\, \beta} \  \, e^{\frac{\sigma^2}{2i} \gamma}\, \, e^{\frac{\sigma^3}{2i}\theta}$. Then $A= d\theta + \cos \gamma \, d
\beta$.

Now we turn to the non Abelian problem, which we formulate in the following way : {\em For a given group H, how can one construct a potential
${\bf A}^a$ such that the non-Abelian Chern-Simons integrand
$CS(A)$ is a total derivative?}

In the solution that we present \cite{JNP}, the ``total derivative''
form for the Chern-Simons density of
$A^a$ is
achieved in two steps.  The \pr, which we find, 
directly leads to an Abelian form of the Chern-Simons
density,
\numeq{eq:286}{
A^a \rd{A^a} + \fract13 f_{abc}
A^a A^b A^c  =
\gamma \rd \gamma,
}
for some $\gamma$.  Then Darboux's theorem (or usual fluid
dynamical theory) ensures that $\gamma$ can be presented
in Clebsch form, so that $\gamma \rd \gamma$ is
explicitly a total derivative.

We begin with a pure gauge $g^{-1}\rd g$ in
some non-Abelian group $G$ (called the Ur-group) 
whose Chern-Simons integral coincides
with the winding number of $g$.
\numeq{eq:287}{
W(g) = \frac1{16\pi^2} \int \rd{^3r} \textrm{CS}(g^{-1}\rd g) 
= \frac1{24\pi^2} \int \tr (g^{-1}\rd g)^3 
}
We consider a normal subgroup $H\subset G$, with
generators $T^a$, and construct a non-Abelian gauge potential for $H$ by
projection.
\numeq{eq:288}{
A^a \propto ~\tr (T^a g^{-1} \rd g )\ 
}
Within $H$, this is not a pure gauge.
We determine the group structure that    ensures
 the Chern-Simons 3-form  of $A^a$
to be proportional to 
$\tr (g^{-1} \rd g)^3$. Consequently, the  
constructed non-Abelian gauge fields, belonging to the group
$H$, carry quantized Chern-Simons number.  Moreover, we
describe the properties of the Ur-group $G$ 
that guarantee that the
projected potential $A^a$ enjoys sufficient generality to
represent an arbitrary potential in $H$.

Since $\tr (g^{-1} \rd g)^3$ is a total derivative for an
arbitrary group \cite{TrJaZuWi} (although this fact cannot in general be
expressed in finite terms) our construction ensures that the
form of $A^a$, which is achieved through the projection \refeq{eq:288},
produces a ``total derivative'' expression (in the limited sense indicated above) for
its Chern-Simons density. 

Conditions on the Ur-group
$G$, which we take to be compact and semi-simple, are
the following. First of all $G$ has to
be so chosen that it has sufficient number of parameters to
make  $\tr (T^a g^{-1} \rd g)$ a generic potential for
$H$.  Since we are in three dimensions, an $H$-potential
$A^a_i$ has $3\times {\rm dim} H$ independent functions; so a
minimal requirement will be
\numeq{eq:289}{
{\rm dim~} G \ge 3 {\rm ~dim~} H \ \ .
}
Secondly we require that the $H$-Chern-Simons form for $A^a$
should coincide with that of $g^{-1}\rd g$.
As we shall show in a moment, this is achieved if $G/H$ is a
symmetric space.  In this case, if we split the Lie algebra of
$G$ into the $H$-subalgebra spanned by $T^a$,
$a=1,\dots, {\rm dim~} H$, and the orthogonal
complement spanned by $S^A$, $A=1,\dots, ({\rm
dim~}G-{\rm dim~}H)$, the commutation rules are of the
form
\begin{subequations} 
\begin{align}
\relax [T^a, T^b{]} &= f_{abc}
T^c,
\label{eq:2.2a} \\
\relax[ T^a, S^A{]} &= h^{a AB} S^B,
\label{eq:2.2b} \\
\relax[S^A, S^B{]} &= N ~h^{a AB} T^a \ .
\label{eq:2.2c}
\end{align}
\end{subequations} 
$(h^a)^{AB}$ form a (possibly reducible)
representation of the
$H$-generators $T^a$. The constant  $N$
depends on 
normalizations. More explicitly, if the structure
constants for the Ur-group $G$ are named ${\bar f}_{abc},~
\{a,b,c\} =1,\dots,{\rm dim} G$, then the conditions (\ref{eq:2.2a}--c) require
that ${\bar f}_{abc}$ vanishes whenever an odd number of indices belongs to
the orthogonal complement labeled by $A,B,..$. Moreover, $f_{abc}$ 
are taken to be the conventional structure constants for $H$ and this may
render them proportional to (rather than equal to) ${\bar f}_{abc}$.

We define the
traces of the generators by
\begin{align}
\tr (T^a T^b) &= -N_1 \delta^{ab} \ \ ,
\quad \tr (S^A S^B) = -N_2 \delta^{AB}
\nonumber  \\
\tr (I^a S^A) &=0 \ \ .
\label{eq:2.3}
\end{align}
We can evaluate the quantity $\tr[S^A ,S^B]T^a =
\tr S^A[S^B,T^a]$ using the commutation rules.  This immediately gives the
relation $N_1N=N_2$. 

Expanding $g^{-1} \rd g$ in terms of generators, we write
\begin{equation}
g^{-1}  \rd g= (T^a A^a + S^A \alpha^A),
\label{eq:2.4}
\end{equation}
which defines the $H$-potential $A^a$. 
Equivalently
\begin{equation}
A^a =-\frac{1}{N_1} \tr(T^a g^{-1}  \rd g).
\label{eq:2.5n}
\end{equation}
From $\rd{(g^{-1} \rd g)} =-g^{-1} \rd g  g^{-1} \rd g$, we get the
Maurer-Cartan relations
\begin{align}
F^a\equiv \rd{A^a} + \fract12
f_{abc}A^b A^c &= -\frac{N}{2}
h^{a A B} \alpha^A \alpha^B,  \nonumber \\
\rd{\alpha^A} + h^{a B A} A^a \alpha^B &= 0 \ \ .
\label{eq:2.6}
\end{align}
Using these results, the following chain of equations shows that
the
Chern-Simons 3-form for the $H$-gauge group is
proportional to $\tr (g^{-1} \rd g)^3$.
\begin{align}
\frac {1}{16\pi^2} (A^a
\rd{ A^a}+\fract13 f^{abc}A^a
A^b A^c) &=\frac{1}{48\pi^2} ( A^a
\rd {A^a}+2~ A^a F^a) \nonumber \\
&=\frac {1}{48\pi^2} ( A^a \rd{ A^a} - N
h^{a AB}A^a \alpha^A\alpha^B) \nonumber \\
&=\frac {1}{48\pi^2} ( A^a
 \rd {A^a}+ N \rd{ \alpha^A} \alpha^A)   \nonumber\\
&=-\frac
{1}{48\pi^2}\bigg(\frac1{N_1} \tr A \rd
A+\frac{N}{N_2}\tr\rd\alpha\alpha\bigg)
 \nonumber\\
&=-\frac {1}{48\pi^2 N_1} \tr(A\rd A + \alpha \rd  \alpha)
\nonumber\\ 
&=-\frac {1}{48\pi^2N_1} \tr g^{-1} \rd g  ~\rd{(g^{-1}
\rd g)}\nonumber \\ 
&=\frac {1}{48\pi^2N_1} \tr(g^{-1} \rd g)^3 \label{eq:2.7n}
\end{align}
In the
above sequence of manipulations, 
we have used the Maurer-Cartan relations
(\ref{eq:2.6}), which rely on the symmetric space structure of
(\ref{eq:2.2a}--c), and the trace relations (\ref{eq:2.3}), along with $N_1N=N_2$

We thus see that $\int \textrm{CS}(A)$ is indeed the winding number of
the configuration $g\in G$.  Since $\tr(g^{-1} \rd g)^3$ is a
total derivative locally on $G$, the potential (\ref{eq:2.5n}),
with the symmetric space structure of (\ref{eq:2.2a}--c), does indeed fulfill
the requirement of making $\textrm{CS}(A)$ a total derivative. 
It is therefore appropriate to call our construction (\ref{eq:2.5n})
a ``non-Abelian Clebsch parameterization".

In explicit realizations, given a gauge group of interest $H$,
we need to choose a group $G$ 
such that the conditions
(\ref{eq:289}), (\ref{eq:2.2a}--c) hold. In general this is not possible.
However, one can proceed recursively. Let us suppose that the
desired result has been established for a group, which we call
$H_2$. Then we form $H\subset G$ obeying (\ref{eq:2.2a}--c)
as $H=H_1 \times H_2$, where
$H_1$ is the gauge group of interest, satisfying ${\rm dim}G \geq
3~{\rm dim} H_1$. For this choice of $H$, the result (\ref{eq:2.7n})
becomes
\numeq{eq:2.8n}{
\textrm{CS} (H_1) +\textrm{CS}  (H_2) =  \frac {1}{48\pi^2 N_1}\tr(g^{-1}
\rd g)^3.
}
But since $\textrm{CS}  (H_2)$ is already known to be a total derivative,
(\ref{eq:2.8n}) shows the desired result: $\textrm{CS} (H_1)$ is a total
derivative.

With $SU(2)$ as the $U r$-group and $H = U(1)$ or $SO(2)$, we achieve Clebsch
parameterization for an Abelian potential, as explained in \refeq{gfield} - \refeq{chaineq}.
$[T^a = \frac{\sigma^3}{2i}; S^A = \{\frac{\sigma^2}{2i}, \frac{\sigma^3}{2i}\}]$

For an explicit working out of a non Abelian case, we consider a potential for $SU(2) \approx
O(3)$, which possesses nine independent functions. We take $G=O(5), H=O(3)\times O(2)$.  We consider the
4-dimensional spinorial representation of $O(5)$.
With the generators normalized  by $\tr(T^aT^b) = -
\delta^{ab}$, the Lie algebra generators of $O(5)$ are given by matrices of Pauli matrices.
\begin{align}
T^a &=\frac{1}{ 2i} 
\left(
\begin{array}{cc}
\sigma^a & 0 \\
0 &\sigma^a
\end{array}
\right)  \nonumber \\
T^0 &=\frac{1}{ 2i} 
\left(
\begin{array}{cc}
-I & 0 \\
0 &I
\end{array}
\right)  \label{eq:3.1n}   \\
S^A &=\frac{1}{ i\sqrt{2}}
\left(
\begin{array}{cc}
0 & 0 \\
\sigma^A &0
\end{array}
\right)
\qquad
\tilde{S}^A = \frac{1}{ i\sqrt{2}}
\left(
\begin{array}{cc}
0 & \sigma^A \\
0 &0
\end{array}
\right)
\nonumber
\end{align}
$T^a$ generate
$O(3)$, with the conventional structure constants $\epsilon_{abc}$,
and $T^0$ is the generator of $O(2)$.  $S,\tilde{S}$ are the coset
generators. 

A general group element in $O(5)$ can be written in the form $g=Mhk$
where $h\in O(3)$, $k\in O(2)$, and
\begin{equation}
M = \frac{1}{\sqrt{1+{\bf \bar{w}} \cdot {\bf w} - \fract{1}{ 4}({\bf
w}\times{\bf {\bar w}})^2 }}
\left(
\begin{array}{cc}
I- \frac{i}{ 2}({\bf w} \times {\bf \bar{w}})\cdot {\bf \sigma} & -{\bf w} \cdot
{\bf \sigma}
\\[2ex]
{\bf \bar{w}} \cdot {\bf \sigma} & I+ \frac{i}{ 2}({\bf w} \times
{\bf\bar{w}})\cdot {\bf \sigma}.
\end{array}
\right)
\label{eq:3.2n}
\end{equation}
$w^a$ is a complex 3-dimensional vector, with the bar denoting complex
conjugation. 
${\bf w}\cdot {\bf
\bar{w}} = w^a \bar{w}^a$ and $({\bf w} \times {\bf\bar{w}})^a =
\epsilon_{abc} w^b \bar{w}^c$. 
The general $O(5)$ group element contains ten independent real functions.
These are collected as six from $M$ (in the three complex functions $w^a$), 
three in $h$, and one in~$k$. 

 The $O(3)$ gauge
potential given by $- \tr (I^a g^{-1} \rd  g)$ reads
\begin{align}
A^a &= R^{ab} (h) ~a^b  + (h^{-1} \rd  h)^a
\nonumber \\
a^a &= \frac{1}{1+ {\bf w} \cdot {\bf\bar{w}} - \fract{1}{ 4}({\bf w} \times
{\bf\bar{w}})^2}
\Bigg\{ \frac{w^a \rd{\bf\bar{w}} \cdot ({\bf w} \times
{\bf\bar{w}}) + \bar{w}^a \rd  {\bf w} \cdot ({\bf \bar{w}}\times {\bf w})}{2} \label{eq:3.3n}\\
&~~~~~~~~~~~~~~~~~~~~~~~~~~~~~~~~~~~~~~~~~~~~~~~~{}+
\epsilon_{abc} (\rd { w^b} \bar{w}^c -
w^b \rd {\bar{w}^c}) \Bigg\}.
\nonumber
\end{align}
where $R^{ab} (h)$ is defined by $hI^a h^{-1} = R^{ab} h^b$ and $k$ does not
contribute. 
$A^a$ is the $h$-gauge transform of $a^a $,  which depends
on six real  parameters $(w^a)$.  The three gauge
parameters of $h\in O(3)$,
along with the six, give the
nine functions needed to parameterize a general $O(3)$- [or $SU(2)$-] 
potential
in three dimensions.  The Chern-Simons form is 
\begin{align}
\textrm{CS} (A) &= \frac{1}{16\pi^2} (A^a \rd{A^a} + \fract13
\epsilon_{abc} A^a A^b A^c)
\nonumber \\
&=  \frac{1}{16\pi^2} (a^a \rd{ a^a} + \fract13
\epsilon_{abc} a^a a^b a^c) -
\rd{\left[\frac{1}{16\pi^2}  (\rd h h^{-1})^a a^a )\right]} 
+
\frac{1}{24\pi^2} \tr (h^{-1} \rd  h)^3.
\label{eq:3.4n}
\end{align}
The second equality reflects the usual response of the Chern-Simons density
to gauge transformations.
Using the explicit form of
$a^a$ as given in (\ref{eq:3.3n}), we can further reduce this.
Indeed we find
\begin{equation}
a^a \rd{a^a} + \fract13
\epsilon_{abc} a^a a^b a^c =
(-2) \frac{({\bf \bar{w}} \times \rd  {\bf\bar{w}}) \cdot {\bf\rho} + ({\bf w} \times 
\rd {\bf w}) \cdot {\bf\bar{\rho}}}{[1+ {\bf w}\cdot {\bf \bar{w}} -\fract{1}{ 4}
({\bf w} \times {\bf{\bar w}})^2 ]^2},
\label{eq:3.5n}
\end{equation}
$$
\rho_k \equiv \fract12 \epsilon_{ijk} \rd {\bar{w}^i} \rd { \bar w^j}.
$$
Defining an Abelian potential
\begin{equation}
a = \frac{{\bf w} \cdot \rd {\bf\bar{w}} -{\bf \bar{w}} \cdot \rd  {\bf w}}{1+
{\bf w}\cdot {\bf \bar{w}} - \fract{1}{ 4}({\bf w} \times {\bf\bar{w}})^2},
\label{eq:3.6}
\end{equation}
we can easily check that $a \rd  a$ reproduces (\ref{eq:3.5n}).
\begin{equation}
\textrm{CS} (A)  = \frac{1}{16\pi^2}  a\rd  a + \rd { \left[ \frac{ (\rd
h h^{-1})^a a^a )}{16 \pi^2} \right] }+ \frac{1}{48\pi^2} \tr(h^{-1} \rd  h)^3
\label{eq:3.7n}
\end{equation}
If desired, the Abelian potential $a$ can now be written in the Clebsch form making
$a \rd  a$ into a total derivative, while the remaining two terms already are total
derivatives, though in a ``hidden'' form for the last expression. This completes our
construction.

\subsubsection*{(ii) Total-derivative form for the Chern-Simons term on a K\"{a}hler manifold} 
\addcontentsline{toc}{subsubsection}{(ii) Total-derivative form for the Chern-Simons term on a K\"{a}hler manifold}

In 4-dimensional space $(x_i, x_2, x_3, x_4)$ we can introduce complex
coordinates (holomorphic and anti-holomorphic) appropriate to Kahler
(even-dimensional) manifold.
\numeq{7.4.26}{
(z, \bar{z}) = (x_1 \pm i x_2), \quad (w, \bar{w}) = (x_3 \pm i x_4)
}
It is then required that the holomorphic and anti-holomorphic components of the
curvature $F_{\mu \nu}$ vanish. \cite{last}
\numeq{7.4.27}{
F_{z w} = F_{\bar{z} \bar{w}} = 0
}
It follows that
\begin{eqnarray}
A_z = U^{-1} \, \partial_z U, \qquad \quad &\ & A_w = U^{-1} \, \partial_w
V,\nonumber\\
A_{\bar{z}} = - (\partial_{\bar{z}} U^\dagger) (U^{-1})^\dagger, &\ & A_{\bar{w}} = -
(\partial_{\bar{w}} V^\dagger) (V^{-1})^\dagger.
\label{7.4.8}
\end{eqnarray}
and after a further complex gauge transformation we may replace (\ref{eq:3.6}) by
\begin{eqnarray}
A_z = g^{-1} \, \partial_z \, g &\ & A_w = g^{-1} \partial_w\, g \nonumber \\
A_{\bar{z}} = 0 = A_ {\bar{w}}
\label{7.4.9}
\end{eqnarray}
with Hermitian $g = U^\dagger U$. When we define $\d_\pm$ as the holomorphic and anti
holomorphic derivatives $d_+ = dz \frac{\partial}{\partial z} + d w
\frac{\partial}{\partial w}, \ d \_ = d \bar{z} \frac{\partial}{\partial \bar{z}} + d \bar{w}
\frac{\partial}{\partial \bar{w}}$, then $ A = A_z dz + A_w dw + A_{\bar{z}} d \bar{z} +
A_{\bar{w}} d \bar{w} = g^{-1} d_+ g  \equiv a$ [with requirement \refeq{7.4.27} and in the
selected gauge (7.4.29). This has the property of Abelianizing the gauge expressions.
\numeq{7.4.30}{
F = d\_ a, \quad t r F^2 =\quad t r (d \_ ad a) = d\_ tr (a d \_ a)
}
The last formula is also consistent with the Chern-Simons formula, since the cubic
contribution in (7.3.1) vanishes, leaving the Abelianized quantity
\numeq{7.4.31}{
CS = t r (a d \_ a).
}
Additionally one can show that
\numeq{7.4.32}{
CS = t r (a d \_ a) = d_+ \Omega + d \_ \Phi.
}
[In fact the above is established by varying $tr(a d \_ a).]$ In this way one arrives at the final
result \cite{last}
\numeq{7.4.33}{
t r F^2 = d \_ d_+ \Omega
}

It is to be emphazized the that \refeq{7.4.32} holds only with the
holomorphic restriction \refeq{7.4.27}, and gauge choice (7.4.29). No explicit
formula for $\Omega$ is available; it is determined by ``intergrating" a variation:
\numeq{7.4.33}{
\delta \Omega = 2 tr a  d_{-} v \quad v \equiv g^{-1}
\delta g
}
For $\Phi$, which however does not contribute to $t r F^2$, but is needed for a
reconstruction of the Chern-Simons term (7.4.32), we find
\numeq{7.4.35}{
\delta \Phi =  tr (a d_+ v) 
}
Evidently this construction does not give a parameterization for an arbitrary vector potential, only one which can be gauged to (7.3.28).

It is instructive to see some explicit expressions. In the Abelian case, with $A= a = d_+ \theta$, we have, $F = d\_ d_+ \theta, \ F^2 = (d\_ d_+ \theta) (d\_
d_+\theta) = d\_ (d_+ \theta d \_ d_+ \theta), CS = A F = d_+ \theta d\_ d_+ \theta =
d_+ (\theta d \_ d_+ \theta),\\ \Omega = \theta d\_ d_+ \theta, \Phi =0.$ Note
that the Abelian connection is parameterized in terms of one function
$\theta$, or two functions if the gauge freedom is included. But a general
connection in 4-space, requires  four functions.

In the non-Abelian $SU(2)$ case, we take $ g = e^{\frac{\sigma^a}{2} \theta^a}$, then
\begin{eqnarray}
A\equiv a = g^{-1} d_+ g = \big(d_+(\hat{\theta}^a \sinh \theta) + (\cosh \theta -1)
(i \varepsilon_{abc} d_+ \hat{\theta}^b \hat{\theta}^c - \hat{\theta}^a d_+
\theta)\big) \frac{\sigma^a}{2}
\end{eqnarray}
with $\theta \equiv \sqrt{\theta^a \theta^a}, \hat{\theta}^a \equiv
\theta^a/\theta$. The Chern-Simons term is reconstructed from
\begin{subequations}
\begin{equation}
\Omega = \frac{1}{2} d_{+} \theta d_{ -} \theta + (\cosh \theta -1) d_+ \hat{\theta}^a
d\_ \hat{\theta}^a 
+ i(\sinh \theta-\theta)  \varepsilon_{abc} \hat{\theta}^a d_{+}  \hat{\theta}^b d_{-} \hat{\theta}^c
\end{equation}
\numeq{lastone}{
 \Phi = -\frac{i}{2} (\sinh \theta - \theta) \varepsilon_{abc} \hat{\theta}^a d_+
\hat{\theta}^b d_{-}
\hat{\theta}^c
}
\end{subequations}
Note that three functions are involved $(\theta ^a)$ in the parameterization of the connection; six if the
gauge freedom is included. However a $SU(2)$ connection on a 4-space requires twelve functions.

% A useful Journal macro
\def\Journal#1#2#3#4{{#1} {\bf #2}, #3 (#4)}
\def\add#1#2#3{{\bf #1}, #2 (#3)}
\def\Book#1#2#3#4{{\em #1}  (#2, #3 #4)}
\def\Bookeds#1#2#3#4#5{{\em #1}, #2  (#3, #4 #5)}
\def\inter#1{(\em #1)}
% \Journal{}{}{}{}
% \Book{}{}{}{}

% Some useful journal names
\def\NPB{{\em Nucl. Phys.}} % put <B> in next field
\def\PLA{{\em Phys. Lett.}} % put <A> in next field
\def\PLB{{\em Phys. Lett.}} % put <B> in next field
\def\PRL{{\em Phys. Rev. Lett.}}
\def\PRD{{\em Phys. Rev. D}}
\def\PR{{\em Phys. Rev.}}
\def\PT{{\em Physics Today}}
\def\ZPC{{\em Z. Phys.} C}
\def\SJNP{{\em Sov. J. Nucl. Phys.}}
\def\AnnP{{\em Ann. Phys.}\ (NY)}
\def\JETPL{{\em JETP Lett.}}
\def\LMP{{\em Lett. Math. Phys.}}
\def\CMP{{\em Comm. Math. Phys.}}
\def\PTP{{\em Prog. Theor. Phys.}}
\def\PNAS{{\em Proc. Nat. Acad. Sci.}}
\def\JMP{{\em J. Math. Phys.}}
\def\JH{{\em JHEP}}
\def\HPA{{\em Helv. Phys. Acta}}
\def\SMM{{\em Sci. Mem. Moscow Univ. Math. Phys.}}


\begin{thebibliography}{1}
\addcontentsline{toc}{section}{References}
%\bibitem [0] {keldysh} J.Schwinger, {\it J. Math. Phys.} {\bf 2}
\bibitem{keldysh} J.Schwinger, {\it J. Math. Phys.} {\bf 2},
407 (1961); P.M.Bakshi and K.Mahanthappa, {\it J. Math. Phys.}
{\bf 4}, 12 (1963); L.V.Keldysh, {\it Zh. Eksper. Teoret. Fiz.} {\bf 47}, 151 (1964) [English translation: {\it Sov. Phys. JETP} {\bf
20}, 1018 (1964)].

\bibitem{LanLif}
A standard physics text on the subject is L. Landau and E. Lifshitz, {\it Fluid Mechanics} (2nd ed.
Pergamon, Oxford UK 1987).
A mathematical treatment is V. Arnold and B. Khesin, {\it Topological Methods in Hydrodynamics}
(Springer-Verlag, Berlin 1988). The relation between Langrange and Euler descriptions of a fluid
is discussed by R. Salmon, {\it Ann. Rev. Fluid. Mech.} {\bf 20}, 225 (1988); see also I. Antoniou and
G. Pronko, arXiv: hep-th/0106119.

\bibitem{ArnKhe}
These brackets were first posited (in somewhat imprecise form) by L. Landau, {\it Zh. Eksper. Teoret.
Fiz.} {\bf 11}, 592 (1941) [English translation: {\it J. Phys. USSR} {\bf 5}, 71 (1941)]; see also P.
Morrison and J. Greene, {\it Phys. Rev. Lett.} {\bf 45}, 790 (1980), (E) {\bf 48} 569 (1982).  

\bibitem{Salm}
S0(2,1) together with the Galileo group, equivalently the ``Schr\"{o}dinger" group,  provides the
maximal symmetry group for the kinetic term of non-relativistic dynamics. Various interactions
in various dynamical systems preserve the S0(2,1) symmetry. In particle dynamics the inverse
square potential is S0(2,1) invariant [R.~Jackiw, \Journal{\PT}{25{\rm(1)}}{23}{1972}; U.~Niederer,
\Journal{\HPA}{45}{802}{1972}; C.R.~Hagen, \Journal{\PRD}{5}{377}{1972}.] In fluid
mechanics the equation of state that follows from (\ref{eqone54b}): pressure =$\propto
 \rho^{1 +2/d}$ enjoys Schr\"{o}dinger group invariance; see M.~Hassa\"{\i}ne and P.~Horvathy,
\Journal{\AnnP}{282}{218}{2000}; L.~O'Raifeartaigh and V.~Sreedhar,
\Journal{\AnnP}{293}{215}{2001}.


\bibitem{Chp} %13
S. Chaplygin, \Journal{\SMM}{21}{1}{1904}.
[Chaplygin was a colleague of  fellow USSR Academician N.~Luzin. Although
accused by Stalinist authorities of succumbing excessively to foreign influences,
unaccountably both managed to escape the fatal consequences of their alleged
actions; see N.~Krementsov, \Book{Stalinist Science}{Princeton University
Press}{Princeton NJ}{1997}.] The same model (\ref{eqone57}) was later put
forward by H.-S.~Tsien, {\it J. Aeron. Sci.}\, {\bf 6}, 399 (1939) and T. von~Karman,
{\it J. Aeron. Sci.}\, {\bf 8}, 337 (1941); see also K. Stanyukovich, 
\Book{Unsteady Motion of Continuous
Media}{Pergamon}{Oxford  UK}{1960}, p.~128. 

\bibitem{DesJac}
For an introduction to the properties of the Abelian and non-Abelian
Chern-Simons terms, see S.~Deser, R.~Jackiw, and S.~Templeton,
\Journal{\AnnP}{140}{372}{1982}, (E)~\add{185}{406}{1985}. In fluid mechanics
or in magnetohydrodynamics the Abelian Chern-Simons term is known as the fluid
or magnetic helicity, and was introduced by L.~Woltier,
\Journal{\PNAS}{44}{489}{1958}, and further studied in M.~Berger and G.~Field,
{\it J. Fluid Mech.}{\bf 147}, 133 (1984); H.~Moffatt and A.~Tsinober,
{\it Ann. Rev. Fluid Mech.}{\bf 24}, 281 (1992).

\bibitem{FadJac}
L.~Faddeev and R.~Jackiw, \Journal{\PRL}{60}{1692}{1988}. For a detailed
exposition, see  R.~Jackiw in \Bookeds{Constraint Theory and Quantization
Methods}{F.~Colomo, L.~Lusanna, and G.~Marmo, eds.}{World
Scientific}{Singapore}{1994},  reprinted in R.~Jackiw, \Book{Diverse Topics in
Theoretical and Mathematical Physics}{World Scientific}{Singapore}{1995}.

\bibitem{bacjac}
D. Bak, R. Jackiw and S.-Y. Pi, \Journal{\PRD}{49}{6678}{1994}, appendix.

\bibitem{VarMech}
See for example, V. Arnold {\it Mathematical Methods of Classical Mechanics} (Springer-Verlag,
Berlin 1989); V. Guillemin and S. Sternberg, {\it Sympletic Techniques in Physics} (Cambridge
University Press, Cambridge UK 1990).

\bibitem{Lin}
C.C.~Lin, \Bookeds{International School of Physics E.~Fermi  (XXI)}{G. Careri,
ed.}{Academic Press}{New York}{1963}.

\bibitem{Cle}
A. Clebsch, \Journal{\it J. Reine Angew. Math.}{56}{1}{1859}.

\bibitem{Eck}
C.~Eckart, \Journal{\PR}{54}{920}{1938}.

\bibitem{Darb}
A constructive discussion of the Darboux theorem can be found it the second
work of Ref.~\cite{FadJac}.

\bibitem{Lam}
H. Lamb, \Book{Hydrodynamics}{Cambridge University Press}{Cambridge UK}{1932}, p.~248.

\bibitem{DesJacPol}  
S.~Deser, R.~Jackiw, and A.P.~Polychronakos, 
\Journal{\PLA}{A279}{151}{2001}.

\bibitem{gravcos}
Note that in conventional treatments, as in S. Weinberg, {\it Gravitation and Cosmology} (Wiley,
New York 1972), $\rho$ is a scalar, rather than the time-component of a Lorentz vector. This
distinction makes no difference in the non-interacting case, but our choice allows introducing
self-interactions in a Lorentz invariant way by using the scalar $\sqrt{j^\alpha j_\alpha} \equiv n$.

\bibitem{Wein} 
See, for example, Landau and Lifshitz, Ref.~\cite{LanLif}, or Weinberg, Ref \cite{gravcos}.

\bibitem{BazJac} 
These symmetry transformations were identified by D.~Bazeia and R. Jackiw,
\Journal{\AnnP}{270}{246}{1998}, on the basis of constants of motion that
were found previously; see Ref.~\cite{BorHop}.

\bibitem{BorHop} 
M. Bordemann and J. Hoppe, \Journal{\PLB}{B317}{315}{1993}; A.~Jevicki,
\Journal{\PRD}{57}{5955}{1988}. These authors consider only the planar, $d=2$,
case. 

\bibitem{Baz} %16a
D.~Bazeia, \Journal{\PRD}{59}{085007}{1999}.

\bibitem{AnDet} %17
Analytic details for  implementing the space-time mixing
transformation on~\refeq{disclat} are presented by Bazeia and
Jackiw Ref~\cite{BazJac}.

\bibitem{Oga} %18
N. Ogawa, \Journal{\PRD}{62}{085023}{2000}, who also presents other solutions,
as does Bazeia, Ref.~\cite{Baz}.

\bibitem{JacPol} %19
R. Jackiw and A.P. Polychronakos, \Journal{\CMP}{207}{107}{1999}. 


\bibitem{ref20} %20
The additional transformation rules were derived in Ref.~\cite{JacPol}, on the basis
of constants of motion identified previously in the
$d=2$ case; see Ref.~\cite{BorHop94}.

\bibitem{BorHop94} %21
M. Bordemann and J. Hoppe, \Journal{\PLB}{B325}{359}{1994}.

\bibitem{Gold} %24
That the theory of a membrane [($d=2$)-brane]  in three spatial dimensions is
equivalent to planar fluid mechanics was known to J.~Goldstone (unpublished) and
worked out by his student Hoppe (sometimes in  collaboration with Bordemann).
The method described in Secs.~2.2(i) and~2.2(ii) was presented for
$d=2$ in J.~Hoppe,
\Journal{\PLB}{B329}{10}{1994}, while a version of the argument in
Sec.~2.2 (iii), specialized to $d=2$, is found in Bordemann and Hoppe,
Ref.~\cite{BorHop}. Generalization to arbitrary~$d$ is given in R.~Jackiw and
A.P.~Polychronakos,
{\it Proc. Steklov Inst. Math.} {\bf 226}, 193 (1999) and Ref.~\cite{JacPol}.

\bibitem{sussk}
L. Susskind, \Journal{\PR}{165}{1535}{1968}.

\bibitem{Nutku} %33
Landau and Lifshitz, Ref.~\protect\cite{LanLif}; 
Y. Nutku, \Journal{\JMP}{28}{2579}{1987}; 
P. Olver and Y. Nutku, \Journal{\JMP}{29}{1610}{1988};
M. Arik, F. Neyzi, Y. Nutku, P. Olver, and J. Verosky, \Journal{\JMP}{30}{1338}{1989};
J. Brunelli and A. Das, \Journal{\PLA}{A235}{597}{1997}. 

\bibitem{BarChe} %34
B. Barbishov and N. Chernikov, \Journal{\em{Zh. Eksp. Theort. Fiz.}}{51}{658}{1966} 
[English translation: \Journal{\em{Sov. Phys. JETP}}{24}{437}{1967}].

\bibitem{grndland}
Other solutions, enjoying group invariance or partial invariance, have been constructed by A
Grundland, A. Hamilton and V. Hussin, \Journal{\JMP}{44}{2874}{2003} and in press.

\bibitem{JevSak} % new 36
A.~Jevicki and B.~Sakita, \Journal{\PRD}{22}{467}{1980} and
\Journal{\NPB}{B165}{511}{1980}; J.~Polchinski,
\Journal{\NPB}{B362}{25}{1991}; J.~Avan and A.~Jevicki,
\Journal{\PLB}{B266}{35}{1991} and
\add{B272}{17}{1991}.


\bibitem{FloJac} %35
R. Floreanini and R. Jackiw, \Journal{\PRL}{59}{1873}{1987}.

\bibitem{JacPol2} %26
R. Jackiw and A.P. Polychronakos, \Journal{\PRD}{62}{085019}{2000}.
Some of these results are described in unpublished papers by J.~Hoppe, Karlsruhe
preprint KA-THEP-6-93, Karlsruhe preprint
KA-THEP-9-93, hep-th/9311059.

\bibitem{BergJac}
Y. Bergner and R. Jackiw, \Journal{\PLA}{A284}{146}{2001}.

\bibitem{DuffFiel}
M.J. Duff, in \Bookeds{Fields, Strings and Duality}{C. Efthimiu and B. Greene, eds.}{World
Scientific}{Singapore}{1997}.

\bibitem{ref:3} %27
  B. de Wit, J. Hoppe, and H. Nicolai, 
\Journal{\NPB}{B305}{[FS23] 525}{1988}.

\bibitem{groonib}
M. Hassaine, \Journal{\PLA}{A290}{157}{2001};
T. Nyawelo, S. Groot Nibbelink and J. van Holten, \Journal{\PRD}{64}{021701}{2001},
\add{68}{125006}{2003}; A. Das and Z. Popowicz, \Journal{\PLA}{A296}{15}{2002}; T. Nyawelo,
\Journal{\NPB}{B672}{87}{2003}.

\bibitem{mclerran} See for example,
T. Ludlam and L. McLerran, {\it Physics Today},
{\bf 56}, 48 (2003).

\bibitem{pisarski}
R.Pisarski, \textit{Physica} \textbf{A158}, 246 (1989),
\textit{Phys. Rev. Lett.} \textbf{63}, 1129 (1989);
E.Braaten and R.Pisarski, \textit{Phys. Rev. D} \textbf{42}, 2156
(1990);
for a recent update and later references, see 
U. Kraemer and A. Rebhan, {\it Rep. Prog. Phys.} {\bf 67}, 351 (2004).
%J. Andersen,
%E. Braaten, E.  Petitgirard and M. Strickland, \Journal{\PRD}{66}{085016}{2002}.

\bibitem{GPY} A.D. Linde, {\it Phys. Lett.} {\bf B96}, 289 (1980) ;
D. Gross, R. Pisarski and L. Yaffe, {\it Rev. Mod. Phys.} {\bf 53}, 43 
(1981).

\bibitem{heinz}
U. Heinz, \textit{Ann. Phys.} \textbf{161}, 48 (1985);
H.-T. Elze and U. Heinz, \textit {Phys. Rep.} \textbf{183}, 81
(1989) for recent update and later references se D. Litin and C. Manuel
%\Journal{\PR}{364}{451}{P2002}.
{\it Phys. Rep.} {\bf 364}, 451 (2002).
\bibitem{wong}
S. K. Wong, \textit{Nuovo Cim.} \textbf{A65}, 689 (1970).

\bibitem{bistro}  B. Bistrovic,
R. Jackiw, H. Li, V.P. Nair and S.-Y. Pi, {\it Phys. Rev. D} {\bf 67},
025013 (2003).

\bibitem{refbib44}
R. Jackiw and K. Johnson, {\it Phys. Rev.} {\bf 182}, 1459 (1969); D. Gross and R. Jackiw, {\it Nucl. Phys.} {\bf B14}, 269 (1970).

\bibitem{bal} A.P. Balachandran, S. Borchardt and A. Stern,
\textit{Phys. Rev. D}
\textbf{17}, 3247 (1978); A. Balachandran, G. Marmo,
B-S. Skagerstam and A. Stern, \textit{Gauge Symmetries and Fibre
Bundles}, Lecture Notes in Physics 188 (Springer-Verlag, Berlin,
1982).

\bibitem{Mad} % new 40
E. Madelung, {\it Z. Phys.} {\bf 40}, 322 (1927); see also E.~Merzbacher, 
\Book{Quantum Mechanics}{3rd ed.,  Wiley}{New York}{1998}.

\bibitem{Bohm55}
Similar investigations are by D. Bohm, R. Schiller and J. Tiomno, {\it Nuovo Cim. Suppl.} {\bf 1} 48
(1995); P. Love and B. Boghosian, {\it Physica} {\bf A332}, 47 (2002).

\bibitem{JackiwPi}
R. Jackiw, S.-Y.Pi and A. Polychronakos, {\it Ann. Phys.} (NY) {\bf 301}, 157 (2002).

\bibitem{sussarv}
L. Susskind, arXiv: hept-th/0101029.

\bibitem{SiebWit}
N. Seiberg and E. Witten, \Journal{\JH}{9909}{032}{1999}.

\bibitem{OkaOo}
Y. Okawa and H. Ooguri, \Journal{\PRD}{64}{046009}{2001}

\bibitem{PiJac}
R. Jackiw and S.-Y. Pi, \Journal{\PRL}{88}{111603}{2002}.

\bibitem{jacKIW}
R, Jackiw, \Journal{\PRL}{41}{1635}{1978}, {\it Acta Phys. Austr. Suppl.} {\bf XXII}, 383 (1980),
reprinted in R. Jackiw, \Book{Diverse Topics in Theoretical and Mathematical Physics}{World
Scientific}{Singapore}{1995}.

\bibitem{sp11}
This approach has been taken to study wave propagation in noncommuting Maxwell theory. Z.
Guralnik, R. Jackiw, S.-Y. Pi and A Polychronakos, \Journal{\PLB}{B517}{450}{2001}; R. Cai,
\Journal{\PLB}{B517}{457}{2001}; J. Zahn, arXiv: hep-th/0405253.

\bibitem{unkn}
J. Madore, S. Schraml, P. Schupp and J. Wess, {\it Eur. Phys. J. C} {\bf 16}, 161  (2000), B. Jurco, S.
Schraml, P. Schupp and J. Wess, {\it Eur. Phys. J. C} {\bf 17}, 521 (2000); D. J. Gross, A. Hashimoto
and N. Itzhaki, {\it Adv. Theor. Math Phys.} {\bf 4}, 893 (2000); D. Bak, K. Lee, and J.-H. Park,
\Journal{\PLB}{B501}{305}{2001}.

\bibitem{Polyc}
A. P. Polychronakos, {\it Ann. Phys.} (NY) {\bf 174} (2002).

\bibitem{Feyn}
R. P. Feynman, \Book{Statistical Mechanics}{Benjamin}{Reading MA}{1972} p. 319.

\bibitem{Dashen}
R. Dashen and D. Sharp, \Journal{\PR}{165}{1857}{1968}; J. Grodnik and D. Sharp,
\Journal{\PRD}{6}{1531, 1546}{1970}; G. Goldin, \Journal{\JMP}{12}{462}{1971}.

\bibitem{Dunne}
G. Dunne, R. Jackiw and C Trugenberger, \Journal{\PRD}{41}{661}{1990}; G. Dunne and R. Jackiw,
{\it Nucl. Phys. B} ({\it Proc. Suppl.}) {\bf 33C}, 114 (1993).

\bibitem{Magro}
G. Magro, arXiv: quant-ph/0302001.

\bibitem{Ger}
Guralnik {\it et. al.}, Ref \cite{sp11}.

\bibitem{Fairlie}
D. Farlie, P. Fletcher and C. Zachos, \Journal{\PLB}{B218}{203}{1989}; D. Farlie and  C. Zachos,
\Journal{\PLB}{B224}{101}{1989}; C. Duval, Z. Horvath and P. Horvathy, {\it Int. J. Mod. Phys.} {\bf
B15}, 3397 (2001).

\bibitem{VPN}
V.P. Nair and R. Ray, \Journal{\NPB}{B676}{659}{2004}.

\bibitem{FHaldane}
R. Balakrishnan and A. Bishop, \Journal{\PRL}{55}{537}{1985}; F. Haldane,
\Journal{\PRL}{57}{1488}{1986}.

\bibitem{JacPi}
R. Jackiw and S.-Y. Pi, \Journal{\PRD}{61}{105015}{2000}.

\bibitem{TrJaZuWi} %12
S. Treiman, R. Jackiw, B. Zumino, and E.~Witten, \Book{Current Algebra and
Anomalies}{Princeton University Press/World
Scientific}{Princeton NJ/Singapore}{1985}.

\bibitem{JNP} %36
 R. Jackiw, V.P. Nair, and S.-Y. Pi, \Journal{\PRD}{62}{085018}{2000}.

\bibitem{last}
R. Bott and S. S. Chern,  {\it Act. Math.} {\bf 114}, 71 (1965).

\end{thebibliography}
\end{document}